\documentclass[11pt,a4paper,twoside,openright]{report}

\usepackage{bbm} 
\usepackage{amsmath} 
\usepackage{amssymb} 
\usepackage{physics} 

\usepackage{graphicx}
\usepackage[rightcaption]{sidecap}  
\usepackage{tikz}
\usetikzlibrary{babel}
\usepackage{subfig}

\usepackage[open=true]{bookmark}
\usepackage{hyperref} 
\usepackage{caratula} 
\usepackage{emptypage} 

\usepackage[english, spanish]{babel} 
\usepackage[utf8]{inputenc} 
\usepackage{cite}
\usepackage{pifont} 
\definecolor{bordo}{HTML}{841026} 
\usepackage{enumitem}

\usepackage{lipsum} 
\usepackage{color}
\usepackage[]{xcolor} 
\usepackage{comment}

\usepackage{array}
\usepackage[margin=1in]{geometry} 
\allowdisplaybreaks  
\usepackage{fancyhdr} 
\sidecaptionvpos{figure}{t}

\usepackage{rotating}
\usepackage[Sonny]{fncychap} 
\ChNumVar{\fontsize{90}{20}\usefont{T1}{pnc}{b}{n}\selectfont}
\ChNameVar{\Large\rm}
\ChTitleVar{\huge\sf}
\definecolor{claro}{HTML}{FCFCFC}
\definecolor{oscuro}{HTML}{808080}
\newcommand{\colornumberchap}{\color{claro}}
\newcommand{\colornamechap}{\color{bordo}}

\makeatletter
\renewcommand{\DOCH}{%
    \hfill
    \begin{turn}{90} 
    {\colornamechap\CNV\FmN{\hspace{-.2cm}\@chapapp}}
    \end{turn}
    \colorbox{bordo}{\colornumberchap\CNoV\hspace{.15cm}\thechapter\hspace{.25cm}}\\%
}
\renewcommand{\DOTI}[1]{%
    \nointerlineskip\raggedright%
    \vskip3ex%
    \textcolor{bordo}{\rule{\textwidth}{5pt}}
    \vskip-2.1ex%
    \rule{\textwidth}{.5pt}
    \vskip3ex%
    \parbox[t]{\linewidth}{\raggedright\CTV\FmTi{#1}}\par\nobreak%
    \vskip 40\p@%
    \centering
}

\renewcommand{\DOTIS}[1]{%
    \nointerlineskip\raggedright%
    \vskip3ex%
    \textcolor{bordo}{\rule{\textwidth}{5pt}}
    \vskip-2.1ex%
    \rule{\textwidth}{.5pt}
    \vskip3ex%
    \parbox[t]{\linewidth}{\raggedright\CTV\FmTi{#1}}\par\nobreak%
    \vskip 40\p@%
    \centering
}
\makeatother
\titulo{Fases geométricas en\\sistemas cuánticos abiertos: }
\subtitulo{análisis y aplicaciones}
\autor{Ludmila Viotti}
\director{Paula I. Villar}
\consejero{Fernando C. Lombardo}
\lugardetrabajo{Departamento de Física, FCEyN, UBA}
\fecha{2023}

\begin{document}
\pagenumbering{roman}

\maketitle

\chapter*{Resumen}
\setcounter{page}{1} 
Esta tesis reúne el estudio de distintos sistemas cuánticos de pocos grados de libertad expuestos a los efectos de un entorno no controlado. En particular, el eje del trabajo es la relación entre los fenómenos de decoherencia y disipación inducidos por el entorno, y el concepto denominado {\em fase geométrica}. La primer mención a estas fases en el contexto de la mecánica cuántica se remonta al trabajo pionero de Berry. En 1984 Berry mostró que la fase acumulada por un autoestado de un Hamiltoniano con dependencia temporal en un ciclo adiabático consta de dos contribuciones distintas: una que denominó 'geométrica', y la bien conocida fase dinámica. 

A partir del trabajo de Berry, la noción de fase geométrica fue generalizada más allá del escenario original, con definiciones que aplican a evoluciones unitarias arbitrarias. 
Estas fases geométricas emergen naturalmente en una descripción geométrica del espacio de Hilbert, donde se manifiestan como holonomías. Resultan significantes no sólo a nivel fundamental y del formalismo matemático de la mecánica cuántica, sino también para la explicación de múltiples fenómenos que incluyen, entre otros, el efecto Hall fraccionario. Más aún, en una mirada más actual, las fases geométricas resultan elementos prometedores con aplicaciones prácticas, como la construcción de compuertas geométricas para el tratamiento de información cuántica.

Sin embargo, un estado puro en evolución unitaria es una idealización y todo experimento e implementación real debe lidiar con la presencia de un entorno que interactúa con el sistema de estudio, lo que requiere una descripción en términos de estados mixtos y evoluciones no-unitarias.
La definición de una fase geométrica que aplique en tal escenario es todavía un problema abierto, y distintas propuestas coexisten. Caracterizar estos objetos se torna de este modo un proyecto con múltiples motivaciones que van desde aspectos fundamentales de la mecánica cuántica hasta aplicaciones tecnológicas.

\chapter*{Abstract}
\textbf{\large Geometric Phases in Open Quantum Systems: Analysis and Applications}
\\
\\\indent
This thesis is composed of several studies performed over different quantum systems of few degrees of freedom exposed to the effect of an uncontrolled environment. The primary focus of the work is to explore the relation between decoherence and dissipative effects induced by the presence of the environment, and the concept known as {\em geometric phases}. The first mention of these phases in the context of quantum mechanics goes back to the seminal work by Berry. He demonstrated that the phase acquired by an eigenstate of a time-dependent Hamiltonian in an adiabatic cycle consists of two distinct contributions: one termed 'geometric' and the other known as the dynamical phase.

Since Berry's work, the notion of geometric phases has been extended far beyond the original context, encompassing definitions applicable to arbitrary unitary evolutions. These geometric phases naturally arise in the geometric description of Hilbert space, where they manifest as holonomies. They possess significance not only in the fundamental understanding of quantum mechanics and its mathematical framework but also in explaining various physical phenomena, including the Fractional Hall Effect. Moreover, in a modern perspective, geometric phases hold promise for practical applications, such as constructing geometric gates for quantum information processing and storage.

However, in practice, a pure state of a quantum system is an idealized concept, and every experimental or real-world implementation must account for the presence of an environment that interacts with the observed system. This interaction necessitates a description in terms of mixed states and non-unitary evolutions. The definition of a geometric phase applicable in such scenarios remains an open problem, giving rise to multiple proposed solutions. Consequently, characterizing these geometric phases becomes a multifaceted task, encompassing motivations that span from fundamental aspects of quantum mechanics to technological applications.

\chapter*{}
Aprovechando que la tesis la escribo yo, y puede entonces ser (más o menos) como me parezca, me tomo este espacio para hablar sobre algunas personas, aunque en realidad no haya forma de reconocer apropiadamente el aporte que hicieron a esta etapa, y el que hacen en mi vida en general. Sin embargo, por hacer el intento y como gesto de voluntad, estampo sus nombres en esta página.

En primer lugar quiero agradecer a mi directora, Paula, y a mi director (o consejero de estudios, según a quién se le pregunte), Fernando, que me formaron no sólo en la praxis de investigar, sino también como docente, y hasta me enseñaron el laborioso arte de lidiar con las tareas administrativas. Su respaldo y su confianza hicieron de mi doctorado una experiencia sumamente enriquecedora, pero, por sobre todas las cosas, humana. Realmente tildan todos los casilleros en cuanto a lo que se puede esperar de un director o directora.
Y, si ya venía con dos grandes victorias en lo que se refiere a directores, la breve tutela de Saro no cortó la racha. Cuando esperaba un investigador testaferro encontré, una vez más, no sólo una persona, sino una que ejerció de educador, para cerrar la etapa-doctorado con un broche de oro.

Quiero además agradecer a los que hicieron el camino conmigo. A Rami por su teoría 'Es que ella es de Paternal', devenida posteriormente en 'Es que ella es de Rosario', que tiene más poder explicativo del que él mismo supone, y a Tiago, por ponerse esa remera de Pescado rabioso, y por los 13 años que vinieron después.

A mamá, que a veces con firme intención, y otras casi sin darse cuenta, me abrió las primeras puertas para que pudiera ser, y hacer, lo que yo quiera. A Yoe, el primer secuaz, y a la familia que nos inventamos en el desarraigo, con esxs tíxs, primxs, y ahora sobrino, que sí elegimos.

Y en esta línea de afectos voluntarios, quiero también agradecer a mis amigos y amigas que están siempre revoloteando, a pesar de que yo escriba, y que ellxs nunca vayan a leer, sobre fases geométricas (en sistemas cuánticos abiertos), y a mis alumnos y alumnas, la verdadera justificación de todo esto.

\vspace{1.5cm}
\begin{center}
   \textcolor{bordo}{\ding{163}}
\end{center}
\vspace{1.5cm}
\begin{center}
    {\em A Martín y a Reynaldo, que estarían chochícimos.}
\end{center}
\clearpage

\tableofcontents
\chapter{Introducción}
\pagenumbering{arabic}

{\em Fases geométricas en mecánica cuántica - }Cuando se considera la dinámica generada por un Hamiltoniano con dependencia temporal explícita bajo las condiciones que definen el llamado régimen adiabático se observa un comportamiento particular: si el sistema cuántico está a tiempo inicial en un autoestado del Hamiltoniano, entonces permanece durante toda la evolución en el autoestado correspondiente del Hamiltoniano instantáneo a tiempos posteriores. Esto, sin embargo, determina el estado del sistema a menos de una fase. En particular, para Hamiltonianos con dependencia periódica en el tiempo $H(t + T) = H(t)$, la ausencia de transiciones implica que el estado final coincida, a menos de una fase, con el estado inicial.
En 1984 Berry mostró, en un trabajo pionero \cite{Berry1983original}, que dicha fase consta, además de la ya por entonces bien conocida fase dinámica, de una componente extra que denominó {\em goemétrica}. Fue Simon \cite{simon1983holonomy} quien, poco tiempo después, sentó el carácter geométrico de la fase de Berry demostrando que puede entenderse como la holonomía de un fibrado lineal sobre el espacio de parámetros. 

Partiendo de este punto, se han propuesto nociones de fase geométrica bien definidas en escenario mucho más generales que aquél considerado por Berry. Por ejemplo, Aharonov y Anandan \cite{aharonov1987phase} eliminaron la restricción de evolución adiabática introduciendo el concepto de fase geométrica no-adiabática, aunque manteniendo la condición de evolución cíclica. Samuel y Bhandari  \cite{samuel1988general} fueron todavía un paso más allá, definiendo la fase geométrica asociada a cualquier evolución en la que el estado permanezca puro, sin la necesidad de que sea cíclica ni preserve norma.
Mukunda y Simon \cite{mukunda1993quantum} recuperaron, mediante su {\em abordaje cinemático}, la expresión para la fase geométrica asociada a la trayectoria unitaria más general posible, explicitando al mismo tiempo la independencia de la fase respecto de la dinámica: la fase geométrica depende exclusivamente del camino que el estado traza en el espacio proyectivo de rayos (esto es, en el espacio de estados físicos) y de la estructura particular del espacio de Hilbert, pero es independiente de la dinámica específica que da origen a dicho camino. 

Las diversas generalizaciones introducidas por estos autores recuperan formalmente los resultados previos si las hipótesis correspondientes se satisfacen, formando de este modo un conjunto de definiciones consistente.
Otras generalizaciones que, a diferencia de las ya mencionadas, no serán tratadas en esta tesis incluyen la definición de fases geométricas para estados mixtos en evolución unitaria \cite{uhlmann1986parallel, uhlmann1989berry, uhlmann1991gauge, sjoqvist2000geometric, singh2003geometric, chaturvedi2004geometric}, las llamadas fases geométricas no-diagonales \cite{maninioffdiagonal,sjoqvistoffdiagonal} y las fases geométricas no-abelianas que surgen en el contexto de evoluciones adiabáticas degeneradas en energía \cite{wilczek1984appearance}.

Emergiendo en la estructura geométrica del espacio de Hilbert y de la curva sobre éste trazada por el estado, las fases geométricas resultan elementos relevantes para el estudio de la mecánica cuántica a nivel fundamental y de su descripción matemática. Pero su relevancia no se agota allí, y también han permitido explicar múltiples fenómenos físicos como el efecto Hall fraccionario~\cite{thouless1982_app_hall} o el comportamiento mostrado por aislantes y superconductores topológicos\cite{bernevig2013_app_supercond, asboth2016_app_supercond}.

El auge de la información cuántica ha abierto una nueva vía para las fases geométricas con aplicaciones prácticas, por ser considerada una herramienta robusta para el procesamiento de información \cite{vedral2003geometric}. Esta nueva línea de interés fue impulsada por el trabajo de Zanardi y colaboradores \cite{Zanardi_1999, pachos1999non, pachos2001quantum}, quienes propusieron utilizar fases geométricas para construir compuertas cuánticas (las operaciones lógicas básicas que componen una computación cuántica).
Sin embargo, la búsqueda de aplicaciones prácticas impone un nuevo escenario que no puede ignorarse. La imagen de un sistema representado por un estado puro cuya evolución está generada por un operador hermítico es una idealización, y los sistemas reales se encuentran siempre sujetos a efectos producto de su interacción con el entorno.
En esta dirección, compuertas cuánticas de alta fidelidad fueron implementadas con iones atrapados~\cite{leibfried2003_app_qi}, mientras que la voluntad de perfeccionar el rendimiento de los dispositivos diseñados para el procesamiento de información cuántica con respecto a los efectos del entorno ha sugerido el uso de esquemas de compuertas cuánticas no-adiabáticas~\cite{xiang2001_app_qi, zhu2002_app_qi, li2020_path,ding2021_path, sjoqvistshortcut, measuringshortcut}. 
\\
\\\indent
{\em Fases geométricas en sistemas abiertos - }La propuesta de aplicaciones prácticas para el procesamiento de información cuántica no es, sin embargo, la única motivación para el estudio de la fase geométrica acumulada por sistemas cuánticos abiertos.

En primer lugar, el estudio de la fase geométrica en un sistema abierto implica necesariamente una definición que generalice las bien establecidas expresiones, aplicables al caso de estados puros, al contexto de estados mixtos en evolución no-unitaria~\cite{breuer2002libro, rivas2012libro}. La búsqueda de tal generalización conduce a nuevos niveles tanto de comprensión e interpretación de la mecánica cuántica a nivel fundamental como del desarrollo matemático asociado a su descripción.
Esta tarea no está definitivamente cerrada, y diversas propuestas y abordajes coexisten. 
Basándose en los trabajos previos para la fase geométrica de mezclas estadísticas de estados en evolución unitaria,
Tong y sus colaboradores~\cite{tong2004kinematic} ampliaron el concepto de fase geométrica al caso de estados mixtos en evolución no unitaria, mediante una forma funcional que se reduce a la fase unitaria acumulada por un estado puro consistentemente.
Por otra parte, Carollo y colaboradores~\cite{Carollo_original, Carollo_review} propusieron un promedio sobre el ensamble de trayectorias que se obtiene al considerar la interacción entre el sistema y el entorno en el contexto de trayectorias cuánticas. Esta propuesta se discutió intensamente en \cite{Sjo_no, bassi2006_no, sjoqvist2010_hidden, buri}.

Otros tratamientos del problema reportan que el efecto de fluctuaciones clásicas en los parámetros de control de una evolución cíclica pueden promediarse mitigando su efecto en la fase de Berry acumulada~\cite{de2003berry}. La presencia de un baño térmico da origen a contribuciones geométricas a la decoherencia~\cite{whitney2003berry, whitney2005geometric}, detectadas experimentalmente en~\cite{ berger2013_noise_cqed, berger2015_noise_cqed}. Estos resultados sugieren un motivo más para el estudio de las fases geométricas acumuladas por sistemas cuánticos abiertos: su relación con otros fenómenos físicos y la posibilidad de comprenderlos o detectarlos a través de ésta. Caracterizar la influencia del entorno sobre la fase geométrica y establecer conexiones con otros efectos del entorno constituye una extensión de los estudios tradicionales que favorece, entre otras cosas, propuestas para la detección indirecta de fenómenos \cite{grupo2006,grupo2013,grupo2014,VILLAR2015246, farias2020towards, oxman2018two}.
\\
\\\indent
{\em Estructura de la tesis - } La estructura de esta tesis es la siguiente. En el capítulo \ref{ch:2} se introducen y discuten aspectos generales de las fases geométricas. Se presentan definiciones para la fase geométrica acumulada por un sistema aislado que muestran creciente grado de generalidad, se examinan sus propiedades y se ilustran y comparan los resultados mediante un ejemplo paradigmático: un sistema de dos niveles en presencia de un campo magnético clásico. Este capítulo se completa introduciendo el problema de la definición de una fase geométrica que resulte aplicable a sistemas abiertos y el desarrollo de la propuesta de Tong et al. \cite{tong2004kinematic}, que será la herramienta clave en los capítulos \ref{ch:3} a \ref{ch:5}.

En el capítulo \ref{ch:3} se estudia en detalle la fase geométrica en un modelo de Jaynes-Cummings disipativo, como caso paradigmático de la electrodinámica cuántica en cavidades. El modelo de Jaynes-Cummings unitario \cite{jaynes1963comparison} describe un sistema formado por un átomo de dos niveles en interacción con un único modo del campo electromagnético en el régimen en que se conserva el número de excitaciones y es considerado el más simple capaz de dar cuenta satisfactoriamente del acoplamiento materia-radiación. En el capítulo, se considera que los principales mecanismos por los cuales el sistema 'átomo + modo' interactúa con el entorno son la pérdida de fotones a través de las paredes de la cavidad y el continuo e incoherente bombeo del sistema de dos niveles \cite{carmichael1989subnatural, yamamoto2003semiconductor, laussy2008strong, vera2009characterization, lodahl2015interfacing}. Primeramente se discuten los contextos y abordajes para los cuáles el caso unitario resulta comparable con el caso disipativo, para posteriormente desarrollar la comparación. Los diversos resultados obtenidos se analizan y justifican en términos de las trayectorias descritas por el estado. Se encuentra, entre otros resultados, que la fase geométrica resulta particularmente robusta para el caso resonante en que la frecuencia del modo del campo coincide con la frecuencia natural del átomo.

En el capítulo \ref{ch:4} se estudia el efecto de las fluctuaciones de vacío del campo electromagnético y las modificaciones introducidas por condiciones de contorno no-triviales. El escenario propuesto consiste de dos partículas de dos niveles (o qubits) originalmente no-interactuantes entre sí, pero acopladas al campo electromagnético cuántico. Este conjunto de un sistema {\em bipartito} sujeto a la acción del entorno resulta favorable para el estudio del efecto del entorno sobre el entrelazamiento. El objetivo en este capítulo es caracterizar la dinámica, el entrelazamiento y la fase geométrica en presencia de fluctuaciones cuánticas del vacío electromagnético, comparando los resultados cuando se introduce una condición de borde que modifique la estructura del entorno. En particular, se considera si la presencia del contorno puede ser utilizada como una herramienta para inferir otros efectos del entorno sobre el sistema. Los resultados observados se interpretan también en términos de interacciones efectivas entre partículas reemplazando los efectos del contorno por partículas imagen ubicadas en la región inaccesible del espacio.

Tras analizar, en el capítulo \ref{ch:4}, el efecto del vacío cuántico del campo electromagnético sobre el sistema en presencia de un contorno conductor perfecto, en el capítulo \ref{ch:5} se reemplaza este contorno ideal por un material semiconductor que permite acceder a efectos originados en el movimiento relativo constante. Entre éstos, nos interesa el fenómeno conocido como {\em fricción cuántica} 
\cite{resultados1, resultados3,resultados4,resultados8, barton2010van, klatt2017quantum, resultados6,resultados7,resultados9}
por lo que se estudia un sistema sujeto a este efecto: un átomo de dos niveles que se mantiene, mediante un forzado externo, en movimiento paralelo a un medio dieléctrico. El conjunto completo se encuentra además inmerso en el estado de vacío del campo electromagnético. La dinámica del sistema, la pérdida de coherencia y las distintas correcciones introducidas a la fase geométrica se investigan y comparan con la bibliografía y los resultados existentes sobre el fenómeno de fricción cuántica. El capítulo finaliza con una propuesta experimental para la detección de correcciones en la fase geométrica y la consecuente detección indirecta de fricción cuántica.

En los capítulos \ref{ch:3} a \ref{ch:5} se estudia el concepto de fase geométrica propuesto por Tong et al., definido para estados mixtos en evolución no-unitaria, en el capítulo \ref{ch:6} el abordaje se modifica a una descripción en términos de trayectorias cuánticas. A esta descripción se accede, por ejemplo, cuando el sistema se monitorea continua e indirectamente.
El objetivo de este capítulo es describir las propiedades de la fase geométrica acumulada a lo largo de trayectorias cuánticas, dónde el carácter estocástico que adquiere la función de onda asociada a cada evolución singular es heredado por la fase geométrica. Para mantener los resultados lo más accesibles posible, se propone un modelo sencillo y paradigmático que en el límite adiabático acumula una fase de Berry. Específicamente, se retoma el modelo tratado en el ejemplo del capítulo \ref{ch:2}, consistente en un átomo de dos niveles inmerso en un campo magnético rotante y en interacción con un baño de osciladores armónicos, y se analiza la distribución de fases geométricas resaltando particularmente la competencia entre los efectos no-adiabáticos y los efectos introducidos por el entorno. 
Con el objetivo de establecer una conexión con los resultados experimentales, se estudia además la distribución de franjas de interferencia obtenidas en un experimento de Eco de Espín.
Finalmente, se investiga la trayectoria específica en la que no se registra ningún cambio abrupto y se muestra que la misma atraviesa una transición topológica como función del acoplamiento con el entorno. 

El capítulo \ref{ch:7} contiene las conclusiones generales de esta tesis. Los apéndices reúnen herramientas teóricas y desarrollos analíticos aplicados a lo largo de la tesis. Específicamente, en el apéndice \ref{apendice1} se resuelve analíticamente la dinámica de un sistema de dos niveles inmerso en un campo magnético clásico rotante para el caso en que este sistema está aislado pero sin imponer condiciones sobre la velocidad de rotación del campo. Por otra parte, el apéndice \ref{sec:ap2} provee una introducción no-exhaustiva al tema de las ecuaciones maestras.

\section{Lista de publicaciones}
\begin{enumerate}
    \item[{[i]}]Viotti, L., Farías, M. B., Villar, P. I., y Lombardo, F. C. (2019). Thermal corrections to quantum friction and decoherence: A closed-time-path approach to atom-surface interaction. Physical Review D, 99(10), 105005.
    \item[{[ii]}]Viotti, L., Lombardo, F. C., y Villar, P. I. (2020). Boundary-induced effect encoded in the corrections to the geometric phase acquired by a bipartite two-level system. Physical Review A, 101(3), 032337.
    \item[{[iii]}] Viotti, L., Lombardo, F. C., y Villar, P. I. (2021). Enhanced decoherence for a neutral particle sliding on a metallic surface in vacuum. Physical Review A, 103(3), 032809.
    \item[{[iv]}] Lombardo, F. C., Decca, R. S., Viotti, L., y Villar, P. I. (2021). Detectable signature of quantum friction on a sliding particle in vacuum. Advanced Quantum Technologies, 4(5), 2000155.
    \item[{[v]}]Viotti, L., Lombardo, F. C., y Villar, P. I. (2022). Geometric phase in a dissipative Jaynes-Cummings model: Theoretical explanation for resonance robustness. Physical Review A, 105(2), 022218.
    \item[{[vi]}] Viotti, L., Gramajo, A. L., Villar, P. I., Lombardo, F. C., y Fazio, R. (2023). Geometric phases along quantum trajectories. Quantum, 7:1029
\end{enumerate}
El capítulo \ref{ch:3} está basado en la publicación [v]. El capítulo \ref{ch:4} está basado en la publicación [ii]. El capítulo \ref{ch:5} Está basado en las publicaciones [i], [iii], y [iv]. El capítulo \ref{ch:6} está basado en las publicaciones [vi]. 

\chapter{Fases geométricas: Aspectos teóricos generales}\label{ch:2}

Este capítulo presenta, en primer lugar, una descripción general de las fases geométricas en el contexto de sistemas aislados, descritos consecuentemente mediante estados puros. Analizar este caso antes de centrar la atención en sistemas cuánticos abiertos permitirá asimilar nociones y ganar intuición sobre las fases geométricas en el marco de una teoría formalmente más simple. Más aún, las fases geométricas fueron propuestas con anterioridad en este escenario, de modo que muchas de sus características e interpretaciones se continúan como imposiciones a la generalización. 
A lo largo del capítulo se introducen distintas expresiones para la fase geométrica acumulada por el estado de un sistema, sus propiedades, e interpretaciones.
Estas expresiones, válidas bajo un determinado conjunto de hipótesis, se organizan en orden cronológico de los trabajos que las presentan, que coincide con el orden creciente de generalidad.

Una vez presentada la fase geométrica en el contexto de los sistemas aislados, se aborda el problema de una definición que aplique al caso de sistemas cuánticos abiertos, la cual no encuentra todavía consenso unánime y plantea diversos desafíos. Hacia el final del capítulo se desarrolla una propuesta particular que será la herramienta de estudio de los próximos capítulos.

\section{Régimen adiabático y fase de Berry}\label{sec:sec2_Berry}
La fase de Berry~\cite{Berry1983original} es un fenómeno fundamental relacionado con el teorema adiabático.
En 1984, Berry mostró que la fase acumulada por el autoestado de un Hamiltoniano $H(t)$ que varía lentamente en un ciclo está relacionada con el circuito descrito por $H(t)$ en un dado espacio de parámetros.

Para reproducir este resultado, se considera un Hamiltoniano $H(R(t))$ que depende explícitamente del tiempo a través de un parámetro multidimensional $R= (R_1, R_2, ...)$, y el conjunto de {\em autoestados instantáneos} de $H(t)$, es decir, de estados $\{\ket{\psi_n(R(t))}\}$ que satisfacen

\begin{equation}
    H(R(t))\ket{\psi_n(R(t))} = E_n(R(t))\ket{\psi_n(R(t))},
    \label{eq:sec2_eigen_inst}
\end{equation}
suponiendo además que los autovalores satisfacen $E_1 < E_2 < ..$ de forma que no existe degeneración.
Por otro lado, se considera también la evolución del estado $\ket{\psi(t)}$ del sistema, dada por la ecuación de Schrödinger

\begin{equation}
    i\hbar |\Dot{\psi}(t)\rangle = H(R(t))\ket{\psi(t)}.
    \label{eq:sec2_sch1}
\end{equation}

Expandiendo el estado del sistema en la base de autoestados instantáneos de $H$ según

\begin{equation}
    \ket{\psi(t)} = \sum_n\, c_n(t)\ket{\psi_n(R(t))},
    \label{eq:sec2_expan}
\end{equation}
la ecuación de Schrödinger se descompone en un conjunto de ecuaciones diferenciales para las funciones $c_n(t) \in \mathbb{C}$ que actúan como coeficientes de la expansión

\begin{equation}
    i\hbar\, \dot{c}_n(t) = \left(E_n - i\hbar\bra{\psi_n}\dot{\psi_n}\rangle\right)\, c_n(t)-i\hbar \sum_{m\neq n}\bra{\psi_n}\dot{\psi}_m\rangle\, c_m(t).
    \label{eq:sec2_sch2}
\end{equation}
Al escribir estas ecuaciones diferenciales se ha omitido explicitar que la dependencia de los autoestados y autovalores en el tiempo es a través del parámetro $R(t)$ con el objetivo de aligerar la notación. Derivando la expresión (\ref{eq:sec2_eigen_inst}), los factores $\bra{\psi_n}\dot{\psi}_m\rangle$ en el lado derecho de la ecuación pueden escribirse en términos de los elementos de matriz $\bra{\psi_n}\Dot{H}\ket{\psi_m}$ del operador $\Dot{H}$. Reemplazando esta relación en las ecuaciones diferenciales (\ref{eq:sec2_sch2}) para los coeficientes, se obtiene

\begin{equation}
    \dot{c}_n(t) = \frac{-i}{\hbar}\left(E_n - i\hbar\bra{\psi_n}\dot{\psi_n}\rangle\right)\, c_n(t)- \sum_{m\neq n}\frac{\bra{\psi_n}\Dot{H}\ket{\psi_m}}{E_m - E_n}\, c_m(t).
    \label{eq:sec2_sch3}
\end{equation}
Si el término con la sumatoria se anula, entonces las soluciones $c_n(t)$ resultan simplemente los valores iniciales $c_n(0)$ multiplicados por un factor de fase y en consecuencia se tiene que $|c_n(t)|=|c_n(0)|$.
En este caso, un sistema preparado en un autoestado $\ket{\psi_n(R(0))}$ del Hamiltoniano a tiempo inicial permanecerá en el autoestado instantáneo correspondiente $\ket{\psi_n(R(t))}$ en todo instante posterior (a menos de una fase).
Las transiciones entre niveles de energía se originan entonces en la sumatoria de la ecuación (\ref{eq:sec2_sch3}). Estas transiciones pueden despreciarse en el {\em régimen adiabático}, caracterizado por una variación del Hamiltoniano $H(R(t))$ lenta con respecto a las escalas propias del sistema $T \sim \hbar/(E_m-E_n)$.

En el régimen adiabático, los coeficientes $c_n(t)$ de la expansión (\ref{eq:sec2_expan}) satisfacen, aproximadamente, la ecuación

\begin{equation}
    \dot{c}_n(t) = \left(-\frac{i}{\hbar}E_n - \bra{\psi_n}\dot{\psi_n}\rangle\right)\, c_n(t),
\end{equation}
de modo que un dado autoestado del Hamiltoniano evoluciona según

\begin{equation}
    \ket{\psi(t)} = e^{-\frac{i}{\hbar}\int_0^t\,dt'\,E(R(t'))}e^{i\phi^n_{\rm a}(t)}\ket{\psi_n(R(t))}.
    \label{eq:sec2_estadoBerry}
\end{equation}

El primer factor exponencial representa la acumulación de {\em fase dinámica} $(1/\hbar)\int_0^t\,dt\, E_n(R(t))$, mientras que el segundo factor es aquél sobre el cual Berry centró su atención.

Cuando el parámetro $R(t)$ se conduce durante un intervalo $t\in[0,T]$ a lo largo de un camino cerrado ${\rm C}$ en el espacio de parámetros, y en consecuencia se tiene que $R(0) = R(T)$, todas aquellas cantidades que pueden escribirse como funciones de $R$ heredan la misma periodicidad. El Hamiltoniano del sistema necesariamente satisface $H(R(0))= H(R(T))$ y lo mismo ocurre con sus autoestados y autoenergías, y en consecuencia con la fase dinámica. La {\em fase de Berry} $\phi^n_{\rm a}$, sin embargo, no puede escribirse como una función del parámetro $R$ y no resulta univaluada sobre el circuito $\phi^n_{\rm a}(0)\neq \phi^n_{\rm a}(T)$ sino que satisface la ecuación

\begin{equation}
    \Dot{\phi}^n_{\rm a}(t) = i\,\bra{\psi_n(R(t))}\nabla_R\ket{\psi_n(R(t))}\,\Dot{R}(t).
    \label{eq:sec2_eqBerry}
\end{equation}
De esta forma, la fase de Berry acumulada al variar el Hamiltoniano a lo largo del circuito, resultado de integrar la ecuación (\ref{eq:sec2_eqBerry}) en el intervalo $t \in [0, T]$, es

\begin{equation}
    \phi^n_{\rm a}({\rm C}) =  i \oint_{\rm C} \bra{\psi_n(R)}\nabla_R\ket{\psi_n(R)}\cdot dR,
    \label{eq:sec2_Berry}
\end{equation}
donde se ha realizado un cambio en la variable de integración $t\rightarrow R$, obteniendo una integral de línea a lo largo del circuito ${\rm C}$ en el espacio de parámetros. 

La expresión (\ref{eq:sec2_Berry}) es independiente de la velocidad a la que se recorre el circuito o, lo que es lo mismo, es independiente frente a reparametrizaciones del circuito que preserven el sentido. Sin embargo, hay que notar que la dependencia en la velocidad con la que varía el parámetro prevalece indirectamente en las hipótesis realizadas. Específicamente, si el parámetro no varía lo suficientemente lento para despreciar las transiciones no-adiabáticas a otros niveles de energía, la evolución del sistema no está dada por la ecuación (\ref{eq:sec2_estadoBerry}). Sin embargo, es posible definir una fase geométrica absolutamente independiente de la tasa de evolución del sistema. Este es precisamente el resultado obtenido por Aharonov y Anandan \cite{aharonov1987phase}.

\section{Fase de Aharonov-Anandan}\label{sec:sec2_AA}
Considérese $\mathcal{H}$ el espacio de Hilbert adecuado para la descripción de un dado sistema cuántico y $\mathcal{N}_0$ el subespacio de vectores no-nulos y norma unitaria de $\mathcal{H}$. Si se describe el estado del sistema mediante un elemento $\ket{\psi} \in \mathcal{N}_0$, el mismo queda entonces definido a menos de un factor de fase. Esto quiere decir que todos aquellos vectores de estado $\ket{\psi} \in \mathcal{N}_0$ que difieren en un factor de fase se consideran {\em equivalentes} ya que recuperan valores de expectación y probabilidades idénticas. Por el contrario, los estados físicos, aquellos en relación 1 a 1 con valores distintos de las propiedades del sistema, están representados por {\em clases de equivalencia} de vectores de $\mathcal{N}_0$, colecciones $\xi =\{e^{i\alpha}\ket{\psi}|0\leq\alpha<2\pi\}$ denominadas {\em rayos} que agrupan en un único elemento (la clase) todos los objetos equivalentes. 

A partir de una noción de evolución cíclica distinta a la de Berry, Aharonov y Anandan \cite{aharonov1987phase} definieron una fase geométrica que resulta idéntica para las infinitas posibles trayectorias $\mathcal{C}\in \mathcal{H}$ en el espacio de Hilbert que se proyecten sobre una misma curva ${\rm C}\in\mathcal{P}$ en el espacio proyectivo de rayos. 
Su propuesta se desarrolla como sigue: la ecuación de Schrödinger define una curva $\mathcal{C}: t \in [0, T] \rightarrow \mathcal{H}$ en el espacio de Hilbert. Mediante un mapa $\Pi: \mathcal{N}_0(\mathcal{H})\rightarrow\mathcal{P}(\mathcal{H})$ esta trayectoria puede proyectarse en una curva ${\rm C} = \Pi(\mathcal{C})$ sobre el espacio de rayos. 

Basándose en estas observaciones es posible construir una definición de evolución cíclica que no hace referencia alguna al Hamiltoniano del sistema, el cual podría incluso ser independiente del tiempo: se define las evoluciones cíclicas como aquellas para las cuales la curva ${\rm C}$ sobre el espacio de rayos asociada a la evolución es una curva cerrada, es decir, evoluciones para las cuales se satisface $\xi(0) = \xi(T)$. 
Los vectores del espacio de Hilbert que representan el estado inicial $\ket{\psi(0)}$ y final $\ket{\psi(T)}$ corresponden en este caso a un mismo estado físico y sólo pueden diferir en un factor de {\em fase total}, lo que puede escribirse como
 \begin{equation}
     \ket{\psi(T)} = e^{i\phi}\ket{\psi(0)}.
 \end{equation}

Por otra parte, cada vector en una misma clase de equivalencia $\xi$ está relacionado con el representante de la clase mediante multiplicación por un factor de fase adecuado

\begin{equation}
    \ket{\psi(t)} = e^{i\,f(t)}\ket{\xi(t)},
    \label{eq:sec2_psiXi}
\end{equation}
y la ecuación de Schrödinger para el estado $\ket{\psi(t)}$ 

\begin{equation}
    i\hbar\,|\dot{\psi}(t)\rangle = H\,\ket{\psi(t)}
    \label{eq:sec2_SchEq}
\end{equation}
implica, cuando se utiliza explícitamente la relación (\ref{eq:sec2_psiXi}), la siguiente ecuación diferencial para el factor de fase $f(t)$

\begin{equation}
    \hbar\Dot{f}(t) = -\bra{\xi(t)}H\ket{\xi(t)} +  i\hbar\langle\xi(t)|\Dot{\xi}(t)\rangle,
    \label{eq:sec2_eqparaft}
\end{equation}
que debe satisfacer la condición $f(T) - f(0) = \phi$. Integrando la ecuación diferencial para el factor de fase, se obtiene que

\begin{equation}
    \phi = -\frac{1}{\hbar}\int_0^T \,dt\,\bra{\xi(t)}H\ket{\xi(t)} +  \int_0^T\,dt\,i\langle\xi(t)|\Dot{\xi}(t)\rangle,
\end{equation}
donde el término en el lado izquierdo de la ecuación es la ya mencionada fase total, el término

\begin{equation}
    -\frac{1}{\hbar} \int_0^T \,dt\,\bra{\xi(t)}H\ket{\xi(t)} = -\frac{1}{\hbar}  \int_0^T \,dt\,\bra{\psi(t)}H\ket{\psi(t)}
    \label{eq_sec2_faseDinamica}
\end{equation}
corresponde a la {\em fase dinámica}, que depende manifiestamente del Hamiltoniano del sistema, y la sustracción de estos dos términos

\begin{equation}
    \phi_{A-A} = \phi + \frac{1}{\hbar}  \int_0^T \,dt\,\bra{\psi(t)}H\ket{\psi(t)} = \int_0^T\,dt\,i\langle\xi(t)|\Dot{\xi}(t)\rangle,
    \label{eq:sec2_AA}
\end{equation}
es la fase geométrica propuesta por Aharonov y Anandan. 
Esta fase depende únicamente de la curva ${\rm C}$ descrita en el espacio de rayos, sin referencia alguna al Hamiltoniano del sistema.
Restringiéndose al caso de un Hamiltoniano $H(t)$ que varía lentamente y expandiendo el estado $\ket{\psi(t)}$ en la base de autoestados del Hamiltoniano instantáneo se recupera el resultado de Berry de la ecuación (\ref{eq:sec2_Berry}).

\section{Interpretación geométrica y caso no-cíclico}\label{sec:sec2_Samuel}
Samuel y Bhandari mostraron \cite{samuel1988general} que la fase geométrica aparece en un contexto todavía más general que aquél para el que aplica la fase de la ecuación (\ref{eq:sec2_AA}). La definición de una fase geométrica no requiere que la evolución sea cíclica ni que se preserve la norma del estado. Para seguir su definición se impone, sin embargo, una descripción geométrica del espacio de Hilbert donde la fase surge como consecuencia de la estructura de este espacio. 
En lo que sigue, se desarrolla el caso unitario, dejando el estudio de una evolución no-unitaria para la sección \ref{sec:sec2_nonUnit}

\subsection{Abordaje en términos de fibrados}\label{sec:sec2_fibrados}

Se considera nuevamente un espacio de Hilbert $\mathcal{H}$ adecuado para la descripción de un dado sistema cuántico y el subespacio $\mathcal{N}_0 \in \mathcal{H}$ de elementos de norma unitaria introducido en la sección anterior \ref{sec:sec2_AA}. Los vectores $\ket{\psi} \in \mathcal{N}_0$ definen el estado del sistema a menos de una fase o, en otras palabras, dos vectores de $\mathcal{N}_0$ relacionados por un factor de fase según $\ket{\psi'} = e^{i \alpha}\ket{\psi}$ son {\em equivalentes}, es decir, describen el mismo estado físico. Las colecciones que agrupan en un único elemento todos los estados equivalentes, reciben el nombre de clases de equivalencia o de rayos y conforman el espacio proyectivo $\mathcal{P}$.
El mapa $\Pi: \mathcal{N}_0\rightarrow\mathcal{P}$ proyecta cada vector en la clase de equivalencia a la cual pertenece.

\begin{SCfigure}[5][ht!]
    \includegraphics[width = .55\linewidth]{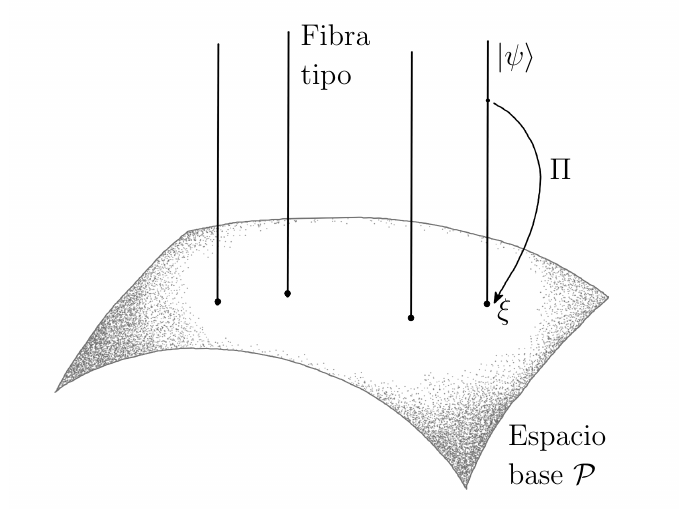}
    \caption{Representación esquemática del subespacio de estados $\mathcal{N}_0$. El espacio proyectivo de rayos $\mathcal{P}$ constituye el espacio base del fibrado. Asociada a cada punto $\xi \in \mathcal{P}$ del mismo hay todo un espacio denominado fibra, es el espacio formado por los vectores equivalentes. El mapa $\Pi$ proyecta los elementos de $\mathcal{N}_0$ en su elemento asociado sobre $\mathcal{P}$.} 
    \label{fig:sec2_hilbert}
\end{SCfigure}

La estructura aquí descrita para el espacio de Hilbert corresponde a un {\em fibrado principal}, en el cual el espacio total o {\em fibrado} se compone de un {\em espacio base} y una {\em fibra tipo} asociada a cada punto del espacio base. 
En este caso, representado en la figura \ref{fig:sec2_hilbert}, el espacio total es el Hilbert $\mathcal{N}_0$, el espacio base es el espacio proyectivo, de rayos o de estados físicos $\mathcal{P}$ y la fibra tipo está formada por todos los vectores equivalentes, relacionados por una transformación de fase. 

En términos generales, los fibrados son variedades que se ven localmente como el producto de dos espacios, el base y la fibra, pero pueden tener topología global no-trivial. Para una introducción a este tema, se refiere a \cite{nakahara2003geometry, chruscinski2004geometric}. El fibrado es {\em principal} cuando la fibra tipo se genera mediante la acción transitiva de un grupo. En el espacio de Hilbert todos los elementos de la fibra tipo se obtienen aplicando transformaciones $U(1)$ al elemento representativo.

Para que las fases geométricas surjan naturalmente en esta descripción, será necesario conferir mayor estructura al espacio. Con este objetivo se introduce a continuación el concepto de {\em conexión}. Una conexión es un objeto que permite comparar elementos pertenecientes a fibras diferentes mediante el establecimiento de una regla de {\em transporte paralelo} entre puntos en una y otra. El transporte paralelo de un elemento cualquiera $\ket{\psi} \in\mathcal{N}_0$ del espacio total a lo largo de una curva ${\rm C}$ en el espacio base consiste en el llamado {\em alzamiento horizontal} de la curva, es decir, la curva $\Pi^{-1}({\rm C})=\mathcal{C} \in \mathcal{N}_0$ cuyo vector tangente satisface, punto a punto, la ecuación

\begin{equation}
    \Im\bra{\psi(t)}\dot{\psi}(t)\rangle = 0
    \label{eq:sec2_paralelo}
\end{equation}
esto es, es perpendicular a la fibra. 

\begin{SCfigure}[5][ht!]
    \includegraphics[width = .55\linewidth]{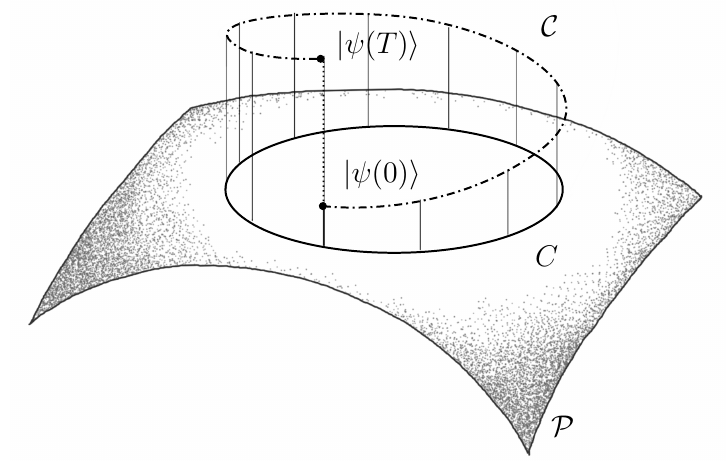}
    \caption{Representación esquemática de una trayectoria cíclica ${\rm C}\in \mathcal{P}$ sobre el espacio proyectivo de rayos (línea sólida). El alzamiento horizontal $\mathcal{C}$ de ${\rm C}$ al espacio total (línea de trazo-punto), transporte paralelo del vector $\ket{\psi(0)}$ a lo largo de ${\rm C}$, no resulta una camino cerrado en $\mathcal{N}_0$. El camino en el espacio total puede cerrarse sobre el rayo aplicando la transformación $U(1)$ conveniente al vector $\ket{\psi(T)}$.} 
    \label{fig:sec2_holonomia}
\end{SCfigure}

Resulta importante observar que el alzamiento horizontal de una curva ${\rm C} \in \mathcal{P}$ cerrada puede ser abierto, o lo que es lo mismo, el transporte paralelo de un vector $\ket{\psi}$ a lo largo de una curva ${\rm C}$ cerrada puede conducir a un vector $\ket{\psi'}\neq \ket{\psi}$ (ver figura \ref{fig:sec2_holonomia}). La transformación $\ket{\psi}\rightarrow\ket{\psi'} = e^{i\alpha}\ket{\psi}$ que une los dos extremos se denomina la {\em holonomía} de la curva $\mathcal{C}$ con respecto a la conexión considerada y proporciona una interpretación geométrica de la fase de Aharonov-Anandan (A-A), que se delinea a continuación.

\vspace{.5cm}
{\em Interpretación geométrica de la Fase de AA - }Considérese una curva $\mathcal{C}: t\in[0,T]\rightarrow\ket{\psi(t)}$ sobre $\mathcal{N}_0$  horizontal, y su vector tangente $|\Dot{\psi}(t)\rangle/\bra{\psi(t)}\ket{\psi(t)}$. La conexión natural

\begin{equation}
    A = \frac{\Im \bra{\psi(t)}\dot{\psi}(t)\rangle}{\bra{\psi(t)}\ket{\psi(t)}},
    \label{eq:se2_conexion}
\end{equation}
transforma, frente a transformaciones $U(1)$ de gauge $\ket{\psi(t)}\rightarrow e^{i \, \alpha(t)}\ket{\psi(t)}$, según 
\begin{equation}
    A \rightarrow A + \Dot{\alpha}(t).
    \label{eq:sec2_tranfGauge}
\end{equation}

Dado que $\mathcal{C}$ es horizontal por definición, la ley de transporte paralelo de la ecuación (\ref{eq:sec2_paralelo}) impone que la conexión se anule a lo largo de la trayectoria del estado que le da origen.
Si el vector de estado $\ket{\psi(t)}$ está además asociado a una evolución cíclica en el sentido de A-A, entonces retorna al rayo inicial en algún instante $T$. 

Considérese, en este escenario, la integral de la conexión $A$ sobre el camino construido a partir de la curva $\ket{\psi(t)}\;;\; t\in [0,T]$, cerrada uniendo $\ket{\psi(T)}$ con $\ket{\psi(0)}$ sobre el rayo. 
Como se ha discutido, la curva $\ket{\psi(t)}$ es horizontal por definición y por lo tanto la conexión se anula $A=0$ sobre ella. Por otra parte, la integral sobre el tramo vertical que cierra el camino da como resultado la diferencia de fase entre $\ket{\psi(T)}$ y $\ket{\psi(0)}$

\begin{equation}
\oint A\, dl_{\mathcal{N}_0} = \int_{\mathcal{C}}A + \int_\text{rayo}A =\arg\bra{\psi(0)}\ket{\psi(T)}.
\end{equation}
En síntesis, la integral total de la conexión en el camino cerrado equivale a la diferencia de fase entre los estados inicial y final.
Por otra parte, la integral de la conexión $A$ sobre una curva cerrada en $\mathcal{N}_0$ es invariante por efecto de la ley de transformación (\ref{eq:sec2_tranfGauge}). En consecuencia puede considerarse como una integral sobre el espacio base.
La holonomía de la curva ${\rm C} \in \mathcal{P}$ asociada a la conexión $A$ es entonces

\begin{equation}
    \mathfrak{g}({\rm C}) = e^{i \oint_{\rm C} A} = e^{i \phi_{A-A}}
\end{equation}
donde puede identificarse el factor de fase de A-A.

En el caso de una evolución general, es decir, no necesariamente cíclica, el vector que describe el sistema no retornará al rayo inicial. Para tratar este caso es necesario establecer un método que permita comparar estados en diferentes fibras. Dicho método recurre a la {\em fase de Pancharatnam} \cite{pancharatnam1956generalized}, definida para dos estados no-ortogonales cualesquiera $\ket{\psi_1}$ y $\ket{\psi_2}$ como
\begin{equation}
    \phi_{\rm P} = \arg\bra{\psi_1}\ket{\psi_2},
    \label{eq:sec2_Pancharatnam}
\end{equation}
y a las curvas {\em geodésicas} del espacio de rayos.

\subsection{Geodésicas del espacio de rayos}\label{sec:sec2_geodesicas}
{\em Concepto de geodésica del espacio de rayos - } Aunque cada elemento de $\mathcal{P}$ es una clase de equivalencia de vectores $\ket{\psi}$, resultará conveniente en lo que sigue trabajar con un elemento representativo, cuidando que las consideraciones resulten invariantes de gauge y apliquen así a todos los elementos de la clase indistintamente. Las geodésicas de un dado espacio son aquellas curvas que resultan el camino más corto (en algún sentido) entre dos puntos. Se podrá definir tales curvas gracias a que el producto interno $\bra{\cdot}\ket{\cdot}$ entre elementos del espacio de Hilbert confiere todavía mayor estructura al espacio proyectivo $\mathcal{P}$, ya que induce sobre éste una métrica (es decir, una noción de distancia) natural. 

Sea $\ket{\psi(t)}$ una curva en $\mathcal{N}_0$ y $|\Dot{\psi}(t)\rangle$ su vector tangente, es posible construir una derivada covariante $\ket{\psi'(t)} = |\Dot{\psi}(t)\rangle - i\,A\,\ket{\psi(t)}$, con $A$ la conexión natural en la ecuación (\ref{eq:se2_conexion}). De esta forma, el producto interno $\bra{\psi'(t)}\ket{\psi'(t)}$ define un intervalo $dl^2 = \bra{\psi'(t)}\ket{\psi'(t)} \,dt^2$ invariante de gauge que contiene, en consecuencia, únicamente información sobre el rayo. La distancia asociada a dicho intervalo $\int \,dt \, \sqrt{\bra{\psi'(t)}\ket{\psi'(t)}}$ puede entonces minimizarse, encontrando la ecuación 

\begin{equation}
    \left(\frac{d}{dt}-i\,A\right)\ket{\psi'(t)}=0.
    \label{eq:sec2_geodesicEq}
\end{equation}
Las curvas sobre el espacio de Hilbert que satisfacen esta ecuación minimizan la distancia invariante de gauge y por lo tanto son curvas que se proyectan, mediante el mapa $\Pi: \mathcal{N}_0\rightarrow\mathcal{P}$, en geodésicas del espacio base.
\\
\\\indent
{\em Regla de las geodésicas - }Las curvas geodésicas son un elemento fundamental para la definición de una fase geométrica que aplique a trayectorias generales.
Esto se debe a que es posible expresar la fase de Pancharatnam (\ref{eq:sec2_Pancharatnam}) como una integral de la conexión $A$ sobre estas curvas. Este resultado se conoce como regla de las geodésicas y se expresa como sigue \cite{samuel1988general, chruscinski2004geometric, mukunda1993quantum}. 

Sean $\ket{\psi_1}$ y $\ket{\psi_2}$ $\in\mathcal{N}_0$ dos estados no-ortogonales pero por lo demás arbitrarios y sea, además, una curva $t \in [0,T]\rightarrow\ket{\psi(t)}$ que conecta estos dos estados con $\ket{\psi(0)} = \ket{\psi_1}$ y $\ket{\psi(T)} = \ket{\psi_2}$, y que se proyecta sobre una geodésica de $\mathcal{P}$, esto es, una curva que satisface la ecuación (\ref{eq:sec2_geodesicEq}). Entonces, la fase $\phi_{\rm P}$ de Pancharatnam (\ref{eq:sec2_Pancharatnam}) entre los estados puede calcularse como la integral de la conexión natural a lo largo de la geodésica
\begin{equation}
    \phi_{\rm P} = \int_{\rm geodesica}\, dt\, A,
    \label{eq:sec2_reglaGeo}
\end{equation}
con la conexión $A$ dada por la ecuación (\ref{eq:se2_conexion}). Esta regla es la pieza fundamental en la construcción de una definición de fase geométrica aplicable al caso de evoluciones no-cíclicas.

\subsection{Generalización al caso no-cíclico}
Los objetos matemáticos descritos en las secciones \ref{sec:sec2_geodesicas} y \ref{sec:sec2_fibrados} dotan de estructura espacio de Hilbert, confiriéndole nociones de distancia, transporte paralelo, etc., y permiten generalizar la noción de fase geométrica más allá del escenario de A-A y abandonar la condición de ciclicidad.
Para ver esto, se considera un vector de estado $\ket{\psi(t)}$ que evoluciona desde un estado inicial $\ket{\psi(0)}$ hasta un dado estado $\ket{\psi(T)}$. Si los estados en los límites del intervalo son no-ortogonales, aunque por lo demás completamente arbitrarios, se puede aplicar el criterio de Pancharatnam para establecer una diferencia de fase $\phi_{\rm P}$ entre ellos. Esta diferencia de fase puede además expresarse como una integral de la conexión $A$ (\ref{eq:se2_conexion}) mediante la regla de las geodésicas (\ref{eq:sec2_reglaGeo}). 

Como se ha discutido, la ley de transformación (\ref{eq:sec2_tranfGauge}) implica que la integral de línea de la conexión $A$ a lo largo de un camino cerrado en el espacio de Hilbert $\mathcal{N}_0$ resulte invariante de gauge $U(1))$. Con el objetivo de construir, a partir de la diferencia de fase $\phi_{\rm P}$ entre los estados inicial y final, una integral de la conexión sobre un camino cerrado, se suma a la misma la integral de $A$ a lo largo de la curva $t\in [0, T]\rightarrow \ket{\psi(t)}$ definida por la ecuación de Schrödinger (\ref{eq:sec2_SchEq})

\begin{equation}
    \int_0^T \,dt\, A = -\frac{1}{\hbar}\int_0^T \,dt \bra{\psi(t)}H\ket{\psi(t)}
    \label{eq:sec2_nocic_aux}
\end{equation}
donde para arribar al lado derecho de la ecuación se ha utilizado la definición de la conexión $A$ de (\ref{eq:se2_conexion}) y la ecuación de Schrödinger (\ref{eq:sec2_SchEq}).
De este modo se obtiene una integral de la conexión $A$ a lo largo de una curva $\mathcal{C}\in \mathcal{N}_0$ cerrada: la construida a partir de la evolución del estado en el intervalo $[0, T]$, con sus extremos $\ket{\psi(T)}$ y $\ket{\psi(0)}$ unidos mediante una curva que se proyecta sobre la geodésica correspondiente. Reemplazando las expresiones (\ref{eq:sec2_reglaGeo}) y (\ref{eq:sec2_nocic_aux}) explícitamente, puede verse que la integral de línea considerada $\phi_{S-B}$ resulta equivalente a la diferencia entre la fase dinámica (\ref{eq_sec2_faseDinamica}) y la fase de Pancharatnam

\begin{equation}
    \phi_{S-B}= \oint_{\mathcal{C}\,\in\,\mathcal{N}_0} A =- \phi_{\rm P} -\frac{1}{\hbar}\int_0^T \,dt \bra{\psi(t)}H\ket{\psi(t)} = \oint_{{\rm C}\, \in\, \mathcal{P}} A.
    \label{eq:sec2_faseSamuel1}
\end{equation}
Para establecer la última igualdad, se he utilizado que la integral sobre el camino cerrado es invariante de gauge y puede por lo tanto considerarse como definida sobre el espacio de rayos $\mathcal{P}$. Esta expresión tiene además un origen puramente geométrico: depende únicamente del camino trazado por el estado físico ${\rm C} \in \mathcal{P}$, y no en la velocidad a la cual dicho camino se atraviesa. 

\begin{SCfigure}[5][ht!]
    \includegraphics[width = .55\linewidth]{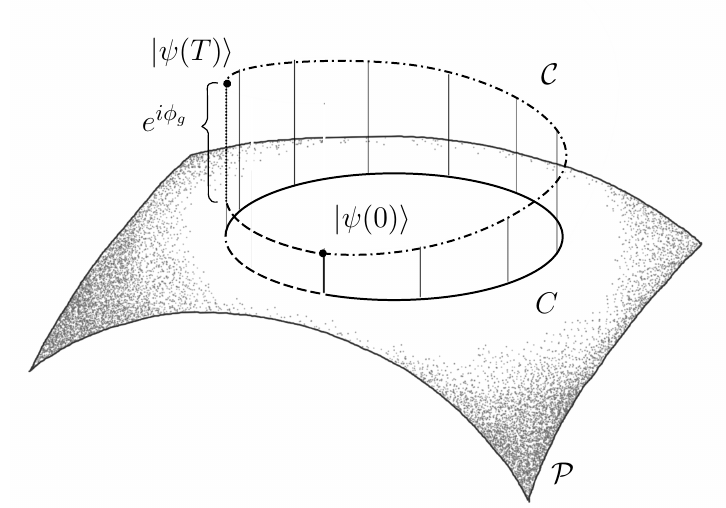}
    \caption{Representación esquemática de una trayectoria no-cíclica ${\rm C}\in \mathcal{P}$ en el espacio proyectivo de rayos (línea sólida). El alzamiento horizontal $\mathcal{C}$ de ${\rm C}$ al espacio total se indica con una línea de trazo-punto. Para definir la holonomía de la curva, la misma se completa con la curva geodésica que une sus extremos (línea de trazos) y que equivale a la diferencia de fase de Pancharatnam.} 
    \label{fig:sec2_abiertas}
\end{SCfigure}
La reducción del resultado (\ref{eq:sec2_faseSamuel1}) a la fase de A-A de la ecuación (\ref{eq:sec2_AA}) para el caso de trayectorias cerradas en el espacio de rayos es inmediata. Ésta además se reduce a la fase de Berry de la ecuación (\ref{eq:sec2_Berry}) cuando se satisfacen las hipótesis correspondientes, de forma que todos los resultados anteriores son casos particulares de la ecuación (\ref{eq:sec2_faseSamuel1}).

\subsection{Generalización al caso no-unitario}\label{sec:sec2_nonUnit}
La generalización de la fase geométrica se puede llevar todavía un paso más allá si se considera un sistema cuántico en evolución no-unitaria en el sentido de evolución de un estado puro que no conserva su norma.
Este tipo de evolución se obtiene, por ejemplo, cuando el sistema se somete a mediciones. De acuerdo con el postulado de colapso, el efecto de una medición en el sistema puede describirse con el operador de proyección $\ket{\psi}\bra{\psi}$ al autoestado correspondiente. Si se considera un estado inicial $\ket{\psi_0}$ sobre el cual se realizan N mediciones sucesivas de forma tal que la N-ésima proyección es otra vez al estado inicial. El estado final del sistema está dado por 

\begin{equation}
    \ket{\psi_0}\bra{\psi_0}\ket{\psi_{N-1}}...\bra{\psi_2}\ket{\psi_1}\bra{\psi_1}\ket{\psi_0},
\end{equation}
y los estados inicial y final tienen una diferencia de fase bien definida según el criterio de Pancharatnam, dada por el argumento del número complejo $\bra{\psi_0}\ket{\psi_{N-1}}...\bra{\psi_2}\ket{\psi_1}\bra{\psi_1}\ket{\psi_0}$. Usando la regla de las geodésicas la fase puede, una vez más, escribirse como una integral de la conexión $A$ sobre una curva cerrada. Dicha curva consiste en una poligonal formada por $N-1$ arcos geodésicos.

\section{Enfoque cinemático}\label{sec:sec2_Mukunda}
En la mayoría de las discusiones sobre la fase geométrica el punto de partida es la ecuación de Schrödinger (\ref{eq:sec2_SchEq}) para algún sistema cuántico particular caracterizado por un dado Hamiltoniano. Sin embargo, la fase geométrica es consecuencia de la cinemática cuántica, esto es, independiente del detalle respecto del origen dinámico de la trayectoria descrita en el espacio de estados físicos.   
Mukunda y Simon~\cite{mukunda1993quantum, mukunda1993quantum2} resaltaron la independencia de la fase geométrica respecto del origen dinámico de la evolución proponiendo un {\em enfoque cinemático} en el cual la trayectoria descrita en el espacio de estados físicos es el concepto fundamental para la fase geométrica.
En su desarrollo, se parte de la consideración de una curva uniparamétrica y suave $\mathcal{C}\subset \mathcal{N}_0$, conformada por una dada secuencia de estados $\ket{\psi(t)}$
 
 \begin{equation}
     \mathcal{C} = \{\ket{\psi(t)} \in \mathcal{N}_0\,|\, t \in [0, T]\subset\mathbb{R}\},
 \end{equation}
donde no se hace ninguna suposición respecto de si $\mathcal{C}$ es una curva abierta o cerrada, ni de el origen dinámico de la secuencia de estados. Se observa luego detenidamente la cantidad $\bra{\psi(t)}\Dot{\psi}(t)\rangle$ construida a partir de de esta curva.
La condición de unitariedad implica que esta cantidad sea imaginaria pura, lo que puede escribirse como
\begin{equation}
   \bra{\psi(t)}\Dot{\psi}(t)\rangle = i\,\Im\bra{\psi(t)}\Dot{\psi}(t)\rangle. 
\end{equation}
Por otra parte, aplicando una transformación $U(1)$ de gauge
\begin{equation}
    \mathcal{C}\rightarrow\mathcal{C'}: \ket{\psi'(t)}= e^{i\,\alpha(t)}\ket{\psi(t)}\;;\; t \in [0,T].
    \label{eq:sec2_U1tranf}
\end{equation}
la cantidad analizada transforma según $\Im\bra{\psi(t)}\Dot{\psi}(t)\rangle\rightarrow \Im\bra{\psi(t)}\Dot{\psi}(t)\rangle + \dot{\alpha}(t)$. A partir de esta ley de transformación se puede construir una funcional de la curva $\mathcal{C}$ que resulta invariante frente a transformaciones $U(1)$ de gauge (\ref{eq:sec2_U1tranf}), es decir, que toma el mismo valor para las curvas $\mathcal{C}$ y $\mathcal{C}'$

\begin{equation}
    \arg\bra{\psi(0)}\ket{\psi(T)} - \Im\int_0^T   \,dt \,\bra{\psi(t)}\Dot{\psi}(t)\rangle.
    \label{eq:sec2_kinGPaux}
\end{equation}

Otra vez, la invarianza de gauge $U(1)$  puede expresarse en términos de los objetos presentados en las secciones \ref{sec:sec2_AA} y \ref{sec:sec2_fibrados} que componen la descripción geométrica del espacio de Hilbert. 
Explícitamente, el mapa $\Pi:\mathcal{N}_0\rightarrow\mathcal{P}$ proyecta las curvas $\mathcal{C}\in \mathcal{N}_0$ y $\mathcal{C}'\in \mathcal{N}_0$, relacionadas mediante una transformación de gauge, en una misma curva ${\rm C} \subset \mathcal{P}$ del espacio $\mathcal{P}$ de rayos, cociente de $\mathcal{N}_0$ con respecto a la acción de $U(1)$.
Esto es,
\begin{align}
    &\mathcal{C}\subset \mathcal{N}_0 \rightarrow \Pi(\mathcal{C}) = {\rm C} \subset \mathcal{P}\\ \nonumber
    &{\rm con}\;\; \Pi(\mathcal{C}') = \Pi(\mathcal{C}).
\end{align}
La invarianza de gauge de la funcional (\ref{eq:sec2_kinGPaux}) significa que ésta es, en realidad, una funcional de la imagen ${\rm C}$ de $\mathcal{C}$. Esta funcional tiene además otra propiedad importante: la invarianza frente a reparametrizaciones monótonamente crecientes 
 \begin{align}
    &\mathcal{C}\rightarrow\mathcal{C'}= \{\ket{\psi'(t')} \in \mathcal{N}_0\,|\, t' \in [0', T']\subset\mathbb{R}\}
    \label{eq:sec2_reparametr}\\[.75em] \nonumber
    &{\rm con}\;\; \ket{\psi'(t'(t))} = \ket{\psi(t)}\;,\;
    \ket{\psi(0'/T')} = \ket{\psi(0/T)}.
\end{align}
Combinando las propiedades de invarianza frente a transformaciones de gauge y reparametrizaciones, se define la funcional 

\begin{equation}
    \phi_u[{\rm C}] \equiv\arg\bra{\psi(0)}\ket{\psi(T)} - \Im\int_0^T   \,dt \,\bra{\psi(t)}\Dot{\psi}(t)\rangle,
    \label{eq:sec2_kinGP}
\end{equation}
que se denomina la {\em fase geométrica} asociada a la curva ${\rm C}$ del espacio de rayos.

Cada término individual del lado derecho de la ecuación (\ref{eq:sec2_kinGP}) queda determinado cuando se toma una curva $\mathcal{C} \subset\mathcal{N}_0$ y por lo tanto depende de la misma. Sin embargo, la sustracción no depende de $\mathcal{C}$ sino de ${\rm C}$, de modo que si se considera esta última curva como el objeto de partida para calcular $\phi_g[{\rm C}]$ puede elegirse cualquier alzamiento $\mathcal{C}$ y el lado derecho de (\ref{eq:sec2_kinGP}) resultará independiente de esta elección. 

Aunque la fase geométrica de la ecuación (\ref{eq:sec2_kinGP}) se obtuvo desde un tratamiento estrictamente cinemático y enfatizando su conexión con la geometría del espacio de Hilbert, cuando se aplica en relación a la ecuación de Schrödinger (\ref{eq:sec2_SchEq}) se recuperan todos los resultados previos, según se satisfagan las condiciones para cada uno de ellos.
Además, el escenario cinemático es particularmente apto para dar una respuesta clara a la pregunta sobre qué le sucede con la fase geométrica frente a una transformación unitaria del estado. 

\subsection{Transformaciones unitarias}\label{sec:sec2_MukundaUnitarias}
Considérese $H(t)$ un dado Hamiltoniano y $\ket{\psi(t)}\,;\, t\in[0,T]$ una solución de la ecuación de Schrödinger (\ref{eq:sec2_SchEq}) para este Hamiltoniano, sin suponer nada respecto de si $H(t)$ o $\ket{\psi(t)}$ son o no cíclicos. En el caso más general la evolución define una curva $\mathcal{C} = \{\ket{\psi(t)}\,;\, t\in[0,T]\}$ con una imagen ${\rm C} = \Pi(\mathcal{C})$ abierta, y 

\begin{equation}
    \phi_u[{\rm C}] = \arg\bra{\psi(t)}\ket{\psi(T)} + \int_0^T \,dt\,\bra{\psi(t)}H(t)\ket{\psi(t)},
\end{equation}
donde se ha usado explícitamente que el estado $\ket{\psi(t)}$ es solución de la ecuación de Schrödinger. Si se aplica una transformación unitaria dependiente del tiempo $U(t)$ sobre el estado $\ket{\psi(t)}$
\begin{align}
    &\;U^\dagger(t)\,U(t)= \mathbb{I}\\ \nonumber
    &\ket{\psi'(t)} = U(t) \, \ket{\psi(t)}
\end{align}
se obtiene, en general, la solución de una ecuación de movimiento que coincide con la ecuación Schrödinger para un Hamiltoniano

\begin{equation}
    H'(t) = U(t)H(t)U^\dagger(t) - i\hbar\,U(t)\dot{U}^\dagger(t) 
\end{equation}
diferente al original que por lo tanto puede asociarse con un sistema diferente. En consecuencia, en el caso general se tiene una curva distinta $\mathcal{C}' = \{\ket{\psi'(t)}\,;\, t\in[0,T]\}$ cuya imagen ${\rm C}' = \Pi(\mathcal{C}')$ es también distinta de la anterior, y la fase de Pancharatnam resulta

\begin{equation}
    \phi_{\rm P} = \arg \bra{\psi'(0)}\ket{\psi'(T)} = \arg \bra{\psi(0)}U(0)^{-1}U(T)\ket{\psi(T)}.
    \label{eq:sec2_unitP}
\end{equation}
Por otro lado, la fase dinámica acumulada en la evolución transformada puede escribirse como

\begin{equation}
    -\int_0^T \,dt\,\bra{\psi(t)}H(t)\ket{\psi(t)} - i\int_0^T \, dt \, \bra{\psi(t)}U(t)^{-1}\dot{U}(t)\ket{\psi(t)}.
    \label{eq:sec2_unitD}
\end{equation}
Observando estas expresiones, se encuentra que en general no sólo difieren la fase de Pancharatnam y la fase dinámica sino que también su diferencia se modifica, y en consecuencia ninguna de las fases conserva su valor ni su expresión funcional al cambiar de $\mathcal{C}$ y ${\rm C}$ a $\mathcal{C}'$ y ${\rm C}'$.

En casos particulares, sin embargo, pueden existir simplificaciones. Uno de estos casos especiales es aquél en que la transformación unitaria es independiente del tiempo. Aún si $H'(t) = U(t)H(t)U^\dagger(t)\neq H(t)$ y en consecuencia generan evoluciones distintas, sin embargo las ecuaciones (\ref{eq:sec2_unitP}) y (\ref{eq:sec2_unitD}) indican que todas las fases coinciden en este caso. {\em Por el contrario, si $U(t)$ depende del tiempo, aún si se satisface que $H'(t) = H(t)$ y por lo tanto $U(t)$ conecta dos soluciones de la misma ecuación de Schrödinger, las cuales las tres fases, dinámica, geométrica, y total, tienen en general valores distintos.}

\section{Ejemplo de aplicación: Sistema de dos niveles en un campo magnético}\label{sec:sec2_ejemplo}
Para ilustrar la descripción teórica con un ejemplo concreto, se estudian las distintas definiciones de fase geométrica presentadas en las secciones anteriores tomando el caso paradigmático de un sistema de espín-$1/2$ inmerso en un campo magnético clásico $\mathbf{B}(t) =\omega\, \hat{\text{\bf{n}}}_\mathbf{B}(t)$ cuya dirección está dada por $\hat{\text{\bf{n}}}_\mathbf{B }=(\sin{\theta}\cos\varphi,\allowbreak \sin{\theta}\sin\varphi, \cos{\theta})$ y puede o no cambiar en el tiempo.
Una evolución de este tipo está generada por el Hamiltoniano 

\begin{equation}
    H(t) = \frac{1}{2}\,\mathbf{B}(t)\cdot \boldsymbol{\sigma},
    \label{eq:sec2_HamiltonianSpin}
\end{equation}
con $\boldsymbol{\sigma} = \sigma_x\,\hat{x} + \sigma_y\,\hat{y} + \sigma_z\,\hat{z}$ un operador vectorial formado por las matrices de Pauli, $\ket{0}$ y $\ket{1}$ los autoestados de la matriz de Pauli $\sigma_z$, y se consideran unidades de $\hbar = 1$.

\subsection{Fase de Berry}\label{sec:sec2_ejemploBerry}
\begin{figure}[ht!]
    \center
    \includegraphics[width = \linewidth]{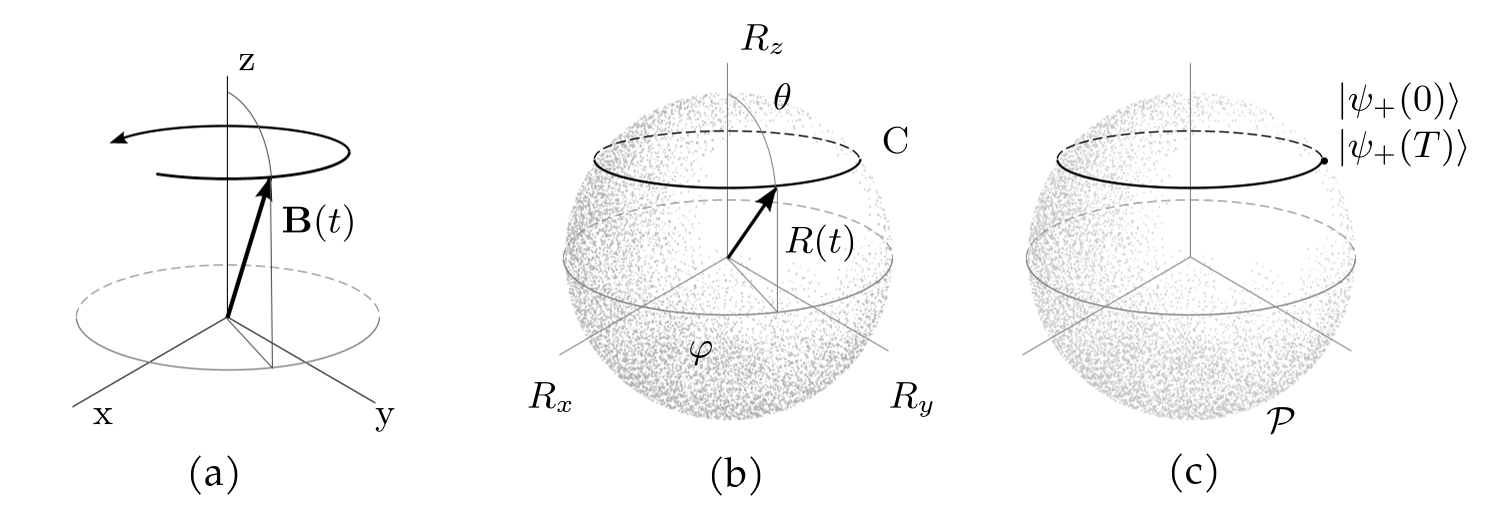}
    \caption{Fase de Berry acumulada por un espín en un campo magnético rotante. (a) Dirección del campo magnético $\mathbf{B}(t)$ en función del tiempo. (b) Espacio de parámetros y trayectoria descrita por el parámetro multidimensional $R(t)$. (c) Trayectoria descrita en la esfera de Bloch $\mathcal{P}$ por la proyección del autoestado $\ket{\psi_+(t)}$.} 
    \label{fig:sec2_EjemploBerry}
\end{figure}

Se examina en primer lugar el caso en el cual la dirección del campo $\mathbf{B}(t)$ varía lentamente en el tiempo con el ángulo polar $\theta$ fijo y ángulo azimutal $\varphi = \Omega\,t$ dependiente del tiempo, como se indica en el panel (a) de la figura \ref{fig:sec2_EjemploBerry}. Cuando el campo rote lo suficientemente lento, el espín de la partícula se alineará instante a instante con el campo, de forma que un autoestado de $H(0)$ permanecerá en el  autoestado instantáneo correspondiente de $H(t)$. Los autoestados instantáneos de $H(t)$ se denotan $\ket{\psi_-(t)}$ y $\ket{\psi_+(t)}$, y pueden escribirse, en términos de los autoestados del operador $\sigma_z$, como
\begin{align}\nonumber
    \ket{\psi_+(t)} =& \cos\frac{\theta}{2}\ket{1} + \sin\frac{\theta}{2}\,e^{i\,\Omega\,t}\,\ket{0}\\
    \ket{\psi_-(t)} =& \sin\frac{\theta}{2}\ket{1} - \cos\frac{\theta}{2}\,e^{i\,\Omega\,t}\,\ket{0},
    \label{eq:sec2_eigenSpin}
\end{align}
con autovalores $E_\pm(t) = \pm\,\omega/2 $ respectivamente.

El Hamiltoniano depende del tiempo a través del campo $\mathbf{B}(t)$, el cual puede ser descrito mediante su norma $|\mathbf{B}|$ y dirección, definida por los ángulos $\theta$ y $\varphi$. En consecuencia, el espacio de parámetros que modelan el Hamiltoniano es idéntico al conjunto de valores posibles para $\mathbf{B}(t)$, que forma, en el caso considerado, una superficie esférica de radio fijo $r = |\mathbf{B}| = \omega$. Recordando la dependencia temporal asumida para el campo, se observa que los posibles circuitos $C$ en el espacio de parámetros corresponden, en este ejemplo, a anillos horizontales sobre una esfera y se ilustran en el panel (b) de la figura \ref{fig:sec2_EjemploBerry}.
El gradiente de los autoestados del Hamiltoniano en coordenadas esféricas $\{r, \theta, \varphi\}$,

\begin{align}\nonumber
    \nabla\ket{\psi_+(t)} =& \frac{-1}{2r}\ket{\psi_-(t)} \hat{\theta} + i\; \frac{\sin(\theta/2)}{r\sin(\theta)}\,e^{i\,\varphi}\,\ket{0} \hat{\varphi}\\
    \nabla\ket{\psi_-(t)} =& \frac{-1}{2r}\ket{\psi_+(t)} \hat{\theta} + i\; \frac{\cos(\theta/2)}{r\sin(\theta)}\,e^{i\,\varphi}\,\ket{0} \hat{\varphi},
\end{align}
da origen a integrandos en la ecuación (\ref{eq:sec2_Berry}) de la forma

\begin{align}\nonumber
    \bra{\psi_+}\nabla\ket{\psi_+} =& i\; \frac{\sin^2(\theta/2)}{r\sin(\theta)}\,\hat{\varphi}= \frac{i}{2}\frac{1-\cos(\theta)}{r\sin(\theta)}\,\hat{\varphi}\\
    \bra{\psi_-}\nabla\ket{\psi_-} =& i\; \frac{\cos^2(\theta/2)}{r\sin(\theta)}\,\hat{\varphi}= \frac{i}{2}\frac{1+\cos(\theta)}{r\sin(\theta)}\,\hat{\varphi}.
\end{align}
Finalmente, por integración directa sobre el circuito ${\rm C}$, se obtiene el valor de la fase de Berry para cada uno de los autoestados

\begin{equation}
    \phi^\pm_{\rm a} ({\rm C}) = i\oint_{\rm C}  \bra{\psi_\pm}\nabla\ket{\psi_\pm}\, r\sin(\theta)d\varphi \, \hat{\varphi} = -\pi(1\mp\cos(\theta)).
    \label{eq:sec2_faseSpinBerry}
\end{equation}

\subsection{Eco de espín}\label{sec:sec2_SpinEcho}
\begin{figure}[ht!]
    \center
    \includegraphics[width = .95\linewidth]{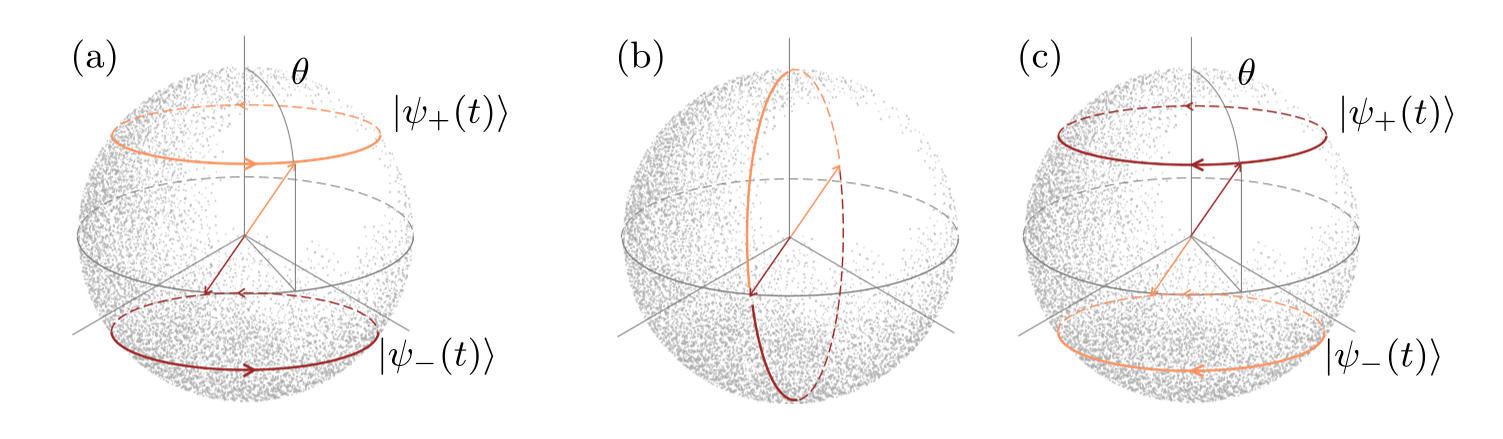}
    \caption{Pasos principales en un protocolo de Eco de espín. (a) Evolución en la esfera de Bloch de cada autoestado durante el primer período $t\in [0,T]$. (b) operación de inversión de espines $\ket{\psi_\pm(T)}\leftrightarrow\ket{\psi_\mp(T)}$, en la que cada autoestado se conduce instantáneamente al autoestado opuesto. (c) Evolución de cada autoestado durante el segundo período $t \in [T, 2T]$, en el cuál el campo magnético rota en dirección opuesta.} 
    \label{fig:sec2_EjemploEcho}
\end{figure}

Un método usual para medir la fase de Berry en un sistema de este tipo consiste en la aplicación del protocolo conocido como {\em Eco de espín} (o Spin Echo, en inglés), el cual está compuesto por la siguiente serie de pasos \cite{hahn} (ver figura \ref{fig:sec2_EjemploEcho}).
En primer lugar, el sistema se prepara en un estado superposición $\ket{\psi(0)}$ que en términos de los autoestados del Hamiltoniano se expresa como $(1/\sqrt{2})(\ket{\psi_+(0)} + \ket{\psi_-(0)})$. A continuación, se conduce el sistema en forma adiabática durante un período $t\in[0,T]$ del Hamiltoniano, causando de esta forma que cada autoestado adquiera tanto una fase dinámica 

\begin{equation}
    -\int_0^T\,dt\, E_\pm(t) = \mp\frac{\omega}{2}\,T,
    \label{eq:sec2_faseSpinDyn}
\end{equation}
como una fase geométrica $\phi^\pm_{\rm a}$

\begin{equation}
    \ket{\psi_\pm(0)}\rightarrow e^{-i\,\int_0^TE_\pm(t) + i \gamma_\pm}\ket{\psi_\pm(T)}.
\end{equation}
La figura \ref{fig:sec2_EjemploEcho}.a muestra cómo evoluciona cada autoestado en la esfera de Bloch durante este paso.
Finalizado ese primer ciclo, se realiza una operación de inversión de espines (conocida como spin flip en inglés) $\ket{\psi_\pm(t)}\rightarrow\ket{\psi_\mp(t)}$, la cual se considera en general instantánea, y se representa en la esfera de Bloch en la figura \ref{fig:sec2_EjemploEcho}.b. 
Finalmente, se conduce el sistema en un segundo ciclo $t\in[T,2T]$ del Hamiltoniano, que se ejecuta esta vez invirtiendo el sentido de rotación del campo mediante $\Omega \rightarrow -\Omega$. La operación de inversión y el segundo ciclo conducen a la cancelación de las contribuciones dinámicas $-\int_0^T\,dt\,E_{\pm}(t)-\int_0^T\,dt\,E_{\mp}(t) = 0$, lo que se observa directamente de su expresión (\ref{eq:sec2_faseSpinDyn}). 

Cada autoestado acumula entonces, a lo largo del protocolo, un factor de fase que es puramente geométrico. La dependencia de la fase de Berry (\ref{eq:sec2_faseSpinBerry}) en el sentido de rotación del campo $\mathbf{B}(t)$ implica contribuciones geométricas que satisfacen $\phi^\pm_{\rm a} - \phi^\mp_{\rm a}  = \mp\,\pi\,(1-\cos\theta)\pm 2\pi$. En síntesis, la evolución total de cada autoestado resulta

\begin{equation}
    \ket{\psi_\pm(0)}\rightarrow e^{\pm\,i (2\phi^+_{\rm a} + 2\pi)}\ket{\psi_\mp(0)},
\end{equation}
donde se ha utilizado que los autoestados $\ket{\psi_\pm(t)}$ respetan la ciclicidad $H(0)=H(T)$ del Hamiltoniano. Esta evolución se traduce en un estado total final

\begin{equation}
    \ket{\psi(2T)} = \frac{1}{\sqrt{2}}(e^{i 2\phi^+_{\rm a}}\ket{\psi_-(0)}+e^{-i 2\phi^+_{\rm a}}\ket{\psi_+(0)}).
\end{equation}
La fase de Berry puede extraerse mediante tomografía~\cite{leek2007_cqed_observation, gasparinetti2016_cqed_observation, cucchietti}  o notando que la probabilidad de que el sistema retorne al estado inicial, es decir, la {\em probabilidad de persistencia}, se relaciona con la fase de Berry según \cite{acotacion_se} 

\begin{equation}
    | \langle \psi(0) | \psi(2\,T) \rangle |^2 = \cos^2(2\,\phi^+_{\rm a}).
\end{equation}

\subsection{Fase de Aharonov-Anandan}\label{sec:sec2_ejemploAA}
\begin{figure}[ht!]
    \center
    \includegraphics[width = .95\linewidth]{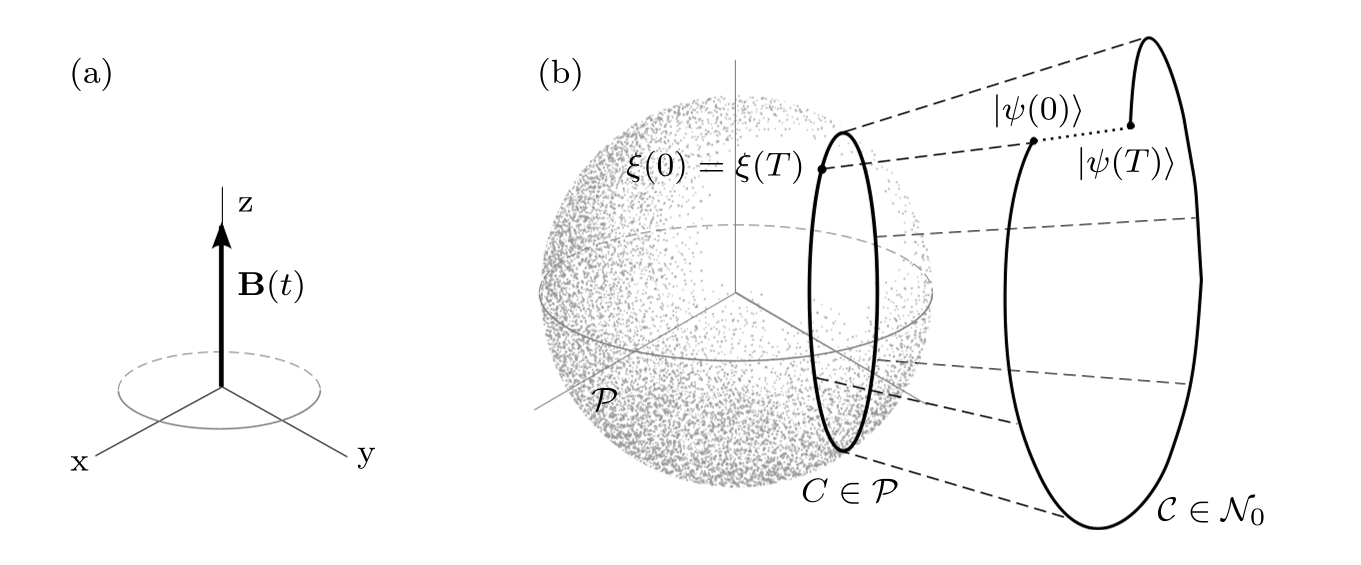}
    \caption{Fase de Aharonov - Anandan acumulada por un espín en un campo magnético fijo en dirección del eje $z$. (a) Dirección del campo magnético $\mathbf{B}$. (b) Trayectoria $\mathcal{C} \in\mathcal{N}_0$ descrita por el estado del sistema y proyección $C\in\mathcal{P}$ sobre la esfera de Bloch. La evolución es cíclica en el sentido de A-A, de modo que los estados físicos $\xi(0)$ y $\xi(T)$ coinciden, generando una curva $C$, cerrada.} 
    \label{fig:sec2_ejAA}
\end{figure}

Con el fin de presentar un ejemplo de cálculo de la fase de Aharonov y Anandan, se considera el caso en el que el campo $\mathbf{B} = \omega\,\hat{z}$ apunta en dirección del eje $z$, como se muestra en la figura \ref{fig:sec2_ejAA}.a. En este caso, el Hamiltoniano de la ecuación (\ref{eq:sec2_HamiltonianSpin}) toma la forma
\begin{equation}
    H = \frac{\omega}{2}\sigma_z.
    \label{eq:sec2_HSpinz}
\end{equation}
El operador de evolución generado por este Hamiltoniano $U(t,0) = e^{i \frac{\omega}{2}\,\sigma_z\,t}$ puede aplicarse directamente sobre un estado inicial superposición $\ket{\psi(0)} = \cos(\theta/2)\ket{1} + \sin(\theta/2)\ket{0}$, para encontrar la expresión del estado a tiempos posteriores $t>0$

\begin{equation}
    \ket{\psi(t)} = \cos\frac{\theta}{2}\,e^{-i\frac{\omega}{2}t}\ket{1} + \sin\frac{\theta}{2}\,e^{i\frac{\omega}{2}t}\ket{0}.
    \label{eq:sec2_ejemploPsiAA}
\end{equation}
La proyección de este estado en el espacio de rayos $\xi(t) =\ket{\psi(t)}\bra{\psi(t)}$ satisface que $\xi(0) = \xi (T)$ para $T = 2\pi/\omega$ y por lo tanto define, para este intervalo temporal, una evolución cíclica según el criterio de A-A. La figura \ref{fig:sec2_ejAA}.b muestra la trayectoria $\mathcal{C} \in\mathcal{N}_0$ descrita por el estado del sistema y proyección ${\rm C}\in\mathcal{P}$ sobre la esfera de Bloch. Estando en presencia de una evolución cíclica es posible utilizar el resultado de A-A y, reemplazando en la ecuación (\ref{eq:sec2_AA}), obtener  
\begin{equation}
    \phi_{A-A} = -\pi(1-\cos(\theta)).
    \label{eq:sec2_ejemploFaseAA}
\end{equation}

Curiosamente, se ha encontrado un valor para la fase geométrica que es idéntico al resultado para la fase adiabática de Berry del ejemplo anterior \ref{sec:sec2_ejemploBerry}. Sin embargo, la dinámica considerada en este caso es completamente diferente, lo que refuerza la idea de un objeto genuinamente geométrico que depende de la curva trazada en el espacio de rayos, y no de la dinámica que origina dicha curva.

\subsection{Enfoque cinemático}\label{sec:sec2_ejemploMukunda}
La fase geométrica definida por la ecuación (\ref{eq:sec2_kinGP}) puede aplicarse a las dos evoluciones consideradas en los ejemplos \ref{sec:sec2_ejemploBerry} y \ref{sec:sec2_ejemploAA}, obtenidas para un qubit que se halla en un campo magnético $\mathbf{B}$ rotante y en un campo fijo en dirección del eje $z$ respectivamente. Las únicas condiciones que esta definición impone sobre la evolución son la pureza del estado inicial y la no-ortogonalidad entre los estados inicial y final, de modo que es además posible abandonar las condiciones de adiabaticidad y ciclicidad, debiendo recuperar los resultados previos cuando estas condiciones sean, de hecho, satisfechas.

\begin{figure}[ht!]
    \center
    \includegraphics[width = .95\linewidth]{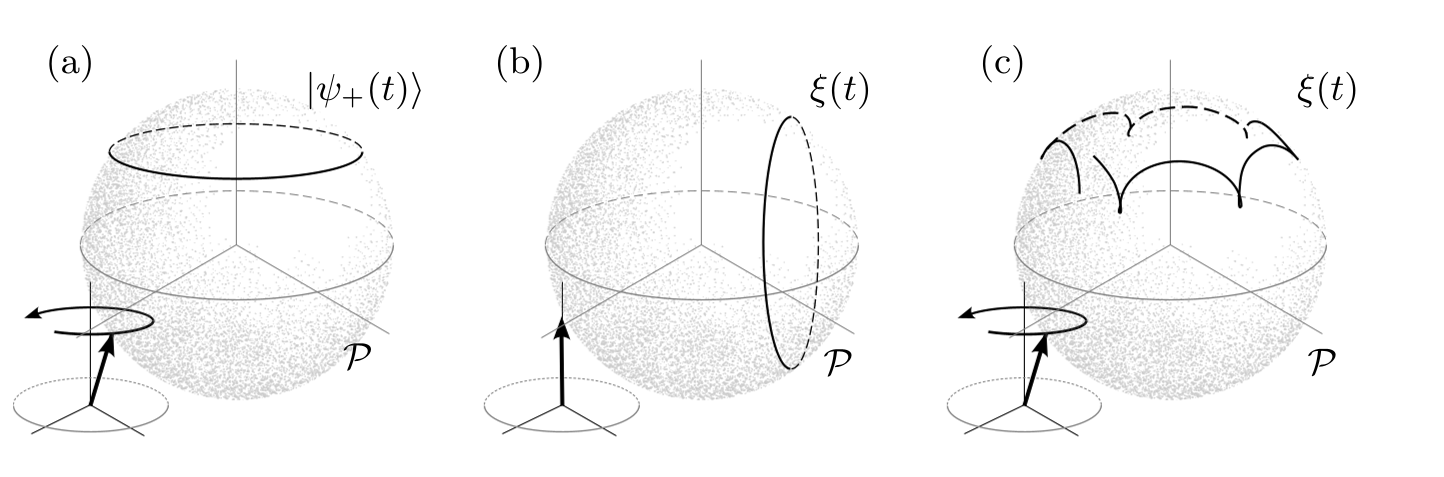}
    \caption{Trayectorias descritas en la esfera de Bloch por distintas evoluciones de un sistema de dos niveles en un campo magnético $\mathbf{B}(t)$ variable. El panel (a) muestra la evolución a partir de un autoestado del Hamiltoniano inicial $\ket{\psi_+(0)}$ en presencia de un campo que rota adiabáticamente, el panel (b) muestra la evolución a partir del estado superposición de la ecuación (\ref{eq:sec2_ejemploPsiAA}) en presencia de un campo magnético estático $\mathbf{B}(t) = B\,\hat{z}$ y, finalmente, el panel (c) corresponde a la evolución a partir de un autoestado del Hamiltoniano inicial $\ket{\psi_+(0)}$ en presencia de un campo que rota con velocidad arbitraria.} 
    \label{fig:sec2_ejMukunda}
\end{figure}

En primer lugar, se continúa tratando el caso de campo magnético fijo $\mathbf{B} = \omega\,\hat{z}$ de la sección \ref{sec:sec2_ejemploAA} para pasar posteriormente al caso del campo magnético rotante estudiado, en el límite adiabático, en la sección \ref{sec:sec2_ejemploBerry}. Dado que el enfoque cinemático hace hincapié en la dependencia exclusiva de la fase en la curva ${\rm C} \in \mathcal{P}$ trazada sobre el espacio proyectivo (que para un sistema de dos niveles es la esfera de Bloch), las curvas correspondientes a cada tipo de evolución se muestran esquemáticamente en la figura \ref{fig:sec2_ejMukunda} para referencia.

\vspace{.2cm}
{\em Campo estático - }Se desea reproducir el estudio de la fase geométrica realizado en la sección anterior \ref{sec:sec2_ejemploAA} en condiciones más generales. Para esto se calcula, recurriendo a la ecuación (\ref{eq:sec2_kinGP}), la fase geométrica acumulada por el estado (\ref{eq:sec2_ejemploPsiAA}) entre un tiempo inicial $t=0$ y un tiempo $t \in [0,T]$ arbitrario. La curva ${\rm C}$ descrita sobre la esfera de Bloch por el estado del sistema resulta un arco de circunferencia abierto para todo $t<T$ o una circunferencia cerrada para $t=T$ (ver figura \ref{fig:sec2_ejMukunda}.b). La fase total, o de Pancharatnam, toma la forma funcional
\begin{equation}
    \arg\bra{\psi(t)}\ket{\psi(t)} =\arg\left\lbrace \cos^2(\theta/2)\,e^{-i\frac{\omega}{2}t} + \sin^2(\theta/2)\,e^{i\frac{\omega}{2}t}\right\rbrace = \tan^{-1}\left\lbrace -\cos(\theta)\,\tan{(\omega\,t/2)}\right\rbrace,
\end{equation}
en términos de los parámetros del problema.
Por otro lado, usando la derivada $|\dot{\psi}(t)\rangle = (-i\omega/2)(\cos(\theta/2)\ket{1}-\sin(\theta/2)\ket{0})$ a la que se accede mediante derivación directa la ecuación (\ref{eq:sec2_ejemploPsiAA}), la fase dinámica acumulada por el estado en $[0,t]$ es

\begin{equation}
    \Im\int_0^t\,dt'\,\bra{\psi(t')}\dot{\psi}(t')\rangle = -\int_0^t\,dt'\, (\omega/2)\cos(\theta).
\end{equation}

La fase geométrica se encuentra sustrayendo las expresiones para estas dos fases

\begin{equation}
    \phi_u[C] = \arg\left\lbrace \cos^2(\theta/2)\,e^{-i\frac{\omega}{2}t} + \sin^2(\theta/2)\,e^{i\frac{\omega}{2}t}\right\rbrace +  (\omega\,t/2)\cos(\theta).
\end{equation}

Si se considera el intervalo $[0,T]$ con $T=2\pi/\omega$, la curva descrita sobre la esfera de Bloch es cerrada y por lo tanto la evolución es cíclica en el sentido de A-A. En este caso particular, la expresión de arriba para la fase geométrica resulta $\phi_u = -\pi(1-\cos(\theta))$, esto es, idéntica al resultado obtenido mediante la ecuación (\ref{eq:sec2_ejemploFaseAA}).

\vspace{.2cm}
{\em Campo magnético rotante - } Se dirige ahora la atención al caso de la evolución descrita en la sección \ref{sec:sec2_ejemploBerry}, pero generalizando el estudio a un escenario donde el campo rota con frecuencia arbitraria y la evolución no es necesariamente cíclica. La relajación de la condición de ciclicidad es tanto en el sentido de Berry como en el de A-A, es decir, no se impone que el Hamiltoniano satisfaga la relación $H(T)=H(0)$ ni que el estado describa una trayectoria cerrada en el espacio de rayos. La curva $C\in\mathcal{P}$ descrita por la proyección del estado en la esfera de Bloch se esquematiza para el caso adiabático en la figura \ref{fig:sec2_ejMukunda}.a, y para el caso en que se han relajado las condiciones de adiabaticidad y ciclicidad, en la figura en la figura \ref{fig:sec2_ejMukunda}.c.

Con el fin de simplificar la comparación entre evolución adiabática y no-adiabática, se considera el mismo estado inicial $\ket{\psi(0)} = \ket{\psi_+(0)}$ de la sección \ref{sec:sec2_ejemploBerry}. 
La dinámica del sistema tiene una solución analítica exacta (ver apéndice \ref{apendice1}) para cualquier estado inicial, lo que permite repetir este estudio para cualquier otra condición inicial de estado puro. Para el caso particular considerado, el estado del sistema $\ket{\psi(t)}$ a tiempo $t$ se describe mediante la expresión

\begin{equation}
    \ket{\psi(t)} = e^{-i\frac{\Omega}{2}\,t}\left\lbrace \left(\cos(\Tilde{\omega}\,t/2)+f(t)\right)\ket{\psi_+(t)}-g(t)\ket{\psi_-(t)}\right\rbrace,
    \label{eq:sec2_estadoUnitaria}
\end{equation}
donde las funciones $f(t)$ y $g(t)$ que aparecen en los coeficientes de la superposición son 
$f(t) = -i\,\sin(\Tilde{\omega}\,t/2)(\omega-\Omega\cos(\theta))/\Tilde{\omega}$ y $g(t) = -i\,\sin(\Tilde{\omega}\,t/2)\,\omega\,\sin(\theta)/\Tilde{\omega}$, y la frecuencia $\Tilde{\omega} = [(\omega\cos(\theta)-\Omega)^2 + \omega^2\,\sin^2(\theta)]^{1/2}$. Una vez en posesión de una forma explícita para el estado del sistema a todo tiempo $t$, el cálculo de la fase geométrica acumulada en un dado intervalo temporal $[0,t]$ es inmediato.
Por derivación directa se obtiene
\begin{align}\nonumber
    |\Dot{\psi}(t)\rangle =& -i(\Omega/2)\ket{\psi(t)}  + i\Omega \bra{0}\ket{\psi(t)}\ket{0}\\[.75em]
    &  + 
    e^{-i\frac{\Omega}{2}\,t}\left\lbrace \left(-(\Tilde{\omega}/2)\sin(\Tilde{\omega}\,t/2)+\Dot{f}(t)\right)\ket{\psi_+(t)}+\dot{g}(t)\ket{\psi_-(t)}\right\rbrace.
\end{align}

Antes de avanzar en el cómputo, sin embargo, se especifica un intervalo temporal que simplifique la comparación directa con la sección (\ref{sec:sec2_ejemploBerry}) y se considera el caso de evolución que tiene lugar durante el intervalo temporal $t\in [0, T]$, con $T = 2\pi/\Omega$, correspondiente a un período de rotación del campo magnético $\mathbf{B}$. Bajo estas condiciones, el sistema satisface la condición de ciclicidad $H(T) = H(0)$ impuesta por Berry mientras que no cumple con aquella definida por A-A. Esto es, el estado del sistema no retorna en general a tiempo $T$ al mismo rayo del cual partió y por lo tanto define una curva ${\rm C}\in\mathcal{P}$ abierta.

Para este intervalo de evolución específico, el primer término de la ecuación (\ref{eq:sec2_kinGP}), la fase total acumulada en la evolución, resulta
\begin{equation}
    \arg\bra{\psi(0)}\ket{\psi(T)}= -\pi -\pi\frac{\Tilde{\omega}}{\Omega} + \arg \left\{1-e^{i\,\frac{\Tilde{\omega}}{\Omega}\,2\pi}\,\frac{\omega\cos(\theta)-\Omega-\Tilde{\omega}}{\omega\cos(\theta)-\Omega +\Tilde{\omega}}\right\},
    \label{eq:sec2_ejemploKin1}
\end{equation}

mientras que la fase dinámica toma la forma 
\begin{align}
   \Im\int_0^T\,dt\,\bra{\psi(t)}\Dot{\psi}(t)  \rangle = - \pi\frac{\omega}{\Omega} + \pi \,\frac{\omega \,\Omega}{\Tilde{\omega}^2}\,\sin^2(\theta)\left(1- \frac{\sin(2\pi\Tilde{\omega}/\Omega)}{2\pi\Tilde{\omega}/\Omega}\right).
   \label{eq:sec2_ejemploKin2}
\end{align}
La sustracción de ambos términos permite encontrar la fase geométrica. 
La expresión exacta que surge de la sustracción de las ecuaciones (\ref{eq:sec2_ejemploKin1}) y (\ref{eq:sec2_ejemploKin2}) resulta poco ilustrativa. Para establecer una comparación con la fase de Berry resulta conveniente realizar una expansión en potencias de la relación entre la frecuencia de rotación del campo y la frecuencia natural del átomo para el caso en que esta relación satisface $\Omega/\omega \ll 1$. En el resultado 

\begin{equation}
    \phi_u[C]= -\pi(1-\cos(\theta)) - \pi\,\sin^2(\theta)\,\frac{\Omega}{\omega} + \mathcal{O}(\Omega/\omega)^2,
\end{equation}
se puede identificar la fase de Berry en el orden $\mathcal{O}(\Omega/\omega)^0$ y la corrección no-adiabática dominante. 

\section{Fases geométricas en sistemas abiertos}\label{sec:sec2_abiertos}
Las secciones anteriores componen una antología de las principales definiciones de fase geométrica para un sistema cuántico aislado, y son aplicables en distintos escenarios de acuerdo con las hipótesis en las que se basan. En cada paso de generalización dado desde la sección \ref{sec:sec2_Berry} a la sección \ref{sec:sec2_Mukunda}, se recuperan las expresiones menos generales si se satisfacen las hipótesis correspondientes, como se corrobora con un ejemplo explícito en la sección \ref{sec:sec2_ejemplo}. 
Por otra parte, todas estas definiciones satisfacen las siguientes propiedades: (i) Son invariantes frente a transformaciones de gauge $U(1)$ y reparametrizaciones monótonas, (ii) Dependen por esto únicamente de la trayectoria descrita por el estado físico en el espacio de rayos y no del Hamiltoniano que genera dicha trayectoria y (iii) son interpretables en términos puramente geométricos como la holonomía de una curva definida sobre un fibrado principal. 

Sin embargo, un estado puro en evolución unitaria es una idealización, y todo experimento e implementación real tiene que lidiar con la presencia de un entorno que interactúa con el sistema de estudio, lo que requiere una descripción en términos de estados mixtos y evoluciones no-unitarias. La definición de una fase geométrica que aplique en tal escenario es todavía un problema abierto. Esfuerzos destacables en esta dirección son aquellos consagrados a definir la fase geométrica acumulada por un estado mixto \cite{uhlmann1986parallel, uhlmann1991gauge, sjoqvist2000geometric, singh2003geometric}, para algunos de los cuales existen incluso reportes de detecciones experimentales \cite{du2003observation}. Otra ruta explorada considera el efecto del entorno como correcciones que permitan mantener las nociones de fase geométrica del caso unitario. Trabajos de este tipo introducen el efecto del entorno mediante un Hamiltoniano no-hermítico \cite{Carollo_original, Carollo_review}, o se concentran en el sistema particular de la sección \ref{sec:sec2_ejemplo}, estudiando las modificaciones a la fase de Berry por ruido clásico en el campo magnético \cite{de2003berry}, o por un entorno cuántico \cite{whitney2003berry,whitney2005geometric}, tanto desde un enfoque teórico, como experimental \cite{berger2013exploring, berger2015geometric}.

El marco en el cual una fase geométrica para sistemas cuánticos abiertos debe definirse es el siguiente: se supone que el efecto del entorno en el sistema de interés es tal que, bajo aproximaciones adecuadas, el sistema puede tratarse {\em efectivamente} como un sistema aislado que experimenta un tipo de evolución lineal no-unitaria

\begin{equation}
    \Sigma: \rho(0) \rightarrow \Sigma_t[\rho(0)]\equiv \rho(t),
\end{equation}
que da cuenta tanto de la dinámica interna del sistema como de su interacción con el entorno, y satisface una {\em ecuación maestra}.

Una consecuencia de este enfoque es que, en el caso general, un estado inicial puro evoluciona en un estado mixto $\rho(t)$. El operador densidad que representa el estado del sistema admite una descomposición $\{\ket{\psi_k(t)}, \omega_k(t)\}$ en estados puros $\ket{\psi_k(t)}$ pesados con probabilidades $\omega_k(t)$, que permite expresarla como

\begin{equation}
    \rho(t) = \sum_{k}\omega_k(t)\ket{\psi_k(t)}\bra{\psi_k(t)}.
    \label{eq:sec2_rhoDecomp}
\end{equation}
La asociación $\rho(t) \rightarrow \{\ket{\psi_k(t)}, \omega_k(t)\}$ entre el operador densidad y el {\em ensamble} de estados $\{\ket{\psi_k(t)}\}$, sin embargo, no es uno-a-uno sino uno-a-muchos, y existen en general diferentes ensambles, con diferentes estados unitarios y diferentes pesos, que pueden asociarse a una misma matriz densidad.

Una estrategia recurrente entre la literatura que aborda el problema de asociar una fase geométrica a un estado mixto $\rho(t)$ es descomponer formalmente la matriz densidad en una mezcla estadística como la de la ecuación (\ref{eq:sec2_rhoDecomp}) y aplicar la fase unitaria (\ref{eq:sec2_kinGP}) sobre cada elemento de la mezcla para asociar una fase geométrica también a $\rho(t)$. Esto fue propuesto, desde una descripción en términos de saltos cuánticos, en  \cite{Carollo_original,Carollo_review} y posteriormente discutido en \cite{Sjo_no, bassi2006_no, buri, sjoqvist2010_hidden}.
En una aproximación diferente al problema, en Tong et al. \cite{tong2004kinematic} se propone una definición de fase geométrica que se vale de una purificación auxiliar del estado como herramienta, pero resulta finalmente independiente de la misma. La siguiente sección desarrolla esta propuesta específica.

\subsection{Enfoque cinemático en sistemas abiertos}\label{sec:sec2_Tong}
Se concluye esta introducción teórica a las fases geométricas delineando la propuesta de Tong et al. \cite{tong2004kinematic} para la fase geométrica acumulada por un sistema abierto. Para esto, se considera un sistema cuántico y el espacio de Hilbert $\mathcal{H}$ de dimensión $N$ asociado al mismo. La evolución del estado puede describirse con la curva ${\rm C} \subset \mathcal{P}$

\begin{equation}
    {\rm C}:t\in[0,T]\rightarrow\rho(t) = \sum_{k=1}^N\omega_k(t)\ket{\psi_k(t)}\bra{\psi_k(t)},
    \label{eq:sec2_rhoDescompTong}
\end{equation}
dónde $\omega_k(t)\geq0$ y $\ket{\psi_k(t)}$ son los autovalores y autoestados, respectivamente, de la matriz densidad $\rho(t)$ del sistema. Todas las funciones no-nulas $\omega_k(t)$ se suponen no-degeneradas en el intervalo $[0, T]$ considerado, y se refiere al trabajo original \cite{tong2004kinematic} para la extensión al caso degenerado. 

Para introducir una noción de fase geométrica se comienza por expresar el estado mixto como la traza parcial de un estado (puro) de un sistema más grande, que se compone del sistema original y otro sistema auxiliar con espacio de Hilbert de igual dimensión. El estado mixto $\rho(t)$ se eleva entonces al {\em estado purificado}

\begin{equation}
    \ket{\Psi(t)} = \sum_{k=1}^N\,\sqrt{\omega_k(t)}\,\ket{\psi_k(t)}\otimes\ket{a_k},
\end{equation}
donde $\ket{\Psi(t)}\in\mathcal{H}\otimes\mathcal{H}_{\rm aux}$ es la purificación de $\rho(t)$, en el sentido de que la matriz densidad se recupera mediante la traza parcial del proyector $\ket{\Psi(t)}\bra{\Psi(t)}$ sobre los grados de libertad auxiliares.

La fase de Pancharatnam entre las purificaciones inicial y final puede escribirse como

\begin{equation}
    \phi_{\rm P} = \arg\left(\sum_{k=1}^N\,\sqrt{\omega_k(0)\omega_k(T)}\bra{\psi_k(0)}\ket{\psi_k(T)}\right).
    \label{eq:sec2_TongPanch}
\end{equation}
Para extraer de esta ecuación una fase geométrica asociada a la curva ${\rm C}$ es necesario eliminar la dependencia en la purificación específica utilizada. Con vistas a este objetivo se observa que, como para cada instante $t\in[0,T]$ $\{\ket{\psi_k(t)}\}$ y $\{\ket{\psi_k(0)}\}$ son bases ortonormales del mismo espacio de Hilbert, existe entonces una transformación unitaria $U(t)$ que lleva de un conjunto al otro $\ket{\psi_k(t)} = U(t)\ket{\psi_k(0)} \forall\, k$. El paso esencial para arribar a una fase puramente geométrica es el de notar que, en realidad, existe una clase de equivalencia de mapas unitarios $\Tilde{U}(t)$ que realizan todos la curva ${\rm C}$. Más en detalle, la expresión (\ref{eq:sec2_rhoDescompTong}) que define la curva ${\rm C}$ es manifiestamente invariante de gauge $U(1)$ de forma que dos transformaciones unitarias $U(t)$ y $U'(t)$ que mapeen $\{\ket{\psi_k(0)}\}$ en $\{\ket{\psi_k(t)}\}$ y en $\{e^{i\alpha_k(t)}\ket{\psi_k(t)}\}$ resultan equivalentes.
Los mapas $U(t)$ que en la clase de equivalencia tienen la forma

\begin{equation}
    \Tilde{U}(t) = U(t)\sum_{k=1}^N\,e^{i\,\alpha_k(t)}\ket{\psi_k(0)}\bra{\psi_k(0)}.
    \label{eq:sec2_equivUni}
\end{equation}

En particular, puede identificarse el mapa $U^\parallel(t)$ que satisface simultáneamente la condición de transporte paralelo para cada $\ket{\psi_k(t)}$, es decir

\begin{equation}
    U^\parallel(t) = U(t)\;|\;\bra{\psi_k(0)}U(t)^\dagger\dot{U}(t)\ket{\psi_k(0)} = 0 \;\;\forall\,k
\end{equation}
y definir la fase geométrica como la diferencia de fase (\ref{eq:sec2_TongPanch}) para este mapa particular. Sustituyendo $U^\parallel(t)=\Tilde{U}(t)$ en la ecuación (\ref{eq:sec2_equivUni}) que describe la relación de equivalencia entre operadores, se encuentra 

\begin{equation}
    \alpha_k(t) = i\,\int_0^t\,dt'\,\bra{\psi_k(0)}U^\dagger(t')\dot{U}(t)\ket{\psi_k(0)},
\end{equation}
y, en consecuencia, la fase geométrica resulta

\begin{equation}
    \phi_g[{\rm C}] = \arg\left(\sum_{k=1}^N\,\sqrt{\omega_k(0)\omega_k(T)}\bra{\psi_k(0)}\ket{\psi_k(T)}\,e^{-\int_0^T\,dt\,\bra{\psi_k(t)}\Dot{\psi}_k(t)\rangle}\right).
    \label{eq:sec2_TongGP}
\end{equation}

La definición arriba propuesta satisface las condiciones que rigen sobre una noción de fase geométrica razonable para un estado mixto. A saber: (i) Es de hecho una fase, dado que su definición a través de la función argumento impone $2\pi$-periodicidad, (ii) es manifiestamente invariante de gauge ya que toma el mismo valor para cualquier operador unitario $U(t)$ en la clase de equivalencia descrita por la ecuación (\ref{eq:sec2_equivUni}), y por lo tanto depende únicamente del camino ${\rm C}$ trazado por el operador $\rho(t)$. Esto puede verificarse inmediatamente sustituyendo la expresión (\ref{eq:sec2_equivUni}) en la ecuación (\ref{eq:sec2_TongGP}), para recobrar que 
\begin{equation}
    \phi_g[{\rm C}]|_{U(t)} = \phi_g[{\rm C}]|_{\Bar{U}(t)}
\end{equation}
La diferencia de fase $\phi_P$, que no es invariante de gauge, coincide con $\phi_g[{\rm C}]$ solamente cuando se toma $U(t) = U^\parallel(t)$ en (\ref{eq:sec2_TongPanch}). (iii) Cuando la evolución es unitaria, se recuperan los resultados esperados \cite{mukunda1993quantum} para estados iniciales puros, y \cite{sjoqvist2000geometric, singh2003geometric} para estados iniciales mixtos y, (iv) es accesible experimentalmente, por ejemplo mediante interferometría.

Será de utilidad para su aplicación en los próximos capítulos especificar (\ref{eq:sec2_TongGP}) para el caso particular en que el sistema se encuentre inicialmente en un estado $\ket{\psi(0)}$ puro. En tal situación, la matriz densidad inicial $\rho(0)$ es un proyector y los autovalores $\omega_k(0)$ en la descomposición espectral (\ref{eq:sec2_rhoDescompTong}) son todos nulos, excepto aquél correspondiente al estado inicial que vale $\omega_+(0)=1$. En consecuencia, la sumatoria de la ecuación (\ref{eq:sec2_TongGP}) posee un único término no-nulo y la fórmula se reduce a la expresión 

\begin{equation}
    \phi_g[C] = \arg\{\bra{\psi(0)}\ket{\psi_+(T)}\}-\Im\int_0^T\,dt\,\bra{\psi_+(t)}\Dot{\psi}_+(t)\rangle.
    \label{eq:sec2_TongGPpuros}
\end{equation}

Notablemente, la expresión hallada bajo estas condiciones se asemeja a la ecuación (\ref{sec:sec2_Mukunda}). Sin embargo, aquella fórmula rige sobre el estado $\ket{\psi(t)}$ del sistema, mientras que en ésta, el cálculo se realiza considerando el autoestado $\ket{\psi_+(t)}$ de $\rho(t)$ que coincida a $t=0$ con el estado inicial, es decir, $\ket{\psi_+(t)} = \ket{\psi_k(t)}|\ket{\psi_k(0)} = \ket{\psi(0)}$. En consecuencia, la fase en (\ref{eq:sec2_TongGPpuros}) admite la interpretación de la fase geométrica unitaria acumulada por el autoestado $\ket{\psi_+(t)}$. 

\vspace{.5cm}
\begin{center}
   \textcolor{bordo}{\ding{163}}
\end{center}
\vspace{.5cm}
En síntesis, en este capítulo se han presentado distintas definiciones para la fase geométrica acumulada por el estado de un sistema en evolución. Las definiciones tratadas difieren en las hipótesis bajo las cuales son válidas. En todos los casos, las definiciones más generales se reducen exactamente a aquellas menos generales si las hipótesis correspondientes son satisfechas.
\clearpage
Para facilitar la comparación, se condensa la información sobre las distintas fases presentadas en un cuadro

\begin{center}
\vspace{0.5cm}
    \begin{tabular}{|>{\centering\arraybackslash}p{.075\linewidth}|>
    {\arraybackslash}p{.42\linewidth}|>
    {\centering\arraybackslash}p{.1\linewidth}|>{\centering\arraybackslash}p{.075\linewidth}|}
    \hline
    Fase &  {\hspace{.75cm} Condiciones sobre la dinámica} & Ecuación & Sección\\
    \hline
    $\phi_{\mathrm{a}}$& Unitariedad, adiabaticidad, ciclicidad (de la dependencia temporal explícita)
    & (\ref{eq:sec2_Berry}) & \ref{sec:sec2_Berry}\\
    $\phi_{A-A}$& Unitariedad, ciclicidad (condición sobre la curva $C$ de estados)& (\ref{eq:sec2_AA}) & \ref{sec:sec2_AA}\\
    $\phi_{S-B}$ & Unitariedad (generalizada a estados puros de norma arbitraria) & (\ref{eq:sec2_faseSamuel1}) & \ref{sec:sec2_Samuel}\\
    $\phi_{u}$ & Unitariedad & (\ref{eq:sec2_kinGP}) & \ref{sec:sec2_Mukunda}\\
    $\phi_g$ & -
    & (\ref{eq:sec2_TongGP}) & \ref{sec:sec2_Tong}\\  
    \hline
\end{tabular}
\vspace{0.5cm}
\end{center}

\vspace{1cm}
Los capítulos \ref{ch:3} a \ref{ch:5} de esta tesis analizan el objeto definido en la ecuación (\ref{eq:sec2_TongGPpuros}) para el caso de diversos sistemas cuánticos en interacción con un entorno, estudiando su relación con otros efectos y características de la evolución, y articulando con diversas potenciales aplicaciones. El objetivo de caracterizar la potencia de la fase geométrica (\ref{eq:sec2_TongGP}) para dar cuenta o resistir la influencia del entorno requerirá de la comparación recurrente con la evolución del sistema aislado. Dado que limitaremos el estudio a evoluciones que parten de estados puros esta comparación será, en el caso general, con la expresión para la fase geométrica de la ecuación (\ref{eq:sec2_kinGP}).

\chapter{Fase geométrica en electrodinámica de cavidades}\label{ch:3}
Este capítulo trata con el escenario particular de la electrodinámica cuántica en cavidades. El {\em modelo de Rabi} es un modelo paradigmático aplicado en múltiples áreas de investigación que van desde la óptica cuántica a la física de la materia condensada. Representa la interacción dipolar entre un átomo con dos niveles de energía bien aislados de los demás y el campo electromagnético presente en una cavidad. Tras realizar sobre el modelo de Rabi la aproximación {\em de onda rotante} se obtiene el {\em modelo de Jaynes-Cummings} (JC), introducido en 1963 con el objetivo de examinar aspectos de la emisión espontánea y revelar la existencia de oscilaciones de Rabi en las probabilidades de excitación atómicas para campos con número de fotones bien definido. Éste puede considerarse el modelo más simple que tiene éxito en describir la interacción materia-radiación y, a pesar de ser extremadamente simple y analíticamente resoluble, logra explicar una gran cantidad de los experimentos conducidos hasta la fecha en el campo de la electrodinámica cuántica y, más recientemente, en arquitecturas de circuitos superconductores.
La literatura que trata el tema de las fases geométricas en los modelos de Jaynes-Cummings y de Rabi cuenta con un considerable número de trabajos, tanto teóricos \cite{fuentes2002vacuum, carollo2003berry, liu2010vacuum, wang2015does} como experimentales \cite{gasparinetti2016_cqed_observation}. La mayor parte de los esfuerzos en esta dirección se realizaron en el contexto de evolución adiabática, calculando la fase de Berry (\ref{eq:sec2_Berry}) que acumulan los autoestados instantáneos de una partícula de espín-$1/2$ que interactúa con un campo magnético externo clásico cuya dirección describe un ciclo variando lentamente. Este límite semiclásico y adiabático corresponde efectivamente al ejemplo de la sección \ref{sec:sec2_ejemploBerry}. En \cite{fuentes2002vacuum}, se generaliza el cómputo al modelo equivalente completamente cuántico, reemplazando el campo rotante clásico por un único modo de un campo cuántico, aunque sin abandonar las condiciones de evolución cíclica y adiabática. 

Siguiendo~\cite{viotti2022geometric}, en este capítulo se estudia en detalle la fase geométrica acumulada en un modelo de JC disipativo, como caso paradigmático dentro del campo de la electrodinámica en cavidades. Se considera que los principales mecanismos por los cuales el sistema 'átomo + modo' interactúa con el entorno son el flujo de fotones a través de las paredes de la cavidad y el continuo e incoherente bombeo del sistema de dos niveles, lo que conforma un escenario frecuente en electrodinámica de cavidades semiconductoras \cite{khitrova2006vacuum, laussy2009luminescence, del2009luminescence}. Para poder establecer comparaciones firmes, se comienza analizando el caso en que el sistema está aislado, para luego expandir el estudio al sistema abierto.

\section{Modelo de Jaynes-Cummings y aproximación de onda rotante}
En 1936 Isidor Rabi investigó el modelo semiclásico que  describe la interacción más simple entre un átomo y un único modo del campo electromagnético contenido en una cavidad~\cite{rabi1936process, rabi1937space}. En su versión completamente cuántica, el Hamiltoniano correspondiente ($\hbar = 1$) es

\begin{equation}
    H_R = \varepsilon\,a^\dagger\,a + \frac{\omega}{2}\sigma_z + g(a^\dagger + a)\,\sigma_x,
    \label{eq:sec3_HRabi}
\end{equation}
donde $\varepsilon$ y $\omega$ son las frecuencias naturales del modo del campo electromagnético y del átomo respectivamente, y $g$ es la constante de acoplamiento efectiva entre el átomo y el campo, considerada real. Los operadores $\sigma_{x,y,z}$ son las matrices de Pauli usuales, que actúan sobre los estados fundamental $\ket{-}$ y excitado $\ket{+}$ del átomo, mientras que $a^\pm$ son los operadores de creación y destrucción de fotones y actúan sobre el campo de la cavidad.

El Hamiltoniano de la ecuación (\ref{eq:sec3_HRabi}) exhibe, en la representación de interacción, dos términos {\em rotantes} que oscilan con frecuencia $\varepsilon - \omega$ y dos términos {\em contrarrotantes} que oscilan con frecuencia $\varepsilon+\omega$. Cuando las frecuencias naturales del sistema de dos niveles y el modo del campo son similares $\varepsilon \sim\omega$ los términos contrarrotantes oscilan mucho más rápidamente que los términos rotantes, de modo tal que en cualquier escala temporal suficientemente prolongada el efecto de los términos rotantes se vuelve dominante en promedio frente al de los términos contrarrotantes. Despreciar los términos contrarrotantes es lo que se conoce como {\em aproximación de onda rotante}. Aplicando esta aproximación al modelo de Rabi, justificada cuando $\varepsilon\sim\omega$ y $g\ll\omega,\varepsilon$, se obtiene el Hamiltoniano de Jaynes-Cummings \cite{jaynes1963comparison}

\begin{equation}
    H_{JC} = \varepsilon \,a^\dagger\,a + \frac{\omega}{2}\,\sigma_z + g(a^\dagger\,\sigma_- + a\,\sigma_+),
    \label{eq:sec3_HJC}
\end{equation}
donde $\sigma_{\pm} = (\sigma_x \pm \sigma_y)/2$ son los operadores de subida y bajada atómicos. Es usual en la literatura aplicar la transformación unitaria $K = \exp\{-i\,\omega\,t(a^\dagger\,a + \sigma_z/2)\}$ sobre el Hamiltoniano $H_{JC}$, que queda de esta forma escrito como

\begin{equation}
    H = \frac{\Delta}{2}\sigma_z + g(a^\dagger\,\sigma_- + a\,\sigma_+),
    \label{eq:sec3_HJC2}
\end{equation}
en términos del {\em detuning} $\Delta = \varepsilon - \omega.$ Una base natural para resolver el problema es la denominada base de estados desnudos $\{\ket{-,n_c},\ket{+,n_c}\,;\, n_c \in \mathbb{N}_0\}$. En el régimen considerado $\varepsilon\sim\omega$, los estados de esta base se acomodan en dobletes estrechamente agrupados en los que la diferencia de energía entre los estados de un mismo doblete es $\Delta$ mientras que una diferencia de energía $\omega \gg \Delta$ los separa de los estados pertenecientes al doblete contiguo. Además, el Hamiltoniano de JC no acopla distintos dobletes, lo que permite resolver la dinámica en el subespacio $\{\ket{+, n_c}\ket{-,n_c+1}\}$ (el acoplamiento entre dobletes introducido por el término de interacción en el modelo de Rabi puede despreciarse con un error de orden $g/\omega$ \cite{grynberg2010introduction}). En esta base restringida los autoestados del Hamiltoniano de la ecuación (\ref{eq:sec3_HJC2}) son
\begin{align}
    \nonumber
    &\ket{\psi^n_+} =  \cos\frac{\theta_n}{2}\ket{+,n_C} + \sin\frac{\theta_n}{2}\ket{-,n_c +1}\\
    &\ket{\psi^n_-} =  \sin\frac{\theta_n}{2}\ket{+,n_c} + \cos\frac{\theta_n}{2}\ket{-,n_c +1}
\end{align}
con autoenergías $E_\pm^n = \pm \,\Omega_n/2$, donde $\Omega_n = (4g^2(n+1)+\Delta^2)^{1/2}$ es la frecuencia de las oscilaciones de Rabi del sistema átomo-modo y con $\cos\theta_n = \Delta/\Omega_n$ modulando la superposición de estados desnudos. Como caso particular, cabe destacar que cuando la condición $\Delta = 0$ de resonancia se satisface, las autoenergías se reducen a $E_\pm^{n, \Delta = 0} = \pm \,g\sqrt{n+1}$ y los autoestados son estados de Bell 

\begin{equation}
    \ket{\psi_\pm^n} = \frac{1}{\sqrt{2}}(\ket{+,n_c} \pm \ket{-,n_c+1}).
\end{equation}

\subsection{Fase geométrica en el modelo de Jaynes-Cummings}\label{sec:sec3_Faseunitaria}
{\em Fase de Berry - } En el innovador trabajo de Fuentes-Guridi et al. \cite{fuentes2002vacuum} se calcula la fase geométrica adiabática para el modelo de JC, posteriormente discutida en \cite{wang2015does} y experimentalmente detectada en \cite{gasparinetti2016_cqed_observation}. En estos trabajos, el sistema se prepara inicialmente en un autoestado del Hamiltoniano $H$, que es posteriormente conducido mediante una transformación de fase adiabática y en completo aislamiento.

Aplicar una transformación (unitaria) de corrimiento de fase $R = \exp\{-i\,\Omega\,a^\dagger\,a\}$ al Hamiltoniano $H$ da origen a un nuevo Hamiltoniano 

\begin{equation*}
    H(\Omega) = \Delta/2\,\sigma_z + g\,(a^\dagger\,\sigma_- \,e^{-i\,\Omega}+a\,\sigma_+ \,e^{i\,\Omega})
\end{equation*} 
que depende del parámetro externo de control $\Omega$. Los autoestados de $H(\Omega)$ se obtienen a partir de los autoestados de $H$ aplicando la misma transformación. Si el parámetro de control $\Omega$ se varía lentamente desde cero hasta $2\pi$, se satisfacen las condiciones impuestas por Berry y se encuentra una fase geométrica

\begin{equation}
    \phi_{\rm a}^n = i \,\oint_{\rm C}\, d\Omega\,\bra{\psi_\pm^n}R(\Omega)^\dagger\frac{d}{d\Omega}R(\Omega)\ket{\psi_\pm^n}]
    = \pi\left(1\mp\cos\theta_n\right),
    \label{eq:sec3_GPadiabatica}
\end{equation}
manifiestamente no-trivial incluso para el caso $n=0$, demostrando que incluso el estado de vacío del campo introduce una corrección en la fase de Berry.
\\
\\\indent
{\em Aproximación cinemática - } Con el objetivo futuro de comparar las dinámicas unitaria y disipativa mediante la inspección de la fase geométrica acumulada en cada caso, cambiamos la perspectiva y abordamos el estudio desde el enfoque cinemático, que permite la considerar escenarios más generales. En particular, nos interesa estudiar las fases acumuladas durante la evolución temporal generada por el Hamiltoniano $H$ de la ecuación (\ref{eq:sec3_HJC2}) sin necesidad de introducir un control externo.

\begin{figure}[ht!]
    \center
    \includegraphics[width = .45\linewidth]{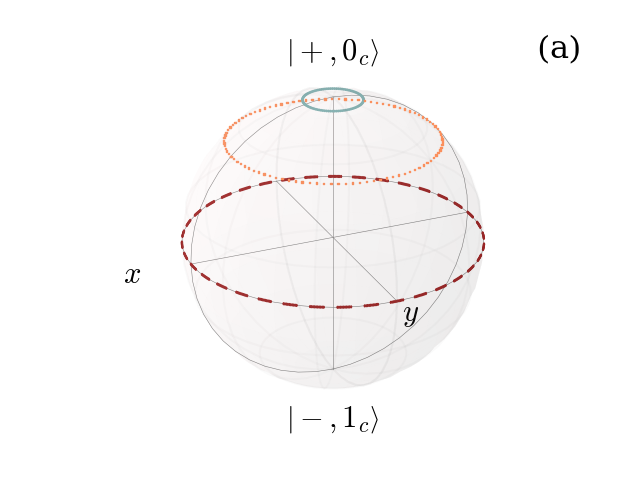}
    \includegraphics[width = .45\linewidth]{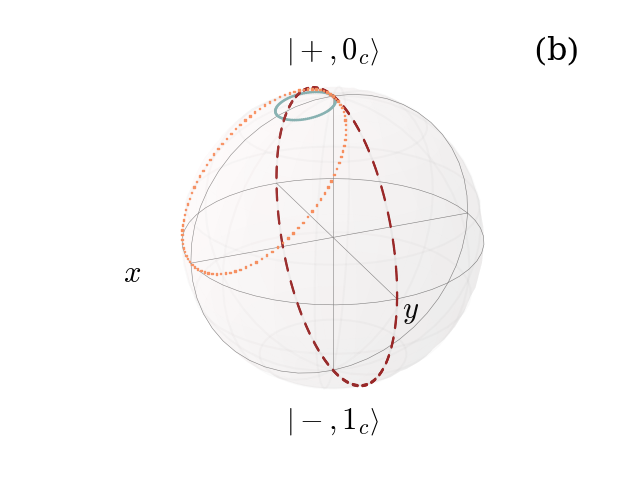}
    \caption{Trayectorias descritas por el estado físico del sistema en la esfera de Bloch (a) cuando se conduce un autoestado del Hamiltoniano inicial en un ciclo adiabático mediante una transformación tipo corrimiento de fase y, (b) para el caso de un estado inicial $\ket{+, 0_c}$ cuya evolución está generada por el Hamiltoniano $H$. Diferentes curvas corresponden a diferentes valores de la relación $\Delta/g$ entre el detuning y el acoplamiento átomo-cavidad según el siguiente código: Las líneas sólidas celestes corresponden al valor $\Delta = 10\,g$, las líneas de puntos anaranjadas describen el caso $\Delta = 2\,g$ y las líneas de trazos bordó señalan el caso resonante. Para cada valor fijo de la relación $\Delta/g$, las trayectorias en una y otra esfera están relacionadas mediante una rotación rígida seguida de una inversión en la parametrización.} 
    \label{fig:sec3_BlochUni}
\end{figure}

Retomando la discusión de la sección    \ref{sec:sec2_Mukunda}, para un estado inicial $\ket{\psi(0)} = \ket{\psi_\pm^n}$ autoestado del Hamiltoniano, la evolución generada por $H$ introduce en el estado únicamente un factor de fase, que es además puramente dinámico $e^{-i\,E_+^n\,t}$. El estado físico del sistema, entonces, permanece estático en el mismo punto del espacio de rayos acumulando fase geométrica nula.
Si por el contrario se considera un estado inicial
$\ket{\psi(0)} = \ket{+,n}$
el estado del sistema a tiempo $t$ resulta

\begin{equation}
    \ket{\psi(t)} = (\cos^2\theta_n\,e^{-i\,E_+^n\,t} + \sin^2\theta_n\,e^{i\,E_+^n\,t})\ket{+,n_c} -i\,\sin\theta_n\sin(E_+^n\,t)\ket{-,n_c+1}.
    \label{eq:sec3_estadoUni}
\end{equation}
La proyección de este estado en el espacio de rayos describe una curva $t'\in[0,t]\rightarrow{\rm C}$ no trivial, que tiene asociada mediante la ecuación (\ref{eq:sec2_kinGP}) una fase geométrica 

\begin{equation}
    \phi_u[{\rm C}] = -\pi(1-\cos\theta_n)\frac{t}{T} + \arg\left\{1+\,e^{2i\,\pi\,\frac{t}{T}}\,\frac{\Omega_n - \Delta}{\Omega_n + \Delta}\right\}.
    \label{eq:sec3_GPunitaria}
\end{equation}
Aquí $T = 2\pi/\Omega_n$ corresponde al período definido por la frecuencia de Rabi. Vale la pena, antes de adentrarse en el caso del sistema con disipación, detenerse en el caso $t=T$ para el cual la ecuación (\ref{eq:sec3_GPunitaria}) se reduce a $\phi_u=-\pi(1-\cos(\theta_n))$. La coincidencia a menos de un signo entre este resultado y aquél de la ecuación (\ref{eq:sec3_GPadiabatica}) se puede explicar comparando las curvas descritas en la esfera de Bloch (es decir, en el subespacio de rayos efectivo) por cada evolución, exhibidas en la figura \ref{fig:sec3_BlochUni}.

La figura \ref{fig:sec3_BlochUni}.a muestra el camino trazado por el estado del sistema cuando el mismo se prepara en un autoestado $\ket{\phi_\pm^n}$ del Hamiltoniano y luego se conduce en un ciclo adiabático bajo la acción de un operador de corrimiento de fase $R(\Omega)$. Es un resultado conocido que en una evolución de este tipo el estado describe círculos de latitud fija, correspondiendo el ecuador al caso $\Delta=0$, o resonante. Por otra parte, la figura \ref{fig:sec3_BlochUni}.b muestra las curvas descritas por el estado en el segundo escenario contemplado, en el que el sistema es preparado en un estado inicial $\ket{+,n}$ y evoluciona por la acción de $H$ durante un período $t'\in[0,T]$. Bajo estas circunstancias el estado describe trayectorias circulares que contienen el polo norte, punto que representa el proyector en el estado $\ket{+, n}$. Para cada valor de la relación $\Delta/g$ entre el detuning y el acoplamiento átomo-cavidad, las curvas descritas en uno y otro caso están relacionadas por una transformación compuesta por una rotación rígida y una reflexión, o equivalentemente, de una rotación rígida y una inversión de la parametrización. Las rotaciones rígidas son isometrías de la esfera de Bloch y se pueden realizar con un operador unitario que actúe sobre el espacio de Hilbert. Si la rotación aplica a toda la curva ${\rm C}$, se sigue que puede describirse con una transformación estática que deja la fase geométrica invariante (ver sección \ref{sec:sec2_MukundaUnitarias}). Por su parte, es inmediato ver en la ecuación (\ref{eq:sec3_GPunitaria}) que una inversión de la parametrización introduce el cambio de signo observado.

\section{Modelo de Jaynes-Cummings disipativo}
Habiendo desarrollado el análisis de la fase geométrica acumulada por el sistema átomo-cavidad en la situación ideal de completo aislamiento, se aborda ahora el estudio para el escenario más realista en el que el mismo sistema se encuentra en interacción con un entorno. El problema se trata para la implementación específica en estructuras semiconductoras, en las que un {\em punto cuántico} (al cuál se sigue, sin embargo, refiriendo como átomo o sistema de dos niveles) se ubica en una nano o micro-cavidad. En este escenario, la fuente principal de disipación es la pérdida de fotones a través de las paredes de la cavidad. Una segunda fuente de disipación y decoherencia es el bombeo incoherente del sistema de dos niveles el cual se asocia, en una descripción microscópica, con la relajación de los pares electrón-hueco en el punto y esta, a su vez, con el bombeo eléctrico externo con el que se excita el sistema \cite{khitrova2006vacuum, laussy2009luminescence, del2009luminescence}. 
A través de una ecuación maestra (\ref{eq:Lindblad}) fenomenológica

\begin{equation}
    \Dot{\rho}(t) = -i\,[H\,,\,\rho(t)] +  \frac{1}{2} \sum_{\alpha} [2\,L_{\alpha}\rho(t) L_{\alpha}^\dagger -  \{L_{\alpha}^\dagger L_{\alpha}, \rho(t)\, \} ] \;,
    \label{eq:sec3_JCEqM}
\end{equation}
se introducen los dos mecanismos mencionados, despreciando otros procesos con menor influencia en la dinámica como el desfasaje puro o el bombeo de fotones en la cavidad, considerando además que el entorno se halla a temperatura cero.
Los operadores de Lindbland
\begin{align}
    L_\gamma =& \sqrt{\gamma}\,a\\
    L_p =& \sqrt{p}\,\sigma_+
\end{align}
representan la pérdida de fotones y el bombeo continuo e incoherente del átomo respectivamente, con los parámetros $\gamma$ y $p$ denominados tasa de pérdida de fotones y la amplitud del bombeo. 

Mientras que el bombeo sobre el átomo es siempre secundario frente a la pérdida de fotones, lo que se traduce en la relación $p/g,\, p/\gamma\ll 1$, la relación entre los parámetros $\gamma$ y $g$ da origen a dos regímenes que se diferencian con claridad \cite{carmichael1989subnatural, yamamoto2003semiconductor, laussy2008strong, vera2009characterization, lodahl2015interfacing}. El régimen de acoplamiento fuerte (SC por sus siglas en inglés) está caracterizado por una constante de interacción $g$ mayor que la tasa de disipación $\gamma$ ($\gamma/g <1 $). Por el contrario, el régimen de acoplamiento débil (WC) refiere a la situación opuesta, en la que la constante de interacción es menor a la tasa de disipación ($\gamma/g >1$). Cabe señalar que cuando se trata con la teoría unitaria de un sistema bipartito, la definición más natural de acoplamiento fuerte y débil refiere a la relación entre la constante caracterizando el acoplamiento entre las partes del sistema ($g$ en este caso) y las cantidades que caracterizan la dinámica interna de cada parte del sistema. En el presente caso, sin embargo, se está considerando un sistema bipartito expuesto además a efectos disipativos que se originan en el acoplamiento de cada subsistema con el entorno y los nombres utilizados para los dos regímenes, que podrían resultar poco intuitivos, son sin embargo recurrentes en la literatura sobre el tema.

Para la descripción de cualquiera de los dos regímenes se asume que el átomo puede estar en el estado fundamental o excitado $\ket{\pm}$, y que el campo fotónico en la cavidad puede tener cero $\ket{0}$ o un fotón $\ket{1}$. En consecuencia, se restringe el estudio a un subespacio de estados, truncando la base de estados desnudos a $\ket{0} = \ket{-,0_c}$, $\ket{1} = \ket{+,0_c}$ y $\ket{2} = \ket{-,1_c}$.
Desarrollando explícitamente el sistema de ecuaciones diferenciales para los elementos de matriz $\rho_{ij}(t)$ condensado en la ecuación (\ref{eq:sec3_JCEqM}), se encuentra que los elementos $\rho_{0i}$ satisfacen
\begin{align}
    \dot{\rho}_{01} &=   - \frac{p}{2} \rho_{01} + i\, \Delta \rho_{01} + i\, g \rho_{02} \nonumber\\
    \dot{\rho}_{02} &=   - \frac{p}{2} \rho_{02} - \gamma\, \rho_{02} + i\, g \rho_{01},
\end{align}
donde es evidente que la dinámica de estos elementos se encuentra desacoplada de los demás, de forma que cualquier condición inicial que implique $\rho_{0i}(0)=0\;;\; i = 1,2$ conducirá a una solución para $\rho(t)$ donde estos elementos se mantienen nulos para todo instante posterior. En consecuencia, la solución para $\rho(t)$ mostrará una estructura diagonal por bloques con un primer bloque $\rho_{00}(t)$ de dimensión $1\times 1$ y un segundo bloque de dimensión $2\times2$ correspondiente al subespacio $\{\ket{1}=\ket{+,0_c}, \ket{2}= \ket{-, 1_c}\}$.

\begin{figure}[ht!]
    \center
    \includegraphics[width = .495\linewidth]{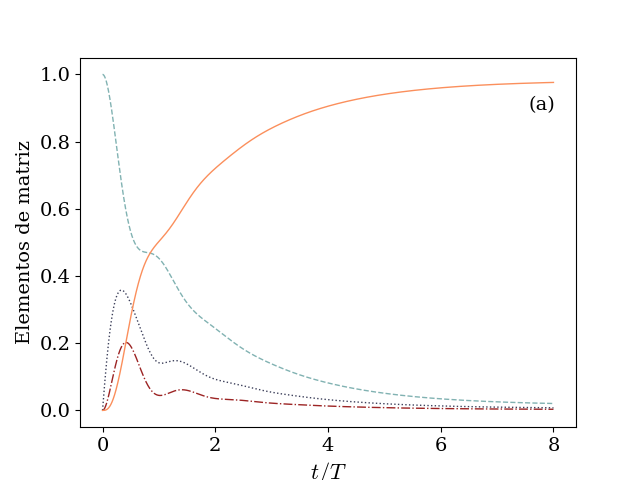}
    \includegraphics[width = .495\linewidth]{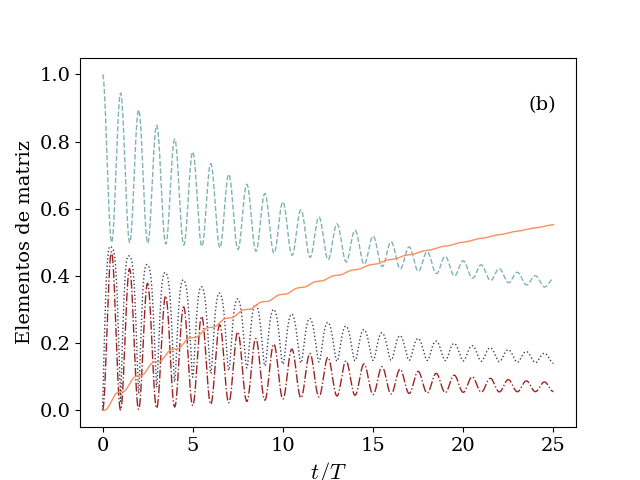}
    \caption{Evolución dinámica de los elementos de matriz $\rho_{ij}$ del operador densidad $\rho(t)$ para un sistema que exhibe una relación entre el detuning y el acoplamiento átomo-modo $\Delta/g= 2.0$. El entorno se encuentra caracterizado por una tasa de bombeo incoherente $p = 0.005\,g$ y una tasa de pérdida de fotones (a) $\gamma= 2.0\,g$ y (b) $\gamma = 0.1\,g$ correspondientes a los regímenes de acoplamiento débil (WC) y fuerte (SC) respectivamente. Las líneas naranja sólida, celeste de trazos y bordó de trazo/punto representan la evolución de las poblaciones $\rho_{00}$, $\rho_{11}$ y $\rho_{22}$ respectivamente, mientras que la línea azul de puntos señala el módulo de las coherencias $|\rho_{12}|$.} 
    \label{fig:sec3_elementos}
\end{figure}

Por analogía con el caso unitario, estudiaremos la evolución del sistema tomando un estado inicial $\rho(0) = \ket{+, 0_c}\bra{+, 0_c}$, que satisface la condición $\rho_{0i}(0)=0\;;\; i = 1,2$ de manera que se espera que el estado $\rho(t)$ a tiempo $t$ exhiba la estructura descrita arriba. Las ecuaciones diferenciales de interés se reducen entonces al sistema 
\begin{align}
    \dot{\rho}_{00} &= - p \rho_{00} + \gamma \rho_{22} \nonumber \\
    \dot{\rho}_{11} &= - i g \left(\rho_{21} - \rho_{12}\right) + p \rho_{00} \nonumber \\
    \dot{\rho}_{22} &=   - i g \left(\rho_{12} - \rho_{21}\right)  - \gamma  \rho_{22} \nonumber \\
    \dot{\rho}_{12} &=   - i g \left(\rho_{22} - \rho_{11}\right) - i \Delta \rho_{12} - \frac{\gamma}{2} \rho_{12}
    \label{eq:sec3_diffEq}
\end{align}
y se resuelven numéricamente para acceder al estado $\rho(t)$ a tiempo $t>0$ arbitrario. La figura \ref{fig:sec3_elementos} muestra el comportamiento de los elementos $\rho_{ij}(t)$ en los regímenes de WC y SC definidos anteriormente. 

En el panel (a) de la figura \ref{fig:sec3_elementos}, donde la relación entre parámetros $\gamma/g$ corresponde al WC en que el acoplamiento con el entorno resulta más fuerte que la constante responsable de las transiciones entre estados, el sistema pierde coherencia en una escala temporal de pocos períodos de Rabi y decae a un estado estacionario $\rho \sim \ket{0}\bra{0}$ vecino al estado de mínima excitación. Por otra parte, el panel (b) de la figura \ref{fig:sec3_elementos} muestra la dinámica de los elementos para una relación $\gamma/g$ que pertenece al SC. En este caso, el estado preserva la coherencia cuántica durante un número considerablemente mayor de pseudo-ciclos $T=2\pi/\Omega_0$ y evoluciona en dirección a un estado asintótico mixto. Las características de la dinámica en cada régimen repercuten en el estudio de la fase geométrica, haciendo del régimen de SC el único escenario conveniente. Se retoma el tema con el análisis que fundamenta esta afirmación en lo que sigue.

\subsection{Caso no unitario: Correcciones de vacío a la fase geométrica}

En esta sección, se muestra cómo la fase geométrica adquirida, calculada siguiendo la definición (\ref{eq:sec2_TongGP}), se corrige respecto del valor unitario cuando el sistema bipartito átomo-modo evoluciona en contacto con el entorno. Recordamos que esta fase geométrica, aplicable al escenario de estados mixtos que evolucionan de forma no-unitaria, se reduce a la ecuación (\ref{eq:sec2_TongGPpuros}) cuando el estado inicial del sistema es, como en este caso, un estado puro. Esta expresión es formalmente idéntica a la fase geométrica para un estado puro en evolución unitaria de la ecuación (\ref{eq:sec2_kinGP}), con la salvedad de que el cálculo se realiza sobre el autoestado $\ket{\psi_+(t)}$ del operador densidad que coincide con el estado del sistema a tiempo $t=0$ en lugar de hacerlo sobre el estado del sistema (que no es un estado puro para $t>0$).

\begin{SCfigure}[][ht!]
    \includegraphics[width = .5\linewidth]{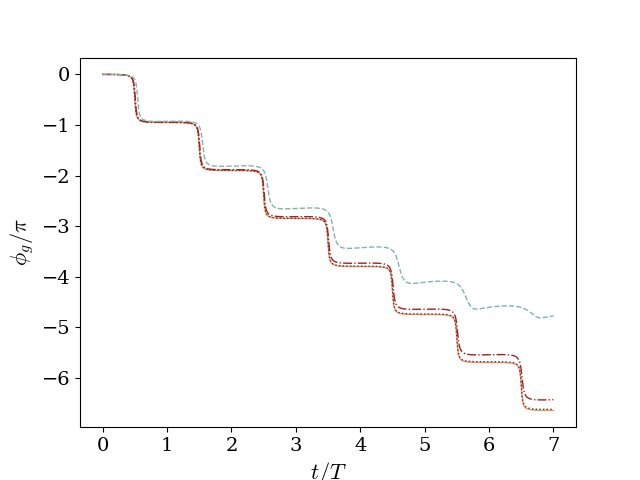}
    \caption{Fase geométrica $\phi_g$ acumulada por un sistema con relación $\Delta = 0.1\,g$ entre el detuning y el acoplamiento átomo-modo. El entorno se encuentra caracterizado por una tasa de bombeo $p = 0.005\,g$ y se muestran 3 casos en que la tasa de pérdida de fotones toma valores  $\gamma = 0.01 \,g$, $\gamma = 0.1 \,g$, y $\gamma = 0.5 \,g$ descritos por la línea azul punteada, bordó de trazo/punto y celeste de trazos respectivamente. Para permitir la comparación se incluye además el caso de evolución unitaria (línea naranja sólida).} 
    \label{fig:sec3_phigamma}
\end{SCfigure}

Los autovalores y autovectores del operador densidad pueden escribirse formalmente en términos de sus elementos de matriz. Para el caso considerado, en el que $\rho(t)$ toma la estructura diagonal en bloques descrita anteriormente, el autoestado de interés tiene la forma

\begin{equation}
    \ket{\psi_+(t)} = \frac{-(\rho_{22} - \epsilon_+)\ket{+,0_c} + \rho_{21}\ket{-,1_c}}{\left((\rho_{22}-\epsilon_+)^2 + \rho_{21}\rho_{12}\right)^{1/2}},
\end{equation}
con $\epsilon_{+} = \frac{1}{2}\left(\rho_{11} + \rho_{22}+((\rho_{11}-\rho_{22})^2 + 4\rho_{21}\rho_{12})^{1/2}\right)$ el autovalor asociado. Recurriendo a este resultado y a la ecuación (\ref{eq:sec2_TongGPpuros}) se encuentra también para la fase una expresión formal en términos de los elementos $\rho_{ij}(t)$

\begin{equation}
    \phi_g(t)=\int_0^t \,dt'\,\frac{\Im \Dot{\rho}_{21}\rho_{12}}{(\rho_{22}-\epsilon_+)^2 + \rho_{21}\rho_{12}}.
    \label{eq:sec3_faseFormal}
\end{equation}
En general, esta fase diferirá de aquella acumulada en una evolución unitaria de forma que puede escribirse, sin pérdida de generalidad, $\phi_g = \phi_u + \delta\phi$, con $\delta\phi$ la diferencia entre la fase unitaria y aquella modificada por la presencia del entorno. Caracterizar la corrección $\delta\phi$ permite relacionar este objeto, perteneciente a la geometría misma del espacio de Hilbert, con los efectos de disipación y decoherencia experimentados por el sistema, así como determinar bajo qué circunstancias $\delta\phi$ resulta despreciable y se puede considerar que la fase geométrica es robusta al efecto del entorno.
\\
\\
{\em Dependencia con el régimen de acoplamiento - } En la figura \ref{fig:sec3_phigamma} se muestra la fase geométrica (\ref{eq:sec3_faseFormal}) acumulada en función de la cantidad de pseudo-períodos $T$ de evolución, comparando tres casos para la relación $\gamma/g$ pertenecientes al régimen SC. Se incluye además como referencia la fase geométrica unitaria (\ref{eq:sec3_GPunitaria}). 

Conforme la relación $\gamma/g$ se incrementa, aumenta la diferencia $\delta\phi$ entre la fase acumulada con aquella correspondiente al caso unitario. Sin embargo, si el valor de la relación aumenta demasiado, la pérdida de coherencia detiene el movimiento del estado en el espacio de rayos y consecuentemente la acumulación de fase. En este sentido, el caso extremo está representado por el régimen de WC, en el cual el estado pierde toda coherencia antes de acumular ninguna fase apreciable. Por este motivo, cualquier estudio sobre la fase geométrica en este sistema debe realizarse en el régimen de SC.

\begin{SCfigure}[][ht!]
    \includegraphics[width = .5\linewidth]{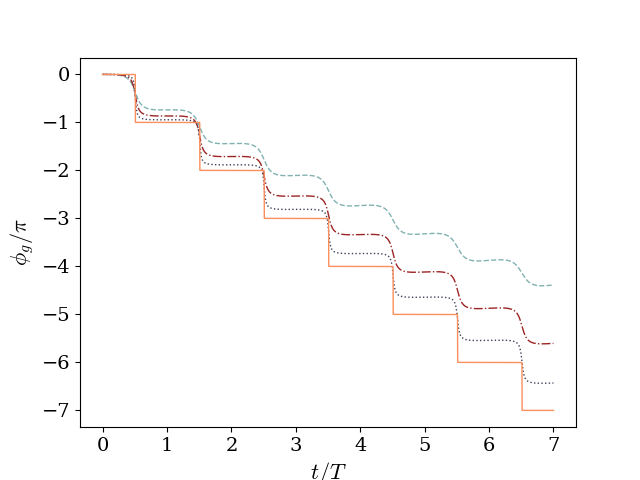}
    \caption{Fase geométrica $\phi_g$ acumulada por el sistema para distintas relaciones $\Delta/g$ entre el detuning y el acoplamiento átomo-modo. El entorno se encuentra caracterizado por una tasa de bombeo incoherente $p = 0.005\,g$ y una tasa de pérdida de fotones $\gamma= 0.1\,g$. La línea naranja sólida describe el caso resonante $\Delta=0$, mientras que las líneas azul de puntos, bordó de trazo/punto y celeste de trazos corresponden a los valores $\Delta = 0.1 \,g$, $\Delta = 0.25 \,g$, y $\Delta = 0.5 \,g$ respectivamente.} 
    \label{fig:sec3_phiDelta}
\end{SCfigure}

\vspace{.2cm}
{\em Dependencia con el detuning - } En la sección \ref{sec:sec3_Faseunitaria} se mostró que en, el contexto de evolución unitaria, la curva uniparamétrica ${\rm C}$ descrita en el espacio de rayos a medida que el sistema evoluciona depende del detuning $\Delta$, dependencia que es heredada por la fase geométrica. Tratando ahora con el sistema abierto, se inicia por observar la fase geométrica acumulada para distintos valores de la relación $\Delta/g$ entre el detuning y el acoplamiento átomo-modo, mientras que el entorno se mantiene en todos los casos igual. En la figura \ref{fig:sec3_phiDelta} se muestra la fase geométrica $\phi_g$ acumulada por el sistema para distintos valores de $\Delta/g$, exponiendo una dependencia manifiesta de la fase en esta relación. En particular, se observa que considerando un entorno idéntico en todos lo casos, conforme aumenta el detuning relativo al acoplamiento átomo-modo, la fase geométrica acumulada es menor (en valor absoluto) y se suaviza. 

\begin{SCfigure}[][ht!]
    \includegraphics[width = .55\linewidth]{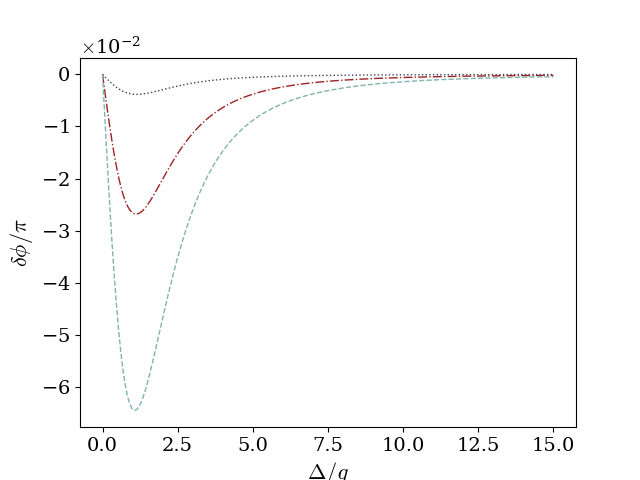}
    \caption{Corrección $\delta\phi = \phi_g - \phi_u$ a la fase unitaria acumulada en un período $t\in[0, 3\,T]$, en función de la relación $\Delta/g$ entre el detuning y el acoplamiento átomo-modo. El entorno está caracterizado por una tasa de bombeo incoherente $p = 0.005\,g$ y se muestran tres casos en que la tasa de pérdida de fotones es distinta. Las líneas azul de puntos, bordó de trazo-punto, y celeste de trazos, corresponden a los valores $\gamma = 0.01\,g$, $\gamma = 0.1\,g$ y $\gamma = 0.25\,g$ respectivamente. } 
    \label{fig:sec3_dphi}
\end{SCfigure}

Como se discutió anteriormente, la fase geométrica exhibida en la figura se puede entender como compuesta por la fase geométrica $\phi_u$ asociada a una hipotética evolución unitaria, y una desviación $\delta\phi$ introducida por el entorno. La dependencia de la componente unitaria en $\Delta/g$ se halla explícita en la ecuación (\ref{eq:sec3_GPunitaria}). Una pregunta que surge naturalmente es qué tipo de dependencia muestra la corrección $\delta\phi$ (que condensa los efectos del entorno sobre la fase geométrica) en la relación $\Delta/g$, o si toda la dependencia en $\Delta/g$ de la fase geométrica se encuentra contenida en la contribución unitaria $\phi_u$.
Para tratar con esta pregunta se analiza la corrección $\delta\phi$ en un instante dado que se mantiene fijo, observándola como función del valor de detuning relativo. El resultado se presenta en la figura \ref{fig:sec3_dphi}, en la cual se contrasta además esta dependencia para tres entornos caracterizados por distintos valores de tasa de pérdida de fotones. 

Puede verse que la corrección $\delta\phi$ en efecto depende de la relación $\Delta/g$, y dos aspectos de esta dependencia se notan inmediatamente. El primer aspecto notable es que la corrección se anula, en todos los casos considerados, cuando se satisface la condición de resonancia $\Delta = 0$. Este hecho resulta prometedor, insinuando que en este caso la fase geométrica $\phi_g$ resultaría robusta a los efectos de un entorno en el régimen de SC. La otra característica de la corrección $\delta\phi$ que se revela en la figura \ref{fig:sec3_dphi} es la no-monotonicidad de la corrección como función de $\Delta/g$, que exhibe un extremo para un valor $\Delta/g$ que depende débilmente tanto en las constantes $\gamma/g$ y $p/g$ que caracterizan el entorno como en el instante $t$ en el que se realiza la observación. 
De esta forma la figura \ref{fig:sec3_dphi} despierta un interés es doble, ya que permite identificar: (i) las condiciones en las cuales el efecto del entorno sobre la fase geométrica es mayor, acercando la posibilidad de una detección experimental y, (ii) las condiciones que mitigan, permitiendo ignorarlo, este efecto o incluso que lo eliminan por completo. Mientras que todas las consideraciones sobre la robustez de la fase geométrica en el caso resonante se presentan en la sección próxima, a continuación se refuerza el estudio de los máximos mediante un abordaje pictórico.

\begin{SCfigure}[50][ht!]
    \includegraphics[width = .35\linewidth]{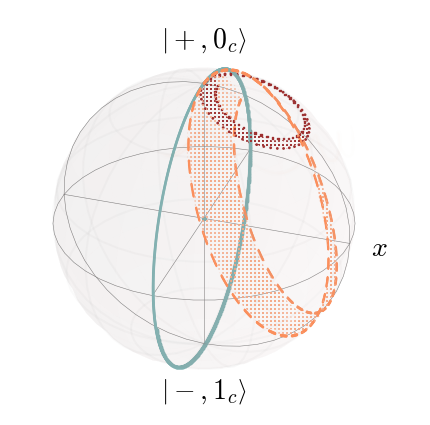}
    \caption{Trayectorias trazadas en la esfera de Bloch por (i) el estado $\ket{\psi(t)}$ del sistema que evoluciona en completo aislamiento y (ii) el autoestado $\ket{\psi_+(t)}$ del estado $\rho(t)$ del sistema expuesto a los efectos del entorno. Se presentan tres casos en los que la relación  $\Delta/g$ toma distintos valores. Las líneas de puntos color bordó, las líneas de trazos anaranjadas y las líneas sólidas celestes representan trayectorias con valores $\Delta = 5\,g$, $\Delta= g$ y $\Delta = 0.01\,g$. Para mayor claridad, se colorea la diferencia entre el área subtendida por la trayectoria unitaria y por la trayectoria del autoestado.} 
    \label{fig:sec3_Bloch_nonu}
\end{SCfigure}

La justificación a continuación se construye a partir de dos ideas. La primera de ellas es de la observación que la fase geométrica acumulada por un sistema de dos niveles en evolución unitaria equivale a la mitad del área encerrada por la trayectoria del estado en la esfera de Bloch o, de forma equivalente, un medio del ángulo sólido encerrado por dicha trayectoria. Aunque esta relación fase-área aplica para el caso de evoluciones unitaria se recuerda, en segundo lugar, que la fase geométrica no-unitaria $\phi_g$ para el caso de un estado inicial puro, dada por la ecuación (\ref{eq:sec2_TongGPpuros}), coincide formalmente con la fase geométrica unitaria adquirida por el autoestado $\ket{\psi_+(t)}$ del operador densidad. Combinando estas dos nociones es posible ganar intuición respecto de la diferencia $\delta\phi = \phi_g - \phi_u$ observando las curvas sobre la esfera de Bloch asociadas a cada fase y comparando el área subtendida por cada una de las curvas, proporcional a la fase geométrica.. 
Las trayectorias descritas en la esfera de Bloch por la proyección de un hipotético estado $\ket{\psi(t)}$ que evoluciona unitariamente y por la proyección del autoestado $\ket{\psi_+(t)}$ se presentan en la figura \ref{fig:sec3_Bloch_nonu} para una evolución en $t'\in[0,T]$ y para tres casos  en los cuales el efecto del entorno influye en la fase geométrica en grado considerablemente distinto, definidos por tres valores distintos de la relación $\Delta/g$.

Para $\Delta = 0.1\,g$, las trayectorias encierran áreas grandes pero casi idénticas, cuya sustracción $\propto \delta\phi$ devuelve un valor pequeño. A medida que el valor de la relación $\Delta/g$ aumenta y las trayectorias se trazan sobre planos más distantes al centro de la esfera de Bloch, la diferencia en el área encerrada por cada una de las curvas aumenta, mientras que el área subtendida por cada una de ellas disminuye. De esta forma, tanto la corrección neta $\delta\phi$ como la corrección relativa $\delta\phi/\phi_u$ se incrementan. Eventualmente, las trayectorias encierran áreas cada vez menores cuya diferencia decrece hasta ubicarse fuera de cualquier rango de detección y finalmente convergen al polo norte para $\Delta/g\gg1$.
\\
\\\indent
{\em Robustez de la fase geométrica en el caso resonante
 - }
Como se ha mencionado, la figura \ref{fig:sec3_dphi} muestra el resultado notable de que la diferencia $\delta\phi$ entre la fase geométrica acumulada en una hipotética evolución unitaria y la fase geométrica acumulada por el sistema abierto se anula cuando se satisface la condición de resonancia $\Delta = 0$. El resultado sugiere que la fase geométrica es robusta a los efectos del entorno en este caso. Esta robustez puede estudiarse y explicarse en términos geométricos analizando la evolución del estado $\rho(t)$ del sistema átomo-modo y se vincula con el salto en $\pi$ que exhibe la fase geométrica unitaria $\phi_u$ del sistema resonante, cuyo estado describe un círculo máximo de la esfera de Bloch. 

Para explicar el salto en $\pi$ de la fase unitaria es necesario recordar, como fue desarrollado en las secciones \ref{sec:sec2_Samuel} y \ref{sec:sec2_Mukunda}, que la fase geométrica asociada a una trayectoria unitaria no-cíclica puede entenderse como la fase geométrica asociada a una curva cerrada específica: aquella construida a partir de la trayectoria abierta original, luego cerrada mediante una curva geodésica que conecte sus extremos. Esta interpretación permite explicar un salto abrupto en $\pi$ que muestra la fase geométrica acumulada en la evolución del sistema cuando el estado recorre un meridiano de la esfera de Bloch. Debido a que las geodésicas de la esfera de Bloch son, precisamente, sus círculos máximos, cuando la evolución recorre uno de éstos sin alcanzar a transitar la mitad de su longitud, la curva geodésica que debe considerarse coincide con la trayectoria de forma que la curva cerrada retorna sobre si misma acumulando fase geométrica nula. Por el contrario, si la trayectoria descrita por el estado supera la mitad del círculo máximo, la geodésica que une sus extremos lo completa encerrando un área de $2\pi$ que corresponde a una fase geométrica $\phi_u = \pi$. 

Para el caso de un sistema en interacción con el entorno, la identidad formal entre la ecuación (\ref{eq:sec2_TongGPpuros}) y la fase geométrica unitaria para el autoestado $\ket{\psi_+(t)}$ de la matriz densisdad demanda el estudio de la curva descrita en la esfera de Bloch por (la proyección de) $\ket{\psi_+(t)}$. 
Lo que se observa es que para el caso resonante $\ket{\psi_+(t)}$ recorre una trayectoria que se superpone con aquella descrita por su análogo unitario pero que por efecto del entorno resulta de menor longitud (para un intervalo temporal idéntico). En un período $t\in[0,T]$ de evolución, entonces, el rayo asociado al estado $\ket{\psi(t)}$ retorna al punto inicial describiendo una trayectoria cíclica, mientras que aquél asociado a $\ket{\psi_+(t)}$ describe una curva abierta. Sin embargo, esto no afecta el valor obtenido para la fase geométrica que resulta $\phi_g = \pi$ siempre que el rayo recorra más de la mitad del círculo máximo. En consecuencia, siempre y cuando los efectos disipativos no sean lo suficientemente destructivos como para impedir que el autoestado supere el polo opuesto en un intervalo $t'\in[0,T]$, la fase geométrica acumulada en un período no se verá afectada. En este sentido, el caso resonante resulta entonces la situación ideal para realizar detecciones experimentales o implementar aplicaciones tecnológicas que requieran un escenario en que se puedan despreciar los efectos del entorno.

\vspace{.5cm}
\begin{center}
   \textcolor{bordo}{\ding{163}}
\end{center}
\vspace{.5cm}
En este capítulo se ha estudiado el efecto que tiene el entorno sobre la fase geométrica en el marco particular que brinda el modelo de Jaynes-Cummings. Este modelo paradigmático, ampliamente aplicado en diversas áreas de investigación, es probablemente el modelo más sencillo que da cuenta de manera satisfactoria de la interacción entre la materia (representada por un sistema de dos niveles) y la radiación electromagnética cuántica (simplificada a un único modo en una cavidad). La extensión del modelo al caso en el cual el sistema bipartito original se encuentra en interacción con el entorno se realizó a través de una ecuación maestra que introduce los efectos ambientales mediante consideraciones fenomenológicas.

En particular, se discutieron los contextos y abordajes para los cuales el caso unitario resulta comparable con el caso disipativo, optando por la interpretación cinemática de la fase geométrica. Esta interpretación resalta la dependencia exclusiva de la fase geométrica en la secuencia de estados físicos atravesada por el sistema, esto es, en la trayectoria descrita en el espacio de rayos. En este marco, los distintos resultados obtenidos fueron detenidamente analizados y justificados en términos de las trayectorias observadas. 

En el capítulo próximo el análisis se modifica de diversas formas. El sistema estudiado consta otra vez de dos partes que en lugar de un sistema de dos niveles y un modo del campo, son dos qubits idénticos. Los mismos estarán inmersos en un campo electromagnético en estado de vacío, que posteriormente se altera introduciendo condiciones de contorno no-triviales. La dinámica del sistema se deriva a partir de una descripción microscópica en lugar de hacerlo mediante consideraciones fenomenológicas.
El estudio, por otra parte, no se limita a la fase geométrica sino que se analiza el efecto del entorno sobre otras propiedades del sistema. 
\chapter{Fluctuaciones del vacío electromagnético}\label{ch:4}
En este capítulo se modifica parcialmente el enfoque respecto del adoptado en el capítulo anterior (capítulo \ref{ch:3}). 
El escenario propuesto consiste en dos partículas de dos niveles (o qubits) originalmente no-interactuantes entre sí, pero acopladas al campo electromagnético cuántico en el que están inmersas. Este modelo de un sistema bipartito que se halla sujeto a la acción del entorno, resulta un contexto propicio para estudiar el efecto del entorno sobre el entrelazamiento entre las partes del sistema. El tipo especial de correlación denominado entrelazamiento no sólo ha sido el foco de discusiones fundacionales \cite{schrodinger1935current, einstein1935can} de la mecánica cuántica sino que también se ha convertido en un {\em recurso} explotable para el almacenamiento y la manipulación de información cuántica \cite{nielsen2000quantum, bennett1992communication, bennett1999entanglement, bennett2000quantum}. Las correlaciones cuánticas decaen en presencia de un entorno, esto es, la degradación del entrelazamiento entre partículas es inevitable \cite{zurek2003decoherence, yu2002phonon, simon2002robustness}. Más aún, se ha encontrado que este decaimiento no siempre sucede de forma asintótica, sino que partículas inicialmente entrelazadas puede decorrelacionarse completamente en un tiempo finito, fenómeno conocido como muerte súbita del entrelazamiento \cite{yu2004finite, yu2009sudden, zyczkowski2001dynamics, rajagopal2001decoherence, diosi2003progressive, dodd2004disentanglement}. Paralelamente, un entorno común puede también producir interacción indirecta entre átomos originalmente independientes, conduciendo al efecto llamado, por analogía, (re)nacimiento del entrelazamiento \cite{mazzola2010interplay, ficek2006dark}. 

Para caracterizar el efecto del entorno sobre el entrelazamiento resulta imprescindible definir y modelar dicho entorno. En particular, las fluctuaciones cuánticas del vacío son un ambiente que no puede ser nunca completamente suprimido de modo que resulta importante, en muchos casos, considerarlo. 
Por otra parte, es sabido que el estado del campo se modifica con la presencia de contornos no-triviales, y que las distorsiones en las fluctuaciones del vacío cuántico se traducen en efectos observables como modificaciones en el corrimiento de Lamb \cite{lamb1947fine} y fuerzas de Casimir \cite{casimir1948attraction, milton2001casimir, lamoreaux2004casimir, passante2018dispersion}.
Considerando: (1) que el vacío cuántico modifica el entrelazamiento y la fase geométrica acumulada por el sistema y (2) que las condiciones de borde modifican el vacío cuántico, surge naturalmente la pregunta sobre en qué medida la presencia de contornos se traduce en nuevas correcciones al entrelazamiento en el sistema bipartito y a la fase geométrica acumulada. Algunos trabajos previos que abordan estas preguntas son \cite{yu2012detecting, yang2016entanglement} respecto de las correlaciones y \cite{cheng2018entanglement} respecto de la fase geométrica. La literatura se extiende también sobre la relación entre el entrelazamiento de sistemas bipartitos y la fase geométrica \cite{oxman2011fractional, khoury2014topological}.

El objetivo en este capítulo es caracterizar la dinámica, el entrelazamiento y la fase geométrica en presencia de fluctuaciones cuánticas del vacío electromagnético, comparando los resultados de espacio libre con aquellos obtenidos cuando se introduce una condición de borde que modifique la estructura del entorno. En particular, se considera si la presencia del contorno puede explotarse para inferir otros efectos del entorno sobre el sistema.

\section{Modelo microscópico y ecuación maestra}
\begin{SCfigure}[5][ht!]
    \includegraphics[width = .5\linewidth]{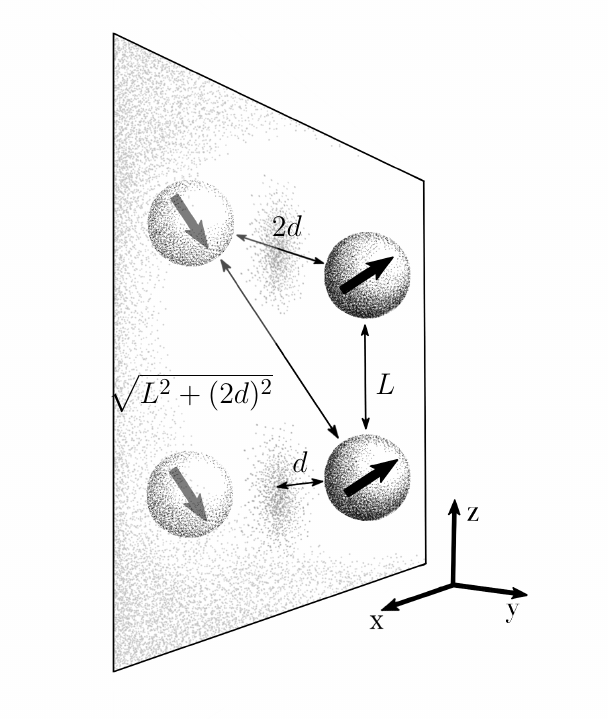}
    \caption{Esquema del escenario propuesto en este capítulo. El sistema de interés se compone de dos qubits (o partículas) neutras con momento dipolar no-nulo y originalmente no-interactuantes entre sí, que están separadas una distancia $L$ en un entorno de campo electromagnético en estado de vacío.  Se estudian los casos (a) de espacio libre, o condiciones de contorno triviales y (b) en presencia de un plano infinito conductor ubicado en $y=0$. En este segundo caso, las partículas se encuentran a una misma distancia $d>0$ del plano, cuyo efecto puede interpretarse en términos de interacciones entre partículas en el espacio libre mediante el método de imágenes.
    } 
    \label{fig:sec4_esquema}
\end{SCfigure}

Como se ha anticipado, se considera un sistema bipartito constituido por dos qubits (a los que se refiere indistintamente de esta forma, o como partículas o sistemas de dos niveles), originalmente desacoplados entre sí. Ambos sistemas de dos niveles están en interacción mínima con el campo electromagnético que los rodea, el cual se encuentra en estado de vacío. El conjunto completo se estudiará (a) en condiciones de contorno triviales y (b) en presencia de un plano infinito de un material conductor perfecto, como se representa esquemáticamente en la figura \ref{fig:sec4_esquema}. 

\vspace{-.5cm}
Ambos casos pueden describirse mediante la misma expresión formal para el Hamiltoniano $H = H_{\rm s} + H_{\rm em} + H_{\rm int}$, donde los dos primeros términos $H_{\rm s}$ y $H_{\rm em}$ corresponden al Hamiltoniano que gobierna la dinámica del sistema bipartito y de la radiación electromagnética libres respectivamente, y $H_{int}$ modela la interacción entre el sistema y el campo de vacío. 

El Hamiltoniano del sistema libre puede escribirse explícitamente como la suma de dos términos

\begin{equation}
    H_{\rm s}^l = \frac{\omega^{l}}{2}\sigma_z^l,
    \label{eq:sec4_Hs}
\end{equation}
donde $l = 1,2$ etiqueta cada partícula, $\omega^l$ representa la frecuencia natural de la partícula $l$ y $\sigma_z^l$ es la matriz de Pauli $\sigma_z$ que actúa sobre la partícula $l$, en producto externo con el operador identidad actuando sobre la otra partícula. Por otro lado, la interacción entre cada átomo y el campo electromagnético se modela en la aproximación dipolar, y el término correspondiente $H_{\rm int}$ del Hamiltoniano también puede expresarse como la suma $H_{\rm int} = \sum_{l=1,2}H_{\rm int}^{l}$ de dos términos 
\begin{equation}
    H_{\rm int}^{l} = -\boldsymbol{\mu}^{l}\cdot\mathbf{E}(t, \mathbf{x}^{l}),
    \label{eq:sec4_Hint}
\end{equation}
dados por el producto interno usual entre vectores del operador (vectorial) dipolo eléctrico $\boldsymbol{\mu}^l$ del qubit y el campo eléctrico $\mathbf{E}(t,\mathbf{x}^l)$ en la posición de cada dipolo. La forma funcional del operador dipolar es $\boldsymbol{\mu}^l = e(\mathbf{r}_+^{l}\,\sigma_+^{l} + \mathbf{r}_-^{l}\,\sigma_-^{l})$, con $e$ la carga del electrón, $\mathbf{r}_\pm^{l}$ la orientación del momento dipolar de la partícula $l$, y $\sigma_\pm = (\sigma_x\pm i\,\sigma_y)/2$. Las orientaciones $\mathbf{r}$ se consideran parámetros clásicos, de forma que todo el carácter operatorial de $\boldsymbol{\mu}$ está contenido en $\sigma_\pm$.

Por su parte, el campo eléctrico puede escribirse como
 
 \begin{equation}
 \mathbf{E}(t, \mathbf{x}) = \sum_{\lambda = 1}^2\int\frac{d^3k}{(2\pi)^3}\,(2\pi\,\omega_\mathbf{k})^{1/2}\,[a_{\lambda,\mathbf{k}} \,A_{\lambda,\mathbf{k}}(\mathbf{x})- a^\dagger_{\lambda,\mathbf{k}} \,A^*_{\lambda,\mathbf{k}}(\mathbf{x})] \, \varepsilon_{\lambda,\mathbf{k}},
 \end{equation}
 donde $a_{\lambda,\mathbf{k}}^\pm$ son los operadores bosónicos de creación y destrucción de fotones para el modo de momento $\mathbf{k}$ y polarización $\lambda$, $\omega_{\mathbf{k}} = k\,c$ es la frecuencia del modo, con $k = |\mathbf{k}|$ y $c=1$ en lo que sigue, y $\varepsilon_{\lambda, \mathbf{k}}$ la $\lambda$-ésima componente del vector unitario $\boldsymbol{\varepsilon}_\mathbf{k}$ que define la dirección de polarización del modo \cite{milonni2013quantum}. Las funciones de modo $A_{\lambda,\mathbf{k}}(\mathbf{x})$ del campo electromagnético deben proponerse de forma que el campo satisfaga las condiciones de borde. En particular, las funciones de modo del espacio libre infinito están dadas por exponenciales complejas 
 
 \begin{equation}
     A_{\lambda, \mathbf{k}}(\mathbf{x}) = e^{-i\,\mathbf{k}\cdot\mathbf{x}},
 \end{equation}
mientras que, para el caso en que se coloca un plano infinito de material conductor perfecto en la posición $y=0$, las funciones que describen el campo eléctrico en la región $y \geq 0$ están dadas por las expresiones

\begin{align}\nonumber
    &A_{1,\mathbf{k}}(\mathbf{x}) = \sqrt{2}\,(\hat{k}_\parallel\times\hat{k}_\perp)\,e^{i\, \mathbf{k}_\parallel\cdot\mathbf{x}_\parallel}\\
    &A_{2,\mathbf{k}}(\mathbf{x})= \frac{\sqrt{2}}{k}[k_\parallel\cos(\mathbf{k}_\perp\cdot\mathbf{x})\cdot\hat{k}_\perp - i\,k_\perp\sin(\mathbf{k}_\perp\cdot\mathbf{x})\hat{k}_\parallel],
\end{align}
donde $\mathbf{k}_\parallel = k_x\,\hat{x} + k_z\,\hat{z}$ y $\mathbf{k}_\perp = k_y\,\hat{y}$ son las componentes paralela  y perpendicular al plano infinito del vector de onda $\mathbf{k}$, y el sombrero indica vectores de norma unitaria. Más concretamente, se considera que los qubits se encuentran separados una distancia $L$ en dirección del eje $z$ entre sí, y se ubican a la misma distancia $d$ de la placa (ver figura \ref{fig:sec4_esquema} para un esquema), de modo que sus posiciones se describen como $\mathbf{x}^1=d\,\hat{y} + L\,\hat{z}$ y $\mathbf{x}^2=d\,\hat{y}$. 

La evolución temporal de este sistema debe abordarse 
mediante una ecuación maestra que se obtiene en este caso desde consideraciones microscópicas siguiendo el desarrollo presentado en el apéndice \ref{sec:ap2}.
Se parte de la ecuación (\ref{eq:ap2_redfieldint}), dónde el primer término se anula al explicitar que el campo se encuentra en estado de vacío, de forma que la ecuación maestra surgirá de especificar para este modelo la expresión

\begin{equation}
    \dot{\rho}_{\rm I}(t) = -\Tr_{\rm em}\int_{0}^t\,dt'\,\bigl[H_{\rm int,I}(t),[H_{\rm int,I}(t'), \rho_{\rm I}(t) \otimes \rho_{\rm em}]\bigr].
    \label{eq:sec4_eqini}
\end{equation}
El Hamiltoniano de interacción entre el campo y cada partícula, en representación de interacción con respecto al Hamiltoniano libre de los qubits y de la radiación electromagnética, 

\begin{equation}
    H_{\rm int, I}(t) = e^{i\,(H_{\rm s} + H_{\rm em})\,t}H_{\rm int}\;e^{-i\,(H_{\rm s} + H_{\rm em})\,t}
\end{equation} 
resulta en una suma de términos que oscilan con frecuencia $\omega_{\mathbf{k}} -\omega^l$ y $\omega_{\mathbf{k}} +\omega^l$ que se obtiene fácilmente aplicando la formula de Baker-Campbell-Hausdorff. Reemplazando la expresión explícita de $H_{\rm int, I}(t)$ en la ecuación (\ref{eq:sec4_eqini}) y retornando a la representación de Schrodinger se obtiene

\begin{equation}
        \Dot{\rho}(t) = -i \,[H_{\rm s}, \rho(t)] - \sum_{l,m =1}^2\,\int_0^t\,dt'\,\Bigl[K_\mp^{l\,m}(t')\Bigl([\sigma_\pm^l,[\sigma_\mp^m, \rho(t)]]+[\sigma_\pm^l,\{\sigma_\mp^m, \rho(t)\}]\Bigr) + {\rm h.c.}\Bigr].
        \label{eq:sec4_eq1}
\end{equation}
Para arribar a la expresión de arriba, se ha realizado además la denominada {\em aproximación secular}, que consiste en despreciar aquellos términos de la ecuación maestra rápidamente oscilantes en la representación de interacción y se vincula estrechamente con la aproximación de onda rotante \cite{fleming2010rotating}.

Los núcleos $K_{\pm}^{lm}(t)$ definidos en la ecuación (\ref{eq:sec4_eq1}) contienen toda la información sobre el efecto del entorno en el sistema bipartito y están dados por

\begin{equation}
    K_{\pm}^{lm}(t) = e^2\,4\pi\,|\mathbf{r}|^2\sum_{n, n' = x,y,z}\,r^l_nr^m_{n'}\bra{0}E_n(t, \mathbf{x}^l)E_{n'}(t, \mathbf{x}^m)\ket{0}\,e^{-i\,(\pm\,\omega)(t-t')}.
    \label{eq:sec4_kernelK}
\end{equation}
Para arribar a esta definición para los núcleos $K_{\pm}^{lm}(t)$ se han realizado algunas suposiciones adicionales: se considera que la frecuencia natural $\omega^l = \omega\, \forall\, l$ y la magnitud del momento dipolar $|\mathbf{r}|$ es la misma para ambas partículas, mientras que se permite todavía que las direcciones de polarización sean distintas. Estas direcciones se representan en la expresión (\ref{eq:sec4_kernelK}) mediante dos vectores unitarios $\hat{r}^l$ de componentes $r_n^l$.
Siguiendo la derivación del apéndice \ref{sec:ap2}, se observa además que los núcleos $K_\pm^{lm}(t)$ decaen en un rango temporal acotado de modo que es válido tomar $t\rightarrow \infty$ en el límite superior de la integral.

Mediante la expansión de los conmutadores y anticonmutadores, la ecuación (\ref{eq:sec4_eq1}) adquiere una expresión mucho más ilustrativa que permite ya en esta instancia identificar efectos del entorno sobre el sistema de distinta naturaleza
\begin{align}
    \Dot{\rho}(t) =& -i \,[H_{\rm s}, \rho(t)]-i\sum_{l, m} \, c_{lm}\,[\sigma_+^l\sigma_-^m,\rho(t)]
    \label{eq:sec4_eq3}\\ \nonumber
    &-\sum_{l,m}\,a_{lm}\bigl(\sigma_+^l\,\sigma_-^m\,\rho(t)+\rho(t)\sigma_+^m\,\sigma_-^l-\sigma_-^m\rho(t)\sigma_+^l-\sigma_-^l\rho(t)\sigma_+^m\bigr),
\end{align}
donde $a_{lm}$ y $c_{lm}$ son coeficientes reales que contienen la información sobre el entorno y sus condiciones de borde, definidos por
\begin{align}
    &a_{lm}(t)=\Re\int_0^\infty\,dt'\,K^{lm}_-(t')\\\nonumber
    &c_{ll}(t)=\Im\int_0^\infty\,dt'\,\bigl[K_+^{ll}(t') - K_-^{ll}(t')\bigr]\\\nonumber
    &c_{lm}(t)\underset{l\neq m}{=}\Im\int_0^\infty\,dt'\,\bigl[K_+^{lm}(t') + K_-^{lm}(t')\bigr],
\end{align}
y resultan constantes como consecuencia de haber tomado el límite $t\rightarrow\infty$ en la integral.

El primer término de la ecuación (\ref{eq:sec4_eq3}) corresponde nuevamente a la evolución unitaria de dos qubits libres no-interactuantes que estaría presente para el sistema en completo aislamiento. Por el contrario, los términos condensados en las sumatorias representan efectos introducidos por las fluctuaciones del vacío cuántico, y son de distinto carácter en dos sentidos: (a) el tipo de evolución que originan y (b) la parte del sistema que involucran.
Los términos agrupados en la primer sumatoria contribuyen a la evolución unitaria, mientras que los términos agrupados en la segunda sumatoria describen efectos no-unitarios. Entre los efectos unitarios que emergen por la presencia del entorno, pueden a su vez reconocerse dos naturalezas. Los términos con coeficientes $c_{ll}$, en los cuales el operador $\sigma_+^l\sigma_-^l$ actúa sobre el subespacio de estados de una única partícula, corresponden a renormalizaciones de la frecuencia natural de cada qubit. Por otra parte, los términos con índices $l\neq m$, en los cuales el operador $\sigma_+^l\sigma_-^m$ actúa sobre el subespacio de estados asociado a una y otra partícula, representa un término de interacción efectiva entre qubits. Considerando los términos disipativos de la ecuación maestra, también en éstos se pueden reconocer distintos efectos según involucren una única partícula o las dos. En particular, el coeficiente $a_{lm}$ con $l\neq m$ se denomina usualmente de {\em decaimiento colectivo}.

Los coeficientes involucrados en la ecuación dinámica del sistema tomarán distintas expresiones explícitas según se propongan condiciones de contorno triviales o no. En lo que sigue, se presentan las expresiones correspondientes al caso del plano infinito de material conductor perfecto emplazado en $y=0$. Sin embargo, es posible extraer las expresiones para el caso de espacio libre tomando el límite en que la distancia $d$ de las partículas al plano tiende a infinito. Por otra parte, las expresiones para los distintos coeficientes permanecerán válidas siempre y cuando se satisfagan las condiciones necesarias para las aproximaciones de Born y Markov (ver apéndice \ref{sec:ap2}) y las funciones de correlación del campo electromagnético decaigan en una escala temporal $\tau_{\rm em}$ mucho más corta que el tiempo característico de relajación del sistema $\tau_{R}$. Esta condición impone una restricción sobre  los valores que pueden tomar los parámetros, a saber, se debe satisfacer que $L\,\omega \gtrsim 1$ y $2d\,\omega \gtrsim 1$. \textcolor{red}{ver}

Todos los coeficientes $a_{lm}$ y $c_{lm}$ pueden a su vez descomponerse en una contribución independiente de la geometría de los contornos y una que surge como consecuencia de la presencia del plano según
\begin{align}
    a_{lm} =&\, \gamma_0\sum_{n=x,y,z}r^l_n\,r^m_n\;\left[f_{1,lm}-f_{2,lm}\right]\\\nonumber
    c_{lm} =&\, \gamma_0\sum_{n=x,y,z}r^l_n\,r^m_n\;\left[h_{1,lm}-h_{2,lm}\right].
    \label{eq:sec4_acDecomp}
\end{align}
En la descomposición de arriba se ha definido el coeficiente $\gamma_0 = e^2|\mathbf{r}|^2\omega^3/2$ proporcional (por un factor $3\pi/2$) a la tasa de decaimiento espontánea para un átomo en vacío. 
A fin de dar una expresión para cada uno de los términos involucrados, se definen además los parámetros adimensionales: $\Tilde{d} = 2\,d\,\omega$, proporcional a la distancia entre las partículas y el plano conductor, $\Tilde{L} = L\,\omega$, proporcional a la distancia entre qubits, y $\Tilde{s} = \omega\,((2d)^2 + L^2)^{1/2}$. 
\\
\\\indent
{\em Efectos de partícula única - } Los términos con $l=m$ representan efectos del campo sobre las partículas tomadas en forma individual. Estos términos estarían presentes también en el caso en que el sistema consistiera de una única partícula. En particular $f_{1,ll}= 1/3$ contiene el efecto disipativo ineludible del entorno como generador de emisión espontánea, mientras que $f_{2,ll}= (\delta_{nx}+\delta_{nz})A_0(\Tilde{d})+2\,\delta_{ny}\,B_0(\Tilde{d})$ da cuenta de un efecto adicional de disipación en cada partícula producto de la presencia del plano. Por otro lado, los factores $h_{1,ll}$ y $h_{2,ll}$ describen los efectos unitarios de partícula única, esto es, corrimientos de las frecuencias producidos por el vacío infinito y por las modificaciones introducidas por el plano respectivamente. Se desprecia $h_{1,ll}$ \cite{hu1992quantum} y se refiere al trabajo original \cite{viotti2020boundary} para la forma explícita del corrimiento en las frecuencias inducido por el plano $h_{2,ll}$.
\\
\\\indent
{\em Efectos emergentes colectivos - }Los efectos de carácter colectivo inducidos por el vacío están modelados por los términos con $l\neq m$ de la descomposición (\ref{eq:sec4_acDecomp}).  El coeficiente $a_{lm}$ de decaimiento colectivo se compone de los factores 
\begin{align}
    f_{1,lm} =& \delta_{nx}A_0(\Tilde{L})+\delta_{ny}A_0(\Tilde{L})-2\delta_{nz}\,B_0(\Tilde{L})\\\nonumber
    f_{2,lm} =& \delta_{nx}A_0(\Tilde{s})+\delta_{ny}\,D_0(\Tilde{s})+\delta_{nz}C_0(\Tilde{s}),\\
    \intertext{mientras que las interacciones efectivas emergentes que contribuyen a la evolución unitaria del sistema se describen con}
    \\
    h_{1,lm} =& \delta_{nx}A_1(\Tilde{L})+\delta_{ny}A_1(\Tilde{L})-\delta_{nz}\,B_1(\Tilde{L})\,2\\\nonumber
    h_{2,lm} =& \delta_{nx}A_1(\Tilde{s})+\delta_{ny}\,D_1(\Tilde{s})+\delta_{nz}\,C_1(\Tilde{s}).
\end{align}
Las funciones $X_{\rm n}(x)\;;\; X = A, ..., D\,,\, {\rm n}= 0,1$ que aparecen en estas expresiones consisten en funciones oscilantes que decaen según 
\begin{align}
    A_{\rm n}(x) =&\, [x\,\cos(x - {\rm n}\,\pi/2) + (x^2-1)\sin(x - {\rm n}\,\pi/2)]/x^3\\\nonumber
    B_{\rm n}(x) =&\, [x\,\cos(x + {\rm n}\,\pi/2) - \sin(x + {\rm n}\,\pi/2)]/x^3\\\nonumber
    C_{\rm n}(x) =& [-(2\Tilde{L}^2+\Tilde{d}^{\,2})x\,\cos(x - {\rm n}\,\pi/2) + (2\Tilde{L}^2+\Tilde{d}^{\,2}(x^2-1))\sin(x - {\rm n}\,\pi/2)]/x^5\\\nonumber
    D_{\rm n}(x) =& [-(2\Tilde{d}^{\,2}-\Tilde{L}^{\,2})x\,\cos(x + {\rm n}\,\pi/2) + (2\Tilde{d}^{\,2}+\Tilde{L}^{\,2}(x^2-1))\sin(x + {\rm n}\,\pi/2)]/x^5.
\end{align}

{\em Interpretación en términos de imágenes - }En síntesis, los efectos que emergen en la dinámica del sistema bipartito como consecuencia de su acoplamiento al entorno muestran distintas naturalezas: (i) Algunos efectos se derivan de las fluctuaciones del vacío cuántico sobre cada qubit y no dependen de ninguna distancia característica del sistema como, por ejemplo, los efectos generados por $f_{1,ll}$.\\ (ii) Otros efectos, también independientes de los contornos, son efectos colectivos emergentes entre los dos qubits del sistema inicialmente no-acoplados, están contenidos en los términos $f_{1,lm}$ y $h_{1,lm}$ ($l\neq m$) y dependen de la distancia $L$ entre partículas. Finalmente (iii), se observan efectos adicionales que dan cuenta de la presencia del plano conductor y afectan tanto a cada partícula individualmente como al sistema bipartito completo. Los coeficientes que modelan el primer tipo, $f_{2,ll}$ y $h_{2,ll}$, dependen del doble de la distancia del qubit al plano, y pueden interpretarse como acciones efectivas ejercidas sobre el átomo por su dipolo imagen, mientras que los coeficientes que modelan el segundo caso, $f_{2,lm}$ y $h_{2,lm}$, exhiben dependencia en la distancia $((2d)^2 + L^2)^{1/2}$ y pueden entenderse como interacciones efectivas ejercidas sobre un qubit por la imagen del otro dipolo.

\section{Análisis de la dinámica del sistema}
En esta sección, se examina la dinámica del sistema bipartito y sus propiedades para los dos casos mencionados de espacio libre y en presencia de un plano conductor infinito, fijo en la posición $y=0$. En ambos casos
los qubits se encuentran separados entre sí una distancia $L$ en dirección $\hat{z}$. Cuando se considere el plano conductor, se denominará $d$ la distancia de las partículas al mismo, que será siempre la misma para los dos qubits. 

Una base naturalmente adecuada para este estudio es la base $\mathcal{B}=\{\ket{+,+}, \ket{+,-}, \ket{-,+}, \ket{-,-}\}$ formada por estados producto $\ket{\psi} = \ket{\pm}\otimes\ket{\pm}$ de autoestados $\ket{\pm}$ del operador $\sigma_z$ para cada qubit. En lo que sigue, se considera esta base y se restringe el estudio a estados iniciales del sistema bipartito de la forma 
\begin{equation}
    \ket{\psi(0)} = \cos(\theta/2)\ket{+,+} + \sin(\theta/2) \ket{-,-}
    \label{eq:sec4_inicial}
\end{equation}
donde $\theta$ determina el grado de entrelazamiento inicial. La restricción a estados iniciales de esta forma se justifica por la intención de analizar los escenarios particulares en los que se observa {\em muerte} y {\em nacimiento} súbito de entrelazamiento. Como se reporta en \cite{mazzola2010interplay}, no se observa muerte súbita del entrelazamiento para ningún estado con un máximo de una excitación. De forma análoga, en dichos estados puede crearse suavemente entrelazamiento, pero no puede aparecer repentinamente. 

La evolución de un estado inicial de la forma (\ref{eq:sec4_inicial}) asume una estructura simplificada en la que algunos elementos de matriz permanecen nulos para todo tiempo. Las poblaciones $\rho_{ii}$ ocupando los estados $\{\ket{1}, \ket{2}, \ket{3}, \ket{4}\}= \{\ket{++}, \ket{+-}, \ket{-+}, \ket{--}\}$ evolucionan según
\clearpage
\begin{align}
    \rho_{11}(t) =& \cos^2(\theta/2)\,e^{-4 \,a_{11}\,t}
    \label{eq:sec4_poblaciones}\\ \nonumber
    \rho_{22}(t) =\rho_{33}(t) =& \cos^2(\theta/2)\,\left\{\frac{a_-}{2\,a_+}(e^{-2 \,a_+\,t}-1)\, e^{-2 \,a_-\,t}+\frac{a_+}{2\,a_-}(e^{-2 \,a_-\,t}-1)\, e^{-2 \,a_+\,t}\right\},
    \intertext{donde se ha empleado la notación $a_\pm = a_{11}\pm a_{12}$ y la población $\rho_{44}(t)$ se deduce de la condición de traza unitaria del operador densidad. Por otro lado, las únicas coherencias no-nulas son $\rho_{41}$ y $\rho_{23}$, o se deducen de éstas, dadas por }
    \rho_{41}(t) =& \sin(\theta)\,e^{-2 \,a_{11}\,t}\,e^{2i\,(c_{11}+\omega)\,t}/2
    \label{eq:sec4_coherencias}\\\nonumber
    \rho_{23}(t) =& \cos^2(\theta/2)\,\left\{-\frac{a_-}{2\,a_+}(1-e^{-2 \,a_+\,t})\, e^{-2 \,a_-\,t}+\frac{a_+}{2\,a_-}(1-e^{-2 \,a_-\,t})\, e^{-2 \,a_+\,t}\right\}.
    \label{eq:sec4_coherencias}
\end{align}

\begin{figure}[ht!]
    \center
    \includegraphics[width = .495\linewidth]{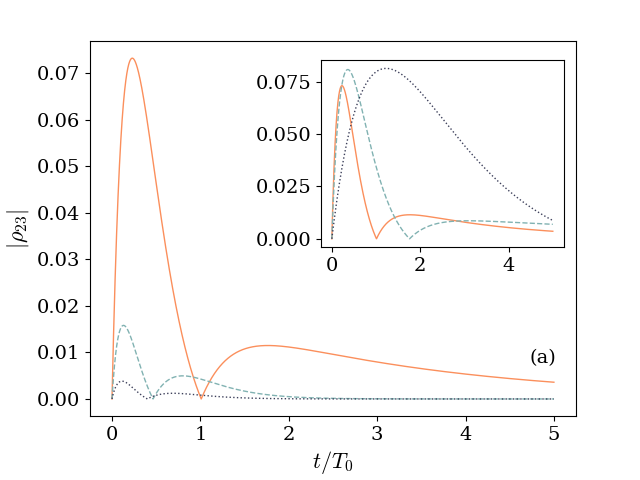}
    \includegraphics[width = .495\linewidth]{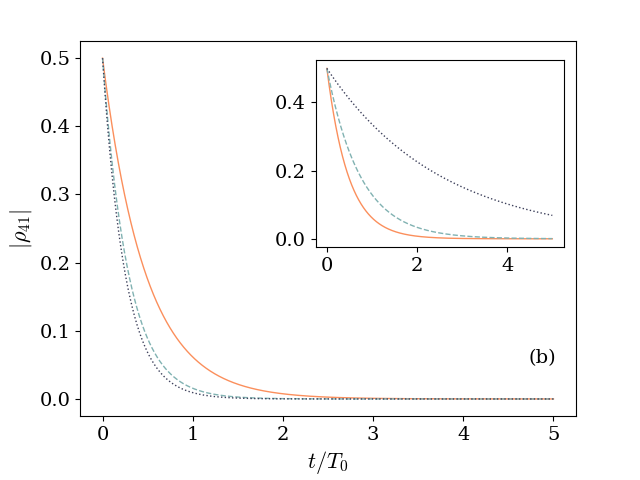}
    \caption{Evolución dinámica en valor absoluto de las coherencias (a) $\rho_{23}(t)$ y (b) $\rho_{41}(t)$ del operador densidad $\rho(t)$ considerando distintas relaciones $d/L$ entre la distancia $d$ del sistema al plano conductor y la distancia $L$ entre las partículas que componen el sistema. El mismo parte de un estado inicial definido por $\theta = \pi/2$, y el efecto del vacío está caracterizado por un parámetro $\gamma_0/\omega = \alpha$, con $\alpha$ la constante de estructura fina. Se incluye también el caso de espacio libre (líneas color naranja, sólidas) para comparación. Las líneas celestes de trazos, y azules de puntos, representan la evolución para valores $d=L$ y $d = 0.5\, L$ respectivamente. Las figuras principales corresponden a partículas con polarización perpendicular al plano, mientras que los insets muestran el caso de partículas con polarización paralela al plano en dirección $\hat{x}$} 
    \label{fig:sec4_coherencias}
\end{figure}

\vspace{.2cm}
{\em Decaimiento de las coherencias - }El decaimiento de las coherencias cuánticas se considera una buena medida del afecto del entorno sobre la dinámica del sistema. La figura \ref{fig:sec4_coherencias} muestra en los paneles (a) y (b) la evolución dinámica del valor absoluto de las coherencias  $\rho_{23}(t)$ y $\rho_{41}(t)$, para distintas relaciones $d/L$ entre la distancia de los qubits al plano y la distancia entre qubits. En ambas figuras se incluye el caso en el que no hay plano conductor $\Tilde{d}\rightarrow\infty$ para permitir la comparación. Cuando se considera la presencia del plano infinito conductor, la orientación del momento dipolar de cada partícula se torna relevante. Para tratar esta cuestión, las figuras principales y los insets muestran el caso de partículas con momento dipolar en direcciones diferentes. Concretamente, las figuras principales corresponden a momento dipolar en dirección perpendicular al plano $\mathbf{r}^l = |\mathbf{r}|\,\hat{y}$, mientras que los insets corresponden al caso de momento dipolar paralelo al plano, pero perpendicular a la separación entre partículas
$\mathbf{r}^l = |\mathbf{r}|\,\hat{x}$

Las coherencias del estado $\rho(t)$ decaen, como es de esperar, con el tiempo adimensional $t\,\omega$. Sin embargo pueden distinguirse en la figura \ref{fig:sec4_coherencias} distintas escalas de decaimiento de las coherencias o {\em decoherencia} para los distintos casos considerados. Si se define la escala de decoherencia para el caso de espacio libre $T_0=\pi/\gamma_0$, se observa que ambas coherencias decaen en una escala temporal $T_\perp<T_0$ más corta que $T_0$ cuando la orientación del dipolo es perpendicular al plano conductor. Por otra parte, la situación se invierte para el caso de polarización en dirección paralela al plano en dirección del eje $x$. En este caso, la escala temporal de decoherencia $T_\parallel>T_0$ resulta más prolongada que para el caso sin contornos. Más aún, este comportamiento se acentúa conforme se reduce la distancia relativa al plano $d/L$.

Esta situación admite una interpretación en términos de interacciones entre partículas (reales e imagen) recurriendo el método de imágenes.
Si un dipolo eléctrico se ubica frente a un plano infinito conductor apuntando en dirección paralela al plano, el efecto de la condición de contorno puede reemplazarse por el efecto de un dipolo de idéntica magnitud y dirección pero de sentido contrario $\mathbf{r}_{\rm im} = - \mathbf{r}$, ubicado en posición opuesta a la misma distancia del contorno que el dipolo real. La diferencia en los sentidos de orientación conduce a que el momento dipolar del conjunto se compense, el dipolo imagen cancela parcialmente el efecto del dipolo real y se observa menos decoherencia. Si por el contrario el dipolo se ubica orientado en dirección perpendicular al plano, entonces el efecto de la condición de contorno se reproduce con otro dipolo en la misma dirección y sentido $\mathbf{r}_{\rm im} = \mathbf{r}$, ubicado a la misma distancia del contorno que el dipolo real. En este caso el efecto total se incrementa y se observa que la decoherencia es más fuerte \cite{mazzitelli2003decoherence}.

\begin{SCfigure}[][ht!]
    \includegraphics[width = .55\linewidth]{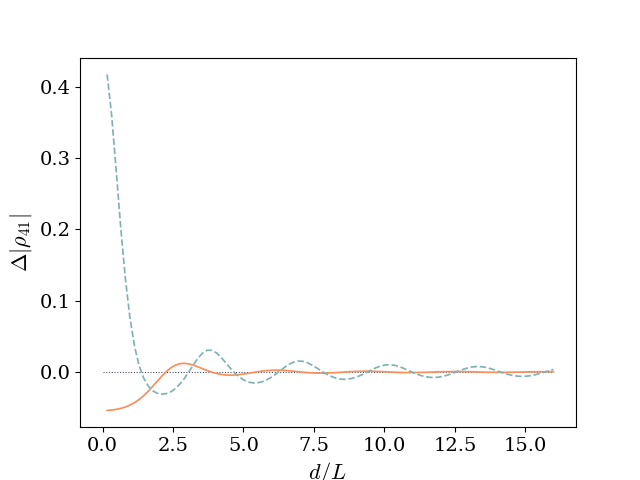}
    \caption{Diferencia entre valores absolutos de $\rho_{41}(T_0)$ a tiempo fijo $t = T_0$ para un sistema en espacio libre y en presencia de un plano conductor, en función de la distancia relativa $d/L$. Se parte de un estado inicial con $\theta = \pi/2$ en un entorno caracterizado por $\gamma_0/\omega = \alpha$, con $\alpha$ la constante de estructura fina. Se presentan los dos casos de polarización perpendicular al plano y paralela en dirección $\hat{x}$ con línea color naranja sólida y color celeste de trazos respectivamente.} 
    \label{fig:sec4_drho}
\end{SCfigure}

Con el objetivo de obtener una imagen más detallada del comportamiento de las coherencias con la distancia qubit-plano $d$, se compara el proceso de caída de las coherencias para el caso de espacio libre con el caso en presencia del plano conductor. A continuación, se toma el ejemplo del elemento $\rho_{41}$ y se refiere al trabajo original \cite{viotti2020boundary} para el análisis del elemento $\rho_{32}$. La evolución de $|\rho_{41}(t)|$ expuesta en el panel (b) de la figura \ref{fig:sec4_coherencias} sugiere que una buena forma de hacerse una idea de la influencia del plano es observar la diferencia $\Delta|\rho_{41}|= |\rho_{41}(t)|_{\rm plano}-|\rho_{41}(t)|_{\rm libre}$ entre los valores absolutos del elemento $\rho_{41}(t)$ a un tiempo fijo $t$ para ambos casos. En la figura \ref{fig:sec4_drho} se muestra esta diferencia en función de la distancia relativa $d/L$ de los qubits al plano a tiempo fijo $t=T_0$. Se presentan nuevamente los dos casos de polarización considerados anteriormente, esto es, el caso en que ambos qubits están polarizados en dirección perpendicular al plano, y en que ambos están polarizados en dirección paralela al plano y perpendicular a la distancia que los separa.  

Esta figura muestra un primer régimen de distancias $d/L$ para el cual el efecto de decoherencia se refuerza si el momento dipolar del qubit es perpendicular al plano, mientras que se suaviza si éste es paralelo. En el primer caso $|\rho_{41}(T_0)|_{\rm plano}<|\rho_{41}(T_0)|_{\rm libre}$, y por lo tanto su diferencia resulta negativa,  mientras que en el segundo caso la relación se invierte y la diferencia resulta positiva. Superado este régimen, lo que ocurre para $d\sim 2\, L$ la relación muestra intervalos de comportamiento inverso, con oscilaciones de amplitud decreciente. Cuando $d/L \gg 1$ se recupera, a partir de ambas polarizaciones, el caso de espacio libre y la diferencia se vuelve nula.

Resulta ilustrativo retomar la interpretación del efecto del plano en términos de momentos dipolares imagen. Cuando los qubits que componen el sistema se encuentran en el primer régimen la distancia que separa las partículas del plano es suficientemente chica como para que la interpretación en términos de dipolos cuyo efecto se cancela o duplica según el caso resulte una buena aproximación. Los momentos dipolares de un dipolo real y su dipolo imagen tienen una acción efectiva que puede pensarse como la suma o la resta de la acción individual, corregida por el hecho de que, si bien próximos, no están realmente superpuestos. Cuando la distancia entre cada partícula y su imagen sea demasiado grande en comparación con las otras escalas del sistema, la acción de los dipolos reales e imagen debe considerarse en detalle y la interpretación simplificada de suma y resta escalar de efectos que supone dipolos ubicados en un mismo punto ya no resulta válida.  

\section{Dinámica del entrelazamiento.}
En esta sección se estudia la dinámica del entrelazamiento en el sistema bipartito expuesto a las fluctuaciones del vacío cuántico poniendo el énfasis, otra vez, en la comparación entre el caso de espacio libre y el caso en que se incluye el contorno no-trivial del plano conductor infinito. En presencia del plano, se retoma el análisis de los dos casos de orientación para el momento dipolar de los qubits.
Entre las muchas medidas de entrelazamiento de estados mixtos motivadas en consideraciones físicas, el {\em entrelazamiento de formación} se propone cuantificar los recursos necesarios para crear un dado estado entrelazado, pero su cálculo implica una minimización sobre todas las descomposiciones posibles en estados puros, volviéndola una medida inconveniente en el caso general.

Sin embargo, para el caso particular de sistemas de dos niveles, la cantidad conocida como {\em concurrencia} se relaciona monótonamente con el entrelazamiento de formación y puede tomarse como una medida del entrelazamiento en sí misma \cite{wootters1998entanglement}.
La concurrencia de un estado $\rho(t)$ se anula si el estado es separable y se extiende monótonamente hasta un valor unitario para el caso de estados maximalmente entrelazados. Puede calcularse mediante la expresión

\begin{equation}
    \mathrm{C}(\rho) = \max(0, \sqrt{\lambda_1}-\sqrt{\lambda_2}-\sqrt{\lambda_3}-\sqrt{\lambda_4})
    \label{eq:sec4_concurrence}
\end{equation}
donde los $\lambda_i$ representan los autovalores de la matriz auxiliar $\Tilde{\rho}=\rho\,(\sigma_y\otimes\sigma_y)\rho^*\,(\sigma_y\otimes\sigma_y)$ en orden decreciente, los cuales están dados para la estructura particular que toma la matriz densidad considerada, por $\{\lambda_i\} =\{(\sqrt{\rho_{11}\,\rho_{44}} \pm |\rho_{41}|)^2, (\rho_{22} \pm |\rho_{23}|)^2 \}$.

\begin{figure}[ht!]
    \center
    \includegraphics[width = .495\linewidth]{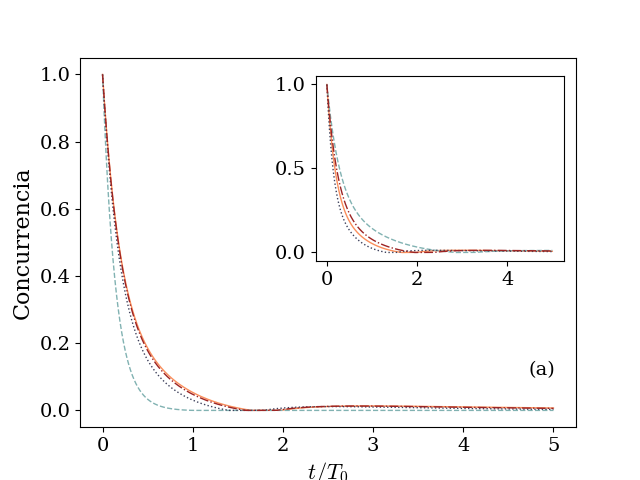}
    \includegraphics[width = .495\linewidth]{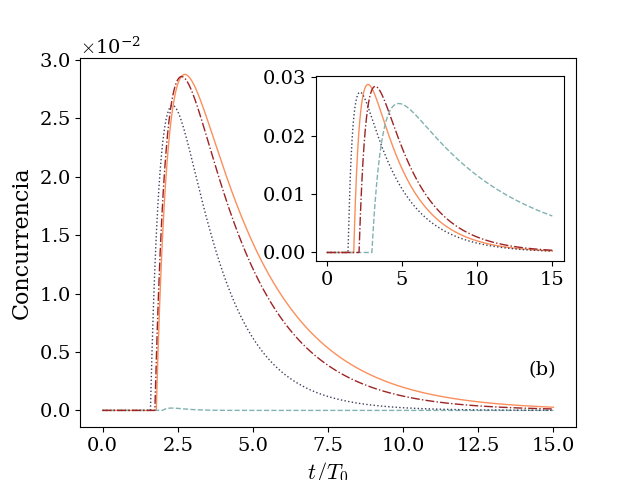}
    \caption{Evolución dinámica de la concurrencia para un estado inicial (a)  maximalmente entrelazado, con $\theta = \pi/2$ y (b) separable, con $\theta = 0$ en un entorno caracterizado por $\gamma_0/\omega = \alpha$, con $\alpha$ la constante de estructura fina. Se consideran distintas relaciones $d/L$ entre la distancia $d$ del sistema al plano conductor y la distancia $L$ entre las partículas que componen el sistema. Se incluye también el caso de espacio libre (líneas color naranja, sólidas) para comparación. Las líneas celestes de trazos, azules de puntos, y bordó de trazo/punto, representan la evolución para valores $d=L$, $d = 2\, L$, y $d = 4\, L$ respectivamente. Las figuras principales corresponden a partículas con polarización perpendicular al plano, mientras que los insets muestran el caso de partículas con polarización paralela al plano en dirección $\hat{x}$}. 
    \label{fig:sec4_concurrence}
\end{figure}

La figura \ref{fig:sec4_concurrence} presenta la evolución dinámica de la concurrencia para los casos extremos de un estado inicial maximalmente entrelazado y un estado inicial separable. El caso de estado inicial maximalmente entrelazado se muestra en el panel (a) mientras que el caso de estado inicial separable, en el panel (b). Se consideran distintas relaciones $d/L$ entre la distancia $d$ del sistema al plano conductor y la distancia $L$ entre las partículas que componen el sistema y se incluye, además, el caso de espacio libre para comparación. Para mayor consistencia con los resultados presentados en la sección anterior, se observan nuevamente las dos direcciones de polarización de las partículas en $\hat{y}$ y en $\hat{x}$, relevantes en presencia del plano conductor.\clearpage

Se observa en primer lugar el caso inicialmente entrelazado del panel (a). En ausencia de contornos, la concurrencia decae en una escala temporal $\sim 1.5 \,T_0$ y muestra, posteriormente un pequeño resurgimiento que alcanza un valor máximo mucho más bajo $\sim 0.015$ y decae asintóticamente. Cuando se incorpora el plano, el decaimiento en el entrelazamiento se comporta cualitativamente igual a las coherencias. Para distancias relativas $d/L$ en el régimen inicial, como el ejemplo $d = L$ representado con línea de trazos, el entrelazamiento se degrada respetando la misma relación entre escalas con que se degradan las coherencias en este régimen de partículas cercanas al plano. El efecto del entorno se acentúa si los momentos dipolares son perpendiculares al plano y se suaviza cuando son paralelos al mismo. A medida que la distancia relativa $d/L$ aumenta, el comportamiento tiende oscilatoriamente al observado para el caso de espacio libre.

En segundo lugar, se examina la dinámica del entrelazamiento cuando el sistema evoluciona a partir de un estado inicial separable, lo que se exhibe en el panel (b) de la figura \ref{fig:sec4_concurrence}. En todos los casos considerados se distingue un período inicial en el que la concurrencia permanece nula, lo que revela que el estado del sistema permanece separable, seguido de una generación súbita de entrelazamiento que muestra siempre la misma forma cualitativa de una cresta que alcanza un valor máximo y luego decae suavemente. Para el sistema en espacio libre, la generación súbita de entrelazamiento tiene lugar en el instante adimensional $\omega\,t\sim 1.8\,T_0$ y crece hasta valores máximos de concurrencia $\mathrm{C} = 0.03$ que decaen prácticamente por completo en una escala $\omega\,t= 15\,T_0$. En presencia del plano conductor, la evolución del entrelazamiento se modifica de distinta forma según las partículas tengan momento dipolar en una u otra de las direcciones consideradas. Para partículas polarizadas en dirección perpendicular al plano, la generación espontánea se observa en instantes similares para cualquier relación de la distancia relativa $d/L$ de las partículas al plano. Sin embargo, los valores máximos de concurrencia alcanzados difieren considerablemente, acercándose al caso de espacio libre para $d=4\,L$ y decreciendo monótonamente conforme se reduce la distancia al plano hasta mostrar un valor prácticamente nulo para $d=0.5\,L$. Por otra parte, para dipolos orientados en dirección $\hat{x}$ paralela al plano y perpendicular a la dirección de separación entre partículas, la generación tiene lugar en instantes adimensionales distintos según sea $d/L$. Específicamente, la generación de entrelazamiento se posterga para distancias relativas en el régimen inicial como $d=L$ y recupera asintóticamente el resultado de espacio libre, oscilando en torno al mismo, cuando $d/L\rightarrow\infty$. Una vez generado el entrelazamiento, éste decae en una escala de tiempo adimensional más prolongada cuanto menor sea la distancia relativa del sistema al plano.

La existencia de resurgimientos y generación de entrelazamiento no debe considerarse un signo de no-markovianidad. En particular, hemos descartado toda contribución de efectos de memoria del entorno al hacer la aproximación de Markov. Este tipo de comportamientos en la evolución del entrelazamiento es consecuencia exclusiva de la dinámica colectiva inducida por presencia del entorno \cite{ficek2003entanglement, ficek2006dark, franco2013dynamics}. El campo irradiado por emisión espontánea de una de las partículas influye sobre la dinámica de la otra mediante el campo de vacío sin que haya retorno de información desde el campo de vacío al mismo qubit.

\section{Fase geométrica}
Por último, se describe la evolución del sistema desde la inspección de la fase geométrica acumulada por su estado, repitiendo las condiciones particulares consideradas en las observaciones de las coherencias y el entrelazamiento con el objetivo de componer un estudio completo de cada uno de dichos escenarios y establecer conexiones y diferencias entre los distintos aspectos indagados.

Cuando el sistema estudiado está acoplado a un entorno como en este caso, se tendrá en general una fase geométrica $\phi_g$, calculada siguiendo la definición (\ref{eq:sec2_TongGP}), que se modifica respecto del valor obtenido en una hipotética evolución unitaria, obteniendo $\phi_g = \phi_u + \delta\phi$.
Recordamos que esta fase geométrica, aplicable al escenario de estados mixtos que evolucionan de
forma no-unitaria, se reduce a la fórmula (\ref{eq:sec2_TongGPpuros}) cuando el estado inicial del sistema es, como en este caso, un estado puro. 
El autoestado de $\rho(t)$ involucrado en el cálculo de la fase geométrica puede expresarse como

\begin{equation}
    \ket{\psi_+(t)} = \frac{-(\rho_{44} - \epsilon_{+})\ket{++} + \rho_{41}\ket{--}}{\left((\rho_{44}-\epsilon_{+})^2 + \rho_{41}\rho_{14}\right)^{1/2}},
\end{equation}
donde $\epsilon_{+} = \frac{1}{2}\bigl(\rho_{11} + \rho_{44}+((\rho_{11}-\rho_{44})^2 + 4|\rho_{41}|^2)^{1/2}\bigr)$ es el autovalor del operador densidad asociado a este autoestado, y satisface $\epsilon_+(0) =1$, mientras que los autovalores restantes se anulan a tiempo inicial.
Mediante el reemplazo de la expresión formal para $\ket{\psi_+(t)}$ en términos de los elementos de matriz $\rho_{ij}$, la ecuación (\ref{eq:sec2_TongGPpuros}) toma la forma 

\begin{equation}
    \phi_g = \arg(\bra{\psi_+(0)}\ket{\psi_+(t)})- \int_0^{t}\,dt'\,\frac{2(\omega + c_{11})\,|\rho_{41}|^2}{(\rho_{44}-\epsilon_{+})^2 + |\rho_{41}|^2}
    \label{eq:sec4_GP4}
\end{equation}

\begin{figure}[ht!]
    \center
    \includegraphics[width = .495\linewidth]{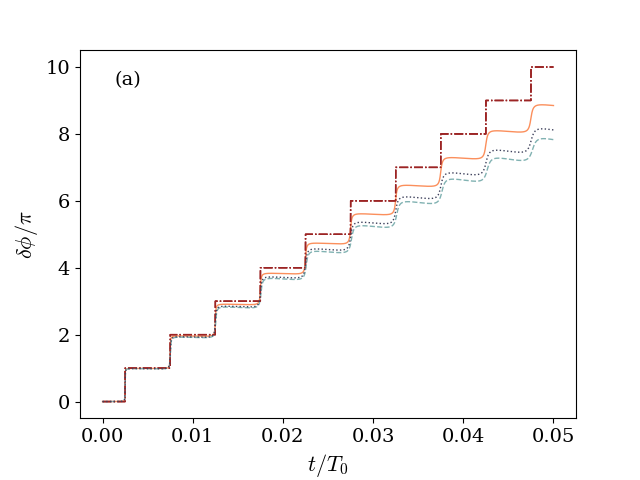}
    \includegraphics[width = .495\linewidth]{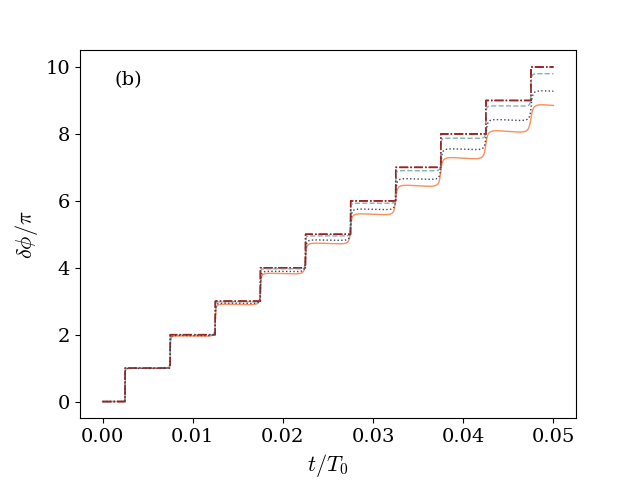}
    \caption{Fase geométrica acumulada a lo largo de la evolución para un sistema cuyas partículas tienen momento dipolar (a) perpendicular al plano y (b) paralelo al plano en dirección del eje $x$, considerando distintas relaciones $d/L$ entre la distancia $d$ del sistema al plano conductor y la distancia $L$ entre partículas. El sistema parte de un estado inicial definido por $\theta = \pi/2$, y el efecto del vacío sobre el mismo está caracterizado por un parámetro $\gamma_0 = 5\times10^{-3}\omega$. Para permitir comparación, se incorpora en los dos gráficos el caso de espacio libre (líneas naranja sólidas) y el caso de evolución unitaria (líneas bordó de trazo-punto), mientras que las líneas celeste de trazos, y azules de puntos, representan la evolución para valores $d=L$ y $d = 0.5\, L$ respectivamente.} 
    \label{fig:sec4_GP}
\end{figure}

\vspace{.2cm}
{\em Evolución dinámica de la fase geométrica - }Se estudia, en primer lugar, la fase geométrica $\phi_g$ acumulada en la evolución dinámica del sistema abierto, y su diferencia $\delta\phi = \phi_g - \phi_u$ respecto de aquella adquirida en la evolución unitaria. En la figura \ref{fig:sec4_GP} se exhibe la fase geométrica acumulada por un estado inicial maximalmente entrelazado ($\theta = \pi/2$). Una vez más, se consideran los casos en que el momento dipolar de las partículas se orienta perpendicular al plano conductor, o paralelo al mismo en dirección $\hat{x}$. Estos dos escenarios se presentan en los paneles  (a) y (b) respectivamente. En cada caso, se muestran dos valores distintos para la relación $d/L$, acompañadas por los casos de espacio libre y de evolución unitaria.
\clearpage

En el caso unitario, en que los dos qubits evolucionan independientemente según el Hamiltoniano (\ref{eq:sec4_Hs}), los escalones corresponden al doble de la bien estudiada fase geométrica de un qubit que traza una trayectoria circular en la esfera de Bloch, resultando $\phi_u = 2\pi(1-\cos(\theta))$ la fase acumulada en un período natural $T_{\rm s}$ de los qubits. Dicho período resulta mucho menor a la escala temporal de decaimiento $T_0$, esto es, satisface la condición $T_{\rm s}\ll T_0$ supuesta en las aproximaciones. 

El efecto del vacío cuántico sin modificar y sus fluctuaciones degrada la fase geométrica que pasa de mostrar escalones perfectos a describir una curva más suave de crecimiento más lento. Eventualmente, el entorno provoca decaimiento total de las coherencias (ver figuras \ref{fig:sec4_coherencias}) y la emergencia de un estado estacionario que detiene su marcha en el espacio de rayos y, en consecuencia, la acumulación de fase geométrica. La incorporación del plano infinito conductor tendrá, una vez más, efectos inversos dependiendo de la dirección de polarización de los qubits. En el panel (a) de la figura \ref{fig:sec4_GP} se observa que, para el caso de polarización perpendicular al plano, el efecto degradante de las fluctuaciones de vacío sobre la fase geométrica se acentúa a medida que la distancia $d$ de las partículas al plano se acorta. Por el contrario, el panel (b) revela que cuando los dipolos se orientan en dirección paralela al plano y perpendicular a la separación entre partículas, el efecto del entorno se atenúa conforme las partículas se acercan al plano, sin superar, sin embargo, el límite establecido por el caso unitario.

Bajo determinadas condiciones, el comportamiento de la fase geométrica puede tratarse analíticamente. En el límite de acoplamiento débil que se considera en este capítulo, es posible realizar una expansión en términos de $\gamma_0/\omega$ que a segundo orden toma, para $t=T_{\rm s}$, la expresión

\begin{align}
    \phi_g(T_{\rm s}) \sim&\, -2\pi(1-\cos\theta)-  4\pi^2\sin^2\theta \frac{a_{11}}{\omega}\label{eq:sec4_GPexpan}\\ \nonumber
    &- \frac{16\pi^3}{3}\sin^2\theta\left((a_{11}^2 +a_{12}^2)(1+\cos\theta) +2\,a_{11}^2\cos\theta\right)\frac{1}{\omega^2}.
\end{align}
El primer término en la expansión corresponde a la fase geométrica $\phi_u$ acumulada durante una evolución unitaria. Resulta notable que la corrección de primer orden para el caso de acoplamiento débil, dependa únicamente del coeficiente $a_{11}$. Esto implica que la presencia de contornos no-triviales en el entorno tiene en la fase geométrica un efecto dominante por sobre la presencia del otro qubit, con la distancia $L$ entre qubits incorporándose al orden siguiente. 

El análisis puede extenderse reemplazando la forma explícita de los coeficientes $a_{lm}$ en la expansión (\ref{eq:sec4_GPexpan}) que, a primer orden en $\gamma_0/\omega$ resulta

\begin{equation}
    \delta\phi(2\,\pi/\omega) \sim -  4\pi^2\sin^2\theta\left(\frac{1}{3}-\sum_{n=x,y,z}(r^1_n)^2\, f_{2,11} \right)\frac{\gamma_0}{\omega},
    \label{eq:sec4_dGPexpan}
\end{equation}
donde se observa que la corrección a primer orden tiene una componente $-  4/3\,\pi^2\sin^2\theta$ de puro vacío y una segunda contribución producto de la presencia del plano conductor. La dependencia $\sin^2(\theta)$ en el estado inicial ocasiona una corrección máxima para los estados maximalmente entrelazados que decrece hasta anularse para estados separables.

\begin{SCfigure}[][ht!]
    \includegraphics[width = .5\linewidth]{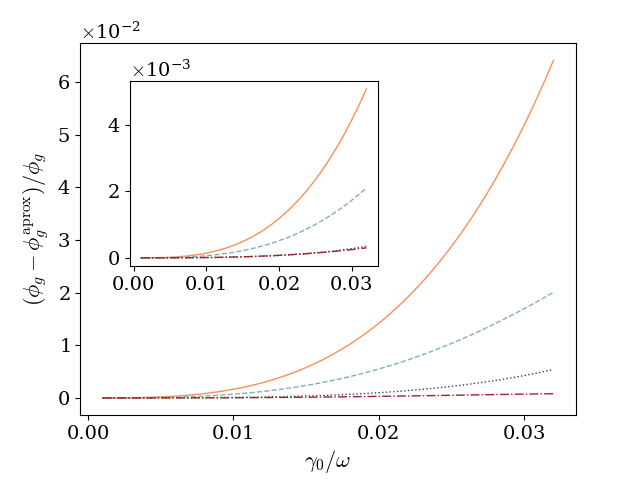}
    \caption{Diferencia relativa $(\phi_g-\phi_g^{\rm aprox})/\phi_g$ entre la fase geométrica (\ref{eq:sec4_GP4}) y el valor aproximado (\ref{eq:sec4_GPexpan}), como función del parámetro $\gamma_0/\omega$ de la expansión, y para una relación $d=L$ entre distancias. Los dos casos de polarización perpendicular al plano y paralela en $\hat{x}$ se presentan en la figura principal y en el inset respectivamente. La línea naranja sólida corresponde al estado separable con $\theta =0$, mientras que las líneas celeste de trazos, azul de puntos, y bordó de trazo-punto, corresponden a estados iniciales con $\theta = \pi/4$, $\pi/2$, y $3\pi/4$ respectivamente. } 
    \label{fig:sec4_dgp}
\end{SCfigure}

Con el objetivo de adquirir una noción del rango de validez de la aproximación, se estudia la diferencia relativa $(\phi_g-\phi_g^{\rm aprox})/\phi_g$ entre la fase geométrica $\phi_g$ calculada mediante la ecuación (\ref{eq:sec4_GP4}) y el valor aproximado $\phi_g^{\rm aprox}$ dado por (\ref{eq:sec4_GPexpan}), como función del parámetro $\gamma_0/\omega$ de expansión, que caracteriza la influencia del entorno sobre el sistema. El resultado de esta comparación de exhibe en la figura \ref{fig:sec4_dgp}. Los dos casos de polarización perpendicular al plano y paralela en $\hat{x}$ se presentan en  en la figura principal y en el inset respectivamente.

Para esta relación $d = L$ entre distancias, la expansión (\ref{eq:sec4_GPexpan}) reproduce el comportamiento de la fase geométrica con un error relativo menor al $10\%$ para valores del parámetro del orden de $\gamma_0/\omega\sim 10^{-2}$. La influencia del contorno en las fluctuaciones de vacío genera variaciones en el rango de $\gamma_0/\omega$ en que la aproximación resulta válida, dependiendo de la distancia del sistema al plano conductor. Por otro lado, el error relativo muestra además una dependencia en el estado inicial del sistema y se incrementa con la probabilidad a tiempo $t=0$ del estado excitado, esto es, aumenta cuanto mayor sea la probabilidad inicial para dicho estado. 

\vspace{.5cm}
\begin{center}
   \textcolor{bordo}{\ding{163}}
\end{center}
\vspace{.5cm}
En resumen, en este capítulo se estudió la dinámica de un sistema compuesto por dos qubits sujeto a los efectos de las fluctuaciones cuánticas del vacío electromagnético tanto en espacio libre como en presencia de un plano infinito conductor. Se observaron tres aspectos particulares, a saber, la pérdida de coherencia, el entrelazamiento entre los dos qubits y la fase geométrica acumulada por el estado del sistema bipartito.

Se definió una escala de decoherencia $T = \pi/\gamma
_0$ asociada a la pérdida de coherencia para el sistema ubicado en el espacio libre y se encontró que existe una jerarquía entre las escalas asociadas a los distintos casos en presencia del plano infinito de material conductor. Específicamente, dicha jerarquía puede expresarse como $T_\parallel > T_0> T_\perp$, donde $T_{\parallel, \perp}$ refiere a la escala de decoherencia para partículas con momento dipolar paralelo y perpendicular al plano conductor respectivamente. 
Este resultado se interpretó en términos de interacciones entre partículas reemplazando los efectos del contorno no-trivial por partículas imagen, concluyendo que el caso en que los dipolos se orientan paralelos al plano origina un entorno más destructivo que conduce a escalas de decoherencia más cortas.
La dependencia de estas escalas con la relación $d/L$ entre la distancia de las partículas al plano y la distancia que separa ambas partículas también fue investigada.

El entrelazamiento entre los qubits que componen el sistema se estudió a través de la concurrencia, analizando la dependencia en la relación $d/L$ y en la orientación de los momentos dipolares. Se puso especial atención en los efectos de muerte y generación espontánea de entrelazamiento, y re-nacimientos del entrelazamiento. Estos fenómenos se interpretaron en términos de interacciones efectivas (esto es, mediadas por el entorno) entre partículas recurriendo al método de imágenes.

Finalmente se consideró la fase geométrica acumulada por el sistema bipartito para un estado inicial maximalmente entrelazado, con particular foco en las modificaciones introducidas por la presencia del entorno en los distintos casos considerados para el estudio de las coherencias y el entrelazamiento. En el límite de acoplamiento débil es posible encontrar una expansión analítica en órdenes del parámetro de acoplamiento que permite reconocer y distinguir las contribuciones del espacio libre y del plano conductor.
\chapter{Fase geométrica y fricción cuántica}\label{ch:5}
Una de las ideas más excitantes que introduce la teoría cuántica de campos es la de que el estado de vacío posee una estructura no-trivial, repleto de partículas de corta vida que aparecen y desaparecen constantemente, llenando el vacío con una {\em energía del punto cero} no nula.
A pesar de las controversias iniciales que despertó esta idea, la evidencia experimental indicando que este vacío vivo tiene consecuencias observables se acumuló en diversos ejemplos como el corrimiento de Lamb \cite{lamb1947fine} y la fuerza de Casimir \cite{casimir1948attraction}. 

En el capítulo \ref{ch:4} se consideró el efecto del vacío cuántico del campo electromagnético sobre el grado de entrelazamiento mostrado por un sistema bipartito y sobre la fase geométrica acumulada por su estado. Las modificaciones en las fluctuaciones de vacío introducidas por la inclusión de contornos no-triviales se trató para el caso particular de un plano infinito de espesor despreciable y material conductor perfecto. Si en tal escenario el sistema se hubiera puesto en movimiento paralelo al plano conductor, esta situación hubiera resultado equivalente al caso estático estudiado, puesto que en un conductor perfecto la carga de redistribuye instantáneamente.

En una propuesta diferente, en este capítulo se compilan y desarrollan los resultados presentados en \cite{viotti2021enhanced, lombardo2022detecting, viotti2019thermal}, y se reemplazan los contornos perfectos por materiales semiconductores que permitan acceder al estudio de los efectos del entorno sobre un sistema en movimiento a velocidad constante. 

\section{Sobre el fenómeno de fricción cuántica}\label{sec:sec5_friccion}
En 1948 Hendrick Casimir demostró que el origen de la fuerza atractiva que tiene lugar entre dos placas paralelas perfectamente conductoras debe buscarse en las fluctuaciones cuánticas del vacío \cite{casimir1948attraction}. Mientras que se han alcanzado numerosas comprobaciones experimentales de este efecto \cite{milton2001casimir, milton2004casimir, lamoreaux1997demonstration, bordag2009advances}, otro fenómeno del mismo origen conocido como {\em Fricción cuántica} elude hasta el momento la detección, lo que se debe principalmente a su rango corto y pequeña magnitud. 
\\
\\\indent
{\em Imagen pictórica - }Empezamos con una descripción introductoria del curioso fenómeno de la Fricción cuántica. Suponiendo que una partícula neutra se encuentra rodeada por un campo electromagnético en estado de vacío, las fluctuaciones en el vacío cuántico inducirán un momento dipolar eléctrico fluctuante $\boldsymbol{\mu}$ en la partícula. 

\begin{figure}[ht!]
    \center
    \subfloat[Dipolo estático]{\includegraphics[height = 4cm]{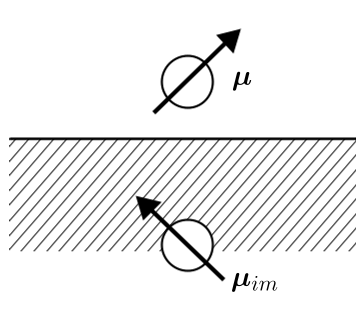}}
    \hspace{1cm}
    \subfloat[Dipolo en movimiento]{\includegraphics[height = 4.15cm]{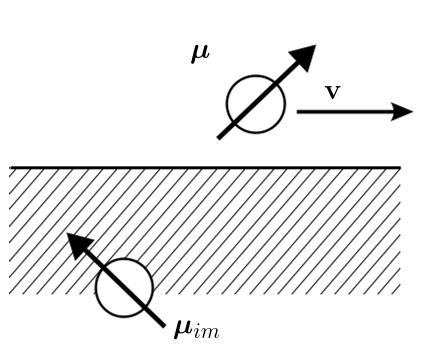}}
    \caption{Representación esquemática de un partícula neutra rodeada por el campo electromagnético en estado de vacío. Si esta partícula se encuentra en las inmediaciones de una superficie conductora neutra, los efectos sobre ella pueden describirse con un dipolo imagen fluctuante en la región inaccesible del espacio. En el caso estático (a) estos efectos corresponden a la Fuerza de Casimir. En el caso dinámico (b) aparece una contribución adicional a esta fuerza paralela a la superficie, la Fricción cuántica.} 
    \label{fig:sec5_Friction}
\end{figure}

Si esta partícula se posiciona en las inmediaciones de una superficie conductora neutra, el campo inducido por el material macroscópico puede describirse, en la región accesible del espacio, con un dipolo imagen fluctuante en la posición espejada del dipolo real. Esta es la situación que se muestra esquemáticamente en el panel (a) de la figura \ref{fig:sec5_Friction}. Cuando el momento dipolar de la partícula real oscila, el dipolo imagen oscila también, lo que resulta en un efecto neto de atracción que empuja la partícula hacia el plano. Esta fuerza inducida por las fluctuaciones se conoce como fuerza de Van der Waals o de Casimir–Polder. 

Por otro lado, las condiciones pueden modificarse poniendo a la partícula en movimiento relativo con respecto a la superficie, lo que se muestra esquemáticamente en el panel (b) de la figura \ref{fig:sec5_Friction}. En este caso las cargas en la superficie conductora se desplazan también, es decir, el dipolo espejado sigue a la partícula real. Si la superficie es un conductor perfecto, las cargas en la superficie se redistribuyen instantáneamente y la situación es completamente análoga al caso estático. Sin embargo, si la partícula viaja por encima de un conductor imperfecto o dieléctrico, el dipolo imagen se retrasa producto de la resistividad del material. Los momentos dipolares fluctuantes real e imagen dan origen a una fuerza con una componente adicional  paralela a la superficie: la Fricción cuántica. 

\subsection{Cálculo de la fuerza mediante un formalismo funcional}
En el marco de esta tesis, e inspirado en el tratamiento de efectos disipativos propuesto en~\cite{fosco2011quantum, Fariasfuncionalcamino, farias2015functional}, se calcula la fuerza disipativa que actúa sobre una partícula en movimiento uniforme y paralelo a una lámina dieléctrica~\cite{viotti2019thermal}. 
El examen se realiza para un modelo simplificado, diferente de aquél estudiado en las secciones \ref{sec:sec5_modelo} a \ref{sec:sec5_exp}. Con el objetivo de identificar modificaciones en la fuerza producidas por la temperatura se utiliza el formalismo de integral de camino cerrada \cite{Schwinger, Keldysh}. A continuación, se describen los pasos principales del desarrollo utilizado para el cálculo de la fuerza inducida por el movimiento relativo.

Considérese una partícula neutra cuyo centro de masa describe una trayectoria rectilínea uniforme de velocidad $v$, paralela a una placa dieléctrica, a una distancia $d$ por sobre la misma. La trayectoria de la partícula se considera determinada por agentes externos. Se considera que la placa dieléctrica se extiende infinitamente en dirección de los ejes $x$ e $y$, ubicada en $z=0$, y que su espesor resulta despreciable. Sin pérdida de generalidad, se supone además que el movimiento de la partícula es en dirección del eje $x$. El conjunto completo se encuentra inmerso en un campo cuántico.
La acción clásica para este sistema puede escribirse, explicitando el origen de las distintas contribuciones, como $S = S^{\rm vac}_0 + S^{\rm pl}_0 + S^{\rm part}_0 + S^{\rm pl}_{\rm int} + S^{\rm part}_{\rm int}$. 
Los primeros tres términos corresponden a las acciones libres del campo de vacío, de la placa y de la partícula respectivamente, mientras que los últimos dos términos representan interacciones entre el campo escalar y la placa, y entre el campo y la partícula. 

Como se ha anticipado, se propone un modelo simplificado en el que el grado de libertad interno de la partícula se describe mediante un oscilador armónico de frecuencia $\omega$, a la vez que el plano dieléctrico, posicionado en $z=0$, se modela como un conjunto de infinitos osciladores armónicos no-interactuantes de idéntica frecuencia $\Omega$, descritos por coordenadas generalizadas $Q(t,\mathbf{x})$. La acción libre que describe una placa de este tipo resulta

\begin{equation}
    S_0^{\rm pl} = \frac{1}{2}\int d^3x\,\delta(z)\left[\dot{Q}(t,\mathbf{x})-(\Omega^2-i\epsilon)Q^2(t,\mathbf{x})\right].
\end{equation}
También el campo electromagnético es reemplazado en este modelo simplificado, considerando que la partícula y la placa se encuentran inmersas en un campo escalar cuántico, cuya acción libre es la acción de Klein-Gordon. 
Por otra parte, el acoplamiento entre la placa y el campo se considera lineal y local, descrito mediante la acción de interacción

\begin{equation}
    S_{\rm int}^{\rm pl} = \lambda\int d^3x\,\delta(z)\,Q(t,\mathbf{x})\phi(t,\mathbf{x}),
\end{equation}
donde $\lambda$ da cuenta de la intensidad de la interacción entre el campo y el plano y $\phi(t,\mathbf{x})$ representa los grados de libertad del campo.
El acoplamiento entre éste y la partícula se describe mediante una corriente $J$ según

\begin{equation}
    S_{\rm int}^{\rm part} = \int d^3x \,\phi(t,\mathbf{x})\,J(t,\mathbf{x}),
\end{equation}
donde la corriente $J(t,\mathbf{x})= g\,q(t)\,\delta(x-v\,t)\,\delta(y)\,\delta(z-d)$ modela una interacción local de intensidad $g$ entre la coordenada generalizada $q(t)$, que describe el grado de libertad interno de la partícula, y el campo.
\\
\\\indent
{\em Propagador efectivo del campo - }El objetivo propuesto es, entonces, hallar una expresión para la fuerza disipativa que actúa sobre la partícula para el caso de temperatura no-nula, específicamente, se considera el caso de {\em equilibrio térmico}, en el cual todos los elementos del sistema se encuentran a la misma temperatura. Esta tarea puede abordarse utilizando el formalismo de Schwinger-Keldysh, también conocido como formalismo de integrales funcionales {\em in-in}, o de integrales cerradas, 
y se considera un contorno de integración específico denominado Contorno de Kadanoff-Baym~\cite{Das_temp, calzetta1988nonequilibrium, calzetta2009nonequilibrium}. Es posible recuperar una representación {\em de tiempo único} duplicando los grados de libertad del sistema. Con este procedimiento se obtiene una funcional generatriz que toma la forma familiar 

\begin{equation}
    \mathcal{Z}_{\rm in-in}[J] = \int \mathcal{D}\phi\,e^{-i\int\,dx\left(\phi\,\frac{K}{2}\,\phi - \phi\,J\right)},
\end{equation}
aunque debe hacerse la salvedad de que, en contraste con la funcional generatriz usual {\em in-out}, el operador diferencial $K\in \mathbb{C}^{2\times2}$ es, en este caso, una matriz. La integral sobre el campo puede hacerse explícitamente, para obtener la conocida expresión 

\begin{equation}
    \mathcal{Z}_{\rm in-in}[J] = \int \mathcal{D}\phi\,e^{-\frac{1}{2}\int\,dx\,dx'\;J_\alpha(t,\mathbf{x})G_{\alpha\beta}(t,\mathbf{x}, t',\mathbf{x'})J_\beta(t',\mathbf{x'})},
\end{equation}
donde se asume sumatoria sobre los índices repetidos, y el propagador de campo libre $G(t,\mathbf{x},t',\mathbf{x'}) \in \mathbb{C}^{2\times2}$ es una matriz con elementos $G_{\alpha\beta}(t,\mathbf{x},t',\mathbf{x'})$.

Integrando sobre los grados de libertad de la partícula y de la placa, se obtiene una funcional que contiene toda la información sobre éstas en una contribución efectiva a la acción del campo $S^{\rm vac}_{\rm int}$ y puede escribirse como

\begin{equation}
    \mathcal{Z}_{\rm in-in} = \int\mathcal{D}\phi\;e^{-\frac{i}{2}\int \,dx\,dx'\, \phi(t,\mathbf{x})\,K'(t,\mathbf{x},t',\mathbf{x'})\,\phi(t',\mathbf{x'})},
\end{equation}
donde $K'(t,\mathbf{x},t',\mathbf{x'}) = K(t,\mathbf{x},t',\mathbf{x'})\delta(\mathbf{x}-\mathbf{x'})\delta(t-t')  - V^{\rm pl}(t,\mathbf{x},t',\mathbf{x'}) - V^{\rm part}(t,\mathbf{x},t',\mathbf{x'})$ es el operador diferencial efectivo para el campo escalar, modificado respecto del operador diferencial $K(t,\mathbf{x},t',\mathbf{x'})$ de campo libre considerado anteriormente por los potenciales $V^{\rm pl/part}(t,\mathbf{x},t',\mathbf{x'})$, que dan cuenta de la presencia de, y de la interacción con la partícula y con la placa. 

Será necesario para el cálculo de la fuerza encontrar la matriz de funciones de correlación del campo $G(t,\mathbf{x},t',\mathbf{x'})$ que satisface

\begin{equation}
    K'(t,\mathbf{x},t',\mathbf{x'})G'(t,\mathbf{x},t',\mathbf{x'}) = \delta(\mathbf{x}-\mathbf{x'})\delta(t-t'),
\end{equation}
para el caso en que el campo está {\em vestido} por la partícula y el plano dieléctrico. Para esto, es útil recordar que los potenciales son proporcionales a constantes de acoplamiento $g$ y $\lambda$ entre el campo electromagnético y la partícula, y entre el campo y la placa, respectivamente. En el límite de acoplamiento débil es posible realizar una expansión perturbativa para el propagador matricial del campo, en la que el elemento $G'_{\alpha\beta}$ puede escribirse, en términos de los correladores $G_{\alpha\beta}$ del campo libre y de los potenciales, según

\begin{equation}
    G'_{\alpha\beta} = G_{\alpha\beta} + G_{\alpha\gamma}\,V^{\rm pl}_{\gamma\delta}\,G_{\delta\mu}\,V^{\rm part}_{\mu\nu}\,G_{\nu\beta} + G_{\alpha\gamma}\,V^{\rm part}_{\gamma\delta}\,G_{\delta\mu}\,V^{\rm pl}_{\mu\nu}\,G_{\nu\beta} + \mathcal{O}(g/\lambda)^4,
    \label{eq:sec5_feynman}
\end{equation}
donde las integrales involucradas en la contracción fueron omitidas para simplificar la notación.
\\
\\\indent
{\em Cálculo de la fuerza - }
Si la partícula y la placa se encuentran en movimiento relativo, existe una transferencia de energía hacia el campo que se evidencia en el hecho de que la acción efectiva del campo escalar adquiere una componente imaginaria positiva, la cual da cuenta de la posibilidad de crear excitaciones de $\phi(t,\mathbf{x})$. La conservación de la energía implica que debe existir alguna fuerza realizando trabajo mecánico cuando la partícula se mueve.
Para encontrar la expresión analítica de la fuerza entre la partícula y el plano dieléctrico, se considera el valor de expectación del tensor de energía-momento en vacío y en el régimen estacionario $\expval{t_{\mu\nu}} = \bra{0_{\rm in}}t_{\mu\nu}\ket{0_{\rm in}}$. La fuerza disipativa que el campo efectivo ejerce sobre la partícula puede obtenerse mediante la técnica de división de puntos, integrando el valor de expectación de la componente $t_{xz}$ en la dirección de movimiento de la partícula en un cubo en torno a la misma. Las dimensiones del cubo se toman luego tendiendo a cero para obtener la fórmula

\begin{equation}
    F = \lim_{z\rightarrow d^+}\expval{t_{xz}(t,x)}-\lim_{z\rightarrow d^-}\expval{t_{xz}(t,x)},
\end{equation}
donde el valor de expectación

\begin{equation}
    \expval{t_{xz}(t,x)} = \lim_{x'\rightarrow x}\bra{0_{\rm in}}\partial_x\phi(t,\mathbf{x})\,\partial'_z\phi(t',\mathbf{x}')\ket{0_{\rm in}}
\end{equation}
puede expresarse como una integral en términos del propagador de Feynman $G_{++}(t,\mathbf{x},t',\mathbf{x'}) = \bra{0_{\rm in}}\mathcal{T}\phi(t,\mathbf{x})\,\phi(t',\mathbf{x}')\ket{0_{\rm in}}$. Expandiendo además el propagador de Faynman siguiendo la ecuación (\ref{eq:sec5_feynman}), es posible calcular cada contribución a la fuerza exactamente.

Para la expresión hallada en el caso más general, se refiere al trabajo original. La fuerza de fricción cuántica, en particular, corresponde al efecto para el caso específico de temperatura nula. En consecuencia, el escenario abordado en la literatura previa es recuperado tomando límite de velocidades no-relativistas y a temperatura nula. En este caso, la magnitud de la fuerza hallada en \cite{viotti2019thermal} resulta

\begin{equation}
    F_{fric} = \frac{\lambda^2g^2}{16\,}\frac{\omega + \Omega}{\omega\,\Omega}\int\frac{dk}{2\pi}\;\frac{\exp\left(-2\frac{d}{v}\sqrt{v^2(\Omega^2 - k^2) - (\omega + \Omega)^2}\right)}{v^2(\Omega^2 - k^2) - (\omega + \Omega)^2},
    \label{eq:sec5_friccionthermal}
\end{equation}
donde se recuerda que $\omega$ y $\Omega$ son las frecuencias características de la partícula y de los infinitos (e idénticos) osciladores que componen la lámina dieléctrica. Con respecto a la expresión (\ref{eq:sec5_friccionthermal}), vale la pena mencionar que el resultado se obtuvo trabajando en unidades de $\hbar = c = 1$, de modo que la velocidad $v$ resulta adimensional, mientras que el resto de los parámetros satisface $[\lambda] = m^{3/2}$, $[g] = m^{1/2}$, $[d] = m^{-1}$, y las frecuencias $[\omega] = [\Omega] = [k] = m$.

\vspace{.5cm}
{\em Pluralidad de expresiones - } La ausencia de verificación experimental ha permitido que diferentes autores obtengan resultados diversos para la fuerza de fricción entre un átomo y una superficie a temperatura nula, los cuales muestran diferentes predicciones respecto de la dependencia en la velocidad del átomo y en la distancia que separa el átomo de la placa.  
La discrepancia se debe en gran medida al uso de múltiples formalismos que involucran distintas aproximaciones y en muchos casos resultan incompatibles, para realizar el cálculo de esta fuerza dependiente de la velocidad. Los formalismos utilizados incluyen la teoría de la respuesta lineal~\cite{resultados1, resultados3}, aproximación de Born-Markov~\cite{resultados4}, teoría de perturbaciones dependientes del tiempo~\cite{resultados8, barton2010van, klatt2017quantum}, el Teorema de fluctuación-disipación~\cite{resultados6,resultados7}, y principios termodinámicos~\cite{resultados9}.
La simultaneidad de abordajes teóricos incompatibles contribuyendo a la controversia en la comunidad de la física de Casimir sólo puede solucionarse con la detección experimental de la fuerza. Por este motivo, en un intento de superar los desafíos involucrados en la implementación de mediciones precisas para la observación de una fuerza de tan pequeña magnitud actuando sobre objetos tan cercanos a una superficie, han surgido una serie de trabajos orientados a determinar las condiciones más favorables para la detección. 

Otros trabajos han seguido un camino alternativo: rastrear indicios de fricción cuántica en otros efectos del vacío cuántico que muestran dependencia en la velocidad pero resulten más accesibles. Siguiendo estas ideas, {\em en el presente capítulo, se estudian la pérdida de coherencia y la fase geométrica de un átomo neutro en movimiento no-relativista con respecto a una superficie dieléctrica.} Se caracteriza el efecto del entorno con énfasis en identificar el comportamiento inducido exclusivamente por el movimiento relativo de la partícula. 

\section{Modelo microscópico}\label{sec:sec5_modelo}
En una elección diferente respecto de aquella realizada en ejemplo de cómputo de la sección anterior, el estudio presentado en lo que sigue propone un modelo notablemente realista. Se considera una partícula neutra que se traslada en un campo electromagnético, el cual se halla en estado de vacío. La partícula se modela como un sistema de dos niveles cuyo centro de masa sigue una trayectoria externamente determinada $\mathbf{x}_{\rm s}(t) = {\rm v}\,t\,\hat{x} + d\,\hat{z}$ a una distancia fija $d$ de un medio dieléctrico semi-infinito de superficie plana.

\begin{SCfigure}[][!ht]
    \includegraphics[width = .55\linewidth]{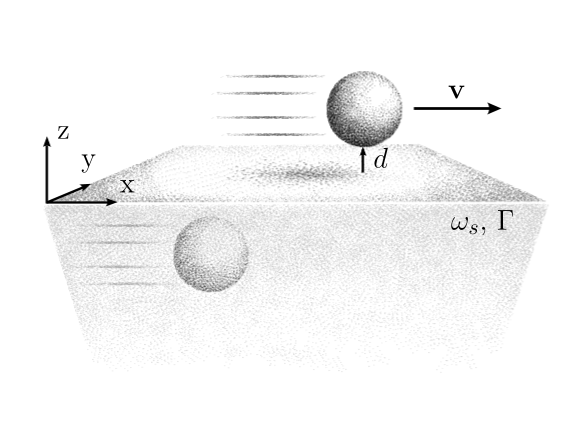}
    \caption{Esquema del escenario considerado. Un sistema de dos niveles (llamado también átomo o partícula) inmerso en el estado de vacío del campo electromagnético, y en movimiento rectilíneo uniforme a velocidad $|\mathbf{v}| = {\rm v}$ y a distancia fija $d$ de un medio dieléctrico. El medio dieléctrico ocupa todo el semiespacio inferior $z<0$ y se describe, según el modelo de Drude-Lorentz, mediante la frecuencia de los plasmones de superficie $\omega_s$ y la tasa de disipación $\Gamma$.} 
    \label{fig:sec5_esquema}
\end{SCfigure}
La dinámica del conjunto completo está generada por un Hamiltoniano $H = H_{\rm s} + H_{\rm em} + H_{\rm int}$, donde el término 

\begin{equation}
    H_{\rm s} = \frac{\omega_{\rm o}}{2}\,\sigma_z,
\end{equation}
con $\omega_{\rm o}$ la diferencia de energía entre los dos niveles, describe una hipotética evolución libre del átomo. El término $H_{\rm em}$ es el Hamiltoniano del campo electromagnético en ausencia del átomo, pero considerando el medio dieléctrico que cubre el espacio $z\leq0$. $H_{\rm int}$ da cuenta de la interacción entre el campo y la partícula, que en la aproximación dipolar, o de longitud de onda larga, está dada por $H_{\rm int} = - \boldsymbol{\mu}\cdot\mathbf{E(\mathbf{x}_{\rm s})}$ y depende explícitamente del tiempo a través de la posición de la partícula, que se trata como una variable clásica contando con que su incerteza resulte irresoluble para la longitud de onda característica del campo.

El estudio se restringe, además, al régimen no-retardado o de campo cercano en el cual la distancia entre la partícula y la superficie es suficientemente pequeña para satisfacer la condición $d\,\omega_{\rm o}\ll1$ (en unidades de $c =1$). En este régimen el tiempo finito que toma a un fotón ser reflejado por la superficie y alcanzar nuevamente la partícula es despreciable en comparación con su escala temporal natural. Como consecuencia, el Hamiltoniano de interacción puede escribirse como $H_{\rm int} = - \boldsymbol{\mu}\cdot\nabla \Phi(\mathbf{x}_{\rm s})$, donde el potencial eléctrico $\Phi$ puede expandirse en la base de ondas planas correspondientes a excitaciones elementales según

\begin{equation}
    \Phi(\mathbf{x}) = \int\,d^2k\int_0^\infty\,d\omega\,\bigl(\,a_{\mathbf{k},\omega}\,A(\mathbf{k}, \omega)\,e^{i\,\mathbf{k}\mathbf{x_{\parallel}}} + {\rm h.c.}\bigr).
\end{equation}
El potencial contiene toda la información sobre el campo eléctrico en la región $z>0$ accesible, incluyendo las modificaciones introducidas respecto del caso de vacío libre por la presencia del semiespacio dieléctrico. Los operadores bosónicos satisfacen las relaciones usuales de conmutación $[a_{\mathbf{k},\omega}, a^\dagger_{\mathbf{k}',\omega'}]= \delta^2(\mathbf{k}-\mathbf{k}')\delta(\omega-\omega')$ y las funciones de la expansión en  modos de excitación simple $A(\mathbf{k},\omega)$ tienen la forma \cite{barton1997van, barton2010van}

\begin{equation}
    A(\mathbf{k},\omega) = \omega_s\sqrt{\frac{1}{4\pi^2\,k}}\,\frac{\sqrt{2\omega\,\Gamma}}{\omega^2-\omega_s^2+i\,\omega\,\Gamma}\,e^{-k\,z}.
    \label{eq:sec5_modefunc}
\end{equation}
En la expresión de arriba $\mathbf{k} = k_x\,\hat{x} + k_y\,\hat{y}$ es un vector de onda paralelo a la superficie del medio, de norma $k = |\mathbf{k}|$, y las frecuencias $\omega_{s}$ y $\Gamma$ parametrizan la constante dieléctrica $\varepsilon(\omega)$ del material que ocupa el semiespacio $z<0$ y que se describe mediante el modelo de Drude-Lorentz. La frecuencia $\omega_s$ define la condición de resonancia con los plasmones de superficie de un hipotético medio no-disipativo ($\Gamma = 0$), mientras que $\Gamma$ representa la tasa de disipación.

La dinámica de este sistema se trata, como en los capítulos anteriores, en el formalismo de ecuaciones maestras. En particular, se deriva una ecuación maestra para el estado reducido del sistema a partir de consideraciones microscópicas, trazando los grados de libertad correspondientes al entorno siguiendo el desarrollo presentado en el apéndice \ref{sec:ap2}. A partir de la ecuación (\ref{eq:ap2_redfieldint}), y observando que el primer término se anula para el campo electromagnético vestido en estado de vacío, se encuentra la ecuación de evolución para el estado del sistema

\begin{equation}
    \dot{\rho}_{\rm I}(t) = -\Tr_{\rm em}\int_{0}^t\,dt'\,\bigl[H_{\rm int,I}(t),[H_{\rm int,I}(t'), \rho_{\rm I}(t) \otimes \rho_{\rm em}]\bigr],
    \label{eq:sec5_eqini}
\end{equation}
donde el subíndice ${\rm I}$ señala que los operadores se encuentran en representación de interacción.

El Hamiltoniano de interacción entre el campo vestido y la partícula, en representación de interacción con respecto a todos los términos de Hamiltoniano libre,

\begin{equation}
    H_{\rm int, I}(t) = e^{i\,(H_{\rm s} + H_{\rm em})\,t}H_{\rm int}\;e^{-i\,(H_{\rm s} + H_{\rm em})\,t}
\end{equation} 
toma la forma de una suma de términos que oscilan con frecuencia $\omega -\omega_{\rm o}$ y $\omega +\omega_{\rm o}$, y se calcula fácilmente aplicando la fórmula de Baker-Campbell-Hausdorff. Reemplazando la expresión explícita de $H_{\rm int, I}(t)$ en la ecuación (\ref{eq:sec5_eqini}) se obtiene, en representación de Schrodinger,

\begin{equation}
        \Dot{\rho}(t) = -i \,[H_{\rm s}, \rho] -i\zeta_{10}(t)\,[\sigma_z, \rho]- \Bigl(\zeta_{00}(t) \, [\sigma_+,\{\sigma_-, \rho\}]+\zeta_{11}(t) \, [\sigma_+,[\sigma_-, \rho]]+ {\rm h.c.} \Bigr),
        \label{eq:sec5_eq1}
\end{equation}
donde se ha realizado, además, la denominada {\em aproximación secular}, que consiste en despreciar términos de la ecuación maestra que resulten rápidamente oscilantes en representación de interacción y se vincula estrechamente con la aproximación de onda rotante \cite{fleming2010rotating}.
Los núcleos $\zeta_{lm}(t)$, con $l,m = \{0,1\}$, introducidos en la ecuación (\ref{eq:sec5_eq1}) contienen toda la información sobre el efecto del entorno en el sistema y son funciones reales del tiempo con parámetros introducidos por la partícula y por el campo modificado por el medio

\begin{equation*}
        \zeta_{lm}(t) = \int_0^tdt'\int d^2k\, d\omega\,|A(\mathbf{k},\omega)|^2\bigl(|\boldsymbol{\mu}.\mathbf{k}|^2+\mu_z^2\,k^2\bigr)\cos(\omega_{\rm o} t' - l\pi/2)\cos(\omega t'-k_x {\rm v} t' - m\pi/2).
\end{equation*}

Si se desea estudiar la dependencia de los distintos efectos que el entorno tiene sobre el sistema con la velocidad de la partícula, es necesario avanzar en una forma interpretable para la ecuación (\ref{eq:sec5_eq1}). Con este objetivo, se desarrollan las múltiples integrales que aparecen en las funciones $\zeta_{lm}(t)$. 

Descomponiendo las funciones trigonométricas en exponenciales, y describiendo el plano $(k_x, k_y)$ en coordenadas polares, las integrales en $k$ tienen solución exacta

\begin{equation}
    \int_0^\infty\,dk\,k^2 e^{-2\,k\,d}e^{\pm i\,k\,{\rm v} \cos(\theta_k)t'} = 2\bigl(2\,d\mp i\,{\rm v}\cos(\theta_k)t' \bigr)^{-3},
    \label{eq:sec5_intk}
\end{equation}
como así también las integrales en la variable angular $\theta_k$
\begin{align}
    \int_0^{2\pi}d\theta_k\, \frac{\cos^2\theta_k}{2\,d\mp i\,{\rm v}\cos(\theta_k)t'} = \pi\frac{4\,d^2- 2{\rm v}^2t'^2}{(4\,d^2+ {\rm v}^2t'^2)^{5/2}}\\\nonumber
    \int_0^{2\pi}d\theta_k\, \frac{\sin^2\theta_k}{2\,d\mp i\,{\rm v}\cos(\theta_k)t'} = \pi\frac{1}{(4\,d^2+ {\rm v}^2t'^2)^{3/2}}.
    \label{eq:sec5_inttheta}
\end{align}
Con estos resultados, los núcleos $\zeta_{lm}(t)$ pueden expresarse según
\begin{equation}
    \zeta_{lm}(t)= \frac{\mu^2}{\pi\,d^3}\int_0^t dt'\int_0^\infty d\omega\,\frac{\Tilde{\Gamma} \omega}{(\omega^2 -1)^2 + \Tilde{\Gamma}^2\omega^2}\cos(\tilde{\omega}_{\rm o} t' - l\pi/2)\cos(\omega t'- m\pi/2)\mathrm{P}(v\,t'),
    \label{eq:sec5_nucleos2}
\end{equation}
donde se han definido las frecuencias $\Tilde{\Gamma}$, $\tilde{\omega}_{\rm o}$ adimensionales como las relaciones 
$\Tilde{\Gamma} = \Gamma/\omega_s$ y $\tilde{\omega}_{\rm o} = \omega_{\rm o}/\omega_s$ entre la tasa de decaimiento y la frecuencia natural del átomo, y la frecuencia de los plasmones de superficie respectivamente. Asimismo, se definió la velocidad $v = {\rm v} /(\omega_s\,d)$ como la relación entre la velocidad de la partícula, y la frecuencia natural de los plasmones de superficie por la distancia entre la partícula y el medio dieléctrico. El valor de la velocidad $\mathbf{v}$ se encuentra acotado por la restricción al caso de velocidades no-relativistas, en que ${\rm v} \ll 1$ (en unidades de $c=1$). La velocidad adimensional $v$ puede, sin embargo, adquirir distintos valores dependiendo de la frecuencia $\omega_s$ de los plasmones de superficie y de la distancia $d$ entre la partícula y el medio. Al escribir la expresión (\ref{eq:sec5_nucleos2}) se han también adimensionalizado las variables de integración según $t\rightarrow \omega_s\,t$ y $\omega \rightarrow \omega/\omega_s$, concentrando todas las dimensiones del núcleo en el factor $\mu^2/(\pi\,d^3)$. Por otra parte, la función algebraica

\begin{equation}
    \mathrm{P}(x) = \frac{4( 1+ n_z^2)+x^2 (-2n_x^2+n_y^2-n_z^2)}{\bigl(4+x^2\bigr)^{5/2}}\sim\frac{1}{8}\mu^{(i)}-\frac{3}{64}\,\mu^{(a)}x^2
    \label{eq:sec5_funcionP}
\end{equation}
contiene toda la información sobre la velocidad de la partícula y sobre su dirección de polarización. Para presentar su expansión en serie de Taylor trunca a segundo orden, se han definido los parámetros $\mu^{(i)}$ y $\mu^{(a)}$ que condensan la dependencia en la dirección de polarización del átomo a cada orden y están dados por
\begin{align}\label{eq:sec5_mues}
    \mu^{(i)} = &1 + n_z^2\\\nonumber
    \mu^{(a)} = & 3n_x^2 + n_y^2 + 4n_z^2.
\end{align} 

A continuación se trata la integral sobre la frecuencia $\omega$, cuya primitiva puede expresarse en términos de la solución $\omega_{\rm r} = \bigl(1 - \Tilde{\Gamma}^2/2 +( \Tilde{\Gamma}^2/4 -1)^{1/2}\,\bigr)^{1/2}$ a la ecuación $(\omega^2 -1)^2 + \Tilde{\Gamma}^2\omega^2 = 0$ que define los polos (complejos) del integrando. Se nota que $\omega_{\rm r}$ es un número complejo en el primer cuadrante para $\Tilde{\Gamma}<2$ y se torna un número imaginario puro para $\Tilde{\Gamma}>2$. El resultado de esta integral difiere para los distintos núcleos, de acuerdo al valor del entero $m$ según
\begin{align}
    \int_0^\infty d\omega\,\frac{\Tilde{\Gamma} \omega\,\cos(\omega t' - m\pi/2)}{(\omega^2 -1)^2 + \Tilde{\Gamma}^2\omega^2} =& \frac{\pi}{\sqrt{4-\Tilde{\Gamma}^2}}\Re{(-i)^m \, e^{i\omega_{\rm r}t'}}
    \label{eq:sec5_intomega}\\[.75em]\nonumber
    +& \frac{1-m}{\sqrt{4-\Tilde{\Gamma}^2}}\Im{e^{i\omega_{\rm r}t'}\mathrm{E}_1(i\omega_{\rm r}t')+ e^{-i\omega_{\rm r}t'}\mathrm{E}_1(-i\omega_{\rm r}t')}
\end{align}
donde $\mathrm{E}_1(z)$ es la función exponencial integral.

Realizar la integral en el tiempo adimensional $t'$ es el paso restante para obtener una expresión para las funciones $\zeta_{lm}(t)$ que modulan la dinámica del sistema. La misma sólo puede realizarse aproximada o numéricamente. En lo que sigue se presentan los resultados analíticos obtenidos para el régimen Markoviano y de velocidades adimensionales bajas, los cuales resultan ilustrativos gracias a la simplicidad de sus expresiones. Para expresiones aproximadas fuera de este régimen ver referencias \cite{lombardo2021detectable, viotti2021enhanced}. 
Para acceder a los regímenes en que estas aproximaciones no resultan válidas, se presentan también resultados numéricos a lo largo del capítulo.

Para tomar el límite Markoviano, resulta conveniente retornar sobre la expresión (\ref{eq:sec5_nucleos2}). La integral en la frecuencia contribuye únicamente para valores de $t$ en un rango acotado. Para tiempos largos la función en $\omega$ se torna un pulso altamente oscilante de integral nula. En consecuencia, para tiempos del orden de la escala de relajación resulta válido extender el límite superior de integración a infinito (para una justificación general de esta aproximación ver el apéndice \ref{sec:ap2}). 

El decaimiento del integrando con la variable $t'$ permite, en ciertos casos, una aproximación adicional, a saber, expandir la función algebraica $P(v\,t)$ en órdenes de la velocidad adimensional según la ecuación (\ref{eq:sec5_funcionP}). Como se ha observado en numerosos trabajos \cite{maghrebi2013quantum, pieplow2015cherenkov, klatt2016spectroscopic, intravaia2016non, svidzinsky2019excitation}, la validez de esta aproximación se encuentra limitada por la relación entre la frecuencia natural del sistema y su velocidad, pudiendo realizarse siempre que se satisfaga la relación $v< \tilde{\omega}_{\rm o}/2$ en términos de los parámetros adimensionales, o ${\rm v} < \omega_{\rm o}\,d/2$ en términos de los parámetros originales del sistema. La velocidad crítica $v_{\rm crit}$ corresponde al umbral a partir del cual el átomo se excita a expensas de su energía cinética, proceso que, de acuerdo con la literatura sobre fricción cuántica, tiene un impacto inmediato en esta fuerza. Se ha reportado que la fricción cuántica entre una partícula sin disipación interna y un semiespacio dieléctrico, a segundo orden en teoría de perturbaciones en el momento dipolar $|\boldsymbol{\mu}|$ de la partícula, es exponencialmente chica antes de este umbral, y sólo una vez superada $v_{\rm crit}$, crece~\cite{intravaia2016non}.   

Desarrollando analíticamente el caso $v<v_{crit}$ en el que la expansión a segundo orden (\ref{eq:sec5_funcionP}) está justificada, la dependencia temporal del factor resultante puede absorberse expresándolo en términos de derivadas parciales en la frecuencia adimensional $\Tilde{\omega}_{\rm o}$.
Con estas aproximaciones, la integral temporal para las funciones $\zeta_{mm}$ con índices $l=m$ impone una condición de resonancia $\delta(\omega- \Tilde{\omega}_{\rm o})$ que implica una forma simple e idéntica 
\begin{equation}
    \zeta_{mm}= \frac{\mu^2}{\pi\,a^3}\left(\frac{1}{8}\mu^{(i)}+\frac{3}{64}\,\mu^{(a)}\,v^2\,\partial^2_{\Tilde{\omega}_{\rm o}}\right)\,h(\Tilde{\omega}_{\rm o},\Tilde{\Gamma})
    \label{eq:sec5_nucleosmarkov}
\end{equation}
para las funciones $\zeta_{00}$ y $\zeta_{11}$ que dan cuenta de los efectos no-unitarios introducidos por el entorno sobre el sistema, en la cual 
\begin{equation}
    h(\Tilde{\omega}_{\rm o},\Tilde{\Gamma}) =\frac{\Tilde{\Gamma} \Tilde{\omega}_{\rm o}}{(\Tilde{\omega}_{\rm o}^2 -1)^2 + \Tilde{\Gamma}^2\Tilde{\omega}_{\rm o}^2}.
    \label{eq:sec5_funcionh}
\end{equation}

\section{Dinámica del sistema}
Mediante la resolución de la ecuación (\ref{eq:sec5_eq1}) se accede a la evolución dinámica del sistema de dos niveles, obteniendo el operador densidad $\rho(t)$ que representa su estado a tiempo $t>0$. Este operador se puede escribir como una matriz de $2\times2$ cuyos elementos diagonales se denominan {\em poblaciones}, mientras que los elementos no-diagonales se refieren como {\em coherencias}. 
La ecuación maestra (\ref{eq:sec5_eq1}) consiste en un sistema de ecuaciones diferenciales para las poblaciones y coherencias
\begin{align}
    \Dot{\rho}_{11}(t) &= -2\, (\zeta_{00}(t) + \zeta_{11}(t))\,\rho_{11} + 2\, (\zeta_{00}(t) - \zeta_{11}(t))\,\rho_{22}\label{eq:sec5_eqdiff}\\\nonumber
    \Dot{\rho}_{12}(t) &= -(2\zeta_{00}(t) + 2\,i\,\zeta_{10}(t) + i\Tilde{\omega}_{\rm o})\,\rho_{12},
\end{align}
donde las ecuaciones restantes se deducen de los vínculos que imponen la condición de hermiticidad $\rho_{21}(t) = \rho_{21}^*(t)$ y de traza unidad $\rho_{11}(t) + \rho_{22}(t) = 1$ que rigen sobre el operador densidad. A lo largo de este capítulo se estudia el escenario en que el estado inicial es puro y está descrito por la superposición 

\begin{equation}
    \ket{\psi(0)} = \cos\left(\frac{\vartheta}{2}\right)\ket{+}+\sin\left(\frac{\vartheta}{2}\right)\ket{-}.
    \label{eq:sec5_psiini}
\end{equation}
Recurriendo al cambio de variables $\rho_-(t) = \rho_{11}(t) - \rho_{22}(t)$ y $\rho_+(t) = \Tr\rho(t) = 1$, se obtiene una solución formal para la evolución de las poblaciones, dada por

\begin{equation}
    \rho_{-}(t) = \cos(\vartheta)\,e^{-4\int\,dt\,\zeta_{00}(t)}-4e^{-4\int\,dt\,\zeta_{00}(t)}\int_0^t\,dt'\,\zeta_{11}(t')e^{4\int\,dt'\,\zeta_{00}(t')}.
    \label{eq:sec5_poblaciones}
\end{equation}
Para la evolución estudiada, las poblaciones tienen la misma forma funcional sea que se aplique la aproximación secular o no. Para este sistema, entonces, dicha aproximación corresponde a desestimar una interacción dinámica entre las coherencias, cuya expresión se encuentra por integración directa de la ecuación correspondiente en (\ref{eq:sec5_eqdiff})

\begin{equation}
    \rho_{12}(t)=\frac{\sin(\vartheta)}{2}\,e^{-2\int_0^t\,dt'\,\zeta_{00}(t')}\, e^{-2i\int_0^t\,dt' \,\zeta_{01}(t')-i\tilde{\omega}_{\rm o}t}.
    \label{eq:sec5_coherencias}
\end{equation}

\begin{figure}[ht!]
    \center
    \includegraphics[width = .495\linewidth]{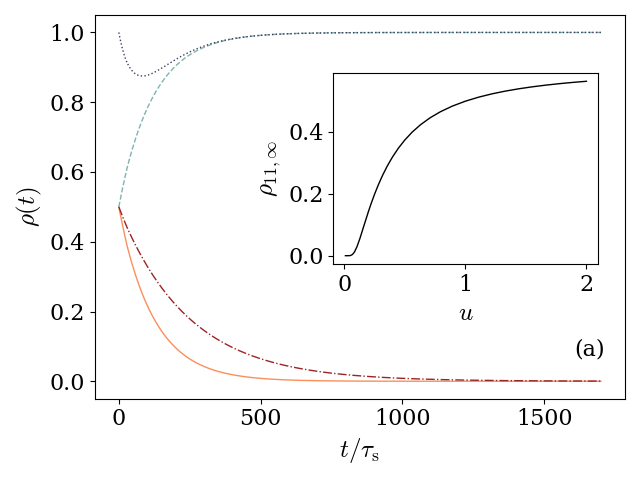}
    \includegraphics[width = .495\linewidth]{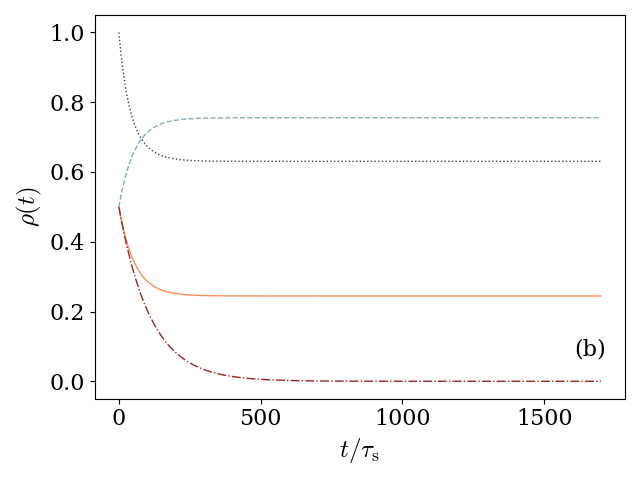}
    \caption{Evolución dinámica de los elementos de la matriz densidad en el régimen de velocidades (a) por debajo de la velocidad crítica, con $v = 0.015\,\Tilde{\omega}_{\rm o}$ y (b) por encima de la velocidad crítica, con $v = 1.5\,\Tilde{\omega}_{\rm o}$. En ambos casos, el eje temporal se normaliza con la escala natural del átomo $\tau_{\rm s} = 2\pi/\omega_{\rm o}$ y se incluye la pureza del estado para completar la descripción. El sistema, de frecuencia natural adimensional $\Tilde{\omega}_{\rm o} = 0.2$ y relación $\mu^2/d^3 = 0.005$ entre su momento dipolar y distancia al semiespacio dieléctrico, parte de un estado inicial definido por $\vartheta = \pi/2$ polarizado en dirección del eje $x$. El campo electromagnético circundante, vestido por medio, está caracterizado por una tasa de disipación adimensional $\Tilde{\Gamma} = 1$. La líneas color naranja sólida, y celeste de trazos representan las coherencias $\rho_{11}$ y $\rho_{22}$ respectivamente, mientras que la línea bordó de trazo-punto describe el valor absoluto de las coherencia. La línea azul punteada corresponde a la pureza del estado. El inset en la figura (a) muestra el valor estacionario alcanzado por la población $\rho_{11}$ en función de la velocidad adimensional del átomo para los mismos parámetros que la figura principal.} 
    \label{fig:sec5_elementos}
\end{figure}

En la figura \ref{fig:sec5_elementos} se presenta la dinámica del sistema, representada por los elementos $\rho_{ij}(t)$ de la matriz densidad que describe su estado. 
Se encuentran comportamientos cualitativamente distintos asociados con dos regímenes de velocidades del átomo, separados por el umbral de la velocidad crítica $v_{\rm crit} = 0.5\, \Tilde{\omega}_{\rm o}$. Un ejemplo correspondiente a cada régimen se muestra en los paneles (a) y (b) de la figura \ref{fig:sec5_elementos}, con velocidades adimensionales $v= 0.015\,\Tilde{\omega}_{\rm o}$ y $v= 1.5\,\Tilde{\omega}_{\rm o}$ $v= 0.015\,\Tilde{\omega}_{\rm o}$.  

Para velocidades adimensionales debajo del umbral definido por $v_{\rm crit}$ el sistema evoluciona en dirección a un estado fundamental estacionario. Sin embargo, cuando la velocidad adimensional supera valor crítico, el sistema tiende a un estado asintótico mixto. Con el objetivo de resaltar la diferencia, en ambos casos se incluye además la evolución de la pureza del estado.

Más en detalle, el panel (a) de la figura \ref{fig:sec5_elementos} presenta el caso con $v = 0.015\,\Tilde{\omega}_{\rm o}$, correspondiente al primer régimen en el cual aplican las aproximaciones que conducen a la expresión (\ref{eq:sec5_nucleosmarkov}) para las funciones $\zeta_{mm}$. Las coherencias y la población $\rho_{11}(t)$ en el estado excitado se suprimen hasta desvanecerse para tiempos suficientemente largos. Iniciando el sistema en un estado puro (\ref{eq:sec5_psiini}), la pureza decrece desde su valor inicial unitario por efecto del entorno. Sin embargo, a medida que todos los elementos, a excepción de $\rho_{22}(t)$, se suprimen, el estado recobra pureza y tiende asintóticamente a un estado que resulta indistinguible del fundamental. Por otra parte, el panel (b) de la figura \ref{fig:sec5_elementos} presenta el caso con $v = 1.5\,\Tilde{\omega}_{\rm o}$. Diferente al caso anterior, en este régimen la población $\rho_{11}(t)$ del estado excitado no se extingue completamente, sino que se suprime parcialmente hasta alcanzar un valor finito. Este comportamiento de las poblaciones, acompañado por la completa anulación de las coherencias, conduce a un estado estacionario mixto de pureza  $\rho_{11,\infty}^2+(1-\rho_{11,\infty})^2$.
El inset en el panel (a) de la figura \ref{fig:sec5_elementos} muestra el valor estacionario alcanzado por el elemento $\rho_{11}$ en función de la velocidad adimensional $v$ del átomo. En esta relación es posible identificar la velocidad crítica que separa los dos regímenes $v_{\rm crit} = \Tilde{\omega}_{\rm o}/2$ (o, en términos de los parámetros dimensionales del problema ${\rm v}_{\rm crit} = d\,\omega_{\rm o}/2$) coincidente con el umbral a partir del cual el átomo puede excitarse a expensas de su energía cinética \cite{resultados8,maghrebi2013quantum, pieplow2015cherenkov, klatt2016spectroscopic, intravaia2016non, svidzinsky2019excitation}.

\begin{figure}[ht!]
    \center
    \includegraphics[width = .55\linewidth]{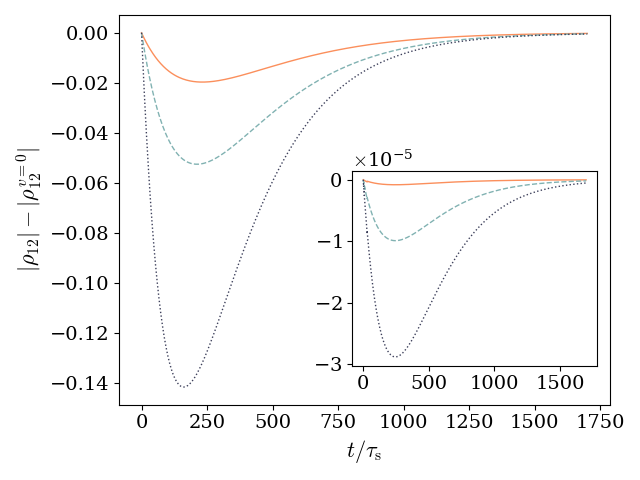}
    \caption{Evolución de la diferencia $|\rho_{12}|-|\rho_{12}^{v=0}|$ entre el valor absoluto de las coherencias a velocidad finita y a velocidad nula, para distintos valores de velocidad adimensional finita. El eje temporal se normaliza con la escala natural del átomo $\tau_{\rm s} = 2\pi/\omega_{\rm o}$ donde $\Tilde{\omega}_{\rm o} = 0.2$. El sistema satisface la relación $\mu^2/d^3 = 0.005$ y parte de un estado inicial definido por $\vartheta = \pi/2$ polarizado en dirección del eje $x$. El entorno está caracterizado por una tasa de disipación adimensional $\Tilde{\Gamma} = 1$. Las líneas color naranja sólida, celeste de trazos, y azul de puntos corresponden a valores $v=0.1$, $v= 0.15$ y $v= 0.3$ en el régimen $v > v_{\rm crit}$ respectivamente en la figura principal, y a los casos $v=0.001$, $v = 0.003$ y $v= 0.005$  en el régimen $v < v_{\rm crit}$ respectivamente en el inset.} 
    \label{fig:sec5_coherencias}
\end{figure}

Respecto del comportamiento de los elementos no-diagonales, se puede realizar algunas observaciones adicionales en lo que refiere a su dependencia con la velocidad adimensional $v$ de la partícula. La figura \ref{fig:sec5_coherencias} muestra la diferencia entre el comportamiento de estos elementos en el caso estático $v=0$ y de velocidad finita $v\neq 0$, para distintos valores de velocidad finita. 

La figura principal presenta curvas en el régimen de velocidades $v>v_{\rm crit}$, mientras que el inset exhibe el comportamiento en el régimen de velocidades $v<v_{\rm crit}$.
En ambos casos se observa que esta diferencia crece más rápidamente conforme la velocidad aumenta. Sin embargo, como se ha notado en el capítulo anterior, la coherencia decae incluso en presencia de fluctuaciones de vacío desnudas del espacio libre. Por este motivo, si bien la presencia del medio dieléctrico y la velocidad relativa del átomo con éste aceleran el proceso, la diferencia entre los casos de velocidad finita y nula alcanza un valor máximo y tiende a anularse posteriormente. En consecuencia, explorar las coherencias a tiempos muy grandes no permitirá identificar efectos inducidos por la velocidad.

La misma observación vale para el estudio de las poblaciones y la pureza en el régimen de velocidades $v<v_{\rm crit}$ (figura \ref{fig:sec5_elementos}.a ). Dado que el sistema converge al estado fundamental independientemente de la velocidad de la partícula, a menos del hecho de que se encuentre en el rango correspondiente, no podrá obtenerse información sobre el valor específico de $v$ observando estos elementos a tiempos prolongados. Para velocidades en el régimen $v>v_{\rm crit}$ (ver figura \ref{fig:sec5_elementos}.b) la situación es diferente en lo que respecta a las poblaciones, dónde el valor de la velocidad se halla codificado en los elementos diagonales del estado estacionario alcanzado. 

Los parámetros que caracterizan el sistema y el material dieléctrico en las figuras \ref{fig:sec5_elementos} y \ref{fig:sec5_coherencias}, y en el resto del capítulo, se eligen recurriendo a datos experimentales. Para el material dieléctrico se considera metales como el oro (Au) o un material de Silicio n-dopado (n-Si). 
Para las partículas, se considera un átomo de rubidio (Rb) o un único centro nitrógeno-vacante (NV) en un diamante como sistema de dos niveles efectivo. Los parámetros correspondientes se resumen en la tabla a continuación para facilitar la lectura
\begin{table}[h!]
\center
\begin{tabular}{|>{\centering\arraybackslash}p{.18\linewidth}|>{\centering\arraybackslash}p{.15\linewidth}|>{\centering\arraybackslash}p{.15\linewidth}|>{\centering\arraybackslash}p{.15\linewidth}|}
    \hline
       & Au & nSi& Valor\\\hline
    $\omega_s \, ({\rm rad}\,s^{-1})$&  $9.7\times 10^{15}$ & $2.47\times 10^{14}$&\\
    $\Tilde{\Gamma}\equiv\Gamma/\omega_s$ &$0.003$ & $1$&\\
    $\Tilde{\omega}^{Rb}_{\rm o} \equiv\omega^{Rb}_{\rm o}/\omega_s$ & 0.2 &  $8$&\\
    $\Tilde{\omega}^{NV}_{\rm o} \equiv\omega_{\rm o}^{NV}/\omega_s$  & $10^{-5}/0.02$ &  $0.0004/0.2$&\\
    $d\, ({\rm m})$ & - & -& $1-5 \times 10 ^{-9}$\\
    \hline
\end{tabular}
\caption{Tabla de valores de los parámetros considerados a lo largo del capítulo para modelar el átomo y el medio dieléctrico.}
\label{tabla}
\end{table}

A continuación, utilizaremos estos resultados para definir una escala temporal en la cual las coherencias del grado de libertad interno del átomo se destruyen por influencia del campo electromagnético circundante vestido por la presencia del material dieléctrico.

\section{Destrucción de las coherencias}\label{sec:sec5_td}

La matriz densidad definida por las ecuaciones (\ref{eq:sec5_poblaciones}) y (\ref{eq:sec5_coherencias}) permite definir una escala temporal de decoherencia $\tau_D$ a partir de la {\em funcional de decoherencia} $\mathcal{D}(t) = \exp(-2\int_0^t\,dt'\,\zeta_{00}(t'))$ como $\mathcal{D}(\tau_D)=\exp(-2)$. 
En el régimen $v<v_{\rm crit}$ (para el que aplica la aproximación (\ref{eq:sec5_nucleosmarkov})) se encuentra, a segundo orden en $v$, una expresión analítica sencilla para esta escala temporal 

\begin{equation}
    \tau_D^{v<v_{\rm crit}}\sim \frac{d^3}{\mu^2}\frac{64}{\mu^{(i)}}\frac{1}{h(\Tilde{\omega}_{\rm o},\Tilde{\Gamma})}\left(1-\frac{3}{8}\frac{\mu^{(a)}}{\mu^{(i)}}\,v^2\,\frac{\partial_{\Tilde{\omega}_{\rm o}}^2\,h(\Tilde{\omega}_{\rm o},\Tilde{\Gamma})}{h(\Tilde{\omega}_{\rm o},\Tilde{\Gamma})}\right),
    \label{eq:sec5_tdmarkov}
\end{equation}
con $h(\Tilde{\omega}_{\rm o},\Tilde{\Gamma})$ dada por la ecuación (\ref{eq:sec5_funcionh}).
En lo que sigue, se estudia la dependencia de la dinámica en la velocidad adimensional del átomo a través de la escala temporal de decoherencia $\tau_D$.
\\
\\\indent
{\em Dependencia en la velocidad - }La expansión en órdenes (\ref{eq:sec5_tdmarkov}) realizada sobre la escala $\tau_D$ muestra un término completamente independiente de la velocidad relativa entre el átomo y el medio, y una corrección por velocidad de orden cuadrático. Esto permite identificar dos contribuciones de distinta naturaleza en el proceso de decoherencia: (a) una contribución {\em estática} generada por las fluctuaciones del vacío electromagnético vestido y (b) una segunda contribución introducida por el movimiento del átomo en este medio.
La competencia entre las dos contribuciones se examina observando la relación $\tau_D/\tau_D^{v=0}$ entre la escala de decoherencia a velocidad finita, y aquella encontrada para velocidad nula. Para el rango de velocidades en que resulta válido el resultado analítico (\ref{eq:sec5_tdmarkov}), esta relación equivale al factor entre paréntesis en dicha expresión, donde se vuelve evidente que la relación satisface $\tau_D/\tau_D^{v=0}\sim1$ si los efectos de la velocidad resultan despreciables. 

\begin{SCfigure}[][ht!]
    \includegraphics[width = .5\linewidth]{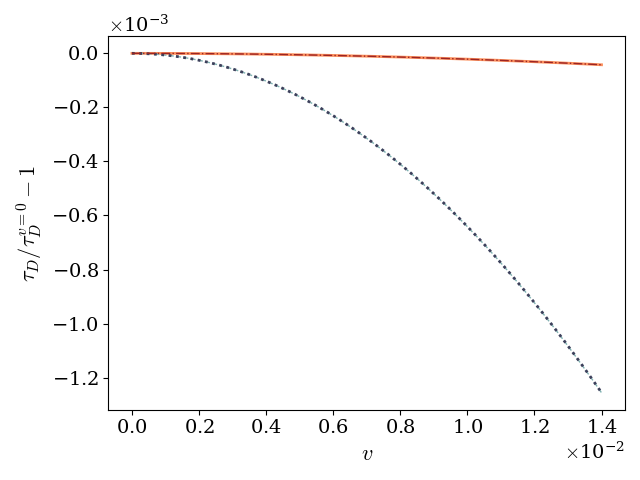}
    \caption{Relación $\tau_D/\tau_D^{v=0}$ entre la escala de decoherencia a velocidad finita y a velocidad nula, como función de la velocidad (adimensional) $v$, para un un átomo de Rb (línea celeste de trazos) y para un centro NV (línea naranja sólida) polarizados en dirección del eje $x$. El campo electromagnético está vestido por un material nSi, y se refiere a la tabla \ref{tabla} para los valores de los parámetros involucrados. Se presenta además la aproximación Markoviana a cada caso en línea azul de puntos, y bordó de trazo-punto respectivamente.} 
    \label{fig:sec5_tdec_u}
\end{SCfigure}

En la figura \ref{fig:sec5_tdec_u} se muestra la relación $\tau_D/\tau_D^{v=0}$ exacta (calculada numéricamente) en función de la velocidad, para dos conjuntos distintos de parámetros adimensionales. Se incluye además el resultado analítico aproximado (\ref{eq:sec5_tdmarkov}) para facilitar la comparación.

Esta figura ilustra una primera consideración: la intensidad de los efectos introducidos por el campo electromagnético vestido por el material dieléctrico y el acceso experimental a los mismos mostrarán una fuerte dependencia en los parámetros que modelan el sistema y el entorno, así como en la relación entre ellos. Por ejemplo, la relación entre la escala de decoherencia a velocidades finita y nula toma valores $\tau_D/\tau_D^{v=0}\sim 1 - 0.072\,\mu^{(a)}/\mu^{(i)}\, v^2$para un átomo de Rb y $\tau_D/\tau_D^{v=0}\sim 1 - 2.14\,\mu^{(a)}/\mu^{(i)}\, v^2$ para un centro NV que se desplazan a una misma distancia de una misma superficie dieléctrica. 
\\
\\\indent
{\em Dependencia en la orientación - }Una segunda observación se desprende de esto último, y es la influencia de la dirección de polarización en el decaimiento de la coherencia cuántica.
En el capítulo anterior, los efectos del entorno sobre el sistema adquirían, en presencia de contornos no-triviales, dependencia en la dirección de polarización de las partículas que componían el sistema. Más aún, esta dependencia se observaba tanto en los efectos colectivos como en aquellos que involucran a cada átomo considerado en forma individual, lo que permite suponer que la dependencia se extiende al caso de un sistema con un único grado de libertad. 

\begin{SCfigure}[10][ht!]
    \includegraphics[width = .5\linewidth, trim = {0 0 25pt 10pt}]{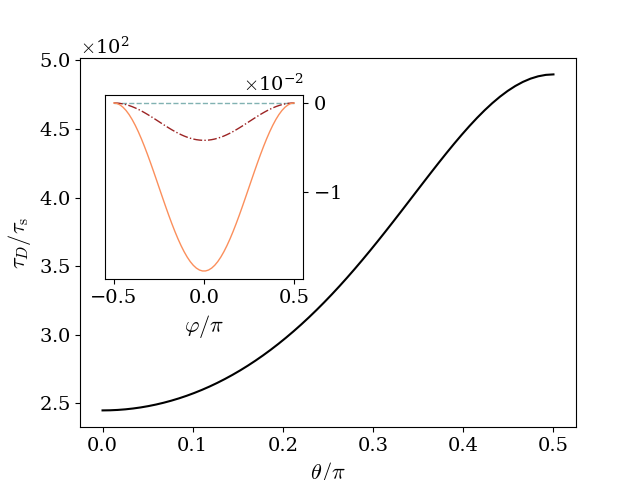}
    \caption{Escala de decoherencia $\tau_D/\tau_{\rm s}$ en función de $\theta$. Se considera un centro NV de frecuencia natural $\Tilde{\omega}_{\rm o} = 0.2$ con velocidad adimensional $v = 0.003$ en un entorno caracterizado por una tasa de disipación adimensional $\Tilde{\Gamma} = 1$
    El inset muestra la escala de decoherencia $\tau_D/\tau_{\rm s}$ sustrayendo el valor para $\varphi = 0$ en función del ángulo azimutal $\varphi$. La líneas color naranja sólida, bordó de trazo-punto, y celeste de trazos, representan valores del ángulo polar $\theta = \pi/2$, $\theta = \pi/4$, y $\theta = 0$ respectivamente.} 
    \label{fig:sec5_polarizacion}
\end{SCfigure}

Para abordar esta cuestión se parametriza la dirección de polarización con variables angulares
\begin{equation}
    \boldsymbol{\mu} = \mu\,\bigl(\cos(\varphi)\sin(\theta)\,\hat{x}+\sin(\varphi)\sin(\theta)\,\hat{y}+\cos(\theta)\,\hat{z}\bigr)
\end{equation}
y se investiga tanto la dependencia en el ángulo polar $\theta$ para valores fijos del ángulo azimutal $\varphi$ como la dependencia en $\varphi$ para distintos valores fijos de $\theta$. Ambas inspecciones se ilustran en la figura \ref{fig:sec5_polarizacion}, que muestra la escala de decoherencia, normalizada con la escala natural del átomo $\tau_{\rm s}$ en función del ángulo polar.
En la misma figura, el inset expone la diferencia 
\begin{equation}
    \frac{\tau_D}{\tau_{\rm s}}-\frac{\tau_D}{\tau_{\rm s}}\Big|_{\varphi =0}
\end{equation}
entre la escala de decoherencia para un ángulo $\varphi$ y su valor para $\varphi = 0$ como función de $\varphi$. La sustracción tiene como objetivo modificar la escala en que se muestran los resultados.

Una característica de la dependencia en la dirección de polarización evidenciada por la figura \ref{fig:sec5_polarizacion} es la  jerarquía entre las dependencias en el ángulo polar y azimutal.
La variación en el ángulo polar $\theta$ se traduce en variaciones de la escala de decoherencia del orden de la propia escala, que se duplica en pasar de $\theta = 0$ a $\theta = \pi/2$. Por el contrario, la variación del ángulo azimutal resulta en variaciones de la escala de decoherencia cuatro órdenes de magnitud más chicas que las primeras. Recordando que el medio ocupa el semiespacio inferior $z<0$ esta dependencia da cuenta de la sutileza de los efectos inducidos exclusivamente por el movimiento relativo con respecto al caso estático. 

El comportamiento descrito para la escala de decoherencia resulta inversamente proporcional al obtenido, en la referencia \cite{intravaia2016non}, para la fricción cuántica (ver figura 5 de la referencia). Esto establece un vínculo no sólo entre el trabajo y la literatura previa, sino también entre los fenómenos, dónde la fricción cuántica acelera la pérdida de coherencia.
Por otro lado, la fuerte diferencia mostrada por la dependencia en los ángulos polar y azimutal puede interpretarse retomando la descripción pictórica en términos de imágenes del capítulo anterior y de la sección \ref{sec:sec5_friccion}. 
\\
\\\indent
{\em Interpretación en términos de imágenes - }Se examina primeramente el caso estático, para introducir posteriormente el movimiento del átomo en el análisis .
Como se ha discutido, cuando el momento dipolar del átomo es perpendicular a  superficie del medio, el átomo imagen que reproduce los efectos de este contorno es, a menos de correcciones por la conductividad finita, un átomo idéntico con polarización en la misma dirección y sentido ubicado en la posición reflejada respecto de la superficie del plano. Cuando la distancia del átomo real (y en consecuencia también de su imagen) es suficientemente corta en comparación con la distancia de observación, el escenario efectivo muestra pequeñas desviaciones respecto del de una partícula con momento dipolar de intensidad $2\mu$. Diferentemente, cuando el átomo real está polarizado en alguna dirección paralela al plano, los efectos de la geometría del contorno pueden reproducirse (a menos de correcciones por la conductividad finita) con una partícula idéntica al átomo real, ubicada en la posición espejada respecto de la superficie. En este caso, sin embargo, el momento dipolar de la partícula imagen tiene sentido contrario al del dipolo real. 

En consecuencia, cuando la distancia del átomo al semiespacio es suficientemente corta en comparación con la distancia de observación, el escenario efectivo es aproximadamente el de una partícula con momento dipolar nulo. Para el caso estático, además, el conjunto formado por el sistema y su entorno es simétrico frente a rotaciones en los planos $z = {\rm cte}$, y como consecuencia los resultados son completamente independientes de $\varphi$.
La velocidad de la partícula $\mathbf{v} = {\rm v}\hat{x}$ rompe esta simetría, implicando una dependencia en $\varphi$. Sin embargo, para velocidades en el rango $v<v_{\rm crit}$, esta dependencia es mucho más débil que la dependencia en $\theta$, insinuando que los efectos disipativos introducidos por el movimiento son mucho más débiles que aquellos ya presentes en el caso estático. 
\clearpage
En lo que sigue, se introduce la fase geométrica como objeto de observación para el análisis de la dinámica con la intención de determinar si los efectos de disipación inducidos por el movimiento resultan más accesibles que a través de la observación de la coherencia cuántica.

\section{Fase geométrica} 
Se recurre una vez más la definición (\ref{eq:sec2_TongGP}) para la fase geométrica $\phi_g$ de un sistema cuántico abierto, que se reduce a la expresión (\ref{eq:sec2_TongGPpuros}) cuando el estado inicial del sistema sea un estado puro, tal como el que se considera en este capítulo. Esta fase difiere en general de la fase geométrica $\phi_u$, dada por la ecuación (\ref{eq:sec2_kinGP}), que acumula el sistema en completo aislamiento de cualquier entorno. En consecuencia, podrá sin pérdida de generalidad expresarse como la suma $\phi_g = \phi_u  + \delta\phi$ entre la fase geométrica acumulada por un hipotético sistema aislado, más un término de corrección.

El autoestado del operador densidad $\ket{\psi_+(t)}$ involucrado en el cálculo de la expresión (\ref{eq:sec2_TongGPpuros}) puede escribirse, para operadores densidad de $2\times2$, como

\begin{equation}
    \ket{\psi_+(t)} = \frac{-(\rho_{22} - \epsilon_+)\ket{+} + \rho_{21}\ket{-}}{\left((\rho_{22}-\epsilon_+)^2 + \rho_{21}\rho_{12}\right)^{1/2}},
\end{equation}
\\
con $\epsilon_{+} = \frac{1}{2}\left(1+((\rho_{11}-\rho_{22})^2 + 4\rho_{21}\rho_{12})^{1/2}\right)$ el autovalor asociado. Con este resultado, y la evolución temporal (\ref{eq:sec5_coherencias}) hallada para las coherencias, se encuentra una expresión formal en términos de los elementos $\rho_{ij}(t)$ para la fase geométrica

\begin{equation}
    \phi_g(t)=\arg\Bigl( \rho_{21}\sin(\vartheta/2)-(\rho_{22} - \epsilon_+)\cos(\vartheta/2)\Bigr)-\int_0^t \,dt'\,\frac{(2\zeta_{01} + \Tilde{\omega}_{\rm o})\,|\rho_{12}|^2}{(\rho_{22}-\epsilon_+)^2 + |\rho_{12}|^2}.
    \label{eq:sec5_faseFormal}
\end{equation}

\begin{figure}[ht!]
    \center
    \includegraphics[width = .5\linewidth]{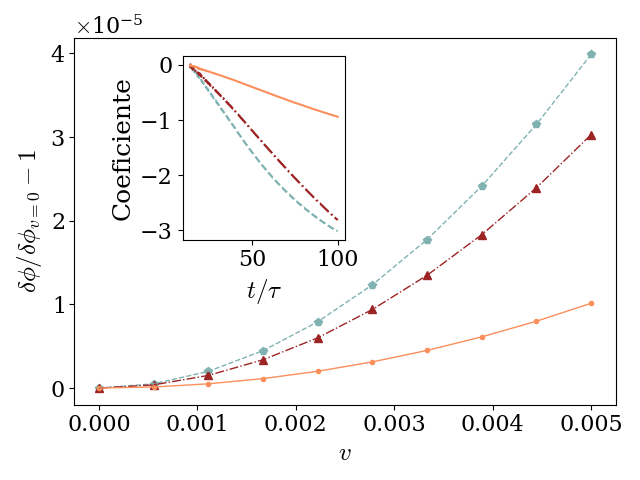}
    \caption{Corrección normalizada $\delta\phi/\delta\phi_{v=0}$ a la fase geométrica acumulada en un intervalo $t=50\,\tau$, para un sistema con momento dipolar orientado en distintas direcciones, y en función de la velocidad adimensional $v$. Para el sistema, se considera la frecuencia natural adimensional $\Tilde{\omega}_{\rm o} = 0.2$, y el efecto del vacío sobre el mismo está caracterizado por $\Tilde{\Gamma} = 1$. Las líneas color naranja sólida, celeste de trazos y bordo de trazo-puntos corresponden a momento dipolar en dirección del eje $y$, del eje $z$, y del eje $x$ respectivamente. En la figura principal, los puntos corresponden a los valores obtenidos, mientras que las líneas corresponden a un ajuste cuadrático que se incorpora en cada caso para facilitar la comparación. El inset muestra la dependencia temporal del coeficiente correspondiente en función del tiempo.} 
    \label{fig:sec5_GP}
\end{figure}

A partir de esta expresión se estudia, en primer lugar, la dependencia de la corrección $\delta \phi$ a la fase geométrica $\phi_u$ acumulada en una evolución unitaria con la velocidad. En este modelo, se espera que esta diferencia exhiba dos contribuciones de origen diverso: (a) una contribución estática $\delta\phi_{v=0}$ debida a la presencia del campo electromagnético vestido por el medio dieléctrico, presente incluso en el caso sin movimiento relativo entre la partícula y el medio, y (b) una contribución adicional generada por el desplazamiento de la partícula que implica, para el caso general, $\delta\phi\neq\delta\phi_{v=0}$. El peso relativo de cada contribución se investiga mediante la relación entre la corrección a velocidad finita y a velocidad nula $\delta\phi_{v}/\delta\phi_{v=0}$.
Esta relación tiende a la unidad cuando los efectos inducidos por el movimiento resulten despreciables. Sin embargo, se apartará del valor unitario mostrando alguna dependencia característica en $v$ a medida que la velocidad adquiera influencia sobre la fase geométrica.
La dependencia de la corrección normalizada $\delta\phi/\delta\phi_{v=0}$ en la velocidad adimensional $v$ se presenta en la figura \ref{fig:sec5_GP}, para distintas orientaciones del momento dipolar del átomo. Se incluye además un ajuste cuadrático para facilitar la comparación.

Más en detalle, la figura muestra la corrección normalizada $\delta\phi_{v}/\delta\phi_{v=0}$, calculada en el instante $t= 50\,\tau$, en función de la velocidad adimensional, para distintas orientaciones del momento dipolar del átomo. 
Para valores de la velocidad $v$ suficientemente chicos, la dependencia en la velocidad resulta despreciable frente a los efectos estáticos y $\delta\phi_{v}/\delta\phi_{v=0}\sim 1$ cualquiera sea la orientación del momento dipolar. Sin embargo, conforme la velocidad adimensional aumenta, la relación se aparta de la unidad en forma cuadrática. Esta dependencia para velocidades bajas, que coincide con la dependencia reportada para el corrimiento de energía y la tasa de disipación del átomo~\cite{klatt2016spectroscopic}, resulta prometedora en comparación con la dependencia mostrada por la fricción cuántica. En este régimen, la mayoría de los resultados analíticos para la fricción predicen una dependencia exponencialmente pequeña, o de orden cuártico en el momento dipolar~\cite{resultados8, klatt2017quantum}. 

Para dar cuenta de la dependencia cuadrática, la figura \ref{fig:sec5_GP} muestra ajustes de este orden acompañando los valores obtenidos. El factor constante que acompaña esta dependencia depende de la orientación del dipolo y del tiempo. Se comparan tres casos con orientaciones distintas. El caso en que el momento dipolar de la partícula se orienta en la dirección del eje $z$, y es entonces perpendicular a la superficie del dieléctrico; el caso en que se orienta en dirección del eje $y$, paralelo a la superficie del dieléctrico y perpendicular a la dirección de movimiento; y finalmente, el caso en que la orientación es paralela a la velocidad de la partícula. El efecto del movimiento sobre la fase geométrica es (relativamente) mínimo para el caso en que el momento dipolar del átomo es paralelo a la superficie dieléctrica y perpendicular a la dirección de desplazamiento.

\section{Propuesta experimental}\label{sec:sec5_exp}
\begin{figure}[ht!]
    \center
    \includegraphics[width = .52\linewidth]{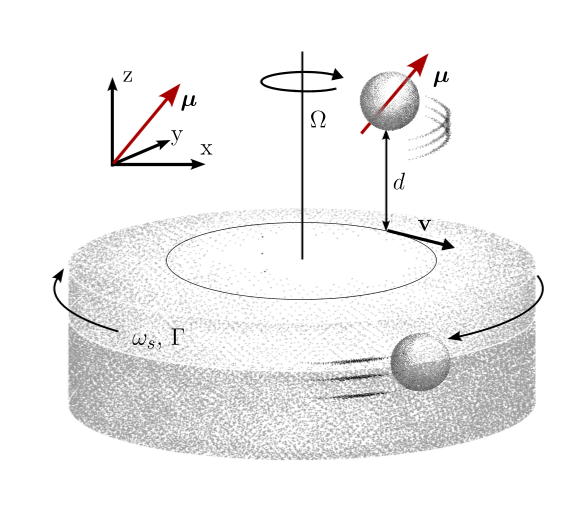}
    \includegraphics[width = .47\linewidth]{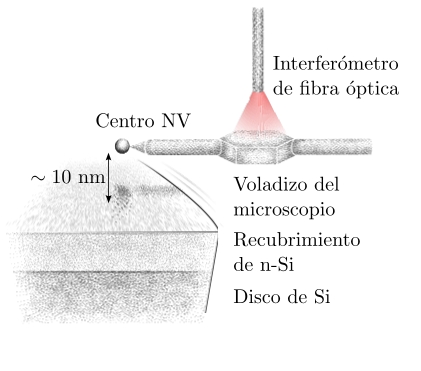}
    \caption{Esquemas conceptuales del arreglo experimental propuesto. La superficie dieléctrica consiste en un disco de silicio recubierto por una capa de silicio n-dopado. La partícula, un centro NV posicionado en la punta de un microscopio de fuerza atómica se mantiene estático en el sistema laboratorio, a una distancia $d\sim10 {\rm nm}$, mediante un microscopio de fuerza atómica modifiado. El movimiento relativo entre el centro NV y la superficie se genera rotando el disco. Los valores de los parámetros considerados en la figura \ref{fig:sec5_materiales} son accesibles con la tecnología actual.} 
    \label{fig:sec5_mesa}
\end{figure}

Producto de la colaboración iniciada en \cite{farias2020towards} y sostenida en~\cite{lombardo2021detectable}, se propone un arreglo experimental factible para medir la corrección inducida en la fase geométrica acumulada, cuyo esquema se muestra en la figura \ref{fig:sec5_mesa}. En el arreglo se utiliza un centro NV como sistema efectivo de dos niveles, posicionado en la punta de un microscopio de fuerza atómica (AFM por sus siglas en inglés) modificado. La distancia de esta punta a algún otro elemento puede controlarse entre algunos nanómetros hasta decenas de nanómetros con resolución sub-nanométrica. El centro NV se presenta como una herramienta prometedora para el estudio de las fases geométricas \cite{maclaurin2012measurable}. Consta de una vacancia, o átomo de carbono faltante en una estructura de diamante ubicada contiguamente a un átomo de nitrógeno que ha sustituido uno de los vecinos inmediatos de la vacancia. El centro NV ofrece un sistema en el cual un único sistema de dos niveles puede ser inicializado, controlado coherentemente y medido. Es además posible mover mecánicamente el centro.

La superficie dieléctrica está compuesta por una placa de silicio n-dopado en la que la densidad de dopado se ajusta para maximizar los efectos. Esta placa cilíndrica de silicio dopado se monta sobre una mesa rotante. Aunque se utilice una mesa rotante, los efectos no-inerciales pueden despreciarse completamente, modelando una partícula que se traslada a velocidad constante sobre la superficie del material.

\begin{figure}[ht!]
    \center
    \includegraphics[height = 6cm]{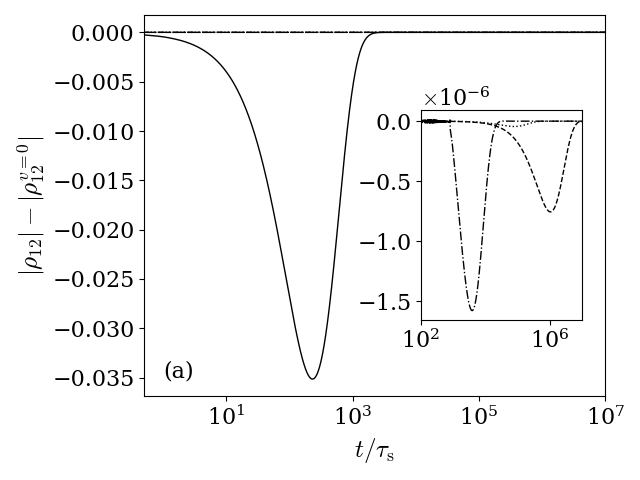}
    \includegraphics[height = 6.25cm]{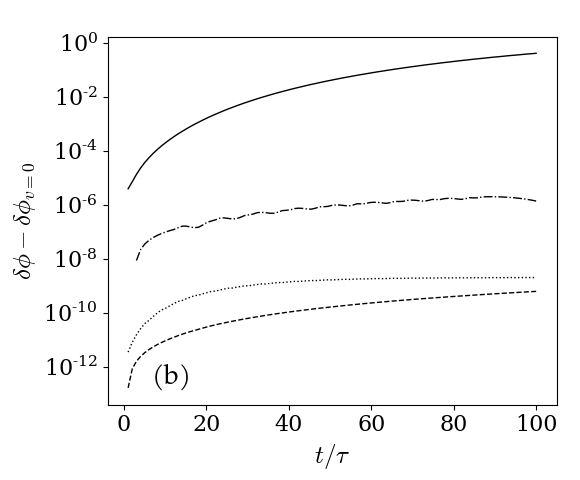}
    \caption{(a) Diferencia $|\rho_{12}|-|\rho_{12}|^{v=0}$ entre el valor absoluto de las coherencias a velocidad finita y a velocidad nula en función del tiempo y (b) diferencia $\delta\phi-\delta\phi_{v=0}$ entre las correcciones a la fase geométrica acumulada a velocidad finita y nula en función del tiempo, para distintas combinaciones de átomo y material dieléctrico. En todos los casos se considera el momento dipolar orientado en dirección del eje $x$. En ambos paneles, las curvas sólida y de trazos corresponden a un centro NV y un átomo de Rb sobre una mesa rotante de nSi respectivamente, mientras que las curvas de trazo-punto y punteadas reproducen ambos casos para una mesa de Au. Para el valor de los parámetros asociados se refiere a la tabla \ref{tabla}.} 
    \label{fig:sec5_materiales}
\end{figure}

Se ha corroborado que un disco de silicio de $12\, {\rm cm}$ de diámetro puede rotarse a frecuencias tan altas como $2\pi\times 7000\, {\rm Hz}$. Esto equivale a velocidades lineales $\sim 3500\,{\rm m}\,{\rm s}^{-1}$ e indica que las velocidades consideradas en los dos paneles de la figura \ref{fig:sec5_materiales}, de ${\rm v} = 370\,{\rm ms}^{-1}$  (equivalente a  $v = 3\times 10^{-4}$) para la mesa de nSi y ${\rm v} = 1100\,{\rm ms}^{-1}$  (equivalente a una velocidad adimensional  $v = 7.5\times 10^{-6}$) son accesibles con la tecnología actual. Dentro del rango de velocidades investigado, el balanceo de la mesa rotante es del orden de $10^{-8}$ radianes, esto es, el movimiento vertical de la superficie es de $1 {\rm nm}$ en los extremos del disco.
Es crítico mantener constante la separación entre el sistema de dos niveles efectivo y la superficie del disco para evitar efectos espurios de decoherencia. 
El experimento puede ser realizado a una distancia entre la punta del microscopio y la superficie $d = 10\,{\rm nm}$ con fluctuaciones en la distancia despreciables en comparación con los efectos de fricción \cite{farias2020towards}.

La estructura de niveles del centro NV y las técnicas para la medición y manipulación del espín electrónico están bien documentadas (ver, por ejemplo~\cite{jelezko2004observation, childress2006coherent, doherty2013n}). 
Los efectos de la fricción cuántica en el sistema de dos niveles pueden observarse realizando el experimento a velocidad relativa nula y finita entre el centro NV y el sistema de silicio dopado. 

Estos efectos, pueden en principio detectarse tanto en las escalas de decoherencia como en la fase geométrica acumulada. Para consideraciones respecto de las posibilidades de medición experimental, la figura \ref{fig:sec5_materiales} muestra la diferencia $|\rho_{12}|-|\rho_{12}|^{v=0}$ entre el valor absoluto de las coherencias, y $\delta\phi-\delta\phi_{v=0}$ entre las correcciones a la fase geométrica acumulada, para los casos a velocidad finita y a velocidad nula. en función del tiempo, y para distintas combinaciones de átomo y material dieléctrico.
La decoherencia en el grado de libertad interno del átomo puede detectarse aplicando un pulso $\pi$ al estado fundamental y midiendo posteriormente la intensidad de la fluorescencia como función del tiempo de retardo. Sin embargo, como se observa en el panel (a) de la figura \ref{fig:sec5_materiales}, la diferencia en la caída de las coherencias es pequeña y difícil de detectar con este método, aún para la propuesta que combina un material nSi con un átomo efectivo conformado por un centro NV, en la cual los efectos son más fuertes. Para esta combinación, por ejemplo, la diferencia tras 200 ciclos naturales $t \sim 200\,\tau_{\rm s}$, correspondientes a aproximadamente $15 \,{\rm ns}$, resulta menor al $5\%$ del valor inicial. Por el contrario, la velocidad tiene un impacto mucho más pronunciado en la fase geométrica, gracias a su carácter acumulativo. Como se observa en el panel (b) de la figura \ref{fig:sec5_materiales}, luego de un intervalo temporal la mitad de corto, la diferencia entre la fase geométrica a velocidad finita y la fase geométrica a velocidad nula es de orden $\mathcal{O}(1)$, un valor experimentalmente accesible.
Se espera que, para los valores de parámetros considerados, el desfasaje geométrico pueda detectarse mediante tomografía del estado.

\vspace{.5cm}
\begin{center}
   \textcolor{bordo}{\ding{163}}
\end{center}
\vspace{.5cm}
En resumen, en este capítulo se estudiaron distintos efectos del entorno para un escenario que incorpora un elemento extra respecto de los capítulos previos, el movimiento relativo entre el sistema y el entorno. En particular, se consideró un sistema de dos niveles que describe una trayectoria recta a distancia fija de un semiespacio dieléctrico, mediante un forzado externo. El conjunto se encuentra inmerso en el vacío del campo electromagnético, de forma que el entorno de la partícula está compuesto por el campo electromagnético vestido por el material. Inspeccionando la dinámica del sistema, se distinguieron dos regímenes distintos delimitados por la velocidad crítica ${\rm v}_{\rm crit} = \omega_o \,d/2$. Por debajo de este umbral de velocidad, el sistema tiende asintóticamente al estado fundamental, mientras que una vez superada la velocidad crítica, el estado asintótico es un estado mixto.
Se estudió con especial énfasis la escala temporal en que decaen las coherencias y la fase geométrica acumulada por el sistema. 

Respecto de la escala de decaimiento de las coherencias, se realizó un estudio minucioso de su dependencia con la velocidad y con la dirección del momento dipolar de la partícula. Para el caso de velocidades por debajo del umbral ${\rm v}<{\rm v}_{\rm crit}$ se encontró una expresión analítica de orden ${\rm v}^2$. Los resultados encontrados se interpretaron, como en el capítulo \ref{ch:4} en términos de interacciones entre partículas, reemplazando los efectos del medio semiconductor por partículas imagen. Estableciendo un vínculo con la literatura previa se encontró que la dependencia de la escala de decoherencia en la orientación del momento dipolar resulta inversamente proporcional a la exhibida por fricción cuántica. Esto sugiere que la presencia de dicha fuerza acelera el proceso de pérdida de coherencia. 

La corrección en fase geométrica acumulada por el sistema se descompuso en dos contribuciones: una estática, debido a la mera presencia del campo electromagnético vestido por el medio dieléctrico, y una dinámica, que depende en la velocidad de la partícula. Esta corrección dinámica es, nuevamente, de orden cuadrático, y se observó la dependencia del coeficiente cuadrático en otros parámetros y variables, como por ejemplo, el instante de observación.

Finalmente, se realizó una propuesta experimental y se comparan los resultados esperados para diversas combinaciones de material dieléctrico y partículas efectivas en la búsqueda de detectar regímenes en los cuales estos efectos resulten experimentalmente accesibles con la tecnología disponible.
\chapter{Fases geométricas a lo largo de trayectorias cuánticas}\label{ch:6}
Como se ha discutido en la sección \ref{sec:sec2_abiertos}, la búsqueda de una noción de fase geométrica que permita tratar con sistemas mixtos en evolución no-unitaria ha resultado en múltiples definiciones.
En particular, en los capítulos \ref{ch:3} a \ref{ch:5} de esta tesis se ha recurrido a la propuesta de Tong y sus colaboradores~\cite{tong2004kinematic}, cuya expresión está dada por la ecuación (\ref{eq:sec2_TongGP}), para abordar el estudio de diversos sistemas paradigmáticos.

Existe, sin embargo, otro nivel de descripción de los sistemas cuánticos abiertos que podría captar características que se desvanecen cuando se atiende exclusivamente a las propiedades del operador densidad reducido. A este nivel se accede, por ejemplo, cuando el sistema es continua e indirectamente monitoreado. En este escenario el estado del sistema está descrito por una función de onda cuya evolución suave se ve interrumpida por saltos cuánticos aleatorios. Esta secuencia de tramos de evolución suave interrumpida por saltos cuánticos se denomina una {\em trayectoria cuántica}. El objetivo de este capítulo es describir las propiedades de la fase geométrica acumulada a lo largo de las trayectorias cuánticas. La aleatoriedad introducida por los saltos que ocurren en una dada trayectoria es heredada por la fase geométrica, que adquiere un carácter estocástico, donde la distribución completa puede reconstruirse con un muestreo aleatorio sobre las trayectorias. 

Dado que la fase geométrica no es un observable, el valor medio de esta distribución de fase no debe necesariamente coincidir con la fase del estado promediado (esto es, la matriz densidad) 
La bibliografía previa al trabajo en que se basa este capítulo~\cite{viotti2023geometric}, a excepción de \cite{gefenWeak}, se restringe al estudio de la dinámica de estados puros con evoluciones suaves, o define valores medios. El estudio de las fluctuaciones en la fase geométrica es todavía en gran medida un territorio por explorar. En este capítulo se da un primer paso en dirección a saldar esa brecha, estudiando esta distribución y en qué medida está (o no) relacionada con la distribución correspondiente de franjas de interferencia en un protocolo de Eco de Espín.

\section{De la ecuación de Lindblad a las trayectorias cuánticas}\label{sec:sec6_trayectorias}
{\em La ecuación de Lindblad - }Con el objetivo de establecer una conexión clara con los capítulos anteriores, se parte de la descripción del estado de un sistema cuántico mediante un operador densidad $\rho(t)$. En este caso, bajo las condiciones que permiten realizar las aproximaciones de Born-Markov, la dinámica está gobernada por una ecuación de Lindblad 

\begin{equation}
   \dot{\rho}_{\rm s}(t) = -i\left[H, \rho_{\rm s}(t) \right] + \frac{1}{2} \sum_{\alpha} [2\,L_{\alpha}\rho_{\rm s}(t) L_{\alpha}^\dagger -  \{L_{\alpha}^\dagger L_{\alpha}, \rho_{\rm s}(t)\, \} ] \;,
   \label{eq:sec6_Lindblad}
\end{equation}
donde el primer término en el lado derecho da cuenta de la evolución unitaria del sistema, mientras que el segundo da cuenta de los efectos no-unitarios introducidos por el acoplamiento con el entorno, y la intensidad y naturaleza de este acoplamiento está codificada en los operadores de Lindblad $L_\alpha$. Para una introducción teórica a las ecuaciones maestras se refiere al apéndice \ref{sec:ap2}, donde la ecuación de Lindblad corresponde a la expresión (\ref{eq:Lindblad}). En este capítulo de considera un Hamiltoniano con dependencia periódica en el tiempo $H(t + 2\pi/\Omega)=H(t)$, con $T= 2\pi/\Omega$ el período de un ciclo en un espacio de parámetros adecuado. Los operadores de Lindblad, si dependen del tiempo, deberán también ser periódicos en el tiempo $L_{\alpha}(t + 2\pi/\Omega)=L_{\alpha}(t)$.
\\
\\\indent
{\em Dinámica monitoreada y trayectorias cuánticas - }La dinámica del sistema cambia radicalmente cuando es posible monitorear continuamente su estado. En este caso el estado del sistema permanece puro y consiste de intervalos de evolución suave (aunque no-unitaria) interrumpidos a tiempos aleatorios por cambios abruptos denominados {\em saltos cuánticos}. Una secuencia de intervalos de evolución suave y un conjunto de cambios bruscos aleatorios conforman lo que se denomina un {\em trayectoria cuántica}. La literatura sobre este tema es vasta, pudiéndose encontrar una descripción general en los siguientes libros y artículos \cite{Molmer:93, manzano, carmichael1993_open, wiseman2009quantum}, y aplicaciones, por ejemplo a sistemas de muchos cuerpos en \cite{daley2014quantum, passarelli2019improving}.
En este abordaje, la evolución del sistema se describe como sigue. 

Si a tiempo $t$ el estado del sistema es $\ket{\psi(t)}$, en un tiempo posterior $t + \delta t$ el estado será

\begin{equation}
|\psi(t+\delta t)\rangle =  \left\{
\begin{tabular}{ ccc } 
 $ \frac{K_o|\psi(t)\rangle}{\sqrt{p_{o}(t)}} $ & con probabilidad  &$p_o(t)$ \\
          &         &         \\ 
 $ \frac{K_{\alpha} |\psi(t)\rangle}{\sqrt{p_{\alpha}(t)}} $ & con probabilidad   &$p_{\alpha}(t)$\\
 \end{tabular}
 \right.
 \label{eq:sec6_evolucion}
\end{equation}
donde $o, \alpha = 1, ...$ etiquetan los distintos operadores $K_{\alpha}$ que inducen los distintos pasos dinámicos

\begin{equation}
    K_{o} = 1- i\,\delta t \left[H-\frac{i}{2}\sum_{\alpha} L^\dagger_{\alpha}L_{\alpha}\right]  \;\;\;\;;\;\;\;\;    K_{\alpha} = \sqrt{\delta t}L_{\alpha}
    \label{eq:sec6_jumpoperators}
\end{equation}
y $p_{o/\alpha}(t) = \bra{\psi(t)}K_{o/\alpha}^\dagger K_{o/\alpha} \ket{\psi(t)}$. Cada elección en el lado derecho de la ecuación (\ref{eq:sec6_evolucion}) representa un paso dinámico de características diferentes. La segunda línea corresponde al acontecimiento de un salto cuántico de tipo $K_{\alpha}$ a tiempo $t$, mientras que la primer línea es una evolución suave (sin saltos), aunque alterada respecto de la unitariedad debido a que adquirir la información de que no ocurrieron saltos modifica la evolución del sistema. El operador de evolución libre de saltos $K_{o}$ puede también entenderse como generado por un Hamiltoniano efectivo de arrastre $H_o$ con el que se relaciona del modo usual $K_{o} = 1 - i \, \delta t\, H_{o}$. 

La evolución completa en un intervalo temporal $[0, t]$ está entonces caracterizada por una secuencia de $N_J$ saltos de tipo $\alpha_i$ que ocurren a tiempo $t_i$. Se denomina la cadena de eventos 

\begin{equation}
	{\cal R} (t, N_{J}) = \{(\alpha_1, t_1), \dots, (\alpha_i, t_i),\dots(\alpha_{N_J}, t_{N_{J}})\},
	\label{eq:R}
\end{equation}
con $0 \ge t_i \ge t \;\; \forall  i$, la {\em trayectoria cuántica}.

Como se ha mencionado arriba, este marco formal emerge naturalmente cuando el sistema se monitorea continua e indirectamente de forma que cada trayectoria puede ser vista como el resultado de mediciones continuas realizadas sobre el entorno en una dada base. Desde esta perspectiva, el monitoreo continuo puede conducir a la mitigación de los efectos de decoherencia producidos por el entorno~\cite{siddiqi2_observing},
y se han propuesto además esquemas de post-selección y corrección de errores~\cite{ahn2002_error, siddiqi1_observation}.  Las propiedades de los operadores de Kraus $K_{o/\alpha}$ garantizan que las probabilidades de obtener un dado resultado se sumen a la unidad, y el paso temporal $\delta\,t$ debe tomarse lo bastante pequeño para que la expansión a primer orden resulte válida, lo que requiere $\sum_{\alpha } p_{\alpha} \ll 1$. Promediando sobre todas las secuencias de saltos cuánticos posibles se recupera la ecuación de Lindblad~\cite{Molmer:93} (\ref{eq:sec6_Lindblad}), mientras que la implicación no es válida en sentido contrario y un número infinito de distintas descomposiciones dan origen a la misma evolución de Lindblad~\cite{manzano}. 

\section{Fase geométrica de un sistema abierto monitoreado}
Se ha mostrado en los distintos capítulos de esta tesis que la acumulación de una fase geométrica en la evolución dinámica de un sistema es un fenómeno que no se restringe a las evoluciones adiabáticas. En particular, para una trayectoria cuántica general consistente en una secuencia de intervalos de evolución suave y un conjunto de saltos cuánticos aleatorios $\mathcal{R}$, una fase geométrica que trata con los dos aspectos de la evolución puede definirse adecuadamente.

\subsection{Fase de Pancharatnam a lo largo de una trayectoria cuántica}

Como se afirmó en la sección \ref{sec:sec6_trayectorias}, la trayectoria cuántica que emerge en una realización de la evolución monitoreada del sistema puede entenderse como una secuencia de intervalos de evolución suave interrumpidos por saltos cuánticos en instantes aleatorios. Tomada de este modo, la evolución en un intervalo $t \in [0, T]$ está caracterizada por una colección ordenada de saltos de tipo $\alpha_i$ que ocurren en instantes $t_i$ según la ecuación (\ref{eq:R}), y el parámetro $t$ es una variable continua dentro de los intervalos delimitados por los $t_i$s. En el enfoque de saltos cuánticos, sin embargo, el algoritmo aplicado para construir las trayectorias es el siguiente~\cite{Molmer:93}. El intervalo temporal $[0, T]$ se discretiza en N pasos de longitud $\delta \,t$ y el estado se actualiza paso a paso consistentemente por la acción de un operador no-hermítico elegido aleatoriamente como se describe en la ecuación (\ref{eq:sec6_evolucion}).
De esta forma, cada trayectoria cuántica puede también entenderse desde un punto de vista algorítmico como una colección ordenada de estados generados por la acción de una secuencia específica de operadores $K_{0, \alpha}$ dados por la ecuación (\ref{eq:sec6_jumpoperators}, y es de esta forma un conjunto discreto de estados.

Para una secuencia de N estados puros discretos, la expresión adecuada para la fase geométrica se construye mediante la diferencia de fase de Pancharatnam entre cada par de estados~\cite{Carollo_original,Carollo_review, mukunda1993quantum}, y está dada por
\begin{equation}
    \phi_P[\psi] = \arg\bra{\psi_1}\ket{\psi_\mathrm{N}}-\arg(\bra{\psi_1}\ket{\psi_{2}}...\bra{\psi_{\mathrm{N}-1}}\ket{\psi_{\mathrm{N}}}).
    \label{eq:sec6_phiDiscreta}
\end{equation}
Esta fase resulta trivialmente invariante de gauge $U(1)$ y no se basa en la condición de adiabaticidad ni requiere unitariedad ya que permite que los estados en la secuencia sean de norma arbitraria siempre y cuando no se anulen perfectamente. Gracias a estas características, se afirma como una definición natural de fase geométrica para ser aplicada a dinámicas monitoreadas en las que la evolución no está generada por operadores hermíticos. La fase definida por la ecuación (\ref{eq:sec6_phiDiscreta}) equivale a la fase geométrica unitaria de la ecuación (\ref{eq:sec2_kinGP}) asociada a la trayectoria construida uniendo estados consecutivos mediante la curva geodésica  del espacio proyectivo correspondiente.
Mientras que esta definición no implica por si misma ninguna condición sobre la cantidad de estados en la secuencia, cuando se aplica en el abordaje de saltos cuánticos a sistemas abiertos el numero N de estados está acotado por debajo como consecuencia de la condición que reina sobre el paso temporal $\delta t$.

La evolución en un intervalo temporal $[0,T]$ consta de $\text{N} =T / \delta\,t \gg1$ estados. Descomponiendo la secuencia de estados $\{\ket{\psi_1} \ket{\psi_2}... \ket{\psi_\text{N}}\}$ en conjuntos delimitadoos por los instantes específicos $t_i$ en los que acontece un salto se construye un puente entre las dos descripciones de la trayectoria brindadas.
Cada sub-intervalo temporal $[t_i, t_{i+1}]$, discretizado en pasos de longitud $\delta t$, consta de un número dado de pasos que depende de los valores específicos de $t_i$ y $t_{i+1}$. Desde un dado instante de salto $t_i$, cualquier paso temporal en el intervalo consecutivo puede hallarse mediante $t_i + k_i\,\delta t$, esto es, sumando una cantidad $k_i \in \mathbb{N}$ de incrementos $\delta t$, hasta alguna cantidad máxima $k_i^*$ que satisface $t_{i+1} = t_i + k^{*}_i\,\delta t$

\begin{figure}[ht!]
    \centering
    \begin{tikzpicture}
        \draw[thick]     (0.0,0) -- (0.6,0);
        \draw[thick]     (1.1,0) -- (5.2,0);
        \draw[thick, ->] (5.7,0) -- (7.0,0);
        
        \filldraw[black] (0.75,0) circle (.4pt);
        \filldraw[black] (0.85,0) circle (.4pt);
        \filldraw[black] (0.95,0) circle (.4pt);
        
        \filldraw[black] (5.35,0) circle (.4pt);
        \filldraw[black] (5.45,0) circle (.4pt);
        \filldraw[black] (5.55,0) circle (.4pt);
        
        \draw[thick]     (0.25,-0.2) -- (0.25, 0.2);
        \draw[thick]     (0.25,-0.2) -- (0.32,-0.2);
        \draw[thick]     (0.25, 0.2) -- (0.32, 0.2);
        \node at         (0.25,-0.5) {$0$};
        
        \draw[thick]     (6.5,-0.2) -- (6.5, 0.2);
        \draw[thick]     (6.5,-0.2) -- (6.43,-0.2);
        \draw[thick]     (6.5, 0.2) -- (6.43, 0.2);
        \node at         (6.55,-0.5) {$T$};
        
        \draw[thick]     (1.5,-0.15) -- (1.5,0.15);
        \node at         (1.5,-0.5) {$t_i$};
        \draw[thick, ->] (1.5, 0.25) -- (1.5,0.8);
        \node at         (1.5, 1.1) {$\ket{\psi(t_i)}$};
        
        \draw[thick]     (4.5,-0.15) -- (4.5,0.15);
        \node at         (4.5,-0.5) {$t_{i+1}$};
        
        \draw     (1.75,-0.1) -- (1.75,0.1);
        \draw     (2.00,-0.1) -- (2.00,0.1);
        \draw     (2.25,-0.1) -- (2.25,0.1);
        \draw     (2.50,-0.1) -- (2.50,0.1);
        \draw     (2.75,-0.1) -- (2.75,0.1);
        \draw     (3.00,-0.1) -- (3.00,0.1);
        \draw     (3.25,-0.1) -- (3.25,0.1);
        
        \draw[thick, ->] (3.5, 0.25) -- (3.5,0.8);
        \node at         (3.5, 1.1) {$\ket{\psi(t_i + k_i\,\delta t)}$};
        \node at         (3.5, 1.5) {\textcolor{white}{.}};
        \draw[thick, ->] (1.75,-0.5) arc (250:310:1.7);
        \node at         (2.5, -1.0) {$+ k_i\,\delta t$};
        
        \draw     (3.50,-0.1) -- (3.50,0.1);
        \draw     (3.75,-0.1) -- (3.75,0.1);
        \draw     (4.00,-0.1) -- (4.00,0.1);
        \draw     (4.25,-0.1) -- (4.25,0.1);
        
        \draw[gray] (0.50,-0.1) -- (0.50,0.1);
        \draw[gray] (1.25,-0.1) -- (1.25,0.1);

        \draw[gray] (4.50,-0.1) -- (4.50,0.1);
        \draw[gray] (4.75,-0.1) -- (4.75,0.1);
        \draw[gray] (5.00,-0.1) -- (5.00,0.1);
        
        \draw[gray] (6.00,-0.1) -- (6.00,0.1);
        \draw[gray] (6.25,-0.1) -- (6.25,0.1);

    \end{tikzpicture}
    \caption{Diagrama ilustrativo del intervalo temporal $[0,T]$. Tanto la discretización en pasos de longitud $\delta\,t$ como la división en los instantes 'de salto' $t_i$ están indicados. La relación entre los instantes temporales y el estado del sistema se representa también. \label{fig:time_interval}}
\end{figure}
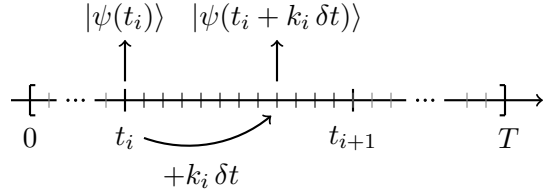

En cada instante, el resultado de la medición realizada en el entorno estará asociado al operador correspondiente de la ecuación (\ref{eq:sec6_jumpoperators}) que actúa sobre el sistema, y al estado generado por dicha acción. En consecuencia, existe una correspondencia uno-a-uno entre el conjunto discreto que conforma el intervalo temporal y la colección de estados que compone la trayectoria. La descomposición del intervalo temporal en los instantes $t_i$ puede entonces mapearse en una descomposición de la trayectoria según

\begin{equation}
    \bigcup_{i=0}^{N_J}\{\ket{\psi(t_i + k_i\,\delta t)} k_i = 0, ..., k^{max}_i -1\}
\end{equation}
con $N_J$ el número de saltos que ocurren en la trayectoria y los índices $i=0$ e $i=N_J +1$ señalando los límites del intervalo temporal completo $t_{0} = 0$ y $t_{N_J +1} = T$.

Introduciendo esta descomposición, la fórmula (\ref{eq:sec6_phiDiscreta}) para la fase geométrica puede reescribirse agrupando las diferencias de fase entre estados consecutivos infinitesimalmente distintos $\ket{\psi(t_i + k_i\delta t)}$ y $\ket{\psi(t_i + k_i\delta t + \delta t)}$  por un lado, y la diferencia de fase entre estados radicalmente distintos $\ket{\psi(t_i)}$ y $K_{\alpha_i}\ket{\psi(t_i)}$ por otro,

\begin{equation*}
    \phi_P = \arg\bra{\psi(0)}\ket{\psi(T)}
    - \sum_{i=0}^{N_J}\sum_{k_i=1}^{k^{*}_i-1} \arg\bra{\psi(t_i + k_i\,\delta t)}\ket{\psi(t_i + k_i\delta t+\delta t)}
    - \sum_{i=0}^{N_J} \arg\bra{\psi(t_i)}K_{\alpha_i}\ket{\psi(t_i)}.
\end{equation*}
Tomando el límite continuo $\delta t/T\rightarrow 0$ dentro de los intervalos de evolución suave~\cite{Carollo_original, Carollo_review}, esta expresión permite derivar una fase geométrica adecuada para la noción de trayectoria cuántica presentada en este capítulo. 

Así, considerando una evolución que transcurre en un intervalo temporal $[0, T]$, parametrizado con $t$, la fase geométrica asociada a una trayectoria en la que se registran $N_J$ saltos en los instantes $t_i$ puede escribirse como

\begin{equation}
    \phi [{\cal R}] = \arg\bra{\psi(0)}\ket{\psi(T)} - {\rm Im}\sum_{i=0}^{N_{J}}\int_{t_i}^{t_{i+1}} dt\frac{\bra{\psi(t)}\ket{\dot{\psi}(t)}}{\bra{\psi(t)}\ket{\psi(t)}} - \sum_{(t_i, \alpha_i) \in \mathcal{R}}\arg\bra{\psi(t_{i})}K_{\alpha_i}\ket{\psi(t_{i})} ,
    \label{eq:sec6_GP_traj}
\end{equation}
donde ${\cal R} = {\cal R}(T, N_J)$ con $t_0 = 0$ y la convención de que $t_{N_{J} +1} \equiv T$ en la sumatoria de integrales.

\subsection{Fase geométrica de una trayectoria cuántica}

La definición de fase geométrica dada por la ecuación (\ref{eq:sec6_GP_traj}) será la base del análisis realizado en este capítulo.
Como resulta evidente por su dependencia en los instantes $t_i$ y la naturaleza de los saltos que tienen lugar a lo largo de la trayectoria, la fase $\phi[{\cal R}(T, N_J)]$ será una variable estocástica dependiente de la trayectoria ${\cal R}(T, N_J)$.

El primer término en la ecuación (\ref{eq:sec6_GP_traj}) es la fase relativa total entre el estado inicial y final. Los términos restantes son de dos tipos distintos, reflejando las propiedades de la dinámica misma. El segundo término representa las fases dinámicas acumuladas a lo largo de los intervalos de evolución suave que tienen lugar antes del primer salto, entre saltos consecutivos y después del último salto registrado, y que debe sustraerse para acceder al objeto puramente geométrico $\phi_{\cal R}$. 
La ocurrencia a tiempo $t_i$ de un salto generado por el operador $K_{\alpha_i}$ incorpora una contribución $\arg\bra{\psi(t_{i})}K_{\alpha_i}\ket{\psi(t_{i})}$ que representa la diferencia de fase entre el estado antes y después del salto. Este término equivale a la fase geométrica asociada a la trayectoria construida uniendo ambos estados con la geodésica más corta en el espacio proyectivo. 

La expresión en la ecuación (\ref{eq:sec6_GP_traj}) es invariante frente a transformaciones de gauge $U(1)$. No requiere que la trayectoria describa una curva cerrada en el espacio de estados ni se basa en la condición de adiabaticidad. Más aún, tampoco demanda unitariedad, ya que está bien definida también para estados $\ket{\psi(t_i)}$ o $\ket{\psi(t')}$ no-normalizados (siempre que sean de norma no-nula).

Adecuada para aplicarse a las trayectorias que emergen en una descomposición de la matriz densidad, la ecuación (\ref{eq:sec6_GP_traj}) se ha empleado en formas limitadas para abordar la definición de fases geométricas que se ajusten a evoluciones no-unitarias. Una ruta inicialmente explorada fue enfocarse en la trayectoria sin saltos~\cite{Carollo_original, Carollo_review}. Este enfoque, que desprecia la posibilidad de los saltos cuánticos restringiéndose a la evolución enteramente suave, preserva las definiciones conocidas para la fase geométrica aplicables a estados puros e introduce los efectos no-unitarios mediante la no-hermiticidad de $H_o$. Si no se registran saltos a lo largo de toda la trayectoria, esto es, si $\mathcal{R}(T, 0) = \emptyset$, la fase geométrica $\phi_{0} \equiv \phi [{\cal R}(T, 0)= \emptyset] $ se lee

\begin{equation}
    \phi_{0}  = \arg\bra{\psi(0)}\ket{\psi(T)}  -{\rm Im}\int_{0}^{T}\frac{\bra{\psi(t)}\ket{\dot{\psi}(t)}}{\bra{\psi(t)}\ket{\psi(t)}}dt
    \label{eq:sec6_GP_nj}
\end{equation}
la cual se reduce trivialmente a la expresión para la fase geométrica acumulada en una evolución unitaria general de la ecuación (\ref{eq:sec2_kinGP}) cuando este sea en efecto el caso y, en consecuencia, los estados estén instantáneamente normalizados haciendo el denominador $\bra{\psi(t)}\ket{\psi(t)}\equiv 1\; \forall \, t$. La ecuación (\ref{eq:sec6_GP_nj}) se reduce además a la expresión (\ref{eq:sec2_AA}) propuesta por Aharonov y Anandan y a la fase adiabática de Berry (\ref{eq:sec2_Berry}) cuando las hipótesis de cada definición se satisfacen, esto es, para evoluciones cíclicas aunque no necesariamente adiabáticas y para evoluciones cíclicas y adiabáticas.
Es importante notar que la fase $\phi_{0}$ está mal definida si alguno de los productos internos en su argumento se anulan. Esta observación resultará de relevancia cuando se discuta la transición topológica en la sección~\ref{sec:sec6_topological}. 

Otros trabajos consideran la descomposición completa de la ecuación de Lindblad, sugiriendo definir la fase geométrica del estado $\rho(t)$ (promedio sobre el ensamble de trayectorias) como un promedio sobre el ensamble de fases $\{\phi_{\mathcal{R}}\} =\phi_{\{\mathcal{R}\}}$ obtenido aplicando la ecuación (\ref{eq:sec6_GP_traj}) a cada trayectoria~\cite{Carollo_original, Carollo_review, buri}. Se ha discutido extensamente si esta es una definición apropiada de fase geométrica para la matriz densidad que representa el estado de un sistema, dado que no permite una relación uno a uno entre el conjunto de matrices densidad posibles y los valores de la fase obtenidos \cite{Sjo_no, bassi2006_no, sjoqvist2010_hidden}.
Todas las propuestas arriba mencionadas, así también como la propuesta (\ref{eq:sec2_TongGP}) realizada desde el enfoque cinemático, se restringen a evoluciones modificadas sobre las que aplican las definiciones para estados puros, o bien, buscan definir una fase geométrica consistente para el operador densidad $\rho(t)$, que representa una descripción en valor medio.

Los procesos estocásticos que emergen en la descomposición de la matriz densidad adquieren, sin embargo, relevancia física en los esquemas de monitoreo continuo.
Dado que la aleatoriedad introducida por la ocurrencia de saltos en una dada trayectoria se refleja en la fase geométrica, este enfoque requiere que el estudio de los efectos inducidos por el entorno en la fase geométrica se realice desde una perspectiva estadística.
La probabilidad asociada con un dado valor de fase geométrica se relaciona con aquella para una dada trayectoria según
\begin{equation}
    P[\phi] = \sum_{\mathcal{R}/\phi[{R}]=\phi} P[\mathcal{R}].
    \label{eq:sec6_phase_probability}
\end{equation}
El valor medio corresponde únicamente al primer momento de dicha distribución
$$
	\Bar{\phi} = \arg\left(\sum \,e^{i\phi} P[\phi]\right).
$$
y en general puede no ser suficiente para caracterizar la dinámica.

\section{El modelo}
Consideramos una evolución unitaria para la cual, en el límite adiabático, la fase geométrica acumulada en un ciclo sería la fase de Berry. 
Es conveniente comentar el límite adiabático de dinámica lenta ya en este punto, puesto que será un tópico fundamental a lo largo de este capítulo.
Si la evolución es unitaria, para Hamiltonianos con dependencia temporal periódica cuyos períodos sean suficientemente largos, un sistema preparado en un autoestado de energía permanece en el autoestado instantáneo a menos de pequeñas correcciones por transiciones de Landau-Zener entre niveles de energía. En otras palabras, la ocupación de un dado autoestado del Hamiltoniano no cambia en el tiempo (ver sección \ref{sec:sec2_Berry}).
La situación cambia completamente en presencia de un entorno. En este caso, un límite adiabático propiamente dicho no está bien definido, dado que el límite de sistema guiado lentamente en el que el límite adiabático se basa, es también el régimen en que las consecuencias del entorno son más severas y el sistema alcanza un estado estacionario. El límite adiabático mismo debe ser reconsiderado en un sistema abierto \cite{Sarandy_2005, thunstrom2005adiabatic, yi2007adiabatic, oreshkov2010adiabatic, venuti2016adiabaticity}, ya que la existencia de un continuo de niveles de energía hace de las diferencias de energía una mala escala de referencia para definir regímenes. Los efectos debidos a la no-adiabaticidad y las correcciones originadas en la presencia del entorno aparecen de esta forma como inevitablemente relacionadas.

Más concretamente, consideramos una partícula de espín $1/2$ en presencia de un campo magnético rotante $\mathbf{B}(t) =\omega\, \hat{\text{\bf{n}}}_\mathbf{B}(t)$, cuya dirección está dada por $\hat{\text{\bf{n}}}_\mathbf{B }=(\sin{(\theta)}\cos(\Omega\,t),\sin{(\theta)}\sin(\Omega\,t), \cos{\theta})$ con ángulo polar $\theta$ fijo y ángulo azimutal variable $\Omega\,t$. 
Una evolución unitaria de este tipo, desarrollada analíticamente en el apéndice \ref{apendice1}, está generada por el Hamiltoniano

\begin{equation}
    H(t) = \frac{1}{2}\,\mathbf{B}(t)\cdot \boldsymbol{\sigma},
    \label{eq:sec6_Hamiltonian}
\end{equation}
con $\boldsymbol{\sigma} = \sigma_x\hat{x} + \sigma_y\hat{y} 
+ \sigma_z\hat{z}$; y $|0\rangle$ y $ |1\rangle $ los autoestados de $\sigma_z$. Los autoestados instantáneos de $H(t)$, por otra parte, se denotan con $\ket{\psi_-(t)}$ y $\ket{\psi_+(t)}$.
Si el sistema pudiera mantenerse perfectamente aislado de su entorno mientras que la dirección del campo $\mathbf{B}(t)$ varía adiabáticamente en un ciclo $t \in [0, T]$ con $T = 2\pi/\Omega$, un autoestado adquiriría una fase adiabática (de Berry) $\Phi_{\rm a}^\pm = -\pi(1\mp\cos\theta)$ donde el signo $\mp$ depende del autoestado de energía en el cual el sistema fue preparado inicialmente. Este resultado se obtiene explícitamente en las secciones \ref{sec:sec2_ejemploBerry} y \ref{sec:sec2_ejemploMukunda}.
\\
\\\indent
{\em Operadores de Lindblad - }Para un sistema que evoluciona según $H(t)$ dado por la ecuación (\ref{eq:sec6_Hamiltonian}) acoplado a un entorno de osciladores armónicos, una ecuación de Lindblad consistente de la forma (\ref{eq:Lindblad}) puede derivarse desde el modelo microscópico siempre y cuando la evolución sea lo suficientemente lenta~\cite{albash2012, albash2015}, con operadores de Lindblad dados por

\begin{eqnarray}
	L_{-}(t)  &= &\sqrt{\gamma_-}\bra{\psi_-(t)}\sigma_x\ket{\psi_+(t)}\ket{\psi_-(t)}\bra{\psi_+(t)}, \nonumber\\  
	L_{+}(t) &= &\sqrt{\gamma_+}\bra{\psi_+(t)}\sigma_x\ket{\psi_-(t)}\ket{\psi_+(t)}\bra{\psi_-(t)}\label{eq:sec6_operators},\\
	L_{d}(t) &=  &\sqrt{\gamma_d}\sum_{i=\pm}\bra{\psi_i(t)}\sigma_x\ket{\psi_i(t)}\ket{\psi_i(t)}\bra{\psi_i(t)},  \nonumber
\end{eqnarray}
que dan cuenta de los procesos de relajación, excitación espontánea y desfasaje respectivamente. Las constantes de acoplamiento consideradas en este capítulo son, en términos de la tasa de disipación $\Gamma$, $\gamma_-=\Gamma\;;\;\gamma_d=0.32\, \Gamma$, mientras que consideramos que $\gamma_+$ es despreciable (todos los resultados que mostraremos son bastante generales y no dependen cualitativamente de los parámetros elegidos). Los saltos generados por los operadores definidos arriba según la ecuación (\ref{eq:sec6_jumpoperators}) llevan, tras tomar el valor medio, a una ecuación  de Lindblad consistente para evoluciones dinámicas lentas~\cite{jumpsAdiabatic}. Los operadores definidos en la ecuación (\ref{eq:sec6_operators}) inducen transiciones entre autoestados instantáneos del Hamiltoniano definido en  la ecuación (\ref{eq:sec6_Hamiltonian}). Con el objetivo de realizar un análisis lo más general posible, se incluye un término adicional en el Lindbladiano que requiere la consideración de un cuarto operador 
\begin{equation}
    L_{z} =  \sqrt{\gamma_z} \sigma_z
    \label{eq:sec6_operator_z}
\end{equation}
a lo largo de una dirección fija en la esfera de Bloch. Esta elección particular de $\sigma_z$ como el operador adicional está motivada por la necesidad de introducir transiciones que no involucren únicamente autoestados de energía. Cualquier otro operador de Lindblad que difiera de aquellos en la ecuación (\ref{eq:sec6_operators}) conduciría a conclusiones cualitativas semejantes.  

\begin{SCfigure}[10][ht!]
    \includegraphics[width = .4\linewidth]{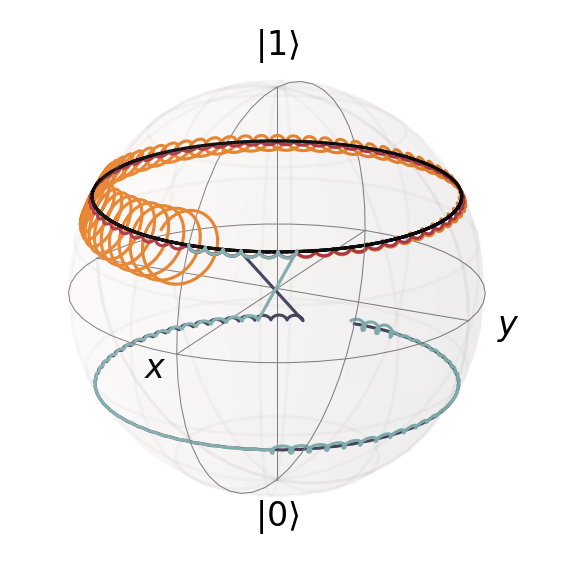}
    \caption{Trayectorias en la esfera de Bloch descritas por el estado del sistema bajo diferentes circunstancias. La línea negra corresponde a la evolución unitaria en el límite adiabático. La línea roja que describe un anillo ondulado corresponde a una dinámica unitaria genérica en la que las correcciones no-adiabáticas comienzan a ser visibles.
    En presencia de un entorno, el estado puede sufrir saltos o ser suavemente conducido durante toda la evolución. La trayectoria naranja corresponde a un arrastre completamente suave. Por otro lado, la curva azul muestra un salto que proyecta el estado en el fundamental y es por lo demás suavemente acarreado. Finalmente, el camino celeste muestra un caso con múltiples saltos.} 
    \label{fig:sec6_bloch}
\end{SCfigure}

Mientras que la evolución unitaria del sistema cerrado sigue la curva enrollada indicada en roja en la figura \ref{fig:sec6_bloch}, la dinámica real seguirá, con alguna probabilidad, el camino azul, esto es, será discontinua y no necesariamente cerrada tras un ciclo del Hamiltoniano incluso en el caso en que el campo varía lentamente. Más aún, cuanto más lenta sea la variación del forzado, más saltos ocurrirán (ver la curva celeste en la figura \ref{fig:sec6_bloch}). La tarea de las próximas secciones será caracterizar las fases geométricas acumuladas bajo estas condiciones.
\\
\\
\subsection{Evolución suave sin saltos}\label{sec:sec6_NoJump}

Una trayectoria cuántica de interés particular es aquella que resulta suave a lo largo de toda la evolución. Antes de abordar la caracterización de las fases geométricas en sistemas indirectamente monitoreados, se provee una introducción a la evolución que da origen a esta trayectoria. Cuando los resultados de las mediciones realizadas sobre el entorno registran cero saltos, la dinámica describe un camino continuo y suave en el espacio de estados y está generada por un Hamiltoniano de arrastre efectivo no-hermítico, que depende tanto del Hamiltoniano hermítico del sistema como de los operadores de Lindblad según lo describe la ecuación (\ref{eq:sec6_jumpoperators}).
En el modelo considerado en este capítulo, el Hamiltoniano de arrastre efectivo $H_o$ que gobierna la dinámica sin saltos [$K_o = 1 - \delta t H_o$ en la Eq.(\ref{eq:sec6_jumpoperators})] está dado por

\begin{equation}
H_o(t) = \left(1 - i\frac{\Gamma}{2\omega} f(t)\right)\,H(t)
\label{eq:sec6_nojump_hamiltonian}
\end{equation}
con $f(t)=\cos^2(\theta) + \sin^2(\theta)\sin^2(\Omega t)$. Nótese que, debido a la unitariedad de las matrices de Pauli, la evolución sin saltos es completamente independiente del cuarto operador de Lindblad $L_z$ introducido ad hoc y, consecuentemente, del parámetro $\gamma_z$. Un ejemplo ilustrativo de la trayectoria generada por la evolución descrita, a la que se refiere como trayectoria sin saltos en lo que sigue, es la curva anaranjada en la figura \ref{fig:sec6_bloch}.

Mientras que esta trayectoria es única, el número de trayectorias posibles (aunque no uniformemente probables) en las que ocurren $N_J > 0$ saltos crece con el número $N_J$ de saltos, y diverge cuando $\delta t$ tiende a cero. Su unicidad hace de la trayectoria sin saltos especialmente adecuada para el análisis de algunas características de las fases geométricas a las que retornaremos en la sección~\ref{sec:sec6_topological}. 
\\
\\\indent
{\em Solución analítica - }
Como se mencionó anteriormente, la evolución sin saltos puede entenderse como la evolución generada por el Hamiltoniano no-hermítico de la ecuación (\ref{eq:sec6_nojump_hamiltonian}), de forma que un estado no-normalizado $|\Tilde{\psi}(t)\rangle$ obedece la ecuación de Schrodinger

\begin{equation}
    i\frac{d}{dt}|\Tilde{\psi}(t)\rangle = H_o(t)  \,|\Tilde{\psi}(t)\rangle
    \label{eq:sec6_D_sch}
\end{equation}
donde $H_o(t)$ es no solamente no-hermítico sino también explícitamente dependiente del tiempo a través de la función $f(t)$. El Hamiltoniano efectivo de arrastre tiene los mismos autoestados que el Hamiltoniano del sistema aislado $H(t)$, pero los autovalores asociados son en este caso complejos y dependientes del tiempo, dados por $\pm \omega/2\,\left[1- i\,\Gamma/(2\omega) f(t)\right]$.

La dinámica del estado normalizado del sistema

\begin{equation}
    \ket{\psi(t)} = \frac{|\Tilde{\psi}(t)\rangle}{\sqrt{\langle\Tilde{\psi}(t)|\Tilde{\psi}(t)\rangle}}
    \label{eq:sec6_D_unitary_state}
\end{equation}
está gobernada por una ecuación no-lineal, considerablemente más complicada que (\ref{eq:sec6_D_sch}) que se encuentra por derivación directa de la ecuación  (\ref{eq:sec6_D_unitary_state}) en combinación con la ecuación (\ref{eq:sec6_D_sch}).

El estado no-normalizado puede expandirse en los autoestados instantáneos del $H_o(t)$, según $|\Tilde{\psi}(t)\rangle = \tilde{c}_{+}\ket{\psi_+(t)} + \tilde{c}_{-}\ket{\psi_-(t)}$. 
El cálculo explícito de la ecuación (\ref{eq:sec6_D_sch}) conduce al siguiente sistema de ecuaciones diferenciales para los coeficientes $\Tilde{c}_\pm(t)$
\begin{equation}
    \Dot{\Tilde{c}}_\pm = \left(\mp i \frac{\omega}{2} - i\frac{\Omega}{2}(1\mp\cos(\theta)) \mp \frac{\Gamma}{4}\,f(t)\right)\Tilde{c}_\pm(t)
    +\,i\frac{\Omega}{2}\sin(\theta)\,\Tilde{c}_\mp(t),
    \label{eq:sec6_D_diff}
\end{equation}
donde el término real $\sim -\Gamma \Tilde{c}_+(t)$ indica que incluso en el caso sin saltos, la presencia del entorno favorece las transiciones entre estados, dado que la amplitud para el autoestado excitado se encuentra suprimida.
Teniendo en cuenta el procedimiento de renormalización involucrado en el traspaso desde el estado no-normalizado hacia el estado real del sistema, esta supresión implica una transferencia de poblaciones desde el autoestado excitado al fundamental.
Como consecuencia, cualquier implementación trivial de la aproximación adiabática está descartada. Una segunda característica observada en la ecuación (\ref{eq:sec6_D_diff}) es que, para los parámetros utilizados en este capítulo, se obtiene una buena coincidencia reemplazando $f(t)$ por el valor medio $f(t)\sim 1 -\sin^2(\theta)/2$. 

Con este reemplazo la dinámica se vuelve fácilmente resoluble en el  sistema rotante. La dinámica del sistema resulta, formalmente, la descrita en el apéndice \ref{apendice1}, con la única diferencia en la forma exacta de los autovalores y coeficientes. La evolución suave sin salto encontrada de este modo está dada por
\begin{equation}
    \ket{\psi_{(\pm)}(t)} = \mathcal{N}_\pm \,e^{-i\Omega/2\,t}
    \left\lbrace \left[\pm(\nu+\varepsilon)e^{-i\varepsilon/2\,t}\mp(\nu-\varepsilon)e^{i\varepsilon/2\,t} \right]\ket{\psi_\pm(t)}-\Omega\sin(\theta)\ket{\psi_\mp(t)}\right\rbrace,
    \label{eq:sec6_D_state}
\end{equation}
donde tanto $\nu = \omega -\Omega\cos(\theta)- i\,\Gamma/2(1-\sin^2(\theta)/2)$ como $\varepsilon = \sqrt{\nu^2+\Omega^2\sin^2(\theta)}$ son cantidades complejas, y
$\mathcal{N}_\pm$ es un factor de normalización. En este punto, debe señalarse que la ecuación(\ref{eq:sec6_D_state}) muestra explícitamente que un autoestado del Hamiltoniano instantáneo inicial de arrastre $H_o(0)$ {\em no} permanecerá en un autoestado instantáneo a tiempo posterior en el caso general.  
\\
\\\indent
{\em Fase geométrica - } La fase geométrica acumulada en una trayectoria sin saltos puede ahora calcularse recurriendo a la ecuación (\ref{eq:sec6_GP_nj}). Mientras que, en el caso general, la expresión obtenida es considerablemente complicada, para valores pequeños de las relaciones $\Omega/\omega \sim \Gamma/\omega$ entre la frecuencia del forzado y la constante de disipación y la amplitud del campo, la fase adopta la forma 
\begin{equation}
    \phi_0 \sim -\pi(1-\cos\theta)- \pi\sin^2\theta\left(\frac{\Omega}{\omega} + \cos\theta\frac{\Omega^2}{\omega^2}\right)
    - \frac{\sin^2\theta}{4}\left(\frac{\Omega}{\omega} + \cos\theta\frac{\Omega^2}{\omega^2}\right)\frac{e^{-4\pi\Im(\nu)/\Omega}-1}{2\Im(\nu)/\Omega},
    \label{eq:sec6_D_GP} 
\end{equation}
donde el primer término en el lado derecho corresponde a la fase adiabática de Berry. El término siguiente es la corrección dominante originada exclusivamente en la no-adiabaticidad de una evolución por lo demás unitaria. El último término da cuenta del efecto no-trivial del entorno en la evolución sin saltos. Cuando la tasa de disipación $\Gamma\rightarrow 0$ tiende a cero, este término se torna una contribución extra debida a la no-adiabaticidad.  

\section{Resultados: distribuciones estadísticas}
\subsection{Distribución de fase geométrica}
\begin{figure}[ht!]
    \center
    \includegraphics[width = .495\linewidth, trim = {1cm 0 .75cm 0}]{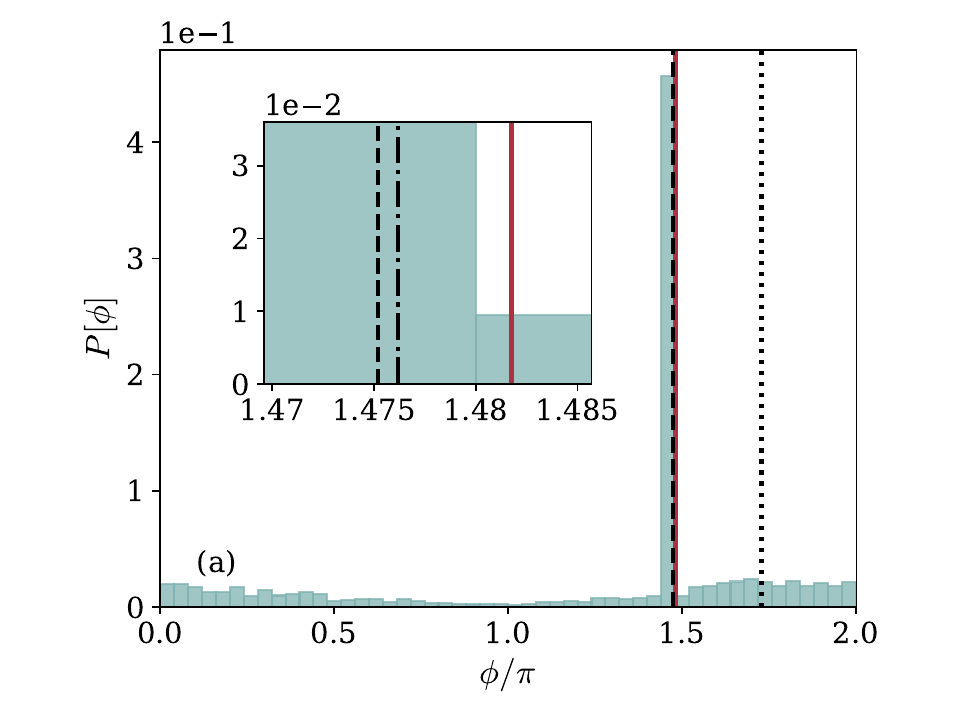}
    \includegraphics[width = .495\linewidth, trim = {.75cm 0 1cm 0}]{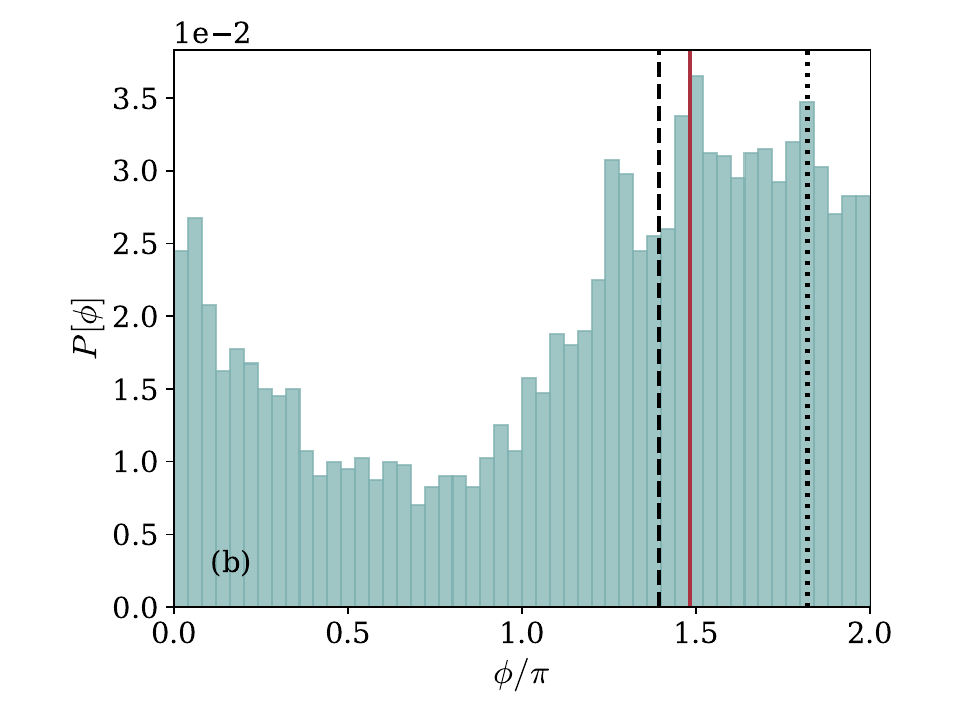}
    \caption{Distribución de probabilidad $P[\phi]$ de fases geométricas para un campo magnético orientado según $\theta = 0.34\pi$ y conducido en un ciclo a frecuencias (a) $\Omega = 5\times 10^{-3}\omega$ y (b) $\Omega = 5\times 10^{-4}\omega$.
    El entorno está caracterizado por una tasa de disipación $\Gamma = 10^{-3}\omega$ y $\gamma_z = 0$. 
    En ambos paneles, la línea sólida roja indica la fase adiabática (de Berry) $\phi^+_\mathrm{a}$, y las líneas negras de trazos y de trazo-punto señalan las fases $\phi_0$ y $\phi_u$ asociadas con una trayectoria sin saltos y con evolución general unitaria. La línea negra punteada indica el primer momento de la distribución $\Bar{\phi}$. 
    El inset en el panel (a) es una ampliación en la cual la diferencia entre estos valores de referencia se hace visible.}
    \label{fig:sec6_hist_phi}
\end{figure}

En esta sección se investiga la distribución del ensamble $\{\phi_{\mathcal{R}}\}=\phi_{\{\mathcal{R}\}}$ de fases geométricas obtenido aplicando la ecuación (\ref{eq:sec6_GP_traj}) a cada realización individual (trayectoria) de la evolución, caracterizada por algún conjunto $\mathcal{R}(T, N_J)$.
En la figura \ref{fig:sec6_hist_phi} se muestran dos casos representativos en los cuales la dinámica de una hipotética evolución unitaria sería o bien rápida (con pequeñas pero finitas correcciones no-adiabáticas); o bien lo suficientemente lenta como para considerarse en el régimen adiabático, mientras que el entorno es igual en ambos casos, caracterizado por una tasa de disipación $\Gamma = 10^{-3}\omega$, que implica $\gamma_- = \Gamma$, $\gamma_d = 0.32\, \Gamma$, y $\gamma_+$ despreciable.

Atendemos en primer lugar el caso con $\gamma_z = 0$, en el cual el entorno induce saltos que involucran los autoestados instantáneos de energía exclusivamente. Las dos situaciones, correspondientes a los dos conjuntos de parámetros indicados anteriormente se muestran en la figura \ref{fig:sec6_hist_phi}, en los paneles (a) y (b) respectivamente. En ambos paneles, se muestra también el valor para la fase adiabática de Berry como referencia, las fases geométricas acumuladas en una trayectoria sin saltos y en evolución unitaria, y el valor medio de la distribución.
Siendo independiente de $\Omega$, la fase de Berry toma el mismo valor $\phi_\mathrm{a} \sim 1.482 \pi$ en una y otra figura. Para los parámetros elegidos, el valor $\phi_{0}$ calculado aplicando la fórmula (\ref{eq:sec6_GP_nj}) sobre la trayectoria sin saltos muestra desviaciones pequeñas respecto de la fase adiabática de Berry $\phi_\mathrm{a}$. 

Mientras que los valores de estas fases geométricas de referencia son similares en ambos casos, la distribución entera es drásticamente diferente en un caso y el otro.
En el primer caso de forzado más rápido, el período $T$ es tal que una cantidad considerable de veces la evolución se completa sin registrar saltos. El valor medio de saltos en la distribución es $\Bar{N}_J=0.63$, y el pico angosto en la figura muestra estos casos de evolución completa suave. La distribución posee, adicionalmente, un pequeño fondo que revela la fase geométrica acumulada en aquellas trayectorias en las que sí ocurrieron saltos.
La composición del ensamble de refleja en el histograma por la presencia de una contribución dominante, correspondiente aproximadamente a $\sim 50\%$ de las realizaciones, del valor de la fase geométrica sin saltos, y el $50\%$ restante de los conteos mostrando una distribución ancha sobre los valores posibles. Esta distribución ancha de fondo puede ser interpretada como la aleatoriedad heredada por la fase geométrica debido al instante (aleatorio) en el que ocurre un salto. Un único término $\bra{\psi(t_i)}K_{-_i}\ket{\psi(t_i)}$ en la ecuación (\ref{eq:sec6_GP_traj}), describiendo la contribución a la fase geométrica de un salto a tiempo $t_i$, da cuenta satisfactoriamente de la distribución de fondo cuando se consideran todos los instantes posibles. 
El pico en la distribución coincide bien tanto con el valor adiabático como con el valor sin saltos. La fase promedio, por otro lado, está ligeramente corrida debido a la pequeña y poco estructurada distribución de fondo, anchamente distribuida sobre $2\pi$. Esto demuestra con claridad que incluso la ocurrencia de un único salto cuántico en un instante aleatorio conduce a fluctuaciones enormes de la fase geométrica acumulada.
En el caso con forzado lento que se muestra en el panel (b) el número medio de saltos en el conjunto de trayectorias es $\Bar{N}_J=1.77$. Esto implica que es mucho más probable que el estado del sistema sufra un cambio abrupto, e incluso más de uno, en cada realización del ciclo. Consecuentemente, la distribución de fases geométricas se ensancha y el pico angosto en torno a $\phi_0$ desaparece. 
Para estudiar esta dinámica son necesarios cumulantes de orden más alto. Las tres líneas que señalan la fase geométrica adiabática, sin saltos, y media no proporcionan información clara sobre la dinámica del sistema monitoreado. 

\begin{figure}[ht!]
    \centering
    \includegraphics[width = .6\linewidth, trim = {1cm 0 0 0}]{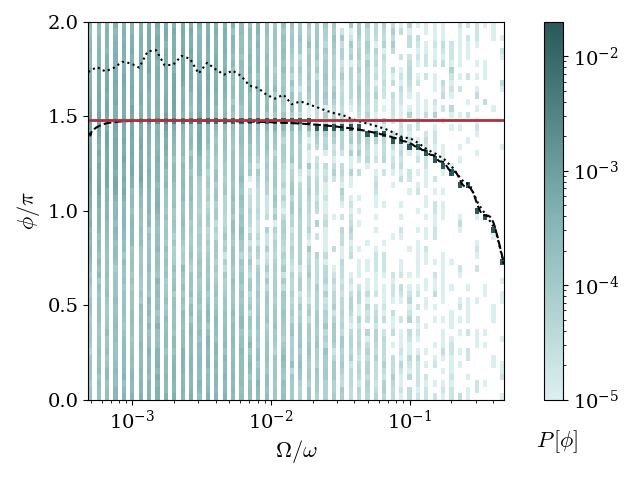}
    \caption{Distribución de probabilidad $P[\phi]$ de las fases geométricas como función de la relación $\Omega/\omega$. El campo está orientado con $\theta = 0.34\pi$ y el entorno está caracterizado por una tasa de disipación $\Gamma = 10^{-3}\omega$ y una amplitud $\gamma_z = 0$ para el cuarto operador de Lindblad. Los valores de fase geométrica se muestran en el eje $y$, mientras que su probabilidad se indica con la intensidad del color de la cuenta. La línea sólida roja indica la fase adiabática (de Berry) $\phi^+_\mathrm{a}$, y la línea negra de trazos señala la fase $\phi_0$ acumulada en una trayectoria suave sin saltos y la línea negra punteada indica el primer momento de la distribución $\Bar{\phi}$.}
    \label{fig:sec6_phi_vs_Omega}
\end{figure}

La relación $\Omega/\omega$ a la cual el campo magnético es rotado tiene, entonces, un impacto directo en la distribución de fases geométricas. Para relaciones más grandes, el sistema se encuentra expuesto al entorno durante un período de tiempo más corto, pero las desviaciones del régimen adiabático se tornan relevantes. Por otro lado, reducir la frecuencia del forzado puede resultar en un sistema expuesto a los efectos ambientales por un período demasiado largo, implicando fuertes correcciones en $\phi_{\mathcal{R}}$ respecto de $\phi^+_\mathrm{a}$. La figura \ref{fig:sec6_phi_vs_Omega} muestra la distribución de valores obtenidos para la fase geométrica a lo largo de un rango de valores para la relación $\Omega/\omega$ que incluye aquellos dos presentados en la figura \ref{fig:sec6_hist_phi}.

Para frecuencias lo suficientemente altas la distribución muestra un pico angosto en torno al valor sin saltos y prácticamente ninguna cuenta de fondo. Por otro lado, este valor 'sin saltos' se desvía considerablemente de la fase de Berry. La distribución ancha de fondo que se aprecia en la figura \ref{fig:sec6_hist_phi}.a se desarrolla a medida que la frecuencia del forzado disminuye, esto es, a medida que el período relativo se extiende. Más allá, la distribución de fondo se torna un segundo pico mientras que el pico en torno al valor sin saltos disminuye. Para valores más pequeños de la relación la distribución muestra el comportamiento exhibido por la figura \ref{fig:sec6_hist_phi}.b, esto es, una distribución ancha con un único pico. Este régimen muestra también efectos del entorno no-despreciables también en el valor sin saltos, que se desvía del resultado adiabático a pesar de que el forzado se realice lentamente. Referimos a la ecuación (\ref{eq:sec6_D_GP}) para una expresión analítica de la dependencia de esta desviación en los distintos parámetros involucrados.

\begin{figure}[ht!]
    \center
    \includegraphics[width = .495\linewidth, trim = {1cm 0 .75cm 0}]{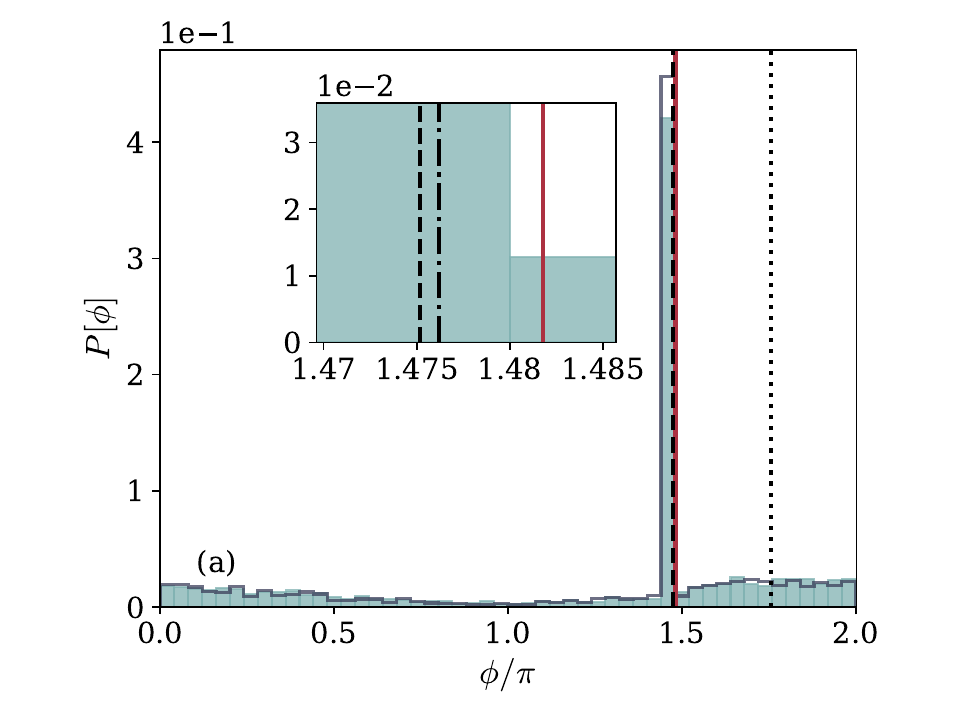}
    \includegraphics[width = .495\linewidth, trim = {.75cm 0 1cm 0}]{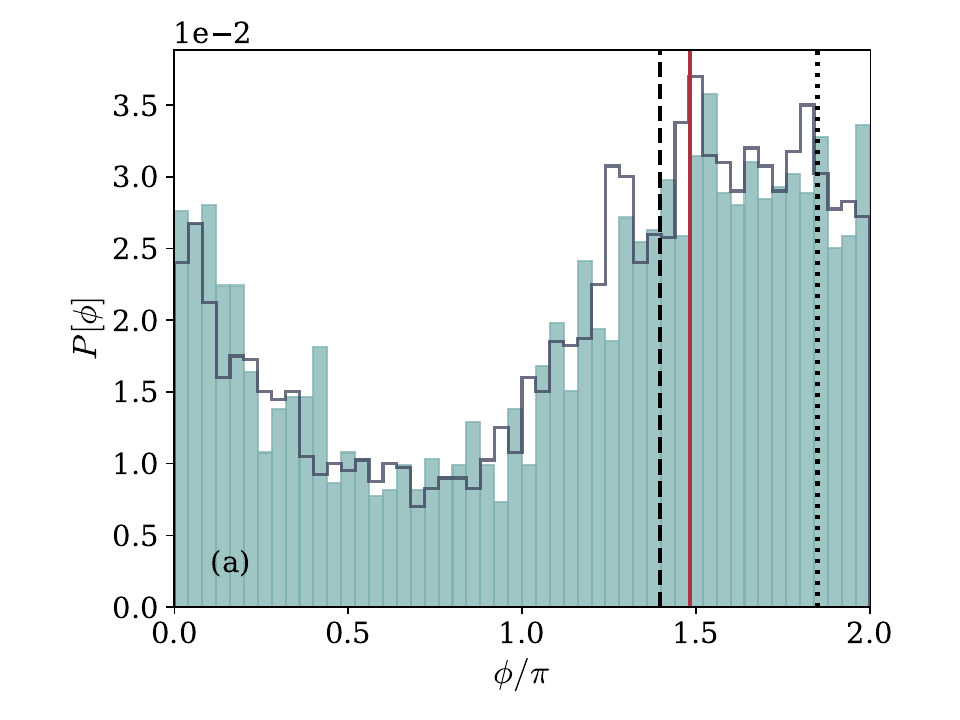}
    \caption{Distribución de probabilidad $P[\phi]$ de fases geométricas para un campo magnético orientado según $\theta = 0.34\pi$ y conducido en un ciclo a frecuencias (a) $\Omega = 5\times 10^{-3}\omega$ y (b) $\Omega = 5\times 10^{-4}\omega$.
    El entorno está caracterizado por una tasa de disipación $\Gamma = 10^{-3}\omega$ y $\gamma_z = 0.1\,\Gamma$. 
    En ambos paneles, un contorno sólido azul indica las distribuciones obtenidas en el caso con $\gamma_z=0$ para comparación. La línea sólida roja indica la fase adiabática (de Berry) $\phi^+_\mathrm{a}$, y las líneas negras de trazos y de trazo-punto señalan las fases $\phi_0$ y $\phi_u$ asociadas con una trayectoria sin saltos y con evolución general unitaria. La línea negra punteada indica el primer momento de la distribución $\Bar{\phi}$. 
    El inset en el panel (a) es una ampliación en la cual la diferencia entre estos valores de referencia se hace visible.}
    \label{fig:sec6_hist_phi_gz04}
\end{figure}

Se concluye esta sección analizando la distribución de fases geométricas cuando $\gamma_z \ne 0$. Un valor no-nulo de $\gamma_z$ induce saltos entre estados que no son autoestados instantáneos del Hamiltoniano y permite, en consecuencia, considerar una clase más amplia de situaciones. La fenomenología resultante depende únicamente cuantitativamente de la elección del operador de Lindblad $L_z$. Específicamente, tomamos $\gamma_z = 0.1\,\Gamma$ y consideramos, como antes, dos valores diferentes de la velocidad a la cual el sistema se conduce en un ciclo. Los resultados se muestran en la figura \ref{fig:sec6_hist_phi_gz04}, con características cualitativas de la distribución muy parecidas a las del caso con $\gamma_z =0$.

La figura \ref{fig:sec6_hist_phi_gz04}.a muestra el caso más rápido. El número medio de saltos  $\bar{N}_J = 0.69$ es ligeramente más grande al obtenido en el caso $\gamma_0=0$. Los saltos adicionales descritos por el operador $K_z$ no son suficientes para modificar cualitativamente la distribución, que continúa mostrando un pico bien definido que surge de las trayectorias suaves sin saltos más una distribución de fondo ancha y pequeña. En la figura \ref{fig:sec6_hist_phi_gz04}.b, que muestra el caso en que el sistema se conduce más lentamente, el número medio de saltos también es ligeramente superior al obtenido para el caso con $\gamma_z = 0$ debido a la presencia de saltos de tipo $\gamma_z$, alcanzando un valor $\bar{N}_J = 2.66$.  

Los casos discutidos arriba contienen un primer mensaje. La naturaleza estocástica de la fase geométrica en dinámicas monitoreadas requiere ser tenida en cuenta y no es posible caracterizarla mediante un único valor. Esto plantea la pregunta adicional de cómo este hecho se refleja en los resultados experimentales. Para abordar esta cuestión, se considera a continuación un protocolo de Eco de Spin (ver sección \ref{sec:sec2_SpinEcho}) y se observa cómo y cuándo la distribución de franjas de interferencia se ve afectada por la aleatoriedad del proceso.

\subsection{Distribución de franjas de interferencia en un protocolo de Eco de Espín}
Como se introdujo en la sección \ref{sec:sec2_SpinEcho}, un mecanismo para medir la fase de Berry es la aplicación de un protocolo de Eco de Espín, el cual consiste en la siguiente serie de pasos. El sistema se prepara inicialmente en un estado superposición $\psi(0)$ que, en términos de los autoestados del Hamiltoniano se expresa como $(1/\sqrt{2})(\ket{\psi_+(0)} + \ket{\psi_-(0)})$. A continuación, se conduce el sistema en un ciclo $t\in[0,T]$ en forma adiabática, causando que cada autoestado adquiera tanto una fase dinámica como una fase geométrica $\phi^\pm_{\rm a}$. Una operación de inversión de espines seguida por un segundo ciclo $t\in[T,2T]$, en sentido inverso conducen a la cancelación de las contribuciones dinámicas, resultando en una fase puramente geométrica. De esta forma, la fase de Berry puede extraerse mediante tomografía o notando que la probabilidad de que el sistema retorne al estado inicial, la {\em probabilidad de persistencia}, se relaciona con la fase de Berry según $| \langle \psi(0) | \psi(2\,T) \rangle |^2 = \cos^2(2\,\phi^+_{\rm a}).$

La relación entre la probabilidad de persistencia y la fase geométrica dada arriba descansa sobre dos factores: el régimen adiabático que impide las transiciones entre autoestados de energía y la cancelación exacta de las fases dinámicas durante el protocolo. Si un experimento de eco se realiza en un sistema que está expuesto al efecto del entorno y continuamente monitoreado, la probabilidad de persistencia conservará su dependencia en la evolución dinámica. Sin embargo, vale la pena entender hasta qué punto es posible conocer características de las fases geométricas a través de un protocolo eco en un sistema monitoreado.

Para cada realización del protocolo, caracterizada por una secuencia de saltos $\mathcal{R}(2\,T, N_J)$, podemos parametrizar la probabilidad de persistencia $\mathcal{P}_\mathcal{R}$ con un ángulo asociado $\varphi_{\cal R}$.
\begin{equation}
    \mathcal{P}_{\mathcal{R}} = | \langle \psi(0) | \psi(2T) \rangle |^2 \equiv \cos^2\left(2\, \varphi_{\cal R}\right).
    \label{eq:sec6_xdefinition}
\end{equation}
Tanto la probabilidad de persistencia como el parámetro $\varphi_{\cal R}$ heredan el carácter estocástico de las trayectorias, y la probabilidad de medir un dado valor $\varphi$ se relaciona con la probabilidad para cada trayectoria según
\begin{equation}
    P[\varphi] = \sum_{\mathcal{R}/\varphi_{\cal R}=\varphi} P[\mathcal{R}].
    \label{eq:sec6_vphase_probability}
\end{equation}
En el caso límite en el cuál la probabilidad de persistencia se aproxima a su valor adiabático, $\varphi$ se aproximará a $\phi^+_{\mathrm{a}}$. 
Fuera de este régimen particular, $\varphi_{\mathcal{R}}$ {\em no} es igual a la fase geométrica $\phi_{\cal R} = \phi[\mathcal{R}]$ sino, como se mencionó anteriormente, una parametrización conveniente de las franjas de interferencia en el eco de espín.

Las desviaciones no-adiabáticas e inducidas por el entorno respecto de $\phi^{+}_\mathrm{a}$ pueden analizarse examinando el ensamble $\{\varphi_\mathcal{R}\} =\varphi_{\{\mathcal{R}\}}$ que se obtiene aplicando la fórmula (\ref{eq:sec6_xdefinition}) a cada realización individual del protocolo. Este estudio permitirá también inspeccionar posibles relaciones, si las hubiera, entre el comportamiento estocástico de las fases geométricas y aquél de los resultados experimentales (notar que $\varphi_{\mathcal{R}}$ está definida módulo $\pi/2$ y a menos de un signo. En consecuencia, cualquier relación entre la distribución de fases y la distribución de resultados experimentales deberá tener esto en cuenta). Otra vez, se espera que la frecuencia $\Omega$ a la que se rota el campo magnético tenga un impacto directo en la distribución~\cite{sjoqvistshortcut, measuringshortcut}. Al aumentar el valor relativo de $\Omega$ el sistema está expuesto a la influencia destructiva del entorno durante períodos temporales más cortos, permitiendo en mayor medida la cancelación parcial de las fases dinámicas. Al mismo tiempo, en este régimen se tornan inevitables desviaciones no despreciables respecto del régimen adiabático. Por otro lado, valores más chicos de $\Omega$ pueden exponer el sistema a los efectos del entorno por períodos largos que conduzcan a fuertes desviaciones en los valores del parámetro-de-eco $\varphi$ respecto de $\phi^+_\mathrm{a}$. 

\begin{figure}[ht!]
    \center
    \includegraphics[width = .495\linewidth, trim = {1cm 0 .75cm 0}]{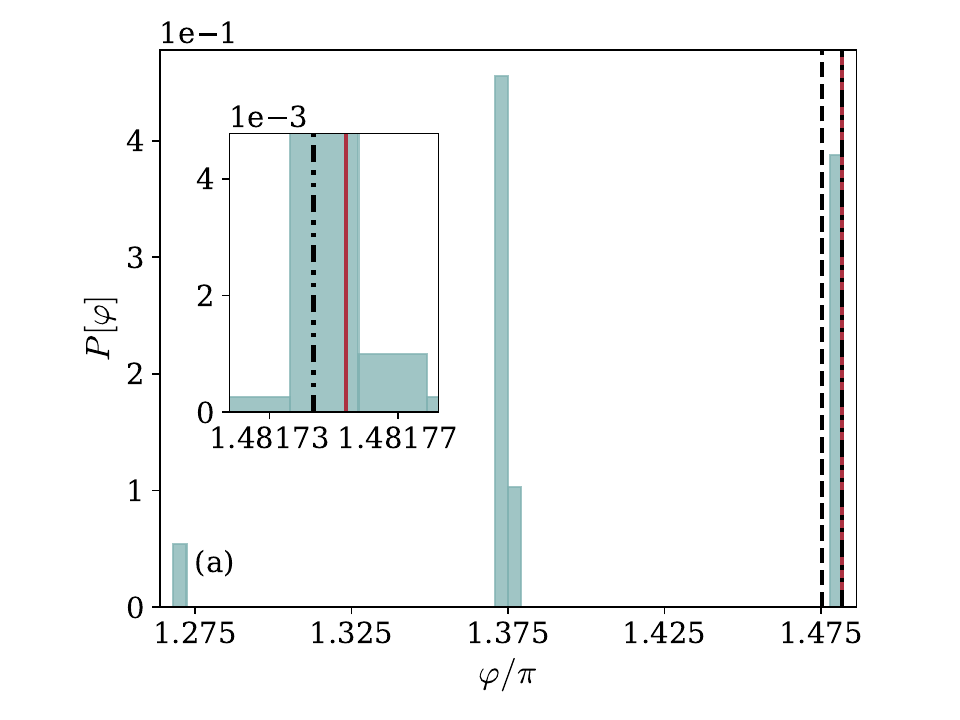}
    \includegraphics[width = .495\linewidth, trim = {.75cm 0 1cm 0}]{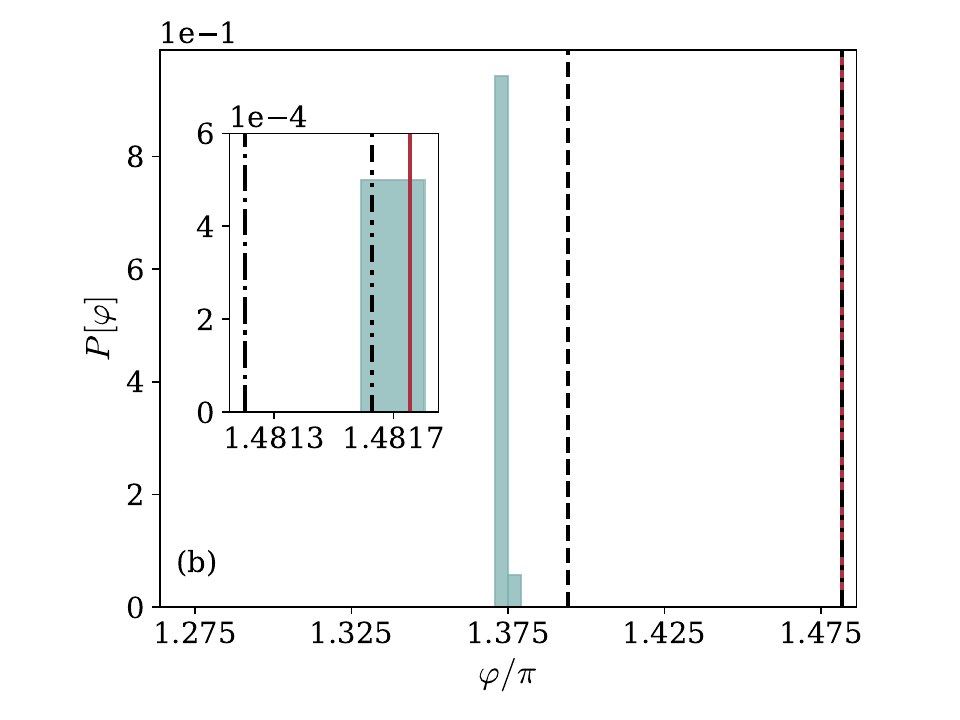}
    \caption{Distribución de probabilidad $P[\varphi]$ obtenida en un protocolo-eco para un campo magnético orientado según $\theta = 0.34\pi$ y conducido en un ciclo a frecuencias (a) $\Omega = 5\times 10^{-3}\omega$ y (b) $\Omega = 5\times 10^{-4}\omega$.
    El entorno permanece igual en ambos casos, caracterizado por una tasa de disipación $\Gamma = 10^{-3}\omega$ y $\gamma_z = 0$. 
    En ambos paneles, la línea sólida roja indica la fase adiabática (de Berry) $\phi^+_\mathrm{a}$, y las líneas negras de trazos y de trazo-punto señalan las fases $\phi_0$ y $\phi_u$ asociadas con una trayectoria sin saltos y con evolución general unitaria. Además, la línea negra de trazo y doble punto indica el valor del parámetro $\varphi$ obtenido en un protocolo en el que no se registran saltos. 
    Los inset en ambos paneles muestran un rango de abscisas en el cuál el resultado obtenido en un experimento de eco realizado suavemente es distinguible de la fase de Berry.}
    \label{fig:sec6_hist_varphi}
\end{figure}

En analogía con el estudio de las fases geométricas, se examina en primer lugar el caso $\gamma_z=0$ y se presenta, en la figura \ref{fig:sec6_hist_varphi} dos casos representativos en los cuales una hipotética evolución unitaria sería rápida, o bien suficientemente lenta para ser considerada dentro del régimen adiabático. Estos casos se muestran en la figura \ref{fig:sec6_hist_varphi}.a y \ref{fig:sec6_hist_varphi}.b respectivamente. En ambos paneles, se indican también la fase adiabática de Berry (que no depende del valor de $\Omega$), la fase $\phi_0$ acumulada en una evolución sin saltos, y la fase $\phi_u$ acumulada en una evolución unitaria general. Se indica también el valor del parámetro $\varphi$ obtenido en un experimento de eco que se completa sin detectar saltos. Para los parámetros elegidos, ambos paneles muestran desviaciones muy pequeñas del valor de $\varphi$ extraído de un protocolo sin saltos respecto de la fase de Berry (ver los inset en la figura \ref{fig:sec6_hist_varphi}). Debe notarse, sin embargo, que la probabilidad de registrar esta trayectoria específica es distinta en un caso y en el otro, como puede verse de las diferencias mostradas entre las distribuciones completas $P[\varphi]$. 

{\em Origen de la estructura de la distribución - }
La primera característica llamativa que aparece es la presencia de tres picos distintos bien definidos. La distribución ancha observada para los valores de fase geométrica desaparece completamente en el eco de espín.
Este comportamiento tiene su origen en el hecho de que, cuando $\gamma_z=0$ sólo son posibles saltos entre autoestados instantáneos de energía lo que conduce, cuando se lo combina con las propiedades de la probabilidad de persistencia, a una distribución de franjas de interferencia cualitativamente diferente a aquella de las fases geométricas. Cada uno de los picos en el panel (a) de la figura \ref{fig:sec6_hist_varphi} surge de un conjunto distinto de trayectorias cuánticas de la siguiente forma: 
\begin{enumerate}
    \item Los protocolos suaves sin saltos generan la acumulación de cuentas en torno al valor $\varphi \sim 1.43\pi$
    \item Protocolos en los que tienen lugar al menos un salto de decaimiento o excitación espontánea y en consecuencia el estado del sistema se proyecta en un autoestado instantáneo $\ket{\psi_\pm(t_i)}$ de $H(t)$ dan origen al pico centrado en $\varphi \sim 1.375\pi$.
    \item Protocolos en los que un único salto de desfasaje tiene lugar dan origen al pico en torno a $\varphi\sim1.275\pi$.
\end{enumerate}

A continuación, se provee una justificación detallada de esta enumeración. Con el objetivo de evidenciar los aspectos cualitativos y los procesos subyacentes, se desprecia en general la no-hermiticidad de la evolución suave entre saltos, pensando en esos intervalos como intervalos de evolución unitaria forzada, ya sea rápida o lentamente. En consecuencia, esta justificación no debe considerarse un análisis cuantitativo.

El pico descrito en el punto (1.) coincide con el valor $\varphi\sim 1.34\,\pi$ extraído en un protocolo sin saltos y puede entenderse como compuesto por estos eventos ya que, siendo esta evolución única, se obtendrá exactamente el mismo valor para el parámetro $\varphi$ en todo experimento que se complete en estas condiciones.
Los picos descritos en los puntos (2.) y (3.) requieren la consideración del caso en que, por el contrario, fueran detectados saltos. 

\begin{figure}[ht!]
    \centering
    \includegraphics[width = .6\linewidth, trim = {1cm 0 0 0}]{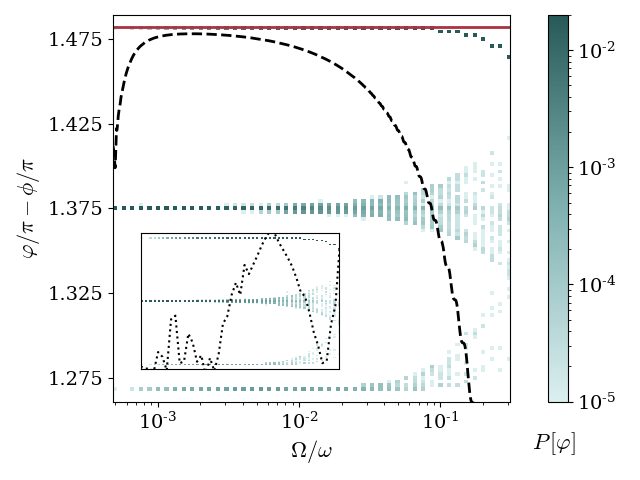}
    \caption{Distribución de probabilidad $P[\varphi]$ de valores para el parámetro $\varphi$ obtenido en un experimento de eco, en función de la relación $\Omega/\omega$. El campo está orientado con $\theta = 0.34\pi$ y el entorno está caracterizado por una tasa de disipación $\Gamma = 10^{-3}\omega$ y una amplitud $\gamma_z = 0$ para el cuarto operador de Lindblad. Los valores de $\varphi$ se muestran en el eje $y$, mientras que su probabilidad se indica con la intensidad del color de la cuenta. La línea sólida roja indica la fase adiabática (de Berry) $\phi^+_\mathrm{a}$, mientras que las líneas negras de trazos y de puntos señalan la fase $\phi_0$ acumulada en una trayectoria suave sin saltos y el primer momento de la distribución de fases geométricas $\bar{\phi}$ respectivamente.}
    \label{fig:sec6_varphi_vs_Omega}
\end{figure}

Cuando $\gamma_z = 0$ tres tipos de saltos son posibles para la descomposición de la ecuación de Lindblad propuesta en este capítulo. De estos tres saltos posibles, dos proyectan el estado del sistema en un autoestado instantáneo de energía, a saber, el decaimiento espontáneo inducido por $L_-$, y la excitación espontánea representada por $L_+$. 
Cuando un primer salto de este tipo tiene lugar en un dado instante $t_i$, inmediatamente después del salto el estado del sistema está dado por
\begin{equation}
    \ket{\Psi(t_i)}= e^{i\,\xi(t_i) + i\,\phi(t_i)}\ket{\psi_\pm(t_i)}
\end{equation}
con $\xi(t_i)$ la fase dinámica y $\phi(t_i)$ la fase geométrica acumuladas desde el inicio de la evolución hasta el instante $t_i$. Si el protocolo finaliza inmediatamente después del salto, la probabilidad de persistencia medida será  $\mathcal{P}_\mathcal{R} = |\bra{\psi(0)}\ket{\psi(2T)}|^2 = 1/2$ y no preserva información alguna sobre la fase geométrica ni dinámica, puesto que son fases globales, ni sobre las características específicas del salto.
Si, por otro lado, el sistema continua evolucionando, la posibilidad de obtener alguna información sobre la fase geométrica posteriormente acumulada o el instante en el que tuvo lugar el salto dependerá de la competencia entre las transiciones no-adiabáticas y la existencia de saltos adicionales.
Si la evolución continúa después del primer salto, esto sucederá de forma suave hasta que, o bien ocurra un nuevo salto cuántico, o bien el protocolo se complete. Diferentes regímenes de $\Omega/\omega$ dan origen a evoluciones suaves de diferentes características. Si el forzado es suficientemente lento la evolución suave está aproximadamente libre de transiciones y $\ket{\psi(t)}\sim e^{i\,\xi(t>t_i) + i\,\phi(t>t_i)}\ket{\psi_\pm(t>t_i)}$, de modo que el resultado obtenido para la probabilidad de persistencia continúa siendo $\mathcal{P}_\mathcal{R} = 1/2$. Más aún, este régimen favorece la ocurrencia de saltos adicionales que refuerzan la pérdida de información re-proyectando el estado del sistema en un autoestado del Hamiltoniano $H(t)$. La completa independencia de los resultados en los instantes $t_i$ en que tienen lugar los saltos hacen que el pico (2.), asociado a esta amplitud de probabilidad, resulte extremadamente definido en el régimen de forzado lento. Por otro lado, si el forzado se ejecuta a frecuencias relativas más altas, a lo largo de la evolución suave posterior al primer salto el estado del sistema adquiere contribuciones del otro autoestado instantáneo producto de los efectos no-adiabáticos, favoreciendo la emergencia de fases relativas. En esta situación, la probabilidad de persistencia adquiere una dependencia en el instante $t_i$ que conduce al ensanchamiento del pico observado en la figura \ref{fig:sec6_varphi_vs_Omega} para forzados rápidos, aunque permanece sin mostrar correlaciones claras con la fase geométrica. Como se fundamenta en los párrafos anteriores, cada salto de excitación o decaimiento espontáneo elimina toda la información previa respecto de la fase geométrica y de los saltos anteriores. La probabilidad para estos eventos de borrado se mitiga en este régimen mediante la reducción de la exposición a los efectos del entorno.

El pico (3.) observado en la distribución, centrado en torno a $\varphi \sim 1.475\pi$ puede entenderse incorporando los saltos de desfasaje a la discusión previa. Un salto de desfasaje tiene el efecto de introducir un corrimiento en $\pi$ en la fase relativa del estado. Si la evolución posterior no involucra transiciones, y no tienen lugar saltos que eliminen información, entonces la dinámica total se asemeja a un experimento de eco adiabático, a menos de correcciones menores, y la probabilidad de persistencia toma el valor $\mathcal{P}\sim\sen^2(2\phi_\mathrm{a})$ (con el coseno reemplazado por seno debido al corrimiento relativo en $\pi$). Esta situación conduce a un valor único y bien definido para $\varphi$ que es independiente del instante $t_i$ en que el salto de tipo $L_d$ tuvo lugar. En consecuencia, el pico descrito en (3.) que surge mediante este procedimiento en el régimen de forzado lento resulta angosto aunque relativamente pequeño, dado que en este régimen los saltos de decaimiento espontáneo son muy probables. A medida que el campo magnético se rota a mayor frecuencia, los efectos no-adiabáticos introducen una dependencia en $t_i$ que es responsable del ensanchamiendo de la distribución que se aprecia en la figura \ref{fig:sec6_varphi_vs_Omega} para valores de $\Omega/\omega$ más largos. 
\\
\\\indent
{\em Análisis de las distribuciones en distintos regímenes - }
En síntesis, para los parámetros considerados en el panel (a), trayectorias con a lo sumo un salto son posibles. Los tres picos corresponden a protocolos sin saltos, protocolos con un salto de tipo $L_\pm$, y protocolos con un salto de tipo $L_d$.
Aquellas trayectorias que permanezcan suaves a lo largo de todo el protocolo inducen el pico a la derecha de la figura \ref{fig:sec6_hist_varphi}.a, esto es, el pico más cercano al resultado sin saltos $\varphi\sim 1.475\pi$ para esta elección de parámetros. El pico en el medio de la figura, centrado en el valor $\varphi\sim\,1.375\pi$  trivialmente asociado a través de la ecuación (\ref{eq:sec6_xdefinition}) con un valor $1/2$ para la probabilidad de persistencia, se construye a partir de aquellos casos en los que el estado del sistema se proyecta, en un dado instante, en un autoestado de $H(t)$. En esas trayectorias, toda la información sobre la fase acumulada antes del salto se pierde. Como consecuencia, inmediatamente después de un salto $L_\pm$, e independientemente tanto de la evolución previa como del instante en que ocurrió el salto, la probabilidad de persistencia toma el valor $1/2$. El tercer pico, ubicado a la izquierda, se debe a las trayectorias en las que ocurre un salto de tipo $L_d$. Este tipo de salto tiene el efecto de introducir un salto en $\pi$ en la fase relativa del estado que se corresponde con la posición del mencionado pico. 

En consecuencia, la distribución de franjas de interferencia muestra tres picos de los cuales dos codifican la misma información, esto es, el valor de $\varphi$ para un protocolo conducido suavemente, mientras que el pico central no contiene prácticamente información.
Más aún, la distribución es considerablemente aguda dado que, para los valores de parámetros elegidos, las trayectorias descritas son todas detectadas mientras que trayectorias cuánticas más complejas son altamente improbables.

En la figura \ref{fig:sec6_hist_varphi}.b los dos picos ubicados en los extremos se han prácticamente desvanecido. Esto revela que cuando el sistema se conduce a frecuencias relativas más bajas, un decaimiento o una excitación espontánea será detectada en casi toda trayectoria. Un efecto similar se obtiene si la tasa de disipación $\Gamma/\Omega$ se aumenta mientras la relación $\Omega/\omega$ se mantiene fija.

Un segundo aspecto de la distribución $P[\varphi]$ es qué características de las fases geométricas captura. En el régimen de forzado rápido de la figura \ref{fig:sec6_hist_varphi}.a, los valores de $\varphi$ obtenidos de los protocolos sin saltos coinciden bien con la fase adiabática de Berry, y estos muestran pequeñas pero finitas desviaciones respecto de la fase geométrica sin saltos. 
El valor de $\varphi$ está más estrechamente relacionado con el valor de la fase adiabática que con el valor de la fase geométrica acumulado en la dinámica de arrastre suave.
Para el régimen de forzado lento de la figura \ref{fig:sec6_hist_varphi}.b, el valor de $\varphi$ sin saltos continúa siendo un buen indicador de la fase adiabática, aún cuando registrar dicho protocolo es en este caso menos probable. Bajo estas condiciones, la mayoría de los experimentos contribuyen al pico central que no está relacionado con ningún valor de fase geométrica.

El examen de la figura \ref{fig:sec6_hist_varphi} sugiere que, como en el caso de las fases geométricas, el compromiso entre las correcciones no-adiabáticas y los saltos inducidos por el entorno se revela más abiertamente cuando la distribución de valores $P[\varphi]$ se analiza en función de la relación $\Omega/\omega$. Esto se muestra en la figura \ref{fig:sec6_varphi_vs_Omega}, cuyo rango de abscisas incluye los dos casos presentados en la figura \ref{fig:sec6_hist_varphi}. La fase de Berry $\phi_{\rm a}$ y los valores $\phi_{0}$ y $\Bar{\phi}$ de la fase geométrica acumulada en trayectorias suaves y del primer momento de la distribución de fases geométricas se dan también como referencia. 

En el régimen de forzado rápido
$\Omega/\omega \gtrsim 0.1$ el valor de $\varphi$ obtenido es, para la mayoría de las realizaciones, aquél que se obtiene en un protocolo sin saltos y muestras desviaciones apreciables, aunque pequeñas de la fase adiabática.Trayectorias con un único salto pueden ser observadas, aunque con menor probabilidad. Si este es el caso, la superposición de autoestados generada por las transiciones no-adiabáticas producirá distribuciones considerablemente anchas en torno a los dos picos restantes, revelando la naturaleza estocástica de los instantes en los que ocurren los saltos. Las correcciones no-adiabáticas tienen un impacto mucho más fuerte sobre $\phi_0$, su comportamiento se descorrelaciona por completo del de la distribución de protocolos de eco. La expresión analítica de la ecuación (\ref{eq:sec6_D_GP}), muestra esta dependencia en $\Omega/\omega$ para la relación de parámetros $\Omega/\omega\sim\Gamma/\omega\ll 1$ en que es válida.

\begin{figure}[ht!]
    \center
    \includegraphics[width = .495\linewidth, trim = {1cm 0 .75cm 0}]{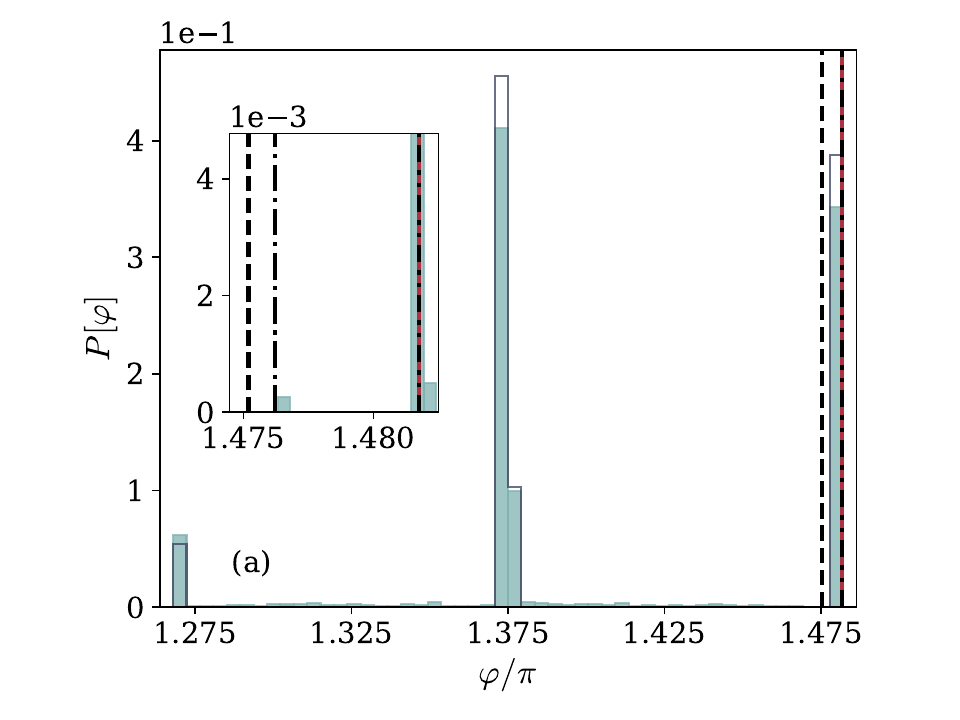}
    \includegraphics[width = .495\linewidth, trim = {.75cm 0 1cm 0}]{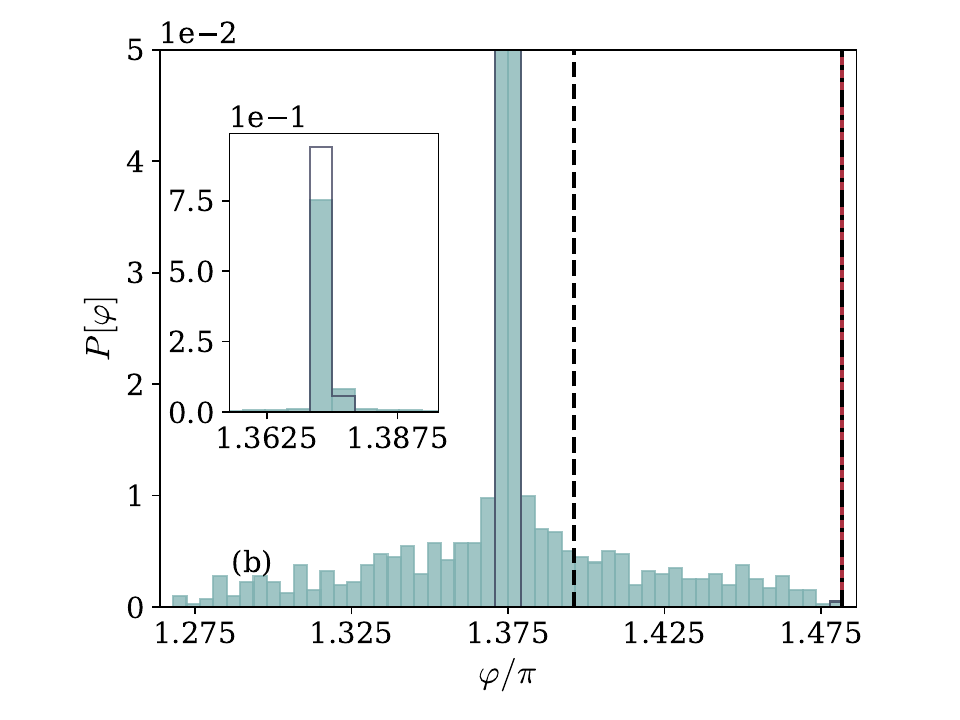}
    \caption{Distribución de probabilidad $P[\varphi]$ obtenida en un protocolo-eco para un campo magnético orientado según $\theta = 0.34\pi$ y conducido en un ciclo a frecuencias (a) $\Omega = 5\times 10^{-3}\omega$ y (b) $\Omega = 5\times 10^{-4}\omega$.
    El entorno permanece igual en ambos casos, caracterizado por una tasa de disipación $\Gamma = 10^{-3}\omega$ y $\gamma_z = 0.1\Gamma$ finita. En ambos paneles, un contorno azul señala las distribuciones obtenidas en el caso con $\gamma_z = 0$. La línea sólida roja indica la fase adiabática (de Berry) $\phi^+_\mathrm{a}$, y las líneas negras de trazos y de trazo-punto señalan las fases $\phi_0$ y $\phi_u$ asociadas con una trayectoria sin saltos y con evolución general unitaria. Además, la línea negra de trazo y doble punto indica el valor del parámetro $\varphi$ obtenido en un protocolo en el que no se registran saltos. 
    Los inset en muestran un rango de abscisas en el cual (a) la diferencia entre los valores de referencia, y (b) la magnitud total del pico central, son visibles.}
    \label{fig:sec6_hist_varphi_gz}
\end{figure}

Por otro lado, a medida que se aproxima al régimen de forzado lento, los tres picos se vuelven más definidos. Este comportamiento se acompaña de un marcado decrecimiento en la altura de los picos laterales y un incremento en las cuentas del pico trivial en el medio. A lo largo de todo el rango se encuentra una región en la que la relación entre los efectos inducidos por el entorno y no-adiabáticos permiten una buena coincidencia entre la fase geométrica acumulada en un protocolo suave no-unitario y el valor de $\varphi$ obtenido. 
De forma diferente a los valores 'sin saltos', que muestran un acuerdo razonable, el inset en la figura \ref{fig:sec6_varphi_vs_Omega} muestra que el (consistentemente trasladado) primer momento de la distribución de fases geométricas $\bar{\phi}$ permanece, a lo largo de todo el rango de frecuencias, completamente descorrelacionado tanto de la distribución de valores $\varphi$ como de todos los valores característicos de eco.
\clearpage
La distribución cambia radicalmente cuando $\gamma_z\neq 0$. En lo que sigue discutimos el caso $\gamma_z = 0.1\,\Gamma$ con $\Gamma = 10^{-3}\omega$. Empezamos reconsiderando los dos casos representados de los paneles (a) y (b) de la figura \ref{fig:sec6_hist_varphi_gz}. El primer aspecto notable es que, aunque los tres picos observados en la figura \ref{fig:sec6_hist_varphi} (indicados ahora con contornos azules) todavía se detectan, ahora coexisten con una distribución ancha.

Como se observa en la figura \ref{fig:sec6_hist_varphi_gz}.a, la altura de los tres picos discutidos previamente decrece en presciencia de $\gamma_z$. La supresión de los picos se ve acompañada de la aparición de una distribución de fondo ancha que cubre el rango completo de fases. La figura \ref{fig:sec6_hist_varphi_gz}.b se ocupa de la situación de forzado lento en la que la probabilidad de que ocurra un salto, o incluso varios a lo largo de la trayectoria, crece.
La inclusión del operador de salto $L_z$ modifica la distribución, que pasa de mostrar de picos bien definidos a cubrir el rango completo de valores posibles para $\varphi$. En particular, los dos picos asociados al protocolo sin saltos desaparecieron. La inclusión de este término en el Lindbladiano induce saltos entre estados que no son autoestados de energía y en ese sentido podemos considerar que los resultados son considerablemente generales y no específicamente dependientes de la elección del operador de Lindblad.

\begin{figure}[ht!]
    \centering
    \includegraphics[width = .6\linewidth, trim = {1cm 0 0 0}]{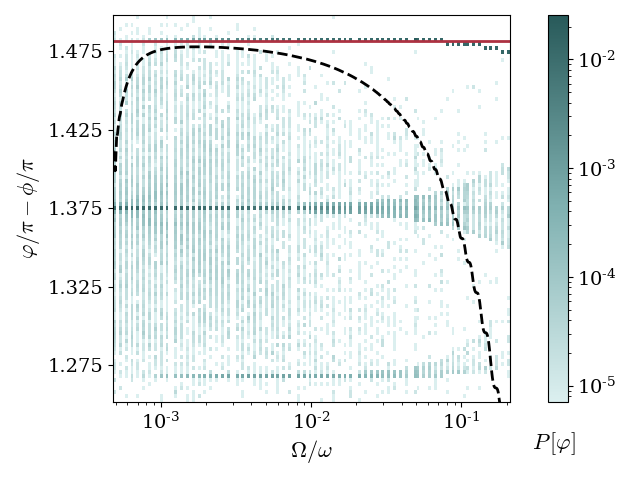}
    \caption{Distribución de probabilidad $P[\varphi]$ de valores para el parámetro $\varphi$ obtenido en un experimento de eco, en función de la relación $\Omega/\omega$. El campo está orientado con $\theta = 0.34\pi$ y el entorno está caracterizado por una tasa de disipación $\Gamma = 10^{-3}\omega$ y una amplitud $\gamma_z = 0.1\Gamma$ finita. Los valores de $\varphi$ se muestran en el eje $y$, mientras que su probabilidad se indica con la intensidad del color de la cuenta. La línea sólida roja indica la fase adiabática (de Berry) $\phi^+_\mathrm{a}$, mientras que la línea negra de trazos señala la fase $\phi_0$ acumulada en una trayectoria suave sin saltos.}
    \label{fig:sec6_varphi_vs_Omega_gz}
\end{figure}

Con el objetivo de obtener una imagen más completa del efecto de un valor finito de $\gamma_z$, la figura \ref{fig:sec6_varphi_vs_Omega_gz} muestra la distribución de valores de $\varphi$ como función de la relación $\Omega/\omega$. 
Para una evolución en el régimen de forzado rápido, en la cual casi no se detectan saltos, el comportamiento exhibido por la distribución es similar al observado para el caso con $\gamma_z=0$. Cuando la velocidad de rotación del campo se reduce, favoreciendo gradualmente la ocurrencia de saltos, el efecto de haber introducido el operador $L_z$ se torna más relevante. Los saltos $L_z$ conducen a valores de $\varphi$ que dan cuenta del instante en que el salto tuvo lugar, y en consecuencia construyen una distribución de fondo ancha.

Resumiendo, mientras que la distribución de franjas de interferencia es, en general, considerablemente distinta de la distribución de fases geométricas acumuladas a lo largo de cada trayectoria, el análisis de un experimento de eco de espín permite extraer información confiable respecto de la fase adiabática (de Berry) en algunos regímenes de parámetros. En la próxima sección, la atención será puesta en la trayectoria sin saltos (el tipo de evolución asociada a los picos laterales de la distribución de probabilidad para el protocolo de eco) y en mostrar que esta evolución sufre una transición topológica como función del acoplamiento con el entorno.

\section{Resultados: transiciones topológicas}\label{sec:sec6_topological}
Como se ha anticipado, se concluye este análisis de las fases geométricas en sistemas monitoreados concentrándose en la trayectoria sin saltos. En esta sección se muestra, siguiendo el espíritu de la referencia~\cite{gefenWeak}, que la dinámica de arrastre sin saltos entraña una transición topológica. Se enfatiza que, aunque el modelo es muy diferente a aquél propuesto en~\cite{gefenWeak}, se cree que la naturaleza de la transición es la misma. Este análisis es un indicio fuerte para la conjetura de que este tipo de transiciones es bastante genérico para los sistemas monitoreados.
\\
\\\indent
{\em Diagrama de fases y puntos singulares - }La fase geométrica $\phi_0$ dada por la ecuación (\ref{eq:sec6_GP_nj}) depende, para cada valor fijo del ángulo $\theta$, de las relaciones $\Omega/\omega$ y $\Gamma/\omega$. Se recuerda que la trayectoria sin saltos, y en consecuencia la fase geométrica asociada a ella, no dependen del valor del parámetro $\gamma_z$. Como función de los parámetros arriba mencionados, la fase geométrica muestra singularidades discretas en ciertos puntos críticos en torno a los cuales da un giro de $2\pi$. 
Recordando que el estado inicial considerado es el autoestado instantáneo excitado $\ket{\psi_+(0)}$ del Hamiltoniano $H(0)$, la condición de ortonormalidad que da origen a un punto singular de la fase implica un estado final  $\ket{\psi(T)} \propto \ket{\psi_-(0)}$. Esto significa que se encontrará un punto singular cuando, exactamente en un período de evolución $t\in[0,T]$, se obtenga una transición de poblaciones completa del estado excitado al fundamental. En la sección \ref{sec:sec6_NoJump} se discutió cómo la dinámica generada por el Hamiltoniano efectivo de arrastre $H_o$ favorece las transiciones del estado excitado al fundamental. Mientras las aproximaciones allí realizadas permanezcan válidas, los puntos singulares de la fase geométrica estarán definidos por la ecuación $(\nu+\varepsilon)-(\nu-\varepsilon) e^{2i\pi\varepsilon/\Omega} = 0$.

\begin{SCfigure}[10][ht!]
    \includegraphics[width = .55\linewidth, trim = {0cm 0 .75cm 0}]{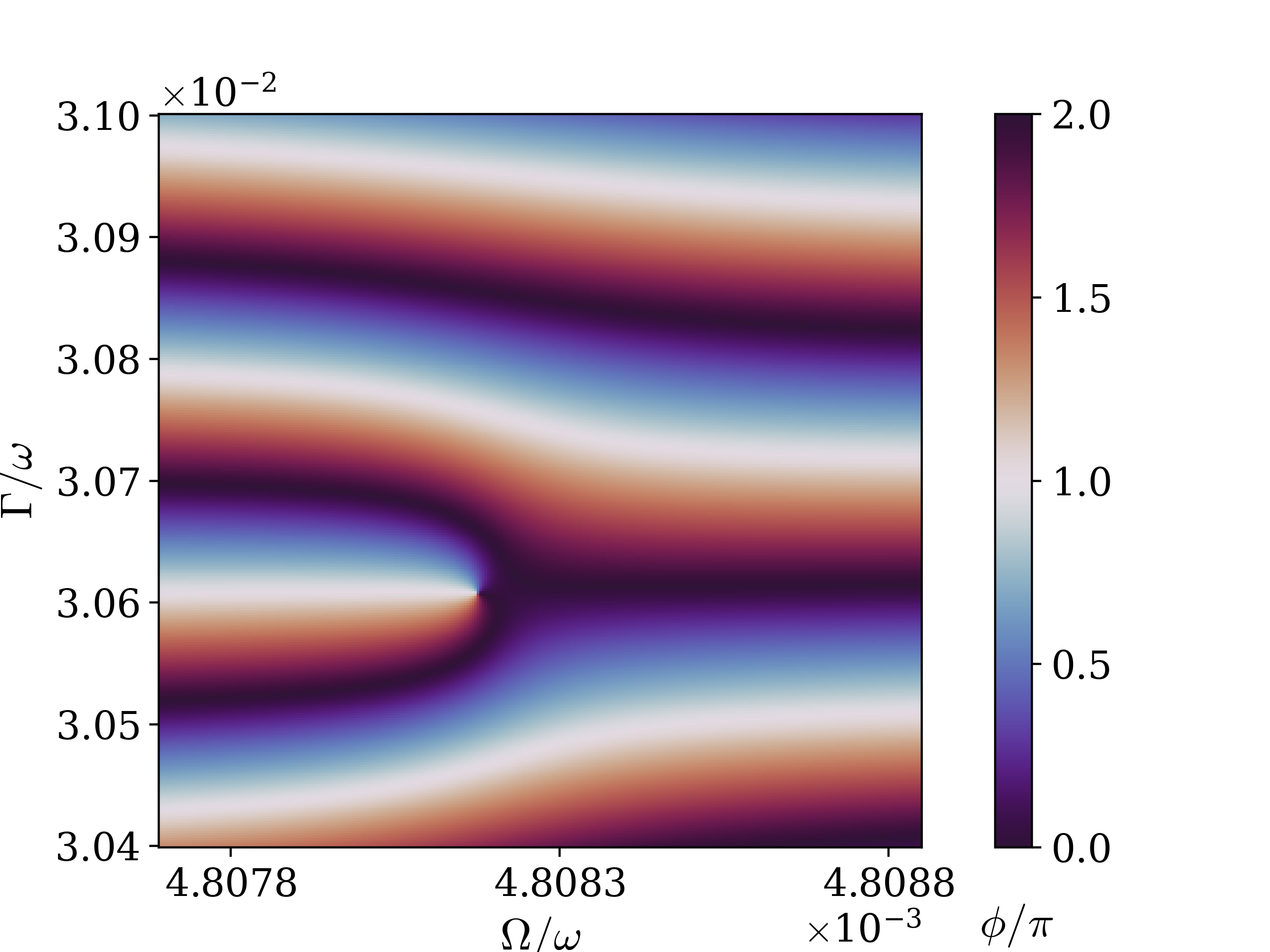}
    \caption{ Fase geométrica en una trayectoria sin saltos, en una región limitada del espacio de parámetros definido por las relaciones $\Omega/\omega$ y $\Gamma/\omega$. El valor de la fase geométrica está indicado por el color según la barra  a la derecha. La dirección del campo está fija en $\theta = 0.34\pi$. Una singularidad de observa en $\Omega/\omega = 4.8082 \times 10^{-3}$ y $\Gamma/\omega = 0.0306$. Las cruces indican puntos ligeramente a izquierda de la singularidad ($\Omega/\omega  = 4.8 \times 10^{-3}$)  y derecha ($\Omega/\omega = 4.8084 \times 10^{-3}$), que se mostrará, pertenecen a distintos sectores.} 
    \label{fig:sec6_Colorplot}
\end{SCfigure}

La figura \ref{fig:sec6_Colorplot} muestra un gráfico de colores de la fase geométrica en el plano $\Gamma - \Omega$ para un valor fijo del ángulo $\theta$. El rango de parámetros se elige para resaltar la singularidad y la vuelta en $2\pi$ de la fase geométrica en torno a la misma.

En esta sección, se mostrará que la colección formada por estos puntos singulares delimita regiones del espacio de parámetros asociadas con distintas clases de evoluciones. Para esto, se definirá un invariante topológico $\mathrm{n} \in \mathbb{Z}$ y se mostrará explícitamente que toma distintos valores en las distintas regiones del espacio de relaciones entre parámetros.

{\em Transición topológica en la trayectoria sin saltos - }Al realizar una inspección directa del Hamiltoniano efectivo de arrastre (\ref{eq:sec6_nojump_hamiltonian}), se encuentra el siguiente comportamiento: si el campo magnético apunta en dirección del eje z, el autoestado excitado de $H(t)$ permanece fijo en el polo de la esfera de Bloch independientemente de los valores tomados por las relaciones entre parámetros $\Omega/\omega$ y $\Gamma/\omega$. En consecuencia, la fase geométrica asociada con la trayectoria sin saltos se anula idénticamente (mod $2\pi)$ para $\theta=0$ y para $\theta=\pi$. Sin pérdida de generalidad, la libertad mod $2\pi$ puede eliminarse de la fase geométrica fijando $\phi_{0}(\theta = 0) =0$ y demandando continuidad simultáneamente.
De esta forma, $\phi_{0}(\theta=\pi)$ está completamente determinada por la evolución, y adquiere un valor 

\begin{equation}
    \phi_{0}(\theta = \pi) = 2\pi\,\mathrm{n},
    \label{eq:sec6_n_definition}
\end{equation}
donde $\mathrm{n}$ es un número entero que caracteriza la dependencia de la fase geométrica con el ángulo $\theta$ para valores fijos de los parámetros.

Siendo un número entero, $\mathrm{n}$ constituye un invariante topológico ya que no puede modificarse mediante deformaciones suaves de $\phi_0(\theta)$. En consecuencia, si la fase geométrica está caracterizada por diferentes valores $\mathrm{n}$ para distintos conjuntos de valores de los parámetros, la misma necesariamente atraviesa una transformación no-suave como el comportamiento singular observado en la figura \ref{fig:sec6_Colorplot}.
De hecho, puntos en el espacio de parámetros ligeramente a la derecha y ligeramente a la izquierda de la singularidad (indicados con cruces en la figura \ref{fig:sec6_Colorplot}) dan origen a evoluciones sin saltos asociadas con valores $\mathrm{n} = 0$ y $\mathrm{n} = 1$ del invariante topológico respectivamente, los cuales rotulan clases topológicas distintas. 

\begin{SCfigure}[20][ht!]
    \includegraphics[width = .5\linewidth, trim = {0 0 0 0}]{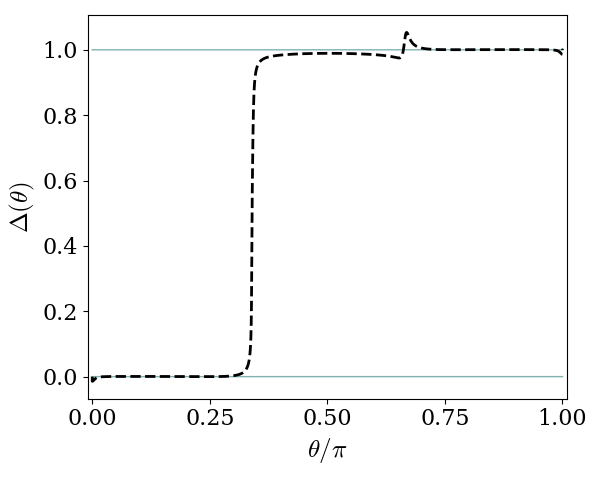}
    \caption{Diferencia $\Delta(\theta)$ entre fases geométricas calculadas para puntos ligeramente a la derecha y ligeramente a la izquierda de la singularidad, indicados en la figura \ref{fig:sec6_Colorplot} con cruces. La fase geométrica está, en cada caso, caracterizada por un valor distintos del invariante topológico ${\rm n}$. Esto se refleja en el hecho de que las fases difieren en $2\pi$ para $\theta = \pi$. \label{fig:sec6_Phi_resta}}
\end{SCfigure}

Para mostrar esto explícitamente, la figura \ref{fig:sec6_Phi_resta} compara el comportamiento como función de $\theta$ de estas fases geométricas mostrando la diferencia $\Delta(\theta)$ entre ellas. Dados dos puntos, llamados por ejemplo (1) y (2) se define $\Delta(\theta)$ como

\begin{equation} 
    \Delta(\theta) = \frac{1}{2\pi} \left[ \phi^{(\Gamma_1,\Omega_1)}_0 -  \phi^{(\Gamma_2,\Omega_2)}_0 \right].
    \label{eq:sec6_phi0_diff}
\end{equation}
Para los puntos considerados, esta diferencia se anula (a menos de pequeñas desviaciones) hasta $\theta = 0.34 \pi$, esto es, hasta el ángulo para el que que se observa la singularidad en el punto específico comprendido entre (1) y (2). En este valor específico de $\theta$ la fase geométrica obtenida por cada conjunto de parámetros se desvía abruptamente de forma que su diferencia muestra un escalón y se estaciona en torno al valor $\Delta = 1$ por el resto del rango. Los valores obtenidos para el invariante topológico $\mathrm{n}$ se traducen en el valor $\Delta(\pi) =1$ para $\theta = \pi$. 

\begin{figure}[ht!]
    \center
    \includegraphics[width = .495\linewidth, trim = {.05cm 0 1.5cm 0}]{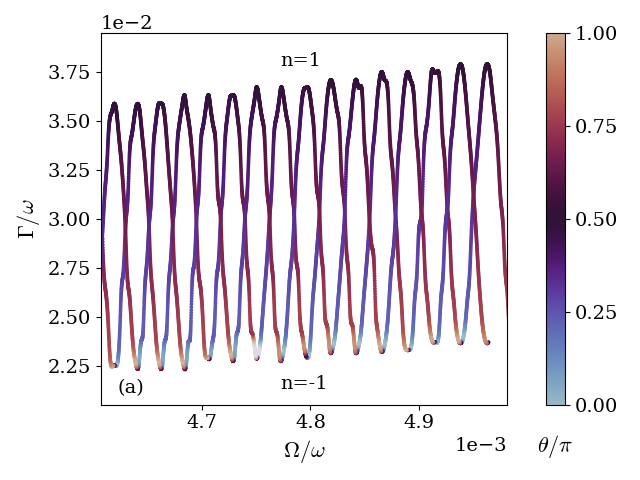}
    \includegraphics[width = .495\linewidth, trim = {1.25cm 0 .3cm 0}]{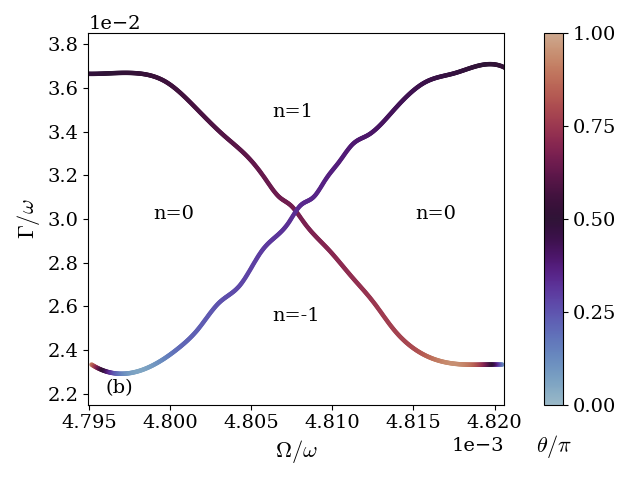}
    \caption{Curvas críticas dividiendo el espacio de parámetros en distintas clases de evolución sin saltos. Las clases se caracterizan por distintos valores del invariante topológico ${\rm n}$. El ángulo crítico $\theta_c$ para el cual se encuentra cada punto singular se indica mediante el color según la barra a la derecha. Los paneles (a) y (b) exhiben distintos rangos para las relaciones $\Omega/\omega$ y $\Gamma/\omega$.} 
    \label{fig:sec6_Singularities}
\end{figure}

Sobre el rango de parámetros completo, la fase geométrica muestra múltiples singularidades cuya ubicación exacta depende del valor de $\theta$. El conjunto de puntos singulares conforma dos curvas contra-oscilantes que definen una cadena de regiones cerradas contiguas y dividen el espacio de parámetros en una región inferior y otra superior. Esto se presenta en la figura \ref{fig:sec6_Singularities}.a.

Las combinaciones de parámetros contenidas en un mismo sector generan evoluciones con el mismo valor ${\rm n}$.  El área debajo de la secuencia de regiones encadenadas está caracterizada por el valor ${\rm n=-1}$. Los puntos
correspondientes de los valores $\Gamma = 0$ y $\Omega/\omega \ll 1$ que definen el régimen adiabático pertenecen a esta región.
Las regiones entre las dos curvas críticas son sectores topológicos triviales con ${\rm n}=0$, mientras que la región superior está caracterizada por un valor ${\rm n}=1$. Vale la pena mencionar que estos sectores topológicos no son igualmente probables. La probabilidad de obtener esta trayectoria aumenta a medida que se reduce $\Gamma$, lo que implica que el sector topológico superior sea menos probable que los demás.
\clearpage

\subsection{Transición topológica en el experimento de eco}
\begin{SCfigure}[20][ht!]
    \includegraphics[width = .55\linewidth, trim = {.5cm 0 0cm 0}]{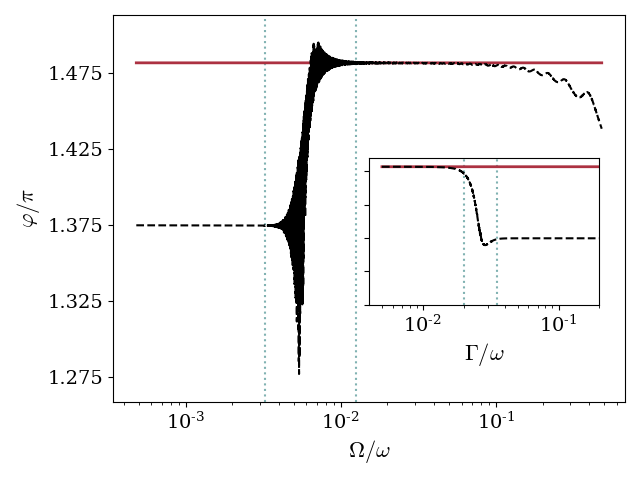}
    \caption{Dependencia del parámetro $\varphi$ (línea negra de trazos) obtenido en un protocolo sin saltos como función de la relación $\Omega/\omega$. El campo está orientado con $\theta = 0.34\pi$ y el entorno está caracterizado por una tasa de disipación $\Gamma = 0.0306\omega$, incluida en los rangos presentados en las figuras \ref{fig:sec6_Colorplot} - \ref{fig:sec6_Singularities}. Se indica también la fase de Berry, mediante una línea sólida roja. El inset muestra el valor de $\varphi$ como función de la relación $\Gamma/\omega$, para $\Omega/\omega = 4.8\times 10^{-3}$, coincidiendo también con los valores considerados en las figuras previas. } 
    \label{fig:sec6_varphi_deviation}
\end{SCfigure}

Con el objetivo de buscar indicios experimentalmente detectables de la transición topológica, se realiza una indagación minuciosa del experimento de eco que se completa sin detectar ningún tipo de salto cuántico. En la figura \ref{fig:sec6_varphi_vs_Omega} el valor de $\varphi$ extraído en este caso mostraba buena coincidencia con la fase adiabática de Berry para un rango amplio de frecuencias. Sin embargo, estrecha coincidencia de $\varphi$ con $\phi_{\rm a}$ no resulta válida para frecuencias arbitrariamente chicas, y se desvía cuando la relación $\Gamma/\Omega$ se torna suficientemente grande. La figura \ref{fig:sec6_varphi_deviation} muestra el valor de $\varphi$ como función de la relación $\Omega/\omega$ entre la frecuencia y la amplitud del campo. Por referencia y para facilitar la comparación, se considera un entorno caracterizado por una constante de disipación $\Gamma/\omega = 0.0306$, contemplada en los rangos exhibidos por las figuras \ref{fig:sec6_Colorplot} a \ref{fig:sec6_Singularities}. 

Para valores altos de la relación $\Omega/\omega$, el valor sin saltos del parámetro $\varphi$ exhibe el comportamiento descrito anteriormente. Sin embargo, a medida que la frecuencia se reduce, muestra un escalón altamente oscilante para finalmente estacionarse en un valor constante $\varphi \sim 1.375\pi$, asociado con una probabilidad de persistencia $1/2$.

Recordando que el estado inicial del experimento de eco puede escribirse, en términos de los autoestados del Hamiltoniano instantáneo $H(0)$ como $(\ket{\psi_+(0)} + \ket{\psi_-(0)})/\sqrt{2}$ y el carácter cíclico de los autoestados instantáneos $\ket{\psi_\pm(t)}$, se deduce que la probabilidad de persistencia toma el valor $\mathcal{P}=1/2$ a tiempo $t=T$ cuando el estado final del sistema coincide exactamente con un autoestado del Hamiltoniano instantáneo $H(T)$. La dinámica suave generada por $H_o$, que asume la expresión analítica de la ecuación (\ref{eq:sec6_D_state}), tiende al decaimiento del estado en el fundamental. Por este motivo se concluye que la probabilidad de persistencia toma el valor $\mathcal{P}=1/2$ y, consecuentemente, $\varphi \sim 1.375\pi$, cuando el arrastre suave suprime la ocupación en el estado excitado dentro del ciclo.

Como se discutió en los párrafos anteriores, la relación entre parámetros que conduce a la transferencia total de poblaciones exactamente en un ciclo $t\in [0,T]$ corresponde a los puntos singulares del diagrama de fases. En consecuencia, la transferencia completa de poblaciones entre los estados excitado y fundamental tiene lugar dentro de un ciclo de rotación del campo si la frecuencia es menor a la del punto singular.
Esta condición establece una conexión entre el valor del parámetro de eco y las clases topológicas de evolución sin saltos, ya que se accede a distintos regímenes de $\varphi$ de uno y otro lado de un punto singular. 

Los límites del rango en el que $\varphi$ realiza el escalón oscilante y pasa de  $\sim \phi_{\rm a}$ al valor central se marcan en \ref{fig:sec6_varphi_deviation} con dos líneas punteadas celestes. La región a la derecha de la primera línea corresponde a evoluciones caracterizadas por el valor $\mathrm{n} = -1$ del invariante topológico. El rango entre líneas celestes corresponde a la secuencia densamente agrupada de sectores topológicamente triviales que se observa en la figura \ref{fig:sec6_Singularities}.a. Finalmente, una vez a la izquierda de la última línea celeste, la evolución asociada está caracterizada por el valor $\mathrm{n} = 1$ del invariante topológico.

El inset en la figura \ref{fig:sec6_varphi_deviation} muestra el parámetro $\varphi$ como función de la tasa de disipación $\Gamma$. En este gráfico, para facilitar la comparación, el valor de la frecuencia de rotación del campo se fija en $\Omega/\omega = 4.8\times 10^{-3}$, también incluido en los rangos de las figuras \ref{fig:sec6_Colorplot} a \ref{fig:sec6_Singularities}. Una vez más, el valor de $\varphi$ muestra una buena coincidencia con la fase adiabática hasta un dado valor crítico de la relación $\Gamma/\Omega$, para el cual se observa un escalón y, posteriormente, se estaciona en $\varphi \sim 1.375\pi$. De la misma forma que en el gráfico principal, líneas punteadas celestes marcan los límites del escalón, separando la figura en tres sectores distinguibles. Los valores de $\Gamma/\omega$ a la izquierda de la primera línea corresponden a evoluciones caracterizadas por $\mathrm{n} = -1$ mientras que los valores a la derecha de la figura, con evoluciones caracterizadas por $\mathrm{n} = 1$. El espacio entre líneas, una vez más, puede asociarse con la región intermedia que es en este caso una sola (ver la figura \ref{fig:sec6_Singularities}.a) y por lo tanto no se observan oscilaciones.

En resumen, la medición de la probabilidad de persistencia en un experimento de eco de espín contiene claros indicadores de la transición topológica. La estructura de picos discutida permite identificar la trayectoria sin saltos. El posterior análisis de este pico, sintetizado en la figura \ref{fig:sec6_varphi_deviation}, es suficiente para capturar la transición topológica.

\vspace{.5cm}
\begin{center}
   \textcolor{bordo}{\ding{163}}
\end{center}
\vspace{.5cm}
En resumen, en este capítulo se abordó el estudio de la fase geométrica acumulada por un sistema abierto dentro de un contexto distinto al propuesto en los capítulos precedentes. Específicamente, se estudió la fase geométrica para un sistema continua e indirectamente monitoreado, escenario en el que el estado del sistema en cada realización de la evolución se torna una variable estocástica. La fase geométrica hereda, por lo tanto, la naturaleza estocástica, haciendo necesario el análisis de la distribución completa de fases geométricas. Este análisis se condujo haciendo hincapié en la competencia entre los efectos no-adiabáticos y los efectos introducidos por el entorno. También se comparó la distribución con diversos valores característicos de fase geométrica, como la fase $\phi_u$ acumulada por una hipotética evolución unitaria y la media sobre la distribución $\bar{\phi}$, estableciendo qué correlaciones existen entre las mismas y en qué rangos de los parámetros.
Con el objetivo de establecer conexión con los resultados experimentales, se estudió además la distribución de franjas de interferencia obtenidas en un experimento de Eco de Espín.
Finalmente, observando la trayectoria específica en la que no se registra ningún salto cuántico, se mostró que la misma atraviesa una transición topológica como función de la tasa de disipación. 
A pesar de las enormes diferencias mostradas por las distribuciones de fase geométrica y de franjas de interferencia de un experimento tipo eco, rastros de esta transición pueden observarse en el comportamiento de dichas franjas de interferencia.
\chapter{Conclusiones}\label{ch:7}
En esta tesis se ha abordado el problema de las fases geométricas acumuladas por sistemas cuánticos abiertos desde una perspectiva amplia que involucró distintas nociones para este objeto y distintos enfoques en el tratamiento de la dinámica. Los diversos sistemas considerados constituyen modelos paradigmáticos extensivamente estudiados en múltiples áreas de investigación, que resultan suficientemente sencillos como para facilitar el acceso a los resultados obtenidos, al mismo tiempo que conservan riqueza para posibilitar un amplio espectro de conclusiones y exámenes.

A lo largo de los distintos capítulos se ha examinado la evolución de estos sistemas relacionando los efectos del entorno, como la pérdida de coherencia, dinámica específica del entrelazamiento, etc., con los efectos observados sobre la fase geométrica acumulada. Por medio de estas comparaciones, se han encontrado escenarios en que la fase geométrica permanece particularmente robusta a pesar del decaimiento de la coherencia cuántica, se ha observado el efecto de las fluctuaciones cuánticas del vacío electromagnético en espacio libre, y las modificaciones introducidas al considerar contornos no-triviales. También se ha extendido este análisis al caso de un campo electromagnético con presencia de medios dieléctricos en el cual el movimiento relativo entre el sistema y el entorno da origen a nuevos fenómenos. Finalmente, se ha considerado el escenario de sistemas continua e indirectamente monitoreados en el cual la descripción del sistema abierto, y en consecuencia también la descripción de la fase geométrica acumulada, se modifican.

Desde su propuesta en el contexto de la mecánica cuántica, las fases geométricas resultaron objetos de interés debido a múltiples razones que se extienden desde los fundamentos mismos de la mecánica cuántica hasta la descripción de fenómenos físicos específicos. En el escenario actual, en que el auge de la información cuántica añade las posibilidad de aplicaciones prácticas en la fabricación de compuertas lógicas, los estudios condensados en esta tesis adquieren interés adicional. 

En el capítulo \ref{ch:3} se ha estudiado el efecto que tiene el  entorno sobre la fase geométrica en el contexto particular del modelo de Jaynes-Cummings. Este modelo paradigmático, ampliamente aplicado en diversas áreas de investigación, es probablemente el modelo más sencillo que da cuenta de manera satisfactoria de la interacción entre la materia (representada por un sistema de dos niveles) y la radiación electromagnética (simplificada a un único modo en una cavidad). La extensión del modelo al caso en el cual el sistema bipartito original se encuentra en interacción con el entorno se realizó a través de una ecuación maestra que introduce los efectos ambientales mediante consideraciones fenomenológicas. Se discutieron los contextos y abordajes para los cuáles el caso unitario resulta comparable con el caso disipativo, optando por la interpretación cinemática de la fase geométrica. Esta interpretación resalta la dependencia exclusiva de la fase geométrica en la secuencia de estados físicos atravesada por el sistema, esto es, en la trayectoria descrita en el espacio de rayos. En este marco, los diversos resultados obtenidos fueron detenidamente analizados y justificados en términos de las trayectorias observadas. Entre los mismos cabe destacar la robustez de la fase geométrica frente a los efectos del entorno en el caso resonante en que la frecuencia del modo en la cavidad coincide con la frecuencia natural del átomo.

En el capítulo \ref{ch:4} se estudió la dinámica de un sistema compuesto por dos qubits sujeto a los efectos de las fluctuaciones cuánticas del vacío electromagnético tanto en espacio libre como en presencia de un plano infinito conductor. Se definió una escala de decoherencia $T = \pi/\gamma_0$ asociada a la pérdida de coherencia para el sistema ubicado en el espacio libre y se encontró que existe una jerarquía entre las escalas asociadas a los distintos casos en presencia del plano infinito de material conductor. Específicamente, dicha jerarquía puede expresarse como $T_\parallel > T_0> T_\perp$, donde $T_{\parallel, \perp}$ refiere a la escala de decoherencia para partículas con momento dipolar paralelo y perpendicular al plano conductor respectivamente. Este resultado se interpretó en términos de interacciones entre partículas reemplazando los efectos del contorno no-trivial por partículas imagen, concluyendo que el caso en que los dipolos se orientan paralelos al plano origina un entorno más destructivo que conduce a escalas de decoherencia más cortas.
La dependencia de estas escalas con la relación $d/L$ entre la distancia de las partículas al plano y la distancia que separa ambas partículas también fue investigada.
El entrelazamiento entre los qubits que componen el sistema se estudió a través de la concurrencia, analizando la dependencia en la relación $d/L$ y en la orientación de los momentos dipolares. Se puso especial atención en los efectos de muerte y generación espontánea de entrelazamiento, y re-nacimientos del entrelazamiento. Estos fenómenos se interpretaron en términos de interacciones efectivas (esto es, mediadas por el entorno) entre partículas recurriendo al método de imágenes.
Finalmente se consideró la fase geométrica acumulada por el sistema bipartito para un estado inicial máximamente entrelazado, con particular foco en las modificaciones introducidas por la presencia del entorno en los distintos casos considerados para el estudio de las coherencias y el entrelazamiento. En el límite de acoplamiento débil es posible encontrar una expansión analítica en órdenes del parámetro de acoplamiento que permite reconocer y distinguir las contribuciones del espacio libre y del plano conductor.

En el capítulo \ref{ch:5} se estudiaron distintos efectos del entorno para un escenario que incorpora un elemento extra respecto de los capítulos previos, el movimiento relativo entre el sistema y el entorno. En particular, se consideró un sistema de dos niveles que describe una trayectoria recta a distancia fija de un semiespacio dieléctrico, mediante un forzado externo. El conjunto se encuentra inmerso en el vacío del campo electromagnético, de forma que el entorno de la partícula está compuesto por el campo electromagnético vestido por el material. Inspeccionando la dinámica del sistema, distinguimos dos regímenes distintos delimitados por la velocidad crítica ${\rm v}_{\rm crit} = \omega_o \,d/2$. Por debajo de este umbral de velocidad, el sistema tiende asintóticamente al estado fundamental, mientras que una vez superada la velocidad crítica, el estado asintótico es un estado mixto.
Se estudia con especial énfasis la escala temporal en que decaen las coherencias y la fase geométrica acumulada por el sistema. 
Respecto de la escala de decaimiento de las coherencias, se realiza un estudio minucioso de su dependencia con la velocidad y con la dirección del momento dipolar de la partícula. Para el caso de velocidades por debajo del umbral ${\rm v}<{\rm v}_{\rm crit}$ se encuentra una expresión analítica de orden ${\rm v}^2$. Los resultados encontrados se interpretan, como en el capítulo \ref{ch:4} en términos de interacciones entre partículas, reemplazando los efectos del medio semiconductor por partículas imagen. Estableciendo un vínculo con la literatura previa se encuentra que la dependencia de la escala de decoherencia en la orientación del momento dipolar resulta inversamente proporcional a la exhibida por fricción cuántica. Esto sugiere que la presencia de dicha fuerza acelera el proceso de pérdida de coherencia. 
La corrección en fase geométrica acumulada por el sistema se descompone en dos contribuciones: una estática, debido a la mera presencia del campo electromagnético vestido por el medio dieléctrico, y una dinámica, que depende en la velocidad de la partícula. Esta corrección dinámica es, nuevamente, de orden cuadrático, y se observa la dependencia del coeficiente cuadrático en otros parámetros y variables, como por ejemplo, el instante de observación.
Finalmente, se realizó una propuesta experimental y se comparan los resultados esperados para diversas combinaciones de material dieléctrico y partículas efectivas en la búsqueda de detectar regímenes en los cuales estos efectos resulten experimentalmente accesibles con la tecnología disponible.

En el capítulo \ref{ch:6} se abordó el estudio de la fase geométrica acumulada por un sistema abierto dentro de un contexto distinto al propuesto en los capítulos precedentes. Específicamente, se estudió las fase geométrica para un sistema continua e indirectamente monitoreado, escenario en el que el estado del sistema en cada realización de la evolución se torna una variable estocástica. La fase geométrica hereda, por lo tanto, la naturaleza estocástica, haciendo necesario el análisis de la distribución de fases geométricas dónde se ha resaltado particularmente la competencia entre los efectos no-adiabáticos y los efectos introducidos por el entorno. También se ha comparado la distribución con algunos valores característicos de fase geométrica, como la fase $\phi_u$ acumulada por una hipotética evolución unitaria, o la media sobre la distribución $\bar{\phi}$, estableciendo en qué rangos, y si, existen correlaciones.
Con el objetivo de establecer conexión con los resultados experimentales, se estudió además la distribución de franjas de interferencia obtenidas en un experimento de Eco de Espín.
Finalmente, se hizo hincapié en la trayectoria específica en la que no se registra ningún salto cuántico y se mostró que la misma atraviesa una transición topológica como función de la tasa de disipación. 
A pesar de las enormes diferencias mostradas por las distribuciones de fase geométrica y de franjas de interferencia de un experimento tipo eco, rastros de esta transición pueden observarse en el comportamiento de dichas franjas de interferencia.

\renewcommand{\theHchapter}{A\arabic{chapter}}
\appendix
\chapter{Evolución unitaria de un sistema de dos niveles en un campo rotante}\label{apendice1}

En este apéndice se desarrolla en forma analítica la dinámica de un modelo paradigmático de la mecánica cuántica: un sistema de dos niveles inmerso en un campo magnético clásico $\mathbf{B}(t) =\omega\, \hat{\text{\bf{n}}}_\mathbf{B}(t)$, cuya dirección está dada por $\hat{\text{\bf{n}}}_\mathbf{B }=(\sin{(\theta)}\cos(\varphi),\allowbreak \sin{(\theta)}\sin(\varphi), \cos{\theta})$ y varía, con el ángulo polar $\theta$ fijo y ángulo azimutal $\varphi = \Omega\,t$ dependiente del tiempo. Este mismo sistema se considera en las secciones(\ref{sec:sec2_ejemplo}) y (\ref{ch:6}) del texto principal. En la primera, se lo estudia como ejemplo de cálculo de fases geométricas en sistemas de evolución unitaria, mientras que en la segunda se investiga el caso en que el sistema se encuentra, además, en interacción con un entorno que se monitorea regularmente.

Una evolución (unitaria) del tipo descrito está generada por el Hamiltoniano de la ecuación (\ref{eq:sec2_HamiltonianSpin})

\begin{equation*}
    H(t) = \frac{1}{2}\,\mathbf{B}(t)\cdot \boldsymbol{\sigma},
\end{equation*}
con $\boldsymbol{\sigma} = \sigma_x\,\hat{x} + \sigma_y\,\hat{y} + \sigma_z\,\hat{z}$ un operador vectorial formado por las matrices de Pauli y, como en el texto principal, se denotan con $\ket{0}$ y $\ket{1}$ los autoestados de la matriz de Pauli $\sigma_z$, y unidades de $\hbar = 1$. La evolución generada por este Hamiltoniano está entonces representada, en el sistema laboratorio, por el operador unitario

\begin{equation}
    U_{\rm lab}(t,t_0) = \mathcal{T}\left\lbrace e^{-i\,\int_{t_0}^{t} dt' H(t')}\right\rbrace
    \label{eq:ap1_Uevollab}
\end{equation}
 
{\em Sistema rotante - }
El problema se simplifica si se aborda desde el sistema de referencia que rota con el campo magnético. La transformación unitaria $U(t)$ que lleva del sistema laboratorio al sistema rotante está descrita por 

\begin{equation}
    U(t) = e^{i\,\frac{\Omega}{2} \,t\,\sigma_z}
    \label{eq:ap1_Urot},
\end{equation}
de modo que los estados y operadores observados desde uno y otro sistema se relacionan según $\ket{\psi_R(t)} = U(t)\ket{\psi(t)}$ y $O_R(t) = U^\dagger(t)\,\,O\,U(t)$.
El Hamiltoniano que gobierna la dinámica en el sistema rotante puede extraerse de la ecuación de Schrödinger en esta representación 

\begin{equation}
     i|\dot{\psi}_R(t)\rangle = H_R(t)\ket{\psi_R(t)},
     \label{eq:ap1_SchRot}
\end{equation}
escribiendo el estado explícitamente como operaciones unitarias sobre el estado inicial en el sistema laboratorio $\ket{\psi_R(t)}= U(t)U_{\rm lab}(t,t_0)\ket{\psi(0)}$. Con este procedimiento se encuentra que la ecuación (\ref{eq:ap1_SchRot}) toma la forma

\begin{equation}
    i\partial_t\left(U(t)U_{lab}(t,t_0)\right)\ket{\psi(0)} = H_R(t)U(t)U_{lab}(t,t_0)\ket{\psi(0)}
\end{equation}
de donde se sigue una expresión para el Hamiltoniano $H_R(t)$ transformado en términos de las operaciones unitarias

\begin{equation}
    i\frac{d}{dt} \left(U(t)U_{\rm lab}(t,t_0)\right)\,U^\dagger_{lab}(t,t_0)U^\dagger(t) = H_R(t).
\end{equation}
Recurriendo a las expresiones explícitas para las transformaciones unitarias $U(t)$ y $U_{\rm lab}(t,t_0)$, dadas en las ecuaciones (\ref{eq:ap1_Urot}) y (\ref{eq:ap1_Uevollab}) respectivamente, se encuentra para el Hamiltoniano en el sistema rotante
\begin{align}\nonumber
    H_R(t) &=  -\frac{\Omega}{2}\,\sigma_z +\; e^{i\,\frac{\Omega}{2}\,t\,\sigma_z}\cdot H(t) \cdot e^{i\,\frac{\Omega}{2}\,t\,\sigma_z}\\
    &=\frac{1}{2}\left[(\omega \cos(\theta)-\Omega )\sigma_z + \omega\sin(\theta)\sigma_x\right].
\end{align}

De esta forma, en el sistema rotante el problema se reduce a resolver la dinámica generada por un Hamiltoniano independiente del tiempo, cuyos autoestados y autovalores están dados por 

\begin{equation}
    \ket{\xi_{R,\pm}} = \mathcal{N}_\pm^{-1}\left[( \omega\cos(\theta)- \Omega \pm \Tilde{\omega})\ket{1} + \omega\sin(\theta)\ket{0}\right], \;\text{con}\; \epsilon_\pm = \pm \frac{\Tilde{\omega}}{2}
    \label{eq:ap1_eigenvecR}
\end{equation}
donde se ha definido la frecuencia  $\Tilde{\omega} = [(\omega\cos(\theta)-\Omega)^2 + \omega^2\,\sin^2(\theta)]^{1/2}$ y $\mathcal{N}_\pm$ indica el factor de normalización correspondiente.
La simplificación que muestra el problema en el sistema que rota con el campo convoca a enfrentar su resolución en dicho escenario, de modo que la solución completa consistirá en la siguiente serie de pasos

\begin{enumerate}
    \item[i.] Aplicar la transformación unitaria $U(0)$ al estado inicial en el sistema laboratorio $\ket{\psi(0)}$ para obtener la forma del estado inicial en el sistema rotante $\ket{\psi_R(0)}$
    \item[ii.] Encontrar, en el sistema rotante, el estado $\ket{\psi_R(T)}$ que ha evolucionado durante el intervalo de interés $t\in[0,T]$
    \item[iii.] Aplicar la transformación inversa $U^\dagger(T)$ en el instante final al estado evolucionado $\ket{\psi_R(T)}$ para obtener el estado correspondiente en el sistema laboratorio $\ket{\psi(T)}$.
\end{enumerate}
Antes de dedicarnos a la serie completa que permita resolver la dinámica de este sistema, nos ocuparemos brevemente en describir con más detalle la evolución en el sistema rotante. 

\vspace{.5cm}
{\em Evolución en el sistema rotante - }
Se desea conocer la evolución de un estado inicial $\ket{\psi_R(0)}$ que puede escribirse, sin pérdida de generalidad, como una descomposición en la base $\{\ket{0}, \ket{1}\}$ de autoestados del operador $\sigma_z$ del sistema laboratorio. Se supone un estado inicial descompuesto en esta base, dado que ésta es una elección usual para la descripción de estados cuánticos puros. 
Por otra parte, otra base natural para la descripción de este problema específico, es la base de autoestados $\{\ket{\xi_{R,\pm}}\}$ del Hamiltoniano $H_R$ cuya evolución 

\begin{equation}
    \ket{\xi_{R,\pm}}\rightarrow e^{\mp i\frac{\Tilde{\omega}}{2}\,t} \ket{\xi_{R,\pm}}
\end{equation}
se conoce inmediatamente. Usando la ecuación (\ref{eq:ap1_eigenvecR}) puede describirse la relación entre las bases $\{\ket{0}, \ket{1}\}$ y $\{\ket{\xi_{R,\pm}(0)}\}$, que puede expresarse como

\begin{align}
    &\ket{1} =\hspace{.3cm}\frac{1}{2\,\Tilde{\omega}}\;(\mathcal{N_+}\ket{\xi_{R,+}}-\mathcal{N_-}\ket{\xi_{R,-}})\\[.75em]\nonumber
    &\ket{0} = - \frac{\mathcal{N_+}}{2\,\Tilde{\omega}}\frac{\omega\cos(\theta) -\Omega -\Tilde{\omega}}{\omega\sin(\theta)}\ket{\xi_{R,+}}+\frac{\mathcal{N_-}}{2\,\Tilde{\omega}}\frac{\omega\cos(\theta)-\Omega +\Tilde{\omega}}{\omega\sin(\theta)}\ket{\xi_{R,-}}, 
\end{align}
para encontrar la evolución de los estados $\ket{0}$ y $\ket{1}$ en un intervalo $t\in[0,T]$.

\begin{align}
    &\ket{1} \rightarrow \left\{\cos(\frac{\Tilde{\omega}}{2}\,t)-i\sin(\frac{\Tilde{\omega}}{2}\,t)\,\frac{\omega\cos(\theta)-\Omega}{\Tilde{\omega}}\right\}\ket{1} -i\sin(\frac{\Tilde{\omega}}{2}\,t)\,\frac{\omega\,\sin(\theta)}{\Tilde{\omega}}\ket{0}\label{eq:ap1_evolRot}\\[.75em]\nonumber
    &\ket{0}\rightarrow \left\{\cos(\frac{\Tilde{\omega}}{2}\,t)+i\sin(\frac{\Tilde{\omega}}{2}\,t)\,\frac{\omega\cos(\theta)-\Omega}{\Tilde{\omega}}\right\}\ket{1} -i\sin(\frac{\Tilde{\omega}}{2}\,t)\,\frac{\omega\,\sin(\theta)}{\Tilde{\omega}}\ket{1}.
\end{align}
Con este resultado resulta inmediato encontrar la evolución, en el sistema rotante, para cualquier estado $\ket{\psi_R(0)}$ descompuesto en la base $\{\ket{0}, \ket{1}\}$.
\\
\\\indent
{\em Evolución en el sistema laboratorio - }Para conocer la evolución de un estado en el sistema laboratorio, es necesario incorporar al resultado para la evolución en el sistema rotante (\ref{eq:ap1_evolRot}), los pasos (i) y (iii) enumerados arriba. En particular, si se considera que el estado inicial en el sistema laboratorio es un autoestado del Hamiltoniano instantáneo a tiempo inicial $H(0)$ 

\begin{equation}
    \ket{\psi_\pm(0)} = \cos(\theta/2)\ket{1} \pm \sin(\theta/2)\ket{0},
    \label{eq:ap1_ini}
\end{equation}
la expresión matemática resulta idéntica para el estado inicial en el sistema laboratorio, puesto que $U(0) = \mathbb{I}$, de modo que el paso (i) se completa inmediatamente con $\ket{\psi_R(0)} = \ket{\psi(0)}$. La evolución en el sistema rotante, listada como paso (ii), se encuentra reemplazando los estados $\ket{0}$ y $\ket{1}$ en (\ref{eq:ap1_ini}) por su evolución a tiempo $t$ en el sistema rotante, obteniendo una expresión para el estado $\ket{\psi_R(t)}$. Realizadas estas dos operaciones, sólo resta retornar al sistema laboratorio mediante la aplicación de la transformación unitaria adjunta $U^\dagger(t)$ sobre el estado $\ket{\psi_R(t)}$ hallado

\begin{equation}
    \ket{\psi(t)} = U^\dagger(t)\ket{\psi_R(t)}.
\end{equation} 

Resultará conveniente, en los distintos exámenes que se realizan en esta tesis, expresar el resultado final en la base de autoestados del hamiltoniano instantáneo $H(t)$, en lugar de hacerlo en la base $\{\ket{0}, \ket{1}\}$ que  resulta finalmente una base auxiliar para desarrollar el cálculo. El resultado obtenido para el estado $\ket{\psi(t)}$ a tiempo $t$ es 

\begin{equation}
    \ket{\psi(t)} = e^{-i\frac{\Omega}{2}\,t}\left\lbrace \left(\cos(\Tilde{\omega}\,t/2)+f(t)\right)\ket{\psi_+(t)}-g(t)\ket{\psi_-(t)}\right\rbrace,
\end{equation}
donde las funciones $f(t)$ y $g(t)$ en los coeficientes que modulan la superposición de autoestados $\ket{\psi_\pm(t)}$ del Hamiltoniano instantáneo tienen la forma $f(t) = -i\,\sin(\Tilde{\omega}\,t/2)(\omega-\Omega\cos(\theta))/\Tilde{\omega}$ y $g(t) = -i\,\sin(\Tilde{\omega}\,t/2)\,\omega\,\sin(\theta)/\Tilde{\omega} $

\chapter{Derivación de ecuaciones maestras}\label{sec:ap2}
Se desea tratar con modelos conformados por dos partes: (i) un sistema de interés al que se refiere como {\em el sistema} y (ii) otro sistema, acoplado al primero, que no se pretende o no es posible caracterizar exactamente y al que se refiere como {\em el entorno}. La dinámica de un  conjunto de este tipo está generada por un Hamiltoniano que puede escribirse formalmente como 

\begin{equation}
    H = H_{\rm s} + H_{\epsilon} + H_{\rm int}
\end{equation}
dónde $H_{\rm s}$ y $H_{\epsilon}$ definen la evolución del sistema y el entorno libres, mientras que $H_{\rm int}$ da cuenta de la interacción entre las dos partes. Inicialmente no se hace ninguna hipótesis respecto de la dependencia temporal de $H$ ni de ninguno de los términos que lo componen. El conjunto completo constituye un sistema cerrado, de modo que su dinámica se describe con la ecuación de Schrödinger o, equivalentemente, según $\ket{\psi(t)} = U(t,t_0)\ket{\psi(t_0)}$ con $U(t,t_0)$ un operador unitario para el cual se tiene la expresión formal

\begin{equation}
    U(t,t_0) = \mathcal{T} \exp(-i\,\int_{t_0}^{t} dt'\,H(t')),
    \label{eq:ap2_Uevol}
\end{equation}
donde $\mathcal{T}$ indica el operador de orden cronológico y $U(t_0, t_0) =\mathbb{I}$. Si se representa el estado del modelo completo con un operador densidad $\rho_{\rm tot}(t)$ en lugar de hacerlo con un vector $\ket{\psi(t)}$ del espacio de Hilbert, la ecuación (\ref{eq:ap2_Uevol}) deviene en 

\begin{equation}
    \dot{\rho}_{\rm tot}(t) = -\frac{i}{\hbar}\,[H(t), \rho_{\rm tot}(t)],
    \label{eq:ap2_VonN}
\end{equation}
conocida como la ecuación de Louville-Von Neumann. Esta ecuación será el punto de partida para la derivación de una ecuación dinámica que describa la evolución del estado $\rho(t)$ del sistema únicamente. Referencias paradigmáticas en las que puede seguirse esta derivación son  \cite{breuer2002libro, rivas2012libro}.

Trabajando todavía con el conjunto, la evolución descrita por $U(t,t_0)$ puede descomponerse y pasar a la {\em representación de interacción} con respecto a un Hamiltoniano $H_0$, en la cual los operadores evolucionan por acción del operador $U_0 = \mathcal{T} \exp(-i\,\int dt'\, H_0)$, mientras que los estados lo hacen según $\ket{\psi_{\rm I}(t)} = U_0^{-1}(t, t_0)\cdot U(t,t_0)\ket{\psi_{\rm I}(t_0)}$ . 

Una elección corriente es pasar a la representación de interacción con respecto a los términos de Hamiltoniano libre, dado que la dinámica que generan es usualmente conocida. Esta elección corresponde a tomar $H_0 = H_{\rm s}+ H_\epsilon$ y consecuentemente $U_0 = U_{\rm s}\otimes U_\epsilon$ e implica que toda la dinámica intrínseca a cada sistema, generada por $H_{\rm s}$ y $H_\epsilon$ esté contenida en los operadores, mientras que la evolución de los estados $\ket{\psi_{\rm I}(t)}$ está dictada exclusivamente por la interacción entre los dos sub-sistemas. 

En lo que sigue se reproduce esta elección suponiendo, en primer lugar, que ninguno de los términos que componen $H_0$ muestra dependencia temporal explícita, de modo tal que el operador de evolución $U_0$ toma la forma

\begin{equation}
U_0(t, t_0) = e^{-i\,H_{\rm s}(t-t_0)}\otimes\, e^{-i\,H_{\epsilon}(t-t_0)}. 
\end{equation}
El caso de Hamiltonianos con dependencia temporal explícita se pospone para la sección \ref{sec:ap2_adiabatica}.

\section{Hamiltoniano libre sin dependencia temporal}\label{sec:apB_1}
Con la elección de representación descrita, los operadores involucrados en la ecuación de movimiento del conjunto son el operador densidad $\rho_{\rm tot, I}(t) = U_0(t)^\dagger\rho_{\rm tot}(t)U_0(t)$, y el Hamiltoniano de interacción $H_{\rm int, I}(t) = U_0(t)^\dagger\rho_{\rm int}(t)U_0(t)$, donde el sufijo ${\rm I}$ indica que el operador correspondiente se encuentra en su representación de interacción. Se toma además $t_0=0$ y, con el objetivo de aligerar la notación, se omite escribirlo en la dependencia funcional. La ecuación que describe la evolución (\ref{eq:ap2_VonN}) toma en esta representación, y con estas elecciones, la expresión

\begin{equation}
    \dot{\rho}_{\rm tot,I}(t) = -\frac{i}{\hbar}\,[H_{\rm int,I}(t), \rho_{\rm tot,I}(t)],
    \label{eq:ap2_VonNI}
\end{equation}
cuya solución formal 

\begin{equation}
    \rho_{\rm tot, I}(t) = \rho_{\rm tot, I}(0)-i\int_{0}^t\,dt'\,[H_{\rm int,I}(t'), \rho_{\rm tot,I}(t')]
\end{equation}
puede sustituirse nuevamente en la misma ecuación (\ref{eq:ap2_VonNI}) y, tras tomar la traza parcial sobre los grados de libertad del entorno, se encuentra

\begin{equation}
    \dot{\rho}_{\rm s, I}(t) = -\frac{i}{\hbar}\,\Tr_{\epsilon}[H_{\rm int,I}(t),\rho_{\rm tot, I}(0)]-\frac{1}{\hbar^2}\,\Tr_{\epsilon}\int_{0}^t\,dt'\,[H_{\rm int,I}(t),[H_{\rm int,I}(t'), \rho_{\rm tot,I}(t')]].
    \label{eq:ap2_eqaux}
\end{equation}
Esta ecuación, aunque exacta, todavía contiene el estado del conjunto completo en el lado derecho. Para eliminarlo se introduce la {\em aproximación de Born}, o aproximación de acoplamiento débil. La misma consiste en suponer que el acoplamiento entre el sistema y el entorno es suficientemente débil para que las correlaciones entre estas dos partes resulten despreciables en la escala temporal de interés. El estado del conjunto completo es de este modo siempre separable $\rho_{\rm tot, I}(t) = \rho_{\rm s, I}(t) \otimes \rho_{\epsilon, I}(t)$. Más aún, se supone que las excitaciones causadas por el sistema en el entorno decaen en una escala temporal $\tau_\epsilon$ irresoluble, de forma que $\rho_{\epsilon}(t)\approx\rho_{\epsilon}(0)$. Aplicando la aproximación de Born a la ecuación (\ref{eq:ap2_eqaux}) se obtiene

\begin{equation}
    \dot{\rho}_{\rm s, I}(t) = -\frac{i}{\hbar}\,\Tr_{\epsilon}[H_{\rm int,I}(t),\rho_{\rm s, I}(0) \otimes \rho_\epsilon]-\frac{1}{\hbar^2}\,\Tr_{\epsilon}\int_{0}^t\,dt'\,[H_{\rm int,I}(t),[H_{\rm int,I}(t'), \rho_{\rm s,I}(t') \otimes \rho_\epsilon]].
\end{equation}
Esta expresión es todavía no-local en el tiempo. Esto es, la variación de $\rho_{\rm I}$ a tiempo $t$ depende no del estado $\rho_{\rm I}(t)$ en dicho instante sino también del estado a todo instante $t'<t$ previo. La no-localidad en el tiempo se traduce en que esta ecuación resulte intratable tanto analítica como numéricamente. 

Para obtener una ecuación local en el tiempo se realiza una segunda aproximación, la {\em aproximación de Markov}. La misma consiste en asumir que la variación del estado $\rho_{\rm s,I}$ es lenta en comparación con la escala en que decaen las excitaciones en el entorno. Es importante recordar que la evolución del operador $\rho_{\rm s,I}$ se debe únicamente a su interacción con el entorno, ya que se ha removido el resto de la dinámica del estado pasando a la representación de interacción. De esta forma, la hipótesis tras la aproximación de Markov supone que $\rho_{\rm I}(t)$ varía apreciablemente en una escala {\em de relajación} $\tau_R \gg \tau_\epsilon$, lo que se justifica cuando el acoplamiento es débil y permite suponer que la variación en el estado del sistema a tiempo $t$ depende del estado en ese instante $\rho_{\rm s, I}(t)$, y no del estado a tiempos anteriores $t'<t$. La ecuación que resulta tras realizar esta segunda aproximación

\begin{equation}
    \dot{\rho}_{\rm s, I}(t) = -\frac{i}{\hbar}\,\Tr_{\epsilon}[H_{\rm int,I}(t),\rho_{\rm s, I}(0) \otimes \rho_\epsilon]-\frac{1}{\hbar^2}\,\Tr_{\epsilon}\int_{0}^t\,dt'\,\bigl[H_{\rm int,I}(t),[H_{\rm int,I}(t'), \rho_{\rm s,I}(t) \otimes \rho_\epsilon]\bigr].
    \label{eq:ap2_redfieldint}
\end{equation}
es local en el tiempo y permite tratamiento analítico o numérico dependiendo de la forma explícita de los operadores involucrados. 
Esta ecuación es la herramienta utilizada para abordar la dinámica de los sistemas físicos en el capítulo \ref{ch:4} y parte del capítulo \ref{ch:5}, mientras que para una ecuación maestra distinta aunque también local en el tiempo se refiere a \cite{paz1999environment}. 

Sin embargo, la expresión (\ref{eq:ap2_redfieldint}) no es la de una ecuación tipo Lindbland \cite{lindblad1976} como las aplicadas en los capítulos \ref{ch:3} y \ref{ch:6}, y no garantiza positividad de la solución en el caso general. Para obtener una ecuación de Lindbland a partir de (\ref{eq:ap2_redfieldint}), es necesario continuar el desarrollo teórico.

\section{Ecuación de Lindbland}\label{sec:apB_2}
Para derivar una ecuación tipo Lindbland a partir de la ecuación (\ref{eq:ap2_redfieldint}), se nota que el Hamiltoniano de interacción en la representación de Schrodinger puede escribirse en términos de un conjunto de operadores hermíticos adimensionales $\{A_\alpha\}$ y $\{B_\alpha\}$ que actúan sobre el estado del sistema y del entorno respectivamente, 

\begin{equation}
    H_{\rm int} = g\sum_\alpha\,A_\alpha\otimes B_\alpha.
    \label{eq:ap2_desc}
\end{equation}
En representación de interacción, estos operadores evolucionan según $A_{\alpha, \rm I}(t) = U_{\rm s}(t)A_\alpha U_{\rm s}(t)$ y  $B_{\alpha, \rm I}(t) = U_{\epsilon}(t)B_\alpha U_{\epsilon}(t)$, de forma que el Hamiltoniano de interacción en representación de interacción puede escribirse como

\begin{equation}
    H_{\rm int,I} = g\sum_\alpha A_{\alpha, \rm I}(t)\otimes B_{\alpha, \rm I}(t).
    \label{eq:ap2_Hintexp}
\end{equation}
Reemplazando esta expresión en la ecuación de evolución para el estado (\ref{eq:ap2_redfieldint}), se tiene, 

\begin{equation}
    \Dot{\rho}_{\rm s,I} = -\frac{g^2}{\hbar^2}\int_{0}^t\,dt'\,\sum_{\alpha, \beta}\expval{B_{\alpha,I}(t)\,B_{\beta,I}(t')}[A_{\alpha, \rm I}(t)A_{\beta, \rm I}(t')\rho_{\rm s,I}(t)- A_{\beta, \rm I}(t')\rho_{\rm s,I}(t)A_{\alpha, \rm I}(t) + {\rm h.c.}],
    \label{eq:ap2_eqaux2}
\end{equation}
donde las $\expval{B_{\alpha,I}(t)\,B_{\beta,I}(t')} = \Tr_\epsilon(B_{\alpha,I}(t)\,B_{\beta,I}(t')\rho_\epsilon)$ son funciones de correlación del entorno. Se ha supuesto, además, que el primer término en la ecuación (\ref{eq:ap2_redfieldint}) se anula. Para que esto en efecto suceda, es condición suficiente que los operadores que actúan sobre el entorno se anulen en valor medio, es decir, que se satisfaga
\begin{equation}
    \expval{B_{\alpha,I}(t)} = 0.
    \label{eq:ap2_condicion}
\end{equation} 

Haciendo el cambio de variables $t\rightarrow t-t'$ en la integral y considerando que el estado del entorno es estacionario, es decir, que se satisface que $[H_\epsilon, \rho_\epsilon]=0$, se tiene que las funciones de correlación del entorno son homogéneas en la variable $t$ y por lo tanto se cumple que

\begin{equation}
    \expval{B_{\alpha,I}(t)\,B_{\beta,I}(t-t')} = \expval{B_{\alpha,I}(t')\,B_{\beta,I}(0)}.
\end{equation}
Como se ha expresado anteriormente, se supone que estas funciones de correlación decaen en una escala temporal $\tau_\epsilon$ que resulta mucho más corta que la escala $\tau_R$ de variación de $\rho_{\rm I}$. Esta diferencia de escalas justifica tomar $t\rightarrow\infty$ en el límite superior de la integral, dado que el integrando será prácticamente nulo fuera del rango delimitado por $\tau_\epsilon$. 

Una última aproximación será necesaria para encontrar una ecuación de movimiento de tipo Lindbland, conocida como {\em aproximación secular}. Esta aproximación consiste en despreciar términos que resulten altamente oscilantes en la escala temporal definida por $\tau_R$ y se vincula estrechamente con la aproximación de onda rotante, frecuente en el campo de la óptica cuántica y en otras áreas de investigación (para un análisis detallado de estas dos aproximaciones y sus implicaciones en el marco de los sistemas cuánticos abiertos se refiere a \cite{fleming2010rotating}).

Para justificar esta última aproximación, se considera la descomposición espectral del Hamiltoniano $H_{\rm s}$. Denominando $\varepsilon$ los autovalores de $H_{\rm s}$ y $\Pi(\varepsilon)$ el proyector al autoespacio asociado a $\varepsilon$, pueden definirse operadores, {\em en representación de Schrödinger},

\begin{equation}
    A(\omega) = \sum_{\varepsilon-\varepsilon' = \omega}\,\Pi(\varepsilon)A_\alpha\Pi(\varepsilon'),
    \label{eq:ap2_Awdef}
\end{equation}
donde la suma se extiende sobre todos los autovalores de energía de $H_{\rm s}$ cuya diferencia tome el valor fijo $\omega$. Sumando sobre todas las diferencias de energía y dada la completitud de la base de autoestados de $H_{\rm s}$, se recupera $A_\alpha$. Una consecuencia inmediata de la definición (\ref{eq:ap2_Awdef}), es que estos operadores resultan, en la representación de interacción, $A_{\alpha, \rm I}(\omega) = \exp(-i\,\omega\,t)A_{\alpha}(\omega)$, de forma que los operadores $A_{\alpha, \rm I}(t)$ en la ecuación (\ref{eq:ap2_eqaux2}) se pueden descomponer según

\begin{equation}
    A_{\alpha, \rm I}(t) = \sum_\omega\,e^{-i\,\omega\,t} \,A_{\alpha}(\omega).
\end{equation}
Introducir explícitamente esta descomposición de los operadores $A_{\alpha, \rm I}(t)$ en la ecuación (\ref{eq:ap2_eqaux2}) resulta en una sumatoria de términos que oscilan con frecuencias $\omega + \omega'$ asociadas a la diferencia entre autoestados de $H_{\rm s}$

\begin{equation}
    \Dot{\rho}_{\rm s,I} = -\frac{1}{\hbar^2}\sum_{\alpha, \beta}\sum_{\omega, \omega'}\,\Bigl\{ \Gamma_{\alpha\beta}(\omega')\,[A_{\alpha}(\omega)A_{\beta}(\omega')\rho_{\rm s,I}(t)- A_{\beta}(\omega')\rho_{\rm s,I}(t)A_{\alpha}(\omega)]e^{-i(\omega + \omega')\,t} + \rm h.c.\Bigr\},
    \label{eq:ap2_eqaux3}
\end{equation}
donde ${\rm h.c.}$ denota el término hermítico conjugado y se han definido las transformadas de Fourier unilaterales de las funciones de correlación

\begin{equation}
    \Gamma_{\alpha\beta}(\omega) = g^2 \int_{0}^\infty dt'\,\expval{B_{\alpha,I}(t')\,B_{\beta,I}(0)}e^{i\omega\,t'}.
\end{equation}
Se denota $\tau_s$ cada escala típica de evolución intrínseca del sistema, definida por $|\omega + \omega'|^{-1}$, esto es, por la inversa de la frecuencia involucrada. Si $\tau_s$ es chico en comparación con la escala de relajación $\tau_R$ del sistema abierto, entonces los términos no seculares presentes en la suma (\ref{eq:ap2_eqaux3}), es decir, aquellos términos para los cuales $\omega + \omega'\neq 0$ pueden despreciarse, dado que oscilan rápidamente durante el intervalo definido por $\tau_R$ en el cuál $\rho_{\rm I}$ varía apreciablemente. En consecuencia se obtiene

\begin{align}
    \Dot{\rho}_{\rm s,I} = -\frac{1}{\hbar^2}\sum_{\alpha, \beta}\sum_{\omega}\,\Bigl\{\Gamma_{\alpha\beta}(\omega)\,[A^\dagger_{\alpha}(\omega)A_{\beta}(\omega)\rho_{\rm s,I}(t)- A_{\beta}(\omega)\rho_{\rm s,I}(t)A^\dagger_{\alpha}(\omega)]  + \rm h.c.\Bigr\},
    \label{eq:ap2_eqaux4}
\end{align}
donde se ha utilizado que $A^{\dagger}(\omega) = A(-\omega)$.
Es conveniente descomponer el factor $\Gamma_{\alpha\beta}(\omega)$ que contiene la información sobre el entorno en partes imaginaria y real según $\Gamma_{\alpha\beta}(\omega)=1/2 \,\gamma_{\alpha\beta}(\omega) + i\,S_{\alpha\beta}(\omega)$, lo que permite, retornando a la representación de Schrödinger, obtener la expresión 

\begin{equation}
    \Dot{\rho}_{\rm s}(t) = - \frac{i}{\hbar}[H_{\rm s}, \rho_{\rm s}(t)] - \frac{i}{\hbar}[H_{\rm LS}, \rho_{\rm s}(t)] + \mathcal{D}[\rho_{\rm s}(t)]
    \label{eq:ap2_Lindbland1}
\end{equation}
donde el operador hermítico $H_{\rm LS} = \sum_{\alpha,\beta} \sum_{\omega}S_{\alpha\beta}(\omega)A_\alpha^\dagger(\omega)A_\beta(\omega)$ da una contribución hamiltoniana a la ecuación y es usualmente llamado {\em corrimiento de Lamb}, puesto que describe una renormalización de las energías del sistema introducido por el acoplamiento al entorno. El último término $\mathcal{D}[\rho_{\rm s}]$ es el {\em disipador} de la ecuación maestra (\ref{eq:ap2_Lindbland1}) 

\begin{equation}
    \mathcal{D}[\rho_{\rm s}] = \frac{1}{2}\sum_{\alpha,\beta} \sum_{\omega}\gamma_{\alpha\beta}(\omega)\left(\bigl\{A_\alpha^\dagger(\omega)A_\beta(\omega), \rho_{\rm s}(t)\bigr\} - 2A_\beta(\omega)\rho_{\rm s}(t)A_\alpha^\dagger(\omega)\right).
\end{equation}
La forma estándar de una ecuación tipo Lindbland 

\begin{equation}
   \dot{\rho}_{\rm s}(t) = -i\left[H, \rho_{\rm s}(t) \right] + \frac{1}{2} \sum_{\alpha} [2\,L_{\alpha}\rho_{\rm s}(t) L_{\alpha}^\dagger -  \{L_{\alpha}^\dagger L_{\alpha}, \rho_{\rm s}(t)\, \} ] \;,
   \label{eq:Lindblad}
\end{equation}
como las que se utilizan para estudiar la evolución de los sistemas físicos en los capítulos \ref{ch:3} y \ref{ch:6}, puede encontrarse diagonalizando las matrices $\gamma_{\alpha\beta}(\omega)$.

\section{Ecuación de Lindbland adiabática}\label{sec:ap2_adiabatica}
Siguiendo el desarrollo propuesto en \cite{albash2012, albash2015} se deriva una ecuación de movimiento para el caso en que el sistema se conduce aplicando un forzado externo, lo que puede representarse escribiendo el Hamiltoniano para el modelo completo como 

\begin{equation}
    H = H_{\rm s}(t) + H_{\epsilon} + H_{\rm int},
\end{equation}
y donde la dependencia temporal del término $H_{\rm s}(t)$ incorpora transformaciones no-triviales en la evolución  del conjunto sistema+entorno.

En este caso, el Hamiltoniano de interacción {\em en representación de Schrödinger} también puede escribirse en términos de un conjunto de operadores hermíticos adimensionales $\{A_\alpha\}$ y $\{B_\alpha\}$ que actúan sobre el estado del sistema y del entorno respectivamente, según la ecuación (\ref{eq:ap2_desc})

\begin{equation*}
    H_{\rm int} = g\sum_\alpha\,A_\alpha\otimes B_\alpha.
\end{equation*} 
Más aun, el desarrollo que conduce, en las secciones \ref{sec:apB_1} y \ref{sec:apB_2}, a la ecuación (\ref{eq:ap2_eqaux2}) 

\begin{equation*}
    \Dot{\rho}_{\rm s,I} = -\frac{g^2}{\hbar^2}\int_{0}^t\,dt'\,\sum_{\alpha, \beta}\expval{B_{\alpha,I}(t)\,B_{\beta,I}(t')}[A_{\alpha, \rm I}(t)A_{\beta, \rm I}(t')\rho_{\rm s,I}(t)- A_{\beta, \rm I}(t')\rho_{\rm s,I}(t)A_{\alpha, \rm I}(t) + {\rm h.c.}],
\end{equation*}
donde $A_{\alpha, \rm I}(t) = U_{\rm s}(t)A_\alpha U_{\rm s}(t)$ indica el operador $A_\alpha$ en la representación de interacción con respecto a los Hamiltonianos libres,
es en realidad absolutamente formal y no difiere si se incorpora el forzado clásico al Hamiltoniano del sistema.
\clearpage
La diferencia fundamental con respecto al desarrollo presentado en la sección anterior está en el operador unitario que conduce a la representación de interacción. En este caso dicho operador está dado por

\begin{equation}
    U_{\rm s}(t,0)\otimes U_B(t,0)= \mathcal{T}\left\{ e^{- i\int_{0}^tdt'\, H_{\rm s}(t')}\right\}\otimes e^{- i H_B\,t}.
\end{equation}
\\
\\\indent
{\em Aproximación adiabática - }Encontrar la forma explícita de los $A_{\alpha, \rm I}(t)$ es en general una tarea imposible. La estrategia, por lo tanto, consiste en reemplazar la forma exacta de $U_s(t,0)$ por una expresión válida en el límite adiabático. Para esto, se considera $\{\ket{\psi_n(t)}\}$ una base de autoestados instantáneos de $H_{\rm s}(t)$, es decir, una base de estados que satisface 

\begin{equation}
    H(t)\ket{\psi_n(t)} = E_n(t)\ket{\psi_n(t)}.
\end{equation}
En el límite adiabático estricto las transiciones entre estados se anulan y en consecuencia se tiene

\begin{equation}
    \ket{\psi_n(t)} = e^{-i\mu_n(t,t')}\ket{\psi_n(t')}.
\end{equation}
Utilizando esta relación, el operador de evolución puede, en el límite adiabático, expandirse en esta base según

\begin{equation}
    U_{\rm s}^{ad}(t,t')=\sum_n \ket{\psi_n(t)}\bra{\psi_n(t')}e^{-i\mu_n(t,t')}.
    \label{eq:ap2_adiabU}
\end{equation}

Por otra parte, resulta útil notar que el operador de evolución satisface

\begin{equation}
    U(t-t',0)=U(t-t', t)U(t,0) = U^{\dagger}(t, t-t')U(t,0).
\end{equation}

Las dos consideraciones realizadas permiten justificar dos aproximaciones sobre el operador $U_s(t,0)$. En primer lugar, se reemplaza $U_{\rm s}(t,0) $ por $ U_{\rm s}^{ad}(t,0)$. Luego, en segundo lugar, se reemplaza también $U_{\rm s}^{\dagger}(t, t-t')$ por $e^{i\,H_{\rm s}(t)\,t'}$, una aproximación que se justifica si las funciones de correlación del entorno decaen lo bastante rápido como para tomar $H_{\rm s}(t')\sim{\rm cte}$ en el intervalo de integración. La combinación de estas aproximaciones conduce a 
\begin{align}\nonumber
    U_{\rm s}(t,0) &\sim U_{\rm s}^{ad}(t,0)\\
    U_{\rm s}(t-t',0) &\sim U_{\rm s}^{ad}(t,0)\,e^{i\,H_{\rm s}(t)\,t'}.
    \label{eq:ap2_approxU}
\end{align}

Introduciendo las formas explícitas dadas por (\ref{eq:ap2_approxU}) y (\ref{eq:ap2_adiabU}) en la ecuación dinámica (\ref{eq:ap2_eqaux2}), se obtiene una ecuación maestra en representación de interacción que resulta válida en el límite adiabático

\begin{equation}
    \dot{\rho}_{\rm s, I} = -g^2\sum_{\substack{\alpha, \beta\\k,l,m,n}} e^{-i[\mu_{nm}(t)+\mu_{lk}(t)]}\;\mathcal{A}_{\alpha}^{mn}(t) \mathcal{A}_{\beta}^{kl}(t)\;\Gamma_{\alpha\beta}(\omega_{lk}(t))\left[\Pi_{mn}(0), \Pi_{kl}(0)\rho_s(t)\right] + {\rm h.c.},
    \label{eq:ap2_EqAdiabInt}
\end{equation}
donde se ha definido la diferencia de fases $\mu_{mn}(t) = \mu_m(0,t) - \mu_n(0,t) $ y $\omega_{mn}(t) = E_m(t) - E_n(t)$, los proyectores $\Pi_{kl}(t) = \ket{\psi_k(t)}\bra{\psi_l(t)}$, y las funciones
\begin{equation}
    \Gamma_{\alpha\beta}(\omega_{mn}(t)) = \int_0^\infty\,dt'\,e^{i\omega_{mn}(t)\,t'}\expval{B_{\alpha,I}(t')\,B_{\beta,I}(0)}.
\end{equation}
De la misma forma que en el caso sin dependencia temporal explícita, en la ecuación (\ref{eq:ap2_EqAdiabInt}) puede realizarse una aproximación secular que permita encontrar una ecuación tipo Lindblad. Siguiendo la discusión que conduce a la ecuación (\ref{eq:ap2_eqaux4}), en este caso dicha aproximación corresponde a tomar $n =m \wedge k = l$ o $k = n \wedge l = m$ (sin contar dos veces sobre el caso particular $k = l = m = n$).

Finalmente, retornando a la representación de Schrodinger mediante operadores que respetan las mismas aproximaciones introducidas al pasar a representación de interacción se encuentra 

\begin{align}
     \dot{\rho}_{\rm s, I} \sim - &\frac{i}{\hbar}[H_{\rm s}(t), \rho_{\rm s}(t)] \\\nonumber
      +&\sum_{\alpha, \beta} \sum_{m\neq n}\gamma_{\alpha\beta}(\omega_{nm}(t))\left[ L_{ mn,\beta}(t) \rho_s(t)L^{\dagger}_{mn,\alpha}(t)+\frac{1}{2}\left\{ L^{\dagger}_{mn,\alpha}(t)L_{mn,\beta}(t), \rho_s(t)\right\} \right]\\\nonumber
      + &\sum_{\alpha, \beta} \sum_{m, n}\gamma_{\alpha\beta}(0)\left[ L_{ mm,\beta}(t) \rho_s(t)L^{\dagger}_{nn,\alpha}(t)+\frac{1}{2}\left\{ L^{\dagger}_{nn,\alpha}(t)L_{mm,\beta}(t), \rho_s(t)\right\} \right] ,
\end{align}
donde se ha definido los operadores de Lindblad $L_{mn}(t) = \mathcal{A}_{\alpha}^{mn}(t)\Pi_{mn}(t)$.

\bibliographystyle{unsrt_lud}
\bibliography{citas, libros, citas_3, citas_4, citas_5, citas_6}

\begin{thebibliography}{100}

\bibitem{Berry1983original}
Berry M.~V.
\newblock {\em Proc. R. Soc. London}, 392(1802):45--57, 1984.

\bibitem{simon1983holonomy}
Simon B.
\newblock {\em Physical Review Letters}, 51(24):2167, 1983.

\bibitem{aharonov1987phase}
Aharonov Y. and Anandan J.
\newblock {\em Physical Review Letters}, 58(16):1593, 1987.

\bibitem{samuel1988general}
Samuel J. and Bhandari R.
\newblock {\em Physical Review Letters}, 60(23):2339, 1988.

\bibitem{mukunda1993quantum}
Mukunda N. and Simon R.
\newblock {\em Annals of Physics}, 228(2):205--268, 1993.

\bibitem{uhlmann1986parallel}
Uhlmann A.
\newblock {\em Reports on Mathematical Physics}, 24(2):229--240, 1986.

\bibitem{uhlmann1989berry}
Uhlmann A.
\newblock {\em Annalen der Physik}, 501(1):63--69, 1989.

\bibitem{uhlmann1991gauge}
Uhlmann A.
\newblock {\em letters in mathematical physics}, 21:229--236, 1991.

\bibitem{sjoqvist2000geometric}
Sj{\"o}qvist E., Pati A.~K., Ekert A., Anandan J.~S., Ericsson M., Oi~D.~K.,
  and Vedral V.
\newblock {\em Physical Review Letters}, 85(14):2845, 2000.

\bibitem{singh2003geometric}
Singh K., Tong D., Basu K., Chen J., and Du~J.
\newblock {\em Physical Review A}, 67(3):032106, 2003.

\bibitem{chaturvedi2004geometric}
Chaturvedi S., Ercolessi E., Marmo G., Morandi G., Mukunda N., and Simon R.
\newblock {\em The European Physical Journal C-Particles and Fields},
  35:413--423, 2004.

\bibitem{maninioffdiagonal}
Manini N. and Pistolesi F.
\newblock {\em Phys. Rev. Lett.}, 85:3067--3071, Oct 2000.

\bibitem{sjoqvistoffdiagonal}
Filipp S. and Sj\"oqvist E.
\newblock {\em Phys. Rev. Lett.}, 90:050403, Feb 2003.

\bibitem{wilczek1984appearance}
Wilczek F. and Zee A.
\newblock {\em Physical Review Letters}, 52(24):2111, 1984.

\bibitem{thouless1982_app_hall}
Thouless D.~J., Kohmoto M., Nightingale M.~P., and den Nijs M.
\newblock {\em Physical review letters}, 49(6):405, 1982.

\bibitem{bernevig2013_app_supercond}
Bernevig B.~A.
\newblock Topological insulators and topological superconductors.
\newblock In {\em Topological Insulators and Topological Superconductors}.
  Princeton university press, 2013.

\bibitem{asboth2016_app_supercond}
Asb{\'o}th J.~K., Oroszl{\'a}ny L., and P{\'a}lyi A.
\newblock {\em Lecture notes in physics}, 919:166, 2016.

\bibitem{vedral2003geometric}
Vedral V.
\newblock {\em International Journal of Quantum Information}, 1(01):1--23,
  2003.

\bibitem{Zanardi_1999}
Zanardi P. and Rasetti M.
\newblock {\em Physics Letters A}, 264(2-3):94--99, dec 1999.

\bibitem{pachos1999non}
Pachos J., Zanardi P., and Rasetti M.
\newblock {\em Physical Review A}, 61(1):010305, 1999.

\bibitem{pachos2001quantum}
Pachos J. and Zanardi P.
\newblock {\em International Journal of Modern Physics B}, 15(09):1257--1285,
  2001.

\bibitem{leibfried2003_app_qi}
Leibfried D., DeMarco B., Meyer V., Lucas D., Barrett M., Britton J., Itano
  W.~M., Jelenkovi{\'c} B., Langer C., Rosenband T., and others .
\newblock {\em Nature}, 422(6930):412--415, 2003.

\bibitem{xiang2001_app_qi}
Xiang-Bin W. and Keiji M.
\newblock {\em Phys. Rev. Lett.}, 87:097901, Aug 2001.

\bibitem{zhu2002_app_qi}
Zhu S.-L. and Wang Z.~D.
\newblock {\em Phys. Rev. Lett.}, 89:097902, Aug 2002.

\bibitem{li2020_path}
Li~K., Zhao P., and Tong D.
\newblock {\em Physical Review Research}, 2(2):023295, 2020.

\bibitem{ding2021_path}
Ding C.~Y., Ji~L.~N., Chen T., and Xue Z.~Y.
\newblock {\em Quantum Science and Technology}, 7(1):015012, 2021.

\bibitem{sjoqvistshortcut}
Gregefalk A. and Sj\"oqvist E.
\newblock {\em Phys. Rev. Applied}, 17:024012, Feb 2022.

\bibitem{measuringshortcut}
Zhang Z., Wang T., Xiang L., Yao J., Wu~J., and Yin Y.
\newblock {\em Phys. Rev. A}, 95:042345, Apr 2017.

\bibitem{breuer2002libro}
Breuer H.-P., Petruccione F., and others .
\newblock {\em The theory of open quantum systems}.
\newblock Oxford University Press on Demand, 2002.

\bibitem{rivas2012libro}
Rivas A. and Huelga S.~F.
\newblock {\em Open quantum systems}, volume~10.
\newblock Springer, 2012.

\bibitem{tong2004kinematic}
Tong D., Sj{\"o}qvist E., Kwek L.~C., and Oh~C.~H.
\newblock {\em Physical review letters}, 93(8):080405, 2004.

\bibitem{Carollo_original}
Carollo A., Fuentes-Guridi I., Santos M.~F., and Vedral V.
\newblock {\em Physical review letters}, 90(16):160402, 2003.

\bibitem{Carollo_review}
Carollo A.
\newblock {\em Modern Physics Letters A}, 20(22):1635--1654, 2005.

\bibitem{Sjo_no}
Sjöqvist E.
\newblock {\em Acta Physica Hungarica B) Quantum Electronics}, 26:195, Dec
  2009.

\bibitem{bassi2006_no}
Bassi A. and Ippoliti E.
\newblock {\em Physical Review A}, 73(6):062104, 2006.

\bibitem{sjoqvist2010_hidden}
Pawlus P. and Sj{\"o}qvist E.
\newblock {\em Physical Review A}, 82(5):052107, 2010.

\bibitem{buri}
Buri\ifmmode~\acute{c}\else \'{c}\fi{} N. and Radonji\ifmmode~\acute{c}\else
  \'{c}\fi{} M.
\newblock {\em Phys. Rev. A}, 80:014101, Jul 2009.

\bibitem{de2003berry}
De~Chiara G. and Palma G.~M.
\newblock {\em Physical review letters}, 91(9):090404, 2003.

\bibitem{whitney2003berry}
Whitney R.~S. and Gefen Y.
\newblock {\em Physical review letters}, 90(19):190402, 2003.

\bibitem{whitney2005geometric}
Whitney R.~S., Makhlin Y., Shnirman A., and Gefen Y.
\newblock {\em Physical review letters}, 94(7):070407, 2005.

\bibitem{berger2013_noise_cqed}
Berger S., Pechal M., Abdumalikov~Jr A.~A., Eichler C., Steffen L., Fedorov A.,
  Wallraff A., and Filipp S.
\newblock {\em Physical Review A}, 87(6):060303, 2013.

\bibitem{berger2015_noise_cqed}
Berger S.~J.
\newblock {\em Geometric phases and noise in circuit QED}.
\newblock PhD thesis, ETH Zurich, 2015.

\bibitem{grupo2006}
Lombardo F.~C. and Villar P.~I.
\newblock {\em Phys. Rev. A}, 74:042311, Oct 2006.

\bibitem{grupo2013}
Lombardo F.~C. and Villar P.~I.
\newblock {\em Phys. Rev. A}, 87:032338, Mar 2013.

\bibitem{grupo2014}
Lombardo F.~C. and Villar P.~I.
\newblock {\em Phys. Rev. A}, 89:012110, Jan 2014.

\bibitem{VILLAR2015246}
Villar P.~I.
\newblock {\em Physics Letters A}, 379(4):246--254, 2015.

\bibitem{farias2020towards}
Far{\'\i}as M.~B., Lombardo F.~C., Soba A., Villar P.~I., and Decca R.~S.
\newblock {\em npj Quantum Information}, 6(1):25, 2020.

\bibitem{oxman2018two}
Oxman L.~E., Khoury A.~Z., Lombardo F.~C., and Villar P.~I.
\newblock {\em Annals of Physics}, 390:159--179, 2018.

\bibitem{jaynes1963comparison}
Jaynes E.~T. and Cummings F.~W.
\newblock {\em Proceedings of the IEEE}, 51(1):89--109, 1963.

\bibitem{carmichael1989subnatural}
Carmichael H., Brecha R., Raizen M., Kimble H., and Rice P.
\newblock {\em Physical Review A}, 40(10):5516, 1989.

\bibitem{yamamoto2003semiconductor}
Yamamoto Y., Tassone F., and Cao H.
\newblock {\em Semiconductor cavity quantum electrodynamics}, volume 169.
\newblock Springer, 2003.

\bibitem{laussy2008strong}
Laussy F.~P., Del~Valle E., and Tejedor C.
\newblock {\em Physical review letters}, 101(8):083601, 2008.

\bibitem{vera2009characterization}
Vera C.~A., Quesada N., Vinck-Posada H., and Rodr{\'\i}guez B.~A.
\newblock {\em Journal of Physics: Condensed Matter}, 21(39):395603, 2009.

\bibitem{lodahl2015interfacing}
Lodahl P., Mahmoodian S., and Stobbe S.
\newblock {\em Reviews of Modern Physics}, 87(2):347, 2015.

\bibitem{resultados1}
Tomassone M. and Widom A.
\newblock {\em Physical Review B}, 56(8):4938, 1997.

\bibitem{resultados3}
Dedkov G. and Kyasov A.
\newblock {\em Technical Physics Letters}, 28:346--348, 2002.

\bibitem{resultados4}
Scheel S. and Buhmann S.~Y.
\newblock {\em Physical Review A}, 80(4):042902, 2009.

\bibitem{resultados8}
Intravaia F., Mkrtchian V.~E., Buhmann S.~Y., Scheel S., Dalvit D.~A., and
  Henkel C.
\newblock {\em Journal of Physics: Condensed Matter}, 27(21):214020, 2015.

\bibitem{barton2010van}
Barton G.
\newblock {\em New Journal of Physics}, 12(11):113045, 2010.

\bibitem{klatt2017quantum}
Klatt J., Far{\'\i}as M.~B., Dalvit D.~A.~R., and Buhmann S.
\newblock {\em Physical Review A}, 95(5):052510, 2017.

\bibitem{resultados6}
Pieplow G. and Henkel C.
\newblock {\em New Journal of Physics}, 15(2):023027, 2013.

\bibitem{resultados7}
Intravaia F., Behunin R., and Dalvit D.~A.
\newblock {\em Physical Review A}, 89(5):050101, 2014.

\bibitem{resultados9}
Reiche D., Intravaia F., Hsiang J.-T., Busch K., and Hu~B.-L.
\newblock {\em Physical Review A}, 102(5):050203, 2020.

\bibitem{nakahara2003geometry}
Nakahara M.
\newblock {\em Geometry, topology and physics}.
\newblock CRC press, 2003.

\bibitem{chruscinski2004geometric}
Chru{\'s}ci{\'n}ski D. and Jamio{\l}kowski A.
\newblock {\em Geometric phases in classical and quantum mechanics}.
\newblock Springer, 2004.

\bibitem{pancharatnam1956generalized}
Pancharatnam S.
\newblock Generalized theory of interference, and its applications: Part i.
  coherent pencils.
\newblock In {\em Proceedings of the Indian Academy of Sciences-Section A},
  volume~44, pages 247--262. Springer India New Delhi, 1956.

\bibitem{mukunda1993quantum2}
Mukunda N. and Simon R.
\newblock {\em Annals of Physics}, 228(2):269--340, 1993.

\bibitem{hahn}
Hahn E.~L.
\newblock {\em Phys. Rev.}, 80:580--594, Nov 1950.

\bibitem{leek2007_cqed_observation}
Leek P.~J., Fink J., Blais A., Bianchetti R., Goppl M., Gambetta J.~M.,
  Schuster D.~I., Frunzio L., Schoelkopf R.~J., and Wallraff A.
\newblock {\em science}, 318(5858):1889--1892, 2007.

\bibitem{gasparinetti2016_cqed_observation}
Gasparinetti S., Berger S., Abdumalikov A.~A., Pechal M., Filipp S., and
  Wallraff A.~J.
\newblock {\em Science advances}, 2(5):e1501732, 2016.

\bibitem{cucchietti}
Cucchietti F.~M., Zhang J.~F., Lombardo F.~C., Villar P.~I., and Laflamme R.
\newblock {\em Physical Review Letters}, 105(24), 2010.

\bibitem{acotacion_se}
Una implementación real del protocolo requiere dos pasos extra. Preparar y
  medir el sistema en el estado superposición $\ket{\psi(0)}$ resulta una
  tarea considerablemente complicada. En cambio, se prepara el autoestado
  fundamental de $\sigma_z$, $\ket{0}$ y se aplica luego un pulso que lo guía
  hasta $\ket{\psi(0)}$. El protocolo usualmente se finaliza con una rotación
  extra de espines que conduzca el estado final a la base $\sigma_z$, donde la
  probabilidad que en efecto se calcula es aquella de que el sistema se halle
  en $\ket{0}$.

\bibitem{du2003observation}
Du~J., Zou P., Shi M., Kwek L.~C., Pan J.-W., Oh~C.~H., Ekert A., Oi~D.~K., and
  Ericsson M.
\newblock {\em Physical review letters}, 91(10):100403, 2003.

\bibitem{berger2013exploring}
Berger S., Pechal M., Abdumalikov~Jr A.~A., Eichler C., Steffen L., Fedorov A.,
  Wallraff A., and Filipp S.
\newblock {\em Physical Review A}, 87(6):060303, 2013.

\bibitem{berger2015geometric}
Berger S.~J.
\newblock {\em Geometric phases and noise in circuit QED}.
\newblock PhD thesis, ETH Zurich, 2015.

\bibitem{fuentes2002vacuum}
Fuentes-Guridi I., Carollo A., Bose S., and Vedral V.
\newblock {\em Physical review letters}, 89(22):220404, 2002.

\bibitem{carollo2003berry}
Carollo A., Santos M.~F., and Vedral V.
\newblock {\em Physical Review A}, 67(6):063804, 2003.

\bibitem{liu2010vacuum}
Liu Y., Wei L., Jia W., and Liang J.
\newblock {\em Physical Review A}, 82(4):045801, 2010.

\bibitem{wang2015does}
Wang M., Wei L., and Liang J.
\newblock {\em Physics letters a}, 379(16-17):1087--1090, 2015.

\bibitem{viotti2022geometric}
Viotti L., Lombardo F.~C., and Villar P.~I.
\newblock {\em Physical Review A}, 105(2):022218, 2022.

\bibitem{khitrova2006vacuum}
Khitrova G., Gibbs H., Kira M., Koch S.~W., and Scherer A.
\newblock {\em Nature physics}, 2(2):81--90, 2006.

\bibitem{laussy2009luminescence}
Laussy F.~P., Del~Valle E., and Tejedor C.
\newblock {\em Physical Review B}, 79(23):235325, 2009.

\bibitem{del2009luminescence}
Del~Valle E., Laussy F.~P., and Tejedor C.
\newblock {\em Physical Review B}, 79(23):235326, 2009.

\bibitem{rabi1936process}
Rabi I.
\newblock {\em Physical Review}, 49(4):324, 1936.

\bibitem{rabi1937space}
Rabi I.~I.
\newblock {\em Physical Review}, 51(8):652, 1937.

\bibitem{grynberg2010introduction}
Grynberg G., Aspect A., and Fabre C.
\newblock {\em Introduction to quantum optics: from the semi-classical approach
  to quantized light}.
\newblock Cambridge university press, 2010.

\bibitem{schrodinger1935current}
Schr{\"o}dinger E.
\newblock {\em science}, 23(50):844--849, 1935.

\bibitem{einstein1935can}
Einstein A., Podolsky B., and Rosen N.
\newblock {\em Physical review}, 47(10):777, 1935.

\bibitem{nielsen2000quantum}
Nielsen M. and Chuang I.
\newblock {\em Quantum Computation and Quantum Information: Cambridge Univ
  Press}.
\newblock American Association of Physics Teachers, 2000.

\bibitem{bennett1992communication}
Bennett C.~H. and Wiesner S.~J.
\newblock {\em Physical review letters}, 69(20):2881, 1992.

\bibitem{bennett1999entanglement}
Bennett C.~H., Shor P.~W., Smolin J.~A., and Thapliyal A.~V.
\newblock {\em Physical Review Letters}, 83(15):3081, 1999.

\bibitem{bennett2000quantum}
Bennett C.~H. and DiVincenzo D.~P.
\newblock {\em nature}, 404(6775):247--255, 2000.

\bibitem{zurek2003decoherence}
Zurek W.~H.
\newblock {\em Reviews of modern physics}, 75(3):715, 2003.

\bibitem{yu2002phonon}
Yu~T. and Eberly J.
\newblock {\em Physical Review B}, 66(19):193306, 2002.

\bibitem{simon2002robustness}
Simon C. and Kempe J.
\newblock {\em Physical Review A}, 65(5):052327, 2002.

\bibitem{yu2004finite}
Yu~T. and Eberly J.
\newblock {\em Physical Review Letters}, 93(14):140404, 2004.

\bibitem{yu2009sudden}
Yu~T. and Eberly J.
\newblock {\em Science}, 323(5914):598--601, 2009.

\bibitem{zyczkowski2001dynamics}
{\.Z}yczkowski K., Horodecki P., Horodecki M., and Horodecki R.
\newblock {\em Physical Review A}, 65(1):012101, 2001.

\bibitem{rajagopal2001decoherence}
Rajagopal A. and Rendell R.
\newblock {\em Physical Review A}, 63(2):022116, 2001.

\bibitem{diosi2003progressive}
Di{\'o}si L.
\newblock {\em Irreversible Quantum Dynamics}, pages 157--163, 2003.

\bibitem{dodd2004disentanglement}
Dodd P. and Halliwell J.
\newblock {\em Physical Review A}, 69(5):052105, 2004.

\bibitem{mazzola2010interplay}
Mazzola L., Maniscalco S., Piilo J., and Suominen K.-A.
\newblock {\em Journal of Physics B: Atomic, Molecular and Optical Physics},
  43(8):085505, 2010.

\bibitem{ficek2006dark}
Ficek Z. and Tana{\'s} R.
\newblock {\em Physical Review A}, 74(2):024304, 2006.

\bibitem{lamb1947fine}
Lamb~Jr W.~E. and Retherford R.~C.
\newblock {\em Physical Review}, 72(3):241, 1947.

\bibitem{casimir1948attraction}
Casimir H.~B.
\newblock On the attraction between two perfectly conducting plates.
\newblock In {\em Proc. Kon. Ned. Akad. Wet.}, volume~51, page 793, 1948.

\bibitem{milton2001casimir}
Milton K.~A.
\newblock {\em The Casimir effect: physical manifestations of zero-point
  energy}.
\newblock World Scientific, 2001.

\bibitem{lamoreaux2004casimir}
Lamoreaux S.~K.
\newblock {\em Reports on progress in Physics}, 68(1):201, 2004.

\bibitem{passante2018dispersion}
Passante R.
\newblock {\em Symmetry}, 10(12):735, 2018.

\bibitem{yu2012detecting}
Yu~H. and Hu~J.
\newblock {\em Physical Review A}, 86(6):064103, 2012.

\bibitem{yang2016entanglement}
Yang Y., Hu~J., and Yu~H.
\newblock {\em Physical Review A}, 94(3):032337, 2016.

\bibitem{cheng2018entanglement}
Cheng S., Yu~H., and Hu~J.
\newblock {\em Physical Review D}, 98(2):025001, 2018.

\bibitem{oxman2011fractional}
Oxman L. and Khoury A.
\newblock {\em Physical Review Letters}, 106(24):240503, 2011.

\bibitem{khoury2014topological}
Khoury A. and Oxman L.
\newblock {\em Physical Review A}, 89(3):032106, 2014.

\bibitem{milonni2013quantum}
Milonni P.~W.
\newblock {\em The quantum vacuum: an introduction to quantum electrodynamics}.
\newblock Academic press, 2013.

\bibitem{fleming2010rotating}
Fleming C., Cummings N., Anastopoulos C., and Hu~B.-L.
\newblock {\em Journal of Physics A: Mathematical and Theoretical},
  43(40):405304, 2010.

\bibitem{hu1992quantum}
Hu~B.~L., Paz J.~P., and Zhang Y.
\newblock {\em Physical Review D}, 45(8):2843, 1992.

\bibitem{viotti2020boundary}
Viotti L., Lombardo F.~C., and Villar P.~I.
\newblock {\em Physical Review A}, 101(3):032337, 2020.

\bibitem{mazzitelli2003decoherence}
Mazzitelli F.~D., Paz J.~P., and Villanueva A.
\newblock {\em Physical Review A}, 68(6):062106, 2003.

\bibitem{wootters1998entanglement}
Wootters W.~K.
\newblock {\em Physical Review Letters}, 80(10):2245, 1998.

\bibitem{ficek2003entanglement}
Ficek Z. and Tana R.
\newblock {\em journal of modern optics}, 50(18):2765--2779, 2003.

\bibitem{franco2013dynamics}
Franco R.~L., Bellomo B., Maniscalco S., and Compagno G.
\newblock {\em International Journal of Modern Physics B}, 27(01n03):1345053,
  2013.

\bibitem{viotti2021enhanced}
Viotti L., Lombardo F.~C., and Villar P.~I.
\newblock {\em Physical Review A}, 103(3):032809, 2021.

\bibitem{lombardo2022detecting}
Lombardo F., Decca R., Liu J., Villar P., and Viotti L.
\newblock Detecting traces of non-contact quantum friction in the corrections
  of the accumulated geometric phase.
\newblock In {\em APS April Meeting Abstracts}, volume 2022, pages F01--044,
  2022.

\bibitem{viotti2019thermal}
Viotti L., Far{\'\i}as M.~B., Villar P.~I., and Lombardo F.~C.
\newblock {\em Physical Review D}, 99(10):105005, 2019.

\bibitem{milton2004casimir}
Milton K.~A.
\newblock {\em Journal of Physics A: Mathematical and General}, 37(38):R209,
  2004.

\bibitem{lamoreaux1997demonstration}
Lamoreaux S.~K.
\newblock {\em Physical Review Letters}, 78(1):5, 1997.

\bibitem{bordag2009advances}
Bordag M., Klimchitskaya G.~L., Mohideen U., and Mostepanenko V.~M.
\newblock {\em Advances in the Casimir effect}, volume 145.
\newblock OUP Oxford, 2009.

\bibitem{fosco2011quantum}
Fosco C.~D., Lombardo F.~C., and Mazzitelli F.~D.
\newblock {\em Physical Review D}, 84(2):025011, 2011.

\bibitem{Fariasfuncionalcamino}
Far\'{\i}as M.~B. and Lombardo F.~C.
\newblock {\em Phys. Rev. D}, 93:065035, Mar 2016.

\bibitem{farias2015functional}
Far{\'\i}as M.~B., Fosco C.~D., Lombardo F.~C., Mazzitelli F.~D., and L{\'o}pez
  A.~E.~R.
\newblock {\em Physical Review D}, 91(10):105020, 2015.

\bibitem{Schwinger}
Schwinger J.~S.
\newblock {\em J. Math. Phys.}, 2:407--432, 1961.

\bibitem{Keldysh}
Keldysh L.~V.
\newblock {\em Zh. Eksp. Teor. Fiz.}, 47:1515--1527, 1964.

\bibitem{Das_temp}
Das A.~K.
\newblock {\em {Finite Temperature Field Theory}}.
\newblock World Scientific, New York, 1997.

\bibitem{calzetta1988nonequilibrium}
Calzetta E. and Hu~B.-L.
\newblock {\em Physical Review D}, 37(10):2878, 1988.

\bibitem{calzetta2009nonequilibrium}
Calzetta E.~A. and Hu~B.-L.~B.
\newblock {\em Nonequilibrium quantum field theory}.
\newblock Cambridge University Press, 2009.

\bibitem{barton1997van}
Barton G.
\newblock {\em Proceedings of the Royal Society of London. Series A:
  Mathematical, Physical and Engineering Sciences}, 453(1966):2461--2495, 1997.

\bibitem{lombardo2021detectable}
Lombardo F.~C., Decca R.~S., Viotti L., and Villar P.~I.
\newblock {\em Advanced Quantum Technologies}, 4(5):2000155, 2021.

\bibitem{maghrebi2013quantum}
Maghrebi M.~F., Golestanian R., and Kardar M.
\newblock {\em Physical Review A}, 88(4):042509, 2013.

\bibitem{pieplow2015cherenkov}
Pieplow G. and Henkel C.
\newblock {\em Journal of Physics: Condensed Matter}, 27(21):214001, 2015.

\bibitem{klatt2016spectroscopic}
Klatt J., Bennett R., and Buhmann S.~Y.
\newblock {\em Physical Review A}, 94(6):063803, 2016.

\bibitem{intravaia2016non}
Intravaia F., Behunin R., Henkel C., Busch K., and Dalvit D.~A.~R.
\newblock {\em Physical Review A}, 94(4):042114, 2016.

\bibitem{svidzinsky2019excitation}
Svidzinsky A.~A.
\newblock {\em Physical Review Research}, 1(3):033027, 2019.

\bibitem{maclaurin2012measurable}
Maclaurin D., Doherty M., Hollenberg L., and Martin A.
\newblock {\em Physical review letters}, 108(24):240403, 2012.

\bibitem{jelezko2004observation}
Jelezko F., Gaebel T., Popa I., Gruber A., and Wrachtrup J.
\newblock {\em Physical review letters}, 92(7):076401, 2004.

\bibitem{childress2006coherent}
Childress L., Gurudev~Dutt M., Taylor J., Zibrov A., Jelezko F., Wrachtrup J.,
  Hemmer P., and Lukin M.
\newblock {\em Science}, 314(5797):281--285, 2006.

\bibitem{doherty2013n}
Doherty M.
\newblock {\em Wrchtrup, and L. C. L. Hollenberg. The nitrogen-vacancy colour
  centre in diamond. Physics Reports}, 528:1--45, 2013.

\bibitem{viotti2023geometric}
Viotti L., Gramajo A.~L., Villar P.~I., Lombardo F.~C., and Fazio R.
\newblock {\em {Quantum}}, 7:1029, June 2023.

\bibitem{gefenWeak}
Gebhart V., Snizhko K., Wellens T., Buchleitner A., Romito A., and Gefen Y.
\newblock {\em Proceedings of the National Academy of Sciences},
  117(11):5706--5713, 2020.

\bibitem{Molmer:93}
M{\o}lmer K., Castin Y., and Dalibard J.
\newblock {\em J. Opt. Soc. Am. B}, 10(3):524--538, Mar 1993.

\bibitem{manzano}
Manzano G. and Zambrini R.
\newblock {\em AVS Quantum Science}, 4(2):025302, 2022.

\bibitem{carmichael1993_open}
Carmichael H.
\newblock {\em An open systems approach to quantum optics}.
\newblock Lecture Notes in Physics Monographs. Springer Berlin, Heidelberg,
  1993.

\bibitem{wiseman2009quantum}
Wiseman H.~M. and Milburn G.~J.
\newblock {\em Quantum measurement and control}.
\newblock Cambridge university press, 2009.

\bibitem{daley2014quantum}
Daley A.~J.
\newblock {\em Advances in Physics}, 63(2):77--149, 2014.

\bibitem{passarelli2019improving}
Passarelli G., Cataudella V., and Lucignano P.
\newblock {\em Physical Review B}, 100(2):024302, 2019.

\bibitem{siddiqi2_observing}
Murch K., Weber S., Macklin C., and Siddiqi I.
\newblock {\em Nature}, 502(7470):211--214, 2013.

\bibitem{ahn2002_error}
Ahn C., Doherty A.~C., and Landahl A.~J.
\newblock {\em Physical Review A}, 65(4):042301, 2002.

\bibitem{siddiqi1_observation}
Vijay R., Slichter D., and Siddiqi I.
\newblock {\em Physical review letters}, 106(11):110502, 2011.

\bibitem{Sarandy_2005}
Sarandy M.~S. and Lidar D.~A.
\newblock {\em Physical Review A}, 71(1), jan 2005.

\bibitem{thunstrom2005adiabatic}
Thunstr{\"o}m P., {\AA}berg J., and Sj{\"o}qvist E.
\newblock {\em Physical Review A}, 72(2):022328, 2005.

\bibitem{yi2007adiabatic}
Yi~X., Tong D., Kwek L., and Oh~C.
\newblock {\em Journal of Physics B: Atomic, Molecular and Optical Physics},
  40(2):281, 2007.

\bibitem{oreshkov2010adiabatic}
Oreshkov O. and Calsamiglia J.
\newblock {\em Physical review letters}, 105(5):050503, 2010.

\bibitem{venuti2016adiabaticity}
Venuti L.~C., Albash T., Lidar D.~A., and Zanardi P.
\newblock {\em Physical Review A}, 93(3):032118, 2016.

\bibitem{albash2012}
Albash T., Boixo S., Lidar D.~A., and Zanardi P.
\newblock {\em New Journal of Physics}, 14(12):123016, 2012.

\bibitem{albash2015}
Albash T., Boixo S., Lidar D.~A., and Zanardi P.
\newblock {\em New Journal of Physics}, 17(12):129501, dec 2015.

\bibitem{jumpsAdiabatic}
Yip K.~W., Albash T., and Lidar D.~A.
\newblock {\em Phys. Rev. A}, 97:022116, Feb 2018.

\bibitem{paz1999environment}
Paz J.~P. and Zurek W.~H.
\newblock {\em Coherent atomic matter waves, Les Houches lectures}, pages
  533--614, 1999.

\bibitem{lindblad1976}
Lindblad G.
\newblock {\em Comm. Math. Phys.}, 48(2):119--130, 1976.

\end{thebibliography}

\end{document}